\documentclass[12pt,a4paper]{book}

\usepackage[utf8x]{inputenc}
\usepackage{ucs}
\usepackage{amsmath}
\usepackage{amsfonts}
\usepackage{amssymb}
\usepackage{array}
\usepackage{graphicx}
\usepackage{color}
\usepackage{nicefrac}
\usepackage[francais]{babel}
\usepackage{wrapfig}
\usepackage{hyperref}
\usepackage{pdfpages}
\usepackage[nottoc, notlof, notlot]{tocbibind}
\usepackage{verbatim} 
\usepackage{times}
\usepackage{caption}
\usepackage{subcaption}
\usepackage{lettrine}

\usepackage{tikz}
	\usetikzlibrary{decorations.markings}
\usepackage{geometry}
\usepackage{fancyhdr}
	\renewcommand{\sectionmark}[1]{\markright{\partname.~Section \thesection{}}}
\AtBeginDocument{

}

\makeatletter	
	\renewcommand\theequation{\thesection.\arabic{equation}}
    \@addtoreset{equation}{part}
    \@addtoreset{equation}{section}
\makeatother

\newcommand{\ie}{\emph{i.e.} }
\newcommand{\eg}{\emph{eg.} }
\newcommand{\cf}{\emph{cf.} }
\newcommand{\refe}[1]{(\hyperref[#1]{\ref*{#1}})}
\newcommand{\refcc}[1]{\hyperref[#1]{\ref*{#1}}}
\newcommand{\tr}{\; \text{Tr} \;}
\newcommand{\Str}{\; \text{STr} \;}

\newcommand{\wt}{\widetilde}
\newcommand{\wh}{\widehat}

\newcommand{\normal}[1]{{}^{{}_\circ}_{{}^\circ} #1 {}^{{}_\circ}_{{}^\circ}}
\newcommand{\bnormal}[1]{{}^{{}_\star}_{{}^\star} #1 {}^{{}_\star}_{{}^\star}}
\newcommand{\ket}[1]{\left| #1 \right>}

\newcommand{\bra}[1]{\left< #1 \right|}
\newcommand{\braket}[2]{\big< #1 \big| #2 \big>}
\newcommand{\eqn}[1]{\begin{equation} #1 \end{equation}}
\newcommand{\eqna}[1]{\begin{eqnarray} #1 \end{eqnarray}}
\newcommand{\eqali}[1]{\begin{align} #1 \end{align}}

\newcommand{\comm}[2]{\left[ #1 , #2 \right]}
\newcommand{\acomm}[2]{\left\{ #1 , #2 \right\}}
\newcommand{\cad}{c'est-\`a-dire }
\newcommand{\im}{\mathfrak{Im}}

\newcommand{\corr}[1]{\left< #1 \right>}
\newcommand{\parent}[1]{\left(#1\right)}

\newcommand{\module}[1]{\left\vert #1 \right\vert}
\newcommand{\norm}[1]{\left\Vert #1 \right\Vert}

\newcommand{\sing}[1]{\left\{ #1 \right\}}
\newcommand{\croch}[1]{\left[ #1 \right]}
\newcommand{\bbar}[1]{\left| #1 \right|}

\newcommand{\Gam}{{\bf \Gamma}}

\newcommand{\symform}[2]{\bbar{\begin{array}{c} #1 \\ #2 \end{array}}}

\newcommand{\asymform}[2]{\parent{\begin{array}{c} #1 \\ #2 \end{array}}}
\newcommand{\Asymform}[2]{\croch{\begin{array}{c} #1 \\ #2 \end{array}}}
\newcommand{\red}[1]{\textcolor{red}{#1}}

\newcommand{\etc}{\emph{etc}}
\newcommand{\di}{\text{d}}
\newcommand{\apriori}{\emph{a priori}~}
\newcommand{\infine}{\emph{in fine}~}
\newcommand{\notref}[1]{(${}^{\text{\ref{#1}}}$)}
\newcommand{\NS}{\text{NS}}
\newcommand{\R}{\text{R}}
\newcommand{\IIA}{\text{IIA}}
\newcommand{\IIB}{\text{IIB}}

\title{Condensation de tachyon dans le syst\`eme brane-antibrane}     
\date{\today} 
\author{Flavien \textsc{KIEFER}}

\begin{document}
\frontmatter

\thispagestyle{empty}

\begin{tabular}{m{2cm} m{2cm} m{2cm} }
\includegraphics[scale=.4]{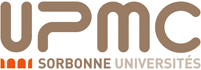}&  \hspace*{3cm}
\includegraphics[scale=.9]{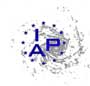} & \hspace*{7cm}
\includegraphics[scale=1]{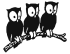} 
\end{tabular}
\vspace*{-.5cm}

{\large

\vspace*{1cm}

\begin{center}

{\bf TH\`ESE DE DOCTORAT DE \\ l'UNIVERSIT\'E PIERRE ET MARIE CURIE}

\vspace*{0.5cm}

Sp\'ecialit\'e \\ [2ex]
{\bf Physique Th\'eorique}\ \\ 

\vspace*{0.5cm}

\'Ecole doctorale ED107 de la R\'egion Parisienne

\vspace*{1cm}

Pr\'esent\'ee par \ \\

\vspace*{0.5cm}

{\Large {\bf Flavien \textsc{KIEFER}}}

\vspace*{0.3cm}
Pour obtenir le grade de \ \\[1ex]
{\bf DOCTEUR de l'UNIVERSIT\'E PIERRE ET MARIE CURIE} \ \\

\vspace*{1cm}

\end{center}

{\flushleft \underline{Sujet de la th\`ese :} \ \\
\ \\
{\Large {\bf Condensation de tachyon dans le syst\`eme brane-antibrane \\}}

\vspace*{1.5cm} 
\flushleft{soutenue le 19 juin 2012}\\[2ex]
\flushleft{devant le jury compos\'e de :  }\\[1ex]
\flushleft{\begin{tabular}{r@{\ }ll}
  & M. Costas {\sc KOUNNAS} & Directeur de th\`ese\\
  & M. Dan {\sc ISRAEL} & Directeur de th\`ese\\
  & M. Volker {\sc SCHOMERUS} & Rapporteur \\
  & M. Massimo {\sc BIANCHI} & Rapporteur  \\
  & M. Vladimir {\sc DOTSENKO} & Examinateur et Président du Jury  \\
  & M. Marios {\sc PETROPOULOS} & Examinateur  \\
\end{tabular}}

}

\normalsize


\cleardoublepage

\vspace*{2cm}

\section*{Laboratoires de rattachement}

\begin{itemize}\itemsep4pt
\item Institut d'Astrophysique de Paris, \\
98bis Bd Arago, 75014 Paris, France \\
\emph{Unit\'e mixte de Recherche 7095, CNRS -- Universit\'e Pierre et Marie Curie} \\

\item LPTENS, \'Ecole Normale Sup\'erieure, \\
24 rue Lhomond, 75231 Paris cedex 05, France \\
\emph{Unit\'e mixte de Recherche du CNRS et de l'\'Ecole Normale Sup\'erieure associ\'ee \`a l'Universit\'e Pierre et Marie Curie, UMR 8549.}
\end{itemize}

\vspace*{2cm}

\section*{Mots-clés}

Th\'eorie des cordes, Tachyon, Brane, Théorie des champs conforme, Fonction de partition, Action effective.

\vspace*{1cm}

\section*{Keywords}

String theory, Tachyon, Brane, Conformal field theory, Partition function, Effective action.

\cleardoublepage

\section*{\centering R\'esum\'e}

En théorie des supercordes de type II, la paire brane-antibrane séparée admet dans son spectre de corde ouverte un tachyon bi-fondamental pour toute séparation $\ell < \pi\sqrt{2\alpha'}$. La dynamique de ce système serait décrite par l'action effective de Garousi~\cite{Garousi:2007fn} en tout $\ell\neq 0$. Notre étude montre que le domaine de validité de cette action concerne exclusivement des condensations de tachyon spatiales. 
Des \'etudes pr\'ec\'edentes menées par Bagchi et Sen~\cite{Bagchi:2008et} ont montr\'e l'existence dans le domaine sous-critique partiel $\ell<\pi \sqrt{\alpha'}$ d'une th\'eorie conforme (CFT) décrivant une condensation dynamique \`a distance \emph{constante}, appel\'ee \emph{tachyon roulant}.
Nous montrons que cette CFT existe en v\'erit\'e sur l'ensemble du domaine tachyonique. Gr\^ace \`a cette d\'emon\-stra\-tion, \emph{i)} nous prouvons que le domaine de validité de l'action de Garousi ne peut pas inclure de condensations de tachyons temporelles et \emph{ii)} nous justifions le calcul de la fonction de partition le long du tachyon roulant pour tout $\ell < \pi\sqrt{2\alpha'}$. En utilisant une méthode proposée par Kutasov et Niarchos~\cite{Kutasov:2003er}, nous déterminons une action effective quadratique pour les champs de tachyon et de distance, au moins valable en tout $\ell$ autour de la solution de tachyon roulant. Cette \'etude a été publiée dans \emph{Physical Review D}~\cite{Israel:2011ut}. 
Par ailleurs, nous avons \'etudié le modèle sigma non linéaire défini comme la déformation perturbative de la CFT du tachyon roulant. Les fonctions b\^eta du groupe de renormalisation, que nous avons obtenues, sont en accord avec les équations du mouvement de l'action effective quadratique proposée, confirmant ainsi son expression.

\section*{\centering Abstract:\\ \centering Tachyon condensation in brane-antibrane system}

In superstring theory of type II, the separated brane-antibrane pair admits a bi-fundamental tachyon in its open string spectrum, for any separation $\ell < \pi\sqrt{2\alpha'}$. The dynamics of this system would be described by the Garousi's effective action~\cite{Garousi:2007fn} for any $\ell\neq 0$. Our study shows however that the domain of validity of this action only includes space-like tachyon condensation. 
Previous studies led by Bagchi and Sen~\cite{Bagchi:2008et} showed the existence, in the partial sub-critical domain $\ell<\pi \sqrt{\alpha'}$ of a conformal field theory (CFT) describing a dynamical condensation at static distance, called \emph{rolling tachyon}. 
We show this CFT actually exists on the whole tachyonic domain. Thanks to this demonstration, \emph{i)} we prove that the domain of validity of Garousi's action excludes time-dependent tachyon condensation and \emph{ii)} we justify the computation of the partition fonction along the rolling tachyon for all $\ell < \pi\sqrt{2\alpha'}$. Using a method proposed by Kutasov and Niarchos~\cite{Kutasov:2003er}, we determine a quadratic effective action for the tachyon and distance fields, at least valid in the whole tachyonic domain around the rolling tachyon solution. This study has led to the publication of an article~\cite{Israel:2011ut} in \emph{Physical Review D}. 
In addition, we studied the non linear sigma model of perturbative deformations along the rolling tachyon CFT. The beta-function of the renormalization group, that we obtained, are in good agreement with the equations of movement derived from the proposed quadratic effective action. This stands as an independent confirmation of its expression.

\cleardoublepage

\section*{\centering Remerciements}

Avant toute chose, je souhaite adresser mes remerciements aux membres du jury~: aux rapporteurs Massimo Bianchi et Volker Schomerus pour l'analyse attentive et minitieuse de mon travail et de ma r\'edaction, aux examinateurs Vladimir Dotsenko et Marios Petropoulos, pour l'attention critique qu'ils ont port\'ee au contenu de ma th\`ese et de ma pr\'esentation. 

Je souhaite exprimer toute ma gratitude à mes co-directeurs de thèse Dan Israel et Costas Kounnas pour leur soutien et leur disponibilité qui m'ont permis de développer en profondeur et avec esprit critique mon travail de thèse. Je suis tr\`es reconnaissant \`a Dan de m'avoir suivi de près pendant ces quatre ans, d'avoir su \^etre pr\'esent et à mon écoute, et d'avoir sauvé ma quatrième année en me conseillant auprès du Pr. Eliezer Rabinovici. Je remercie Costas pour ses recommandations et ses encouragements, quand bien m\^eme les échanges furent moins réguliers.

Je voudrais remercier très sincèrement Laurent Vigroux ainsi que tout le personnel de l'institut d'astrophysique de Paris qui m'ont fait un accueil chaleureux et m'ont accepté au sein du Conseil du Laboratoire, puis m'ont soutenu et encouragé. En particulier, je resterai toujours reconnaissant pour le soutien financier du laboratoire au début de la quatrième année sans lequel je n'aurais jamais pu continuer la thèse puis la soutenir. Un grand merci au GReCO pour l'accueil dans l'équipe. 

Je remercie également l'équipe du laboratoire de physique théorique de l'Ecole Normale, et en particulier les professeurs dont j'ai suivi les cours en master qui m'ont guidé vers cette thèse. Je suis également très reconnaissant pour le soutien financier que le LPT m'a accordé pour mes diverses missions. \\

{\it 
I would like to acknowledge all my gratitude to Pr. Eliezer Rabinovici for having invited me at the Racah Institute of Jerusalem, and supported me, financially and scientifically. These six months were hard and workful, with my time shared between writing the thesis, working with Stefano on the project you supervised, and also visiting a bit Israel. But the quality of the results was, definitely, worth the effort.

I would like also to address a consequent part of my thanks to Stephano, Bjarke, Roberto, Latif and Ofek for this very nice time in Jerusalem. I am especially grateful to Roberto for having hosted me at the beginning of my stay in Jerusalem. 

My thanks go also to Mathias Gaberdiel, Jan Troost and Vladimir Dotsenko for the very enlighting discussions that helped us completing our researches.}  \\

Je souhaite également adresser mes remerciements à David Langlois et Sébastien Renaux-Petel suite à nos nombreuses discussions, qui auraient pu aboutir à un projet de collaboration. La tournure des résultats nous a cependant amené à développer indépendamment le contenu de cette thèse. Mais l'apport de nos discussions à ce travail est cons\'equent. \\

Par ailleurs, je voudrais remercier l'ensemble des \'equipes d'organisateurs b\'en\'evoles de la conf\'erence de jeunes chercheurs franco-anglaise SCGSC (initialement SCSC) qui se d\'eroule tous les ans alternativement \`a Paris et \`a Londres. En particulier, merci \`a Marc pour avoir transmis le "b\'eb\'e" \`a Blaise et moi, merci aussi \`a Enrico et Francesco pour avoir repris le flambeau l'ann\'ee suivante et l'avoir transmis ensuite aux g\'en\'erations suivantes. Au nom des organisateurs, j'adresse ma gratitude \`a Jean-Bernard Zuber, pr\'esident de la FRIF qui accepte chaque ann\'ee de financer cette conf\'erence, ainsi qu'aux ambassades fran\c caise et anglaise de Londres et Paris qui accordent r\'eguli\`erement des financements aux organisateurs. \\

Je suis en outre reconnaissant envers mes \'etudiants des LM100, LP104 et LP111b qui furent mes cobayes de monitorat ; merci \'egalement aux \'equipes d'enseignement de ces UE. Au passage, je voudrais exprimer ma gratitude \`a l'\'equipe de l'association Paris-Montagne qui organise la Science Acad\'emie chaque ann\'ee pour les lyc\'eens motiv\'es par la science en les invitant \`a effectuer des stages dans les laboratoires. Ce fut une belle d\'ecouverte et un excellent exercice de vulgarisation que d'accueillir, avec l'aide de mes coll\`egues, ces quelques lyc\'eens. \\

Ce manuscrit de thèse serait bien pauvre dans sa rédaction si tous les membres de mon équipe de relecteurs assidus ne s'étaient pas désignés volontaires -- d'office pour certains -- pour corriger des fichiers-texte minés de codes mystérieux~: mes parents, Aline et Jacques, mes soeurs, Eloïse, avec son mari Vianney, et Violaine, ma cousine Charline ainsi que parmi mes amis, Amandine et Amélie. Merci d'avoir été particulièrement attentifs sur toutes ces petites (et grosses) fautes d'orthographe et de syntaxe qui se sont glissées un peu partout dans ce manuscrit. Un exercice difficile, bravo ! 

Le soutien psychologique et financier de mes parents m'a permis de garder le cap durant ces quatre dernières années, mais aussi toutes les précédentes qui me menèrent jusqu'au doctorat. Je ne saurais jamais vous en \^etre suffisamment reconnaissant. Sans vous, autant dire que je ne serais pas là à présenter cette thèse. 

Mes soeurs m'ont \'egalement apport\'e un soutien remarquable et sans faille pendant la th\`ese. Je vous en remercie infiniment, ainsi qu'aux p'tits bouts d'choux d'Eloïse et Vianney~: mon neveu Grégoire 2 ans et ma nièce Armelle née le 1er mars 2012, qui sont arrivés en cours de thèse pour rayonner de vie et redonner espoir et sourire à leur tonton pendant des moments difficiles de son travail. 

Je suis aussi particulièrement reconnaissant à ma cousine Charline qui m'a chaleureusement supporté, encouragé et réconforté dans les nombreux moments de doutes, à grand renfort de joie et de positivisme très efficaces. \\

J'adresse un grand merci à tous mes collègues thésards de l'IAP pour l'excellente ambiance au laboratoire garantie par leur bonne humeur quotidenne, mais également les YMCA (et YTA), les pauses-déjeuner, les pauses-café, les soirées animées, les nombreuses bières absorbées -- on est étudiant ou on ne l'est pas -- et enfin les nombreuses discussions physico-philosophico-politico-économico-idiotico-alcolo-scientifiques. D'abord mes chers co-bureaux dans l'ordre chronologique depuis 2008~: Larissa, Komiko, Nicolas, Raphael, Anne et finalement Sylvain. Puis mes chers co-muraux du \emph{Buralland}~: Camila, Isabelle, Typhaine, Alexandre, Jacoppo et pas beaucoup plus loin Ophélia. Toute ma gratitude va aussi à Typhaine qui a bien voulu m'aider \`a l'impression de la thèse à distance et qui a brillament rempli sa mission ! 

Je souhaite remercier très chaleureusement Florence pour m'avoir choisi comme confident de ses mésaventures thésardes (et autres) en plus d'être une excellente amie, en particulier pour avoir aussi écouté les miennes. Merci aussi à Mélody, Romain S., Romain L. et Sophie. Enfin, je remercie bien fort Camila et Isabelle avec Jérome, Benjamin, Hakim et Jean, pour tous les moments incroyables passés ensemble durant ces quatre ans de thèse, des couloirs du labo aux quais de Seine. Une pensée et une larme (pas forcément de tristesse) pour les (trop) nombreuses photos qui ont immortalisé certains clichés. 

Je suis également très reconnaissant envers Alejandro pour la co\"incidence incroyable qui a fait que tu as eu besoin d'un appartement exactement au moment où je recherchais un sous-locataire. Je te remercie aussi pour tous les services que tu m'as rendu pendant ces six mois, malgré les quelques problèmes de connexions internet. 

Merci aussi \`a tous mes anciens camarades \'etudiants du master de physique th\'eorique qui se reconna\^itront. \\

Tous mes amis de longue date m'ont soutenu durant ces quatre ans et je leur en suis infiniment reconnaissant. Vous avez eu la bonne idée de vous installer à Paris au cours de ma thèse et de me distraire de mon travail assidu à base de pique-niques de quais et de parcs, de session boeuf guitare-basse, de balades parisiennes, de soirées tziganes, de raclettes, de barbecue, et j'en passe !... 

Merci en particulier à mon ex-coloc' bassiste Samuel et au clan féminin composé de Caroline, Ga\"elle, Amélie, Anne-sophie, Julie et M\"adeli. Un grand merci surtout à Caroline qui m'a hébergé lors de mon passage d'un mois à Paris en avril. J'ai eu un grand plaisir à te recevoir en retour à Jérusalem. C'était tellement incroyable de marcher ensemble sur les routes de Galilée et dans les rues de la vieille ville.

Merci aussi aux aulnaysiens partis bien loin de Paris~: Olivier, Nicolas, Florent et Yohan.

Merci à l'ensemble du groupe de Bouafle : Alexandre, Amandine, Charlotte, Florie, Maximilien, Cécile et Quentin.

Merci aux Volcanic Butterflies (Jules, Nico' et Mat') pour ces concerts mémorables et ces répet' ma foi fort bien calées -- hell yeah !

Merci à Nicolas, Mickael, Raphael et Charline pour ces moments de purs délices passés à discuter économie de marché, politiques et autres fadaises portugolistiques.

Et enfin merci à la lyrique G\'eraldine qui m'a fait d\'ecouvrir les lumi\`eres du Memphis... \\

Pour finir, je souhaite adresser moultes remerciements à tout le groupe de la section escalade de l'ASP6 et GCN. Je ne vais pas tous vous citer, mais j'espère rester encore longtemps en contact avec vous tous. Pendant ces trois dernières années, nous avons vécu tant d'aventures rocailleuses qui ont été essentielles à mon équilibre mental et physique. J'ai pu découvrir en m\^eme temps que mon travail, un sport qui rassemble l'ensemble des qualités et défauts d'une thèse : concentration et destabilisation, altitude et vertige, endurance et fatigue, rigueur et laisser-aller, travail et repos, mais aussi persévérance et donc patience et impatience. Ce sont aussi des instantanés gravés à jamais dans la t\^ete et dans la pellicule. 

Je suis d'ailleurs très reconnaissant envers Jean-François N'Guyen, mon cher kinésithérapeute, qui a su me garder en un seul morceau.

\cleardoublepage

\vspace*{8cm}

\begin{flushright} 
\`A ma nièce et mon neveu, \\
Armelle et Grégoire,
\end{flushright}

\tableofcontents

\mainmatter

\part{Pr\'eambule~: de la th\'eorie quantique des champs \`a la th\'eorie des cordes}
\label{part:preamb}

	\lettrine{D}{epuis son introduction} dans les ann\'ees 60, sous l'impulsion de scientifiques tels que Vene\-zia\-no, Gross, Scherk, Schwarz ou Polyakov, la th\'eorie des cordes a vu son d\'eveloppement s'intensifier. D'abord appliqu\'ee \`a la mod\'elisation des interactions fortes puis plus tard imagin\'ee comme th\'eorie fondamentale de l'univers, elle est peu \`a peu devenue incontournable dans le paysage de la physique th\'eorique. C'est depuis les ann\'ees 90 que la th\'eorie recoit un gain d'int\'er\^et ph\'enom\'enal, depuis que l'on a d\'ecouvert les D-branes. Depuis lors combien de dualit\'es les mettant en jeu a-t-on d\'ecouvert~? Faisant d'elles des \'el\'ements d'une importance colossale. Et pour cause, la plupart des mod\`eles cosmologiques cordistes actuels se basent sur leur dynamique afin de trouver une origine coh\'erente et unifiante \`a l'univers. Mais nous d\'ecouvrons de jour en jour de nouvelles particularit\'es, de nouveaux objets, de nouvelles g\'eom\'etrie et notre connaissance actuelle de la th\'eorie et de toutes ses implications semble encore tr\`es faible par rapport \`a l'\'etendue de ses ramifications.
	
	Nous allons à présent partir des années 1900, période de grande ferveur scientifique, qui a vu se développer les théories grandioses que sont la \emph{mécanique quantique} et la \emph{relativité restreinte} et \emph{générale}, survoler la période des années 30 à 70 pendant laquelle ont été élaborées les théories des particules relativistes, les \emph{théories quantiques des champs}, pour arriver finalement \`a l'\'epoque s'\'etalant des années 70 \`a nos jours en vue d'un des chefs d'oeuvre de la physique théorique  \emph{et expérimentale} moderne, le \emph{Modèle Standard des particules}. Nous conclurons pourtant sur ses limitations et nous verrons avec quel naturel la théorie des cordes s'immisce dans le paysage de la physique fondamentale pour devenir aujourd'hui la théorie de premier plan. Nous l'introduirons très brièvement, puis nous présenterons dans ce cadre la problématique de la condensation de tachyon et enfin de cette thèse.

\section*{M\'ecanique quantique et relativit\'e}

	L’histoire de la mécanique quantique et celle de la relativité ont débuté quasi simultanément au d\'ebut du si\`ecle dernier. Les avancées impressionnantes dans chacun de ces domaines ont amené les physiciens à s'interroger sur l'unification de ces deux théories.

\subsection*{Mécanique quantique}

	La première explique comment les particules de matière telles que les électrons, protons ou neutrons, émettent des ondes lumineuses -- photons -- tout en conservant l'unité de l'atome. Les travaux conjoints de Schrödinger, Heisenberg, Bohr, Dirac, De Broglie et Einstein -- malgré leurs divergences d'opinion -- convergent vers une formulation des lois physiques extrêmement mathématisée et usant d'un formalisme probabiliste et de théorie des groupes -- espace de Hilbert, états, opérateurs et algèbres, fonctions d'onde\ldots. La puissance de ce développement réside pourtant dans sa concision et son apparente clarté, malgré les mystères qu'il dévoile. 

Les théoriciens découvrent que les particules de matière sont portées dans l'espace suivant la figure d'interférence d'un paquet d'ondes -- c'est la dualité onde-corpuscule. La position de la particule, supposée alors ponctuelle, n'est ainsi jamais déterminée exactement au sein de ce paquet au point qu'on ne peut affirmer que la particule est vraiment portée par le paquet ou totalement dématérialisée dans les ondes. Pour cette raison, les lois physiques s'appliquent non pas à une particule suivant une trajectoire déterminée, mais à un ensemble d'ondes. Les pendants mathématiques de ces lois sont appelés des \emph{équations d'onde}. Ainsi, la "particule", mesurable dans cet ensemble d'ondes sous la forme d'un agrégat ponctuel d'énergie, ne suivrait en réalité aucune trajectoire particulière entre son point d’émission et son point de réception. La résolution des équations d'onde montre une discrétisation -- ou quantification -- des valeurs des observables telles que l'énergie ou les moments cinétiques -- \emph{spin} et moment orbital par exemple. Le système quantique est alors décrit par un nombre fini ou infini d'états classés suivant la valeur de ces observables et représentés par des vecteurs (bra et ket). Dans ce formalisme, les lois physiques s'expriment à l'aide d'opérateurs linéaires agissant sur ces vecteurs d'états. 

Testée théoriquement et expérimentalement, la théorie quantique est une révolution, mais laisse cependant de larges zones d'ombres. En particulier, comment concilier les visions antagonistes d'une matière à la fois onde et corpuscule~? Pourtant la description de l'atome en noyau et couches électroniques donne des accords expérimentaux excellents, jusqu'à l'observation directe. Et il a fallu pour cela abandonner l'intuition \emph{classique} d'un corpuscule parfaitement localisé et de trajectoire déterminée~; mais abandonner aussi l'intuition que les mesures de la position et de l'impulsion d'une particule peuvent être simultanées. 

Face à ces mystères, Dirac insistera pour ne pas chercher de représentation conforme à l'expérience humaine~; les lois atomiques défiant l'intuition, il faut suivre en aveugle la route que les mathématiques nous ouvre. Le lecteur pourra ouvrir l'excellent ouvrage de Cohen-tannoudji, Diu et Lalo\"e~\cite{cohen1973mecanique}.
	
\subsection*{Relativité restreinte et générale}
	La th\'eorie de la relativit\'e, introduite par Lorentz, Poincaré et Einstein -- qui la développera considérablement -- se base sur le constat expérimental que la valeur mesurée de la vitesse de la lumière est indépendante du référentiel galiléen d'observation. A cela s'ajoute le postulat que les lois de la physique ne doivent pas non plus dépendre du choix de ce référentiel. Autrement dit, par changement de référentiel galiléen, les lois physiques observées et la valeur de la vitesse de la lumière doivent être invariantes. Mathématiquement, Lorentz puis Poincaré d\'emontrent que de tels changements de référentiel galiléen prennent la forme de transformations dites \emph{spéciales} appelées aussi \emph{boost de Lorentz}. Les lois physiques sont ensuite déterminées avec la contrainte de \emph{covariance}, \cad d'invariance de forme par transformation spéciale. Einstein développe le formalisme en introduisant les notions de simultanéité et de causalité, concepts extrêmement importants ayant permis d’unir  l'espace et le temps en une seule entité géométrique : l'espace-temps, doté d'une structure causale. L'ensemble du formalisme constitue la \emph{relativité restreinte}.

Einstein souhaitera généraliser le postulat de covariance des lois à l'ensemble des référentiels, y compris accélérés, passant ainsi de la relativité \emph{restreinte} à la relativité \emph{générale}. Constatant que la gravité peut être annulée par le truchement d'une accélération -- principe d'équivalence -- Einstein montrera que la gravité est localement un effet relativiste. Cet aspect de localité le conduira à considérer que l'espace-temps est localement déformé et que la gravité est totalement encodée dans cette géométrie. Toute la puissance de la géométrie Riemannienne vient renforcer cette nouvelle approche et conduit à mathématiser la physique à un niveau supérieur. Mieux encore, les mesures de la précession du périhélie de Mercure et les mesures d'Eddington sur la déformation des rayons lumineux appuient cette révolution ! La géométrie est réconciliée avec la physique et occupera désormais une place de maître dans la physique moderne. On pourra lire les ouvrages de Wald pour la relativité générale~\cite{Wald:1984rg} et de Nakahara pour l'approche mathématique de la géométrie riemannienne~\cite{Nakahara:1990th}.

\section*{Théorie quantique des champs et théories de jauge~: la construction du Modèle Standard}

\subsection*{Théorie quantique des champs}
	La théorie quantique des champs (TQC) est la fusion de ces deux concepts fondamentaux, à savoir la mécanique quantique et la relativité restreinte. Elle nous a permis de comprendre comment un photon se propage  et comment un électron émet ces quanta d'onde électro-magnétique. C'est en 1925 que cette recherche est initiée par Heisenberg, Born et Jordan, plus tard complétée par Dirac. La description de ces particules nécessite de tenir compte à la fois de la mécanique quantique -- les calculs doivent dépendre de la constante de Planck $\hbar$ -- et de la relativité restreinte\footnote{D'un point de vue microscopique l'espace-temps est plat en première approximation. Une généralisation appelle une théorie microscopique de la gravitation, ce qui sera la théorie des cordes.} puisque tout phénomène physique faisant intervenir des particules de vitesse quasi-luminique doit être modélisé par des lois covariantes de Lorentz. Ces recherches ont mené à exprimer en un formalisme très complet l'\emph{électrodynamique quantique}. Grossièrement, l'électron est couplé au champ électromagnétique par sa charge électrique~; il émet et absorbe des quantas d'onde lumineuse (photon) et il est lui-même un quanta de champ -- électronique ici. Ainsi, le processus dominant l'interaction entre des particules de matière chargées consiste en des échanges de photons\footnote{Cependant cette visualisation sera vraiment intégrée lorsque Feynman introduira ses diagrammes et lorsque la convergence perturbative de cette approche sera vérifiée.}. Face au succès de cette théorie, les physiciens ont imaginé de l'adapter à tous les autres types d'interaction connus : interactions faible, forte et gravitationnelle. 
	
	Il était déjà très clair dés 1925 que les photons et les électrons revêtaient des natures bien différentes, ne satisfaisant pas à la même \emph{statistique}. Le premier est un \emph{boson}, \cad il peut se superposer à un photon identique ; le deuxième est un \emph{fermion}, \cad il subit une règle d'exclusion -- de Pauli -- qui interdit toute superposition de particules rigoureusement identiques. En outre, le photon est un champ dont les degrés de liberté peuvent être représentés par ceux d'un quadrivecteur -- appelé \emph{boson vecteur}. Par ailleurs, les degrés de liberté du champ électronique étant régis par une statistique non triviale, il faut introduire un nouvel objet mathématique, le \emph{spineur}. La généralisation de l'électrodynamique quantique à la théorie quantique des champs doit donc regrouper les bosons (scalaires, vecteurs, tenseurs\ldots) et les fermions (spineurs) puis formaliser leur quantification en terme de particules -- \emph{seconde quantification}. Ainsi, dans le \emph{formalisme hamiltonien} les champs sont décrits par des oscillateurs harmoniques et sont exprimés en termes d'opérateurs de création et d'annihilation de quantum à telle ou telle impulsion ou telle ou telle position. Suite aux travaux de Feynman dans les années 1950, le \emph{formalisme lagrangien}, à travers l'introduction des intégrales de chemins, est généralisé au cas quantique et une approche plus statistique est envisagée, en même temps qu'un développement visuel fort -- les diagrammes. Les interactions entre particules sont alors soumises à des poids statistiques qui privilégient tel ou tel processus, tel ou tel échange de particule. Cette approche renoue en quelque sorte avec la vision classique et déterministe dans le sens où -- au niveau perturbatif -- l'intégrale de chemins est une façon de moyenner sur tous les chemins possibles que pourrait prendre une particule pour se rendre d'un point A à un point B. La trajectoire classique est alors celle qui minimise le temps propre de propagation de la particule, mais les effets quantiques indiquent que la particule empreinte tous les chemins possibles à la fois. La théorie des perturbations directement associée à cette approche conduit à comprendre les interactions fondamentales comme relevant d'échanges de quanta de champ d'interaction entre les diverses particules de matière. Nous recommandons la lecture des ouvrages de Itzykson-Z\"uber et de Peskin-Schroeder~\cite{Itzykson:1980rh,Peskin:1995ev} pour une introduction complète à la théorie quantique des champs. En complément, le lecteur pourra trouver d'importants développements dans les ouvrages de Weinberg~\cite{Weinberg:1995mt,Weinberg:1996kr}.
	
\subsection*{Théories de jauge, modèle standard et au-delà}
	La généralisation de l'électrodynamique quantique à l'ensemble des interactions a conduit les physiciens à analyser en profondeur les concepts liés aux interactions dites "de jauge" telles que les interactions électro-magnétiques. En effet, outre sa qualité de boson-vecteur, le photon possède une symétrie interne, nommée \emph{symétrie de jauge}. Or il s'avère indispensable que la théorie entière vérifie cette symétrie de façon à exprimer de manière cohérente les interactions du photon avec la matière fermionique. Dans le dessein final d'unifier l'ensemble des interactions, le choix a été fait de généraliser le concept de "jauge" à tous les bosons d'interaction, devenant ainsi des \emph{bosons de jauge}, introduisant respectivement les \emph{théories de jauge}, chacune associée à un groupe de symétrie de jauge. Des résultats expérimentaux dans les grands accélérateurs de particules appuient cette intuition et fondent le Modèle Standard des interactions electro-faibles et fortes $SU(3)\times SU(2) \times U(1)$. Du côté théorique, les avancées mathématiques en géométrie différentielle permettent d'interpréter géométriquement la covariance de jauge d'une théorie en termes de fibrés. C'est une constatation importante dans le sens où, s'il ne faut pas forcément comprendre que l'origine de toutes les interactions est purement géométrique, pour le moins l'interaction gravitationnelle établie plus haut comme fondamentalement géométrique -- \cad associée à des transformations purement géométriques -- peut être dérivée en terme d'interaction de jauge donc en terme de \emph{graviton}. Sans aller aussi loin, les théories de jauge possèdent déjà des propriétés fantastiques, posant des contraintes de cohérence sur la théorie elle-même et introduisant de nouveaux objets, topologiques et donc non-perturbatifs : solitons, instantons \etc. En ce qui concerne les théories de jauge et les aspects géométriques, le lecteur pourra lire la revue d'Eguchi \emph{et al.} ainsi que les ouvrages de Nakahara et de Binetruy~\cite{Eguchi:1980jx,Nakahara:1990th,Binetruy:2006ad}.
	
	Le modèle standard offre une correspondance assez nette avec les résultats expérimentaux obtenus dans les grands accélérateurs de particules. Ultime particule hypothétique à découvrir, le Higgs est retors et semble se cacher indéfiniment des expérimentateurs -- à moins que le LHC ne finisse par mettre très bientôt la main dessus. Mais il faudrait le trouver, car lui seul semble pouvoir expliquer les différences de masses entre les particules~: entre les leptons, entre les bosons de jauge et entre les quarks. Bien qu'indispensable, il n'offre cependant pas la propriété la plus adéquate : sa masse est ajustée beaucoup trop précisément et sa masse nue extrêmement élevée, ce qui pose un problème de naturalité -- \emph{naturalness problem}. Il faut aller au-delà du modèle standard pour expliquer cette intrigante propriété, sans nécessairement mettre au rebut les théories de Higgs -- ici intervient la \emph{supersymétrie} qui résout en même temps le problème de naturalité et le problème de hiérarchie. Outre cet aspect, il faut de toute façon aller plus loin, car il semblerait que le modèle standard n'est qu'une théorie effective à basse énergie. En effet, trop de paramètres sont à ajuster, entre autre les constantes de couplages, dont celles entre le boson de higgs et les autres particules. Enfin, devant le succès de la théorie électro-faible d'unification de l'électromagnétisme et de l'interaction faible, tout porte à croire que les interactions électro-faibles et fortes doivent également être unifiées à un niveau d'énergie plus élevée -- échelle de \emph{Grande Unification} GUT.

\section*{Gravit\'e quantique, supersym\'etrie et supergravit\'e~: vers la th\'eorie des cordes}

\subsection*{Le problème de la gravité}
	Finalement, que penser de la gravité ? Serait-ce, comme suggéré plus haut, une interaction de jauge ? Peut-on unifier la théorie quantique des champs et la relativité générale ? Malheureusement, un naïf modèle de gravité quantique ne fonctionne pas, car il présente des calculs pathologiques. En effet, les échanges de graviton à échelle infinitésimale ne sont pas contrôlés et donnent des probabilités infinies -- \emph{non-renormalisables}. 
	
	Des divergences UV similaires apparaissent dans le cadre du modèle standard. Le traitement appliqu\'e pour traiter ces divergences se nomme \emph{renormalisation}~\cite{Wilson:1973jj,Itzykson:1980rh}. La théorie -- les paramètres -- est ajustée de telle sorte que les résultats physiques sont finis, ce qui revient souvent à introduire des paramètres \emph{infinis} -- ce fameux ajustement fin. Il s'agit d'une méthode efficace, mais finalement peu satisfaisante dans le cadre du modèle standard~: cet ajustement très précis pour des valeurs extrêmement élevées des paramètres de la théorie est trop artificiel. Cela va dans le sens de l'existence d'une théorie plus fondamentale. Ajoutons que ces paramètres ajust\'es sont des constantes de couplage ou des masses -- dites \emph{nues}. Tant que leur dimension est positive ou nulle, la théorie est dite \emph{renormalisable}. Or c'est toujours le cas des constantes de couplages et des masses nues du modèle standard. Par conséquent, le modèle peut être testé expérimentalement, non pas dans ses paramètres fondamentaux --  puisque ceux-ci sont ajustés -- mais dans sa forme, son expression ; et il l'a été, positivement. 
	
	Ce faisant, la \emph{renormalisation} n'est pas uniquement reli\'ee \`a l'existence de résultats infinis. En particulier, dans les th\'eories renormalisables, la renormalisation traduit un vrai processus \emph{physique}. Ceci est mis en valeur par la méthode de renormalisation introduite par Wilson. Elle consiste à imposer une échelle UV physique $\Lambda_{phys}$ à la théorie, au-dessus de laquelle tous les effets quantiques perturbatifs ou non-perturbatifs doivent \^etre int\'egr\'es. En vérité cette intégration-ci tient, pour sa part, compte d'un cut-off UV $\Lambda_{max}$ qui pourrait être la limite à partir de  laquelle telle théorie est suppos\'ee fausse donc les corrections non pertinentes. Dans le modèle standard, cette limite est typiquement $M_{GUT}\sim 10^{15} \, GeV$ l'échelle d'énergie de la \emph{grande unification}.  L'échelle $\Lambda_{phys}$ en revanche correspond en général à l'énergie maximale atteinte dans un accélérateur, ou dans une expérience de collision particulière, par exemple, et elle est par définition inférieure à $\Lambda_{max}$. En ce sens, les couplages de la théorie -- on parle de \emph{théorie effective} -- définie en-dessous de l'échelle maximale dépendent réellement de l'échelle physique, et cela est vérifi\'e expérimentalement. \\

Dans la théorie des champs de la gravité, la constante de couplage ajustable serait la constante de Newton $G_N$. Or celle-ci est de dimension négative $G_N=M_p^{-2}$ en fonction de la masse de Planck $M_p = 1,22.10^{19} ~ \text{GeV}$. La théorie associée est donc \emph{non-renormalisable}, \cad qu'il existe une infinité de diagrammes UV-divergents à tous les ordres. Ceci indique que la théorie perturbative est effectivement mal définie pour les impulsions élevées ou de manière équivalente aux petites distances. \\

Il existerait deux solutions -- voire une troisième -- à ce problème~:\\

\begin{itemize}\itemsep4pt
\item[\emph{i)}] Est-ce une pathologie du calcul perturbatif ? Auquel cas, peut-être faut-il envisager le problème dans une approche non-perturbative pour le résoudre. Il s'agit de la direction dans laquelle s'est engagée la \emph{gravité à boucle}, qui connaît quelques succès. 

\item[\emph{ii)}] Ou bien est-ce une problème plus fondamental de la physique quantique ? C'est l'option qu'a choisi d'explorer la \emph{théorie des cordes}, en proposant de résoudre le problème de divergence des interactions gravitationnelles en délocalisant les particules. Elles admettraient une structure microscopique de la forme d'une corde, de taille fixée $\ell_s \sim \ell_p$ autour de la longueur de Planck $\ell_p \approx 1,616 . 10^{-34} \, m$, échelle caractéristique de la gravité quantique. La théorie quantique des champs qui en découle est ainsi dotée d'une longueur, donc d'une échelle, minimale d'interaction. Autrement dit, la limite infinitésimale (UV) de la théorie est contrôlée naturellement. 

\item[\emph{iii)}] A-t-on réellement besoin d'une gravité quantique pour décrire nos observations~ ? Jusqu'à quel point la mathématisation des phénomènes est possible~ ? Autrement dit, faut-il bien que les modèles mathématiques soient cohérents d'un bout à l'autre et en dehors du champ d'observations ? Je n'irais pas plus loin dans cette problématique, car je ne connais aucune réponse si ce n'est des questions. Mais je la conçois personnellement comme une option. 
\end{itemize}	 
	 
\subsection*{La supersymétrie et la supergravité}

L'apparente non-naturalité du modèle standard, en particulier l'ajustement fin des constantes de couplages et des masses -- problème de hiérarchie -- associée à la non-finitude de l'énergie du vide, ont amené les théoriciens à tendre vers un modèle mathématique mieux défini et plus naturel. Jusqu'à présent la théorie du modèle standard fonctionnait plutôt bien et l'accord aux expériences de collision très correct. Cependant, les observations astrophysiques donnent d'autres contraintes qui posent la question de la complétude du Modèle Standard. Il y a par exemple le problème de la \emph{matière noire}~: est-elle faite de neutrinos~? Il semble pourtant que les neutrinos soient observationnellement rejetés pour ce rôle-ci. Il faut donc ajouter de nouvelles particules qui ne couplent pas aux photons -- on en trouverait dans l'extension supersymétrique du modèle standard MSSM par exemple. Il y a aussi le problème de la constante cosmologique  qui d'après les dernières observations est infinitésimale, mais non nulle et positive. Dans une théorie couplée à la gravité cette constante correspond à l'énergie du vide. Or dans le cadre du modèle standard, sa valeur n'est absolument pas contrôlée et diverge.

En outre, il y a ce courant de pensée qui guide les théoriciens vers une explication purement mathématique de l'ensemble des choses. C'est une question vraiment profonde qu'il faut avoir en tête en allant plus en avant dans la théorisation. En se basant sur ce principe, que la Nature est mathématique, ou plus faiblement quasi-parfaitement décrite par les mathématiques, alors il faut que ce cadre soit tout à fait bien défini et que dynamiquement les choses y soient " causes et conséquences " les unes des autres sans que jamais il ne faille faire surgir quoique ce soit du néant. Le modèle standard ne satisfait pas à ce principe pour les raisons que nous avons dites~: nombre de paramètres fondamentaux, ajustement fin et infinité de l'énergie du vide -- sans compter les autres problèmes cosmologiques.   \\ 

En étudiant la théorie des cordes et afin d'introduire des objets fermioniques dans l'espace-temps, les cordistes ont été amenés à introduire une \emph{supersymétrie}. Brièvement, les cordes sont décrites par des surfaces sur lesquelles les coordonnées d'espace-temps deviennent des champs scalaires, des bosons. Or en introduisant naturellement des champs fermioniques -- formulation de Green-Schwarz ou formulation RNS -- on découvre qu'il existe une symétrie mélangeant ces fermions avec les bosons, alors appelée supersymétrie. Il s'en suit que le spectre de particules décrites par les cordes hérite -- d'une certaine façon et nous verrons cela plus en détail par la suite -- de cette supersymétrie qui se trouve être une propriété d'un univers cordiste contenant bosons et fermions~ ; chaque boson admet un partenaire fermionique de même masse et vice-versa. 

L'une des caractéristiques essentielles d'une théorie supersymétrique est l'annulation de l'énergie du vide en espace plat\footnote{C'est un premier pas vers son contrôle en espace courbe.} entre les contributions des fermions et celles des bosons et une autre est la résolution du problème de hiérarchie gr\^ace au théorème de non-renormalisation appliqué au boson de Higgs. Mais elle vient aussi avec son lot de problèmes et en particulier prédit l'existence de plus de particules qu'observées. La force du principe de mathématisation est que si un moyen de résoudre mathématiquement tel problème est d\'ecouvert alors \c c'en est une solution sérieusement envisageable. Pour ce cas-ci, on propose que la supersymétrie soit brisée à notre échelle, de telle sorte que les partenaires qui ne sont pas observés -- jusqu'à aujourd'hui -- sont suffisamment massifs pour ne pas être observables dans les gammes d'énergie accessibles dans la plupart des accélérateurs -- mais peut-être pas au LHC ? 

Bien que ce modèle nécessite encore de l'ajustement pour contrôler le degré de brisure de supersymétrie, ce n'est pas le même niveau d'ajustement que celui du modèle standard et le contrôle que l'on a dessus en fait un candidat vraiment sérieux. Mais encore, il faut aller plus loin et comprendre comment dynamiquement arriver à cet ajustement car on ne souhaite pas voir les choses surgir du néant et c'est légitimement dans ce contexte que la théorie des cordes s'inscrit. Pour une introduction à la supersymétrie, nous conseillons la lecture des ouvrages de Binetruy, de Derendinger et de Weinberg~\cite{Binetruy:2006ad,Derendinger:1990tj,Weinberg:2000cr}.
	\\ 

Maintenant, il faut aussi comprendre comment coupler la gravité à ce modèle supersymétrique et si par hasard une gravité quantique ne serait pas mieux définie au sein d'une théorie plus symétrique donc mieux contrainte. En fait, la supergravité qui est l'extension supersymétrique de la gravité quantique vient assez naturellement. En effet, pour l'instant nous n'avons parlé que d'une supersymétrie rigide, globale. Qu'en serait-il si les générateurs de la supersymétrie étaient définis uniquement localement~ ? En ouvrant cette voie, nous nous pla\c cons d'emblée dans une théorie de gravité~: les générateurs de supersymétrie, des spineurs, not\'es $Q_\alpha$ vérifient l'anti-commutateur suivant~:

\begin{align}
\acomm{Q}{\overline Q} = 2 \gamma^\mu P_\mu
\end{align} 

avec $P_\mu$ le générateur des translations. Par conséquent, le simple fait de rendre les générateurs de supersymétrie locaux rend inévitablement les générateurs de translations locaux également. Or cela n'est rien d'autre que rendre l'espace courbe pour la particule concernée par cette impulsion. Alors, nous avons fatalement affaire à une théorie de gravité quantique. L'application de la supersymétrie sur le champ correspondant de spin $2$ exige donc l'existence d'au moins un partenaire supersymétrique de type spin $3/2$ appelé \emph{gravitino}. Les types de supergravités sont d\'enomm\'ees en fonction du nombre de supersymétries ${\mathcal N}$ ou de façon équivalente du nombre de gravitinos. Par exemple, les supergravités de type II en dimension $10$ (pour correspondre à la théorie des supercordes) ont ${\mathcal N}=2$ et deux gravitinos tandis que les supergravités de type I en dimension $10$ toujours n'ont que  ${\mathcal N}=1$ et un seul gravitino. 

Le problème de base de la gravité quantique n'est pourtant pas résolu dans ce modèle, car bien que ce théorème de non-renormalisation existe, la supergravité souffre encore d'être non-renormalisable -- sauf peut-être la supergravité ${\mathcal N}=8$ en dimension $4$. Mais c'est un point positif pour la théorie des cordes~ : dans une théorie des supercordes la gravité quantique est naturelle et s'inscrit immédiatement dans un cadre de supergravité, tout en étant contrôlée dans l'UV.

\section*{La th\'eorie des cordes}

L'idée fondamentale de l'application de la théorie des cordes à la gravité quantique est la suivante. Puisque les divergences des calculs gravitationnels sont toutes ultra-violettes, en introduisant une régularisation UV tout en conservant la symétrie de Poincaré -- boost et translations -- le problème devrait être réglé. En introduisant des objets de dimension non nulle -- par exemple une corde de longueur $\ell_s \sim 10^{-34} m$ -- leurs interactions sont délocalisées le long de leur extension spatiale, de telle sorte qu'il n'existe plus de singularité de vertex. On peut montrer que la théorie dans laquelle ces objets sont des cordes est la théorie la plus fondamentale\footnote{En fait, c'est un peu incorrect, car la théorie que l'on pense fondamentale n'est pas encore connue entièrement mais serait une supergravité à 11 dimensions faisant interagir des membranes et non des cordes~: c'est la théorie M.}~: des objets de dimension supérieure, des membranes par exemple, sont plus énergétiques donc probablement moins fondamentaux. Dans le cadre de la théorie des cordes ce ont des objets de plus grande dimension pouvant être créés à partir de cordes et dans certains cas se désintégrant en cordes. A l'heure actuelle il s'agit de la seule théorie capable de conserver la symétrie de Poincaré tout en imposant une régularisation ultra-violette \emph{physiquement} justifiée. 

Les objets en théories des cordes peuvent être interprétés, en les sondant à basse énergie, en tant que particules. Les cordes ont des masses, qui sont grossièrement fonctions de leur fréquence d'oscillation et proportionnelles à $1/\ell_s^2$ donc rapidement très massives. Les plus importantes sont les cordes non-massives puisque les autres sont tellement massives que la probabilité pour les créer est infinitésimale et leur probabilité de désintégration presque $1$. En outre, afin de distinguer entre ces cordes (massives et non-massives) des particules bosoniques et des particules fermioniques, il faut se concentrer sur la théorie des \emph{supercordes} dans laquelle le spectre de particules admet fermions et bosons et est éventuellement supersymétrique. Il est aussi possible de définir une théorie de cordes bosoniques -- c'est d'ailleurs la première inventée -- mais évidemment puisqu'elle n'admet pas de fermions, elle n'est pas intéressante \infine pour décrire une théorie de notre univers et de nos particules. Il existe 5 types de théories de supercordes IIA, IIB, I, hétérotique $E_8\times E_8$ et $SO(32)$. Chacune a ses propriétés mais aussi et surtout elles sont toutes reliées les unes aux autres par ce qu'on appelle des \emph{dualités} et toutes reliées à la théorie M. Elles décrivent toutes à basse énergie des supergravités. 

Toute théorie des supercordes, pour être mathématiquement cohérente, doit habiter dans un espace-temps\footnote{La théorie M est une supergravité de dimension $11$.} de dimension $10$. Cette contrainte implique de traiter le cas de ces $6$ dimensions supplémentaires. Pour cela, on propose de \emph{compactifier} l'espace le long de ces directions, \cad que chaque dimension est refermée sur elle-même de sorte qu'en allant tout droit on finit par retourner à son point de départ. Cela s'exprime mathématiquement par une relation d'équivalence $X \sim X + 2\pi R$ avec $R$ le rayon de compactification qui caractérise la longueur de l'espace compact dans la direction $X$. En l'appliquant indépendamment sur toute les $6$ directions, il s'agit du schéma le plus simple de compactification, la figure géométrique décrite est un tore à $6$ dimensions noté $T^6$. Le volume de l'espace compact est en général supposé très petit, de sorte que les effets physiques relevant de ces dimensions sont imperceptibles. En fait, il est techniquement envisageable d'appliquer une compactification plus générale sur toute variété de dimension $6$. Cependant, il existe des contraintes, en particulier la conservation d'une certaine quantité de supersymétrie, qui imposent à l'espace de respecter certaines géométries. Par exemple, la conservation d'au moins une supersymétrie et une torsion nulle imposent que la variété compacte doive avoir une holonomie $SU(3)$. Cela se traduit pour une variété de dimension paire avec une métrique réelle en~: variété \emph{Ricci-plate}\footnote{Le tenseur de Ricci $R_{mn} = 0$ doit être nul.} et \emph{K\"ahler}\footnote{La forme de Kähler $J = i G_{I \bar J} dz^I \wedge d\bar z^{\bar J}$ doit être fermée.}, ce qui est appelé une variété de \emph{calabi-Yau}, en l’occurrence ici $CY_6$. Ne rentrons pas plus dans les détails, mais mentionnons simplement que le tore $T^6$ est Calabi-Yau et conserve toutes les supersymétries d'espace-cible à 10 dimensions. Le nombre de degrés de liberté – appelés \emph{modules} -- associés à la création d'un espace de Calabi-Yau peut être très grand et laisse beaucoup de possibilités. Par exemple, le tore a déjà $6$ degrés de libertés. Or, l'objectif de la théorie des cordes est de retrouver dans une limite de basse énergie une théorie de gravité quantique couplée au modèle standard et permettant aussi de satisfaire aux contraintes cosmologiques. Nous avons donc un important problème qui est celui de retrouver la théorie de la Nature dans ce qu'on appelle depuis quelque temps le \emph{paysage} de la théorie des cordes.  \\

Nous venons brièvement de décrire la théorie des supercordes. Dans cette théorie, comme dans la bosonique, les cordes sont des objets compacts unidimensionnels de petite taille qui peuvent soit se refermer sur eux-mêmes, on parle de \emph{corde fermée} soit simplement avoir deux extrémités distinctes, on parle dans ce cas de \emph{corde ouverte}. Les cordes fermées se propagent dans l'espace-temps en décrivant des surfaces \emph{tubulaires} et les cordes ouvertes sous la forme de \emph{nappes}. Elles forment alors des \emph{surfaces d'univers} – par analogie aux lignes d'univers introduites en mécanique classique du point. Les cordes fermées se propagent en général librement dans l'espace-temps. A l'inverse, les cordes ouvertes doivent être attachées par leurs extrémités à des défauts d'espace-temps, des hypersurfaces. Ces défauts se révèlent en fait être des objets eux-mêmes dynamiques, appelés \emph{branes} et en général $Dp$-brane. La lettre $D$ provient des conditions de bord de Dirichlet que vérifient les cordes à leurs extrémités et la variable $p$ indique leur dimension spatiale~: les branes sont des objets de $p+1$ dimension, dont une dimension temporelle. Ainsi les cordes ouvertes se propagent le long de ces branes et elles admettent un spectre de masse dépendant de la configuration des branes auxquelles elles sont attachées. En particulier, leur spectre peut ne pas être supersymétrique mais aussi contenir des niveaux de masse carrée négative, ce qu'on appelle des tachyons. \\

Enfin, l'ensemble de ces cordes et leurs excitations forment à basse énergie un zoo de particules, massives, non-massives ou tachyoniques, bosoniques ou fermioniques. Dans le cadre de la théorie quantique des champs, qui utilise un formalisme de seconde quantification, chaque particule est interprété comme le quanta d'un champ dont la dynamique est régie par une action ou un hamiltonien. Par analogie et au moins à basse énergie, les "particules cordistes" pourraient donc \^etre interprétées comme des quantas de champs et leur dynamique décrite à l'aide d'une action. Le terme d'\emph{action effective de basse énergie} est introduit. Nous en verrons des exemples dans cette thèse. Il est aussi possible de décrire exactement, quoique difficilement,  les cordes en tant que quanta d'une théorie des champs. Ce domaine de recherche majeur constitue ce qu'on nomme la \emph{théorie des champs de corde} dénotée SFT pour \emph{string field theory} et est encore en progrès. Voici quelques références de lectures pour une introduction à la théorie des cordes~\cite{Polchinski:1998rq,Polchinski:1998rr,Kiritsis:2007zz,Green:1987sp,Green:1987mn}. \\

\section*{Le tachyon et la condensation}

Dans ce cadre, nous nous sommes intéressés pendant cette thèse à la description du tachyon de corde ouverte apparaissant dans le spectre des cordes tendues entre une brane et une antibrane (voir définition dans section~\refcc{sec:SCFT}) parallèles et séparées. Ce système admet un tachyon dans ce secteur interbranaire lorsque la distance séparant les branes est inférieur à une certaine distance critique, que nous noterons ici $r_c$. L'évolution du tachyon est particulièrement importante, car le champ dont il est le quanta est instable si bien que le système de branes lui-même est instable. La résolution de cette instabilité s'appelle la \emph{condensation de tachyon} et est le sujet de cette thèse. La condensation de tachyon est tout un domaine de recherche et a été largement étudié depuis les années 80 dans toutes sortes de systèmes, aussi bien supersymétriques que bosoniques. La question essentielle est celle d'obtenir une description effective du système et de comprendre de quelle façon le tachyon évolue et condense, mais surtout de savoir s'il atteint une valeur à laquelle il se stabilise. Un tachyon stable implique un système stable. Quel est sa nature ? De quoi est-il composé ?

Donnons brièvement un exemple. En théorie bosonique, une brane admet naturellement un tachyon de corde ouverte. On sait aujourd'hui que ce tachyon, notons-le $T$ est décrit par un potentiel $V(T)$ qui est minimisé en $T\to \infty$ et s'annule. Cela implique que le système au minimum du potentiel est une théorie de cordes fermées, où les cordes ouvertes, donc aussi la brane initiale, ont disparu ! Quand il évolue temporellement dans son potentiel pour rejoindre ce minimum, le tachyon induit sur la brane une évaporation de sa matière sous forme de cordes fermées ainsi qu'une désintégration en cordes fermées. A la fin, au minimum, la brane s'est totalement dissipée en cordes fermées. 

Nous verrons que la description du tachyon interbranaire dans le système brane-antibrane séparé n'est pas aisé et qu'il n'est pas évident que le schéma ci-dessus soit celui qu'il suit. Avant tout, l'action effective, et donc aussi le potentiel effectif, couplant le tachyon $T$ et la distance $\ell$ qui est aussi représentée par un champ, n'est pas connue exactement. En fait, son expression a été conjecturée par Garousi, mais le domaine de validité de l'action qu'il a déterminée semble restreint en dehors des solutions de condensation dynamique, ce que nous montrons par notre étude. La dynamique du système ou l'issue de la condensation, sont donc assez m\'econnues, bien qu'à l'inverse le cas des branes coincidentes est très bien décrit, connu et même exploité. Néanmoins, nous savons qu'il existe un mode de condensation dynamique, nommé \emph{tachyon roulant}, à distance constante. Bagchi et Sen avaient montré que ce mode de condensation existait pour un domaine de distance restreint à $\ell<\ell_c/\sqrt 2$.

Nous avons démontré qu'en réalité ce mode existe aussi pour $\ell_c/\sqrt 2<\ell<\ell_c$ donc en somme sur l'ensemble du domaine sous-critique $\ell<\ell_c$. Puis nous avons déterminé l'action effective du tachyon à l'ordre quadratique en utilisant une méthode proposée par Kutasov et Niarchos qui l'avaient utilisé avec succès pour obtenir l'action effective du tachyon sur la brane non-BPS (voir section~\refcc{sec:SCFT} pour la définition). L'existence du mode de condensation à distance constante  pour tout $\ell<\ell_c$ est importante, car elle ouvre la voie à une description exacte de l'évolution dynamique du système séparé. Nous mentionnerons brièvement dans la conclusion de futures pistes de recherche~: la description exacte de la condensation d'un tachyon interbranaire à distance constante sur un système séparé en dimension compacte, et une piste de calcul du potentiel effectif du tachyon en fonction de la distance de séparation par identification du modèle off-shell de OSFT au modèle Kondo.

\part{Introduction}
\label{part:intro}

\chapter{Motivations et plan de th\`ese}
\label{chap:mot}

	La condensation de tachyon est un phénomène important en théorie des cordes et a été la cible d'un considérable intérêt ces dernières années. A la fois parce qu'il a été possible de décrire exactement des phases de condensation et aussi parce que le problème que pose son apparition a nécessité de développer les théories de champs de corde. 

En théorie bosonique, cordes fermées et cordes ouvertes admettent généralement un fondamental tachyonique dans leur spectre de masse. En revanche, en théorie des supercordes, la projection GSO tronque le tachyon du spectre des cordes ouvertes et fermées. Toutefois, il existe des exceptions: les systèmes brane-antibrane parallèles et brane non BPS par exemple. Dans ces systèmes, la projection GSO y est telle qu'ils admettent un ou plusieurs tachyons dans leur spectre de cordes ouvertes. Nous nous intéresserons dans cette thèse au tachyon de corde ouverte dans le système brane-antibrane. \\

En théorie quantique des champs, les tachyons sont des champs localisés en un maximum de leur potentiel. En ces points les champs sont instables par conséquent la quantification n'y a aucun sens. En d'autres points, le potentiel peut toutefois admettre un ou plusieurs minimums, locaux ou globaux, où à l'inverse une théorie quantique est très bien définie. On parle ainsi de vide stable si le minimum est global, ou métastable si le minimum est strictement local. La condensation du tachyon est le phénomène de stabilisation du champ dans un de ces minimums. Celle-ci peut être globale aussi bien que locale en espace ou en temps. 

En théorie des cordes, il est communément admis que le potentiel du tachyon de corde ouverte admet au moins un vide stable. On peut montrer que chaque vide stable correspond à un vide de corde fermée, \cad sans corde ouverte donc sans la brane initiale. Si le potentiel admet plus d'un vide stable, les condensations spatiales du tachyons peuvent être locales et les configurations topologiquement non triviales. Peuvent donc apparaître des murs de domaine, interpolant entre deux vides bien distincts, mais aussi des cordes cosmiques, entourées d'une configuration en vortex. Ces défauts topologiques, des solitons, sont identifiés à des branes généralement stables. Ainsi par condensation une brane mère donne naissance à, au moins, une brane fille. Puisque ces dernières sont de dimension inférieure à la brane initiale, on parle de relation de \emph{descente}. Ce processus s'inscrit directement dans le cadre de la théorie K -- voir par exemple l'article de Witten~\cite{Witten:1998cd}. \\

Le tachyon peut aussi se condenser temporellement, \cad dynamiquement. En théorie des cordes, ce mode de condensation peut \^etre d\'ecrit exactement~; c'est un point très important. En cosmologie, par exemple, ce genre de processus existe~: pendant l'expansion, alors que l'univers refroidit, le Higgs posé en son maximum local se condense dynamiquement dans la gouttière du chapeau mexicain tout en brisant spontanément la symétrie électrofaible. Une réalisation explicite et une description exacte de ce phénomène serait très précieuse. En particulier, la condensation est-elle accompagnée d'une phase d'inflation ou de production de particules, dont les effets pourraient se révéler mesurables ? Or en théorie des cordes, la condensation de tachyon temporelle sur une brane instable induit l'évaporation progressive de cette derni\`ere sous la forme de cordes fermées massives -- par exemple identifiables au flux d'énergie nécessaire au (p)reheating dans les scénarios d'inflation -- et l'éventuelle création de branes inférieures – par exemple des cordes cosmiques.  \\

Deux approches complémentaires sont généralement utilisées pour décrire la condensation du tachyon~: \\

\begin{itemize}\itemsep4pt
\item La première est historiquement l'approche de CFT qui consiste à inclure un fond tachyonique, soit dynamique soit statique, et à étudier les dépendances de ces observables en fonction de ce fond. Celle-ci a été d'une grande efficacité pour mettre en valeur la nature des solitons et des résidus de condensation. Malheureusement nous n'avons dans ce contexte aucune recette pour construire l'ensemble des CFTs.
\item La deuxième est l'approche de théorie des champs effective. Dans ce cadre, les CFT constituent des solutions aux équations du mouvement, dérivées d'une certaine action effective. Donc cette démarche est censée permettre de déceler l'ensemble des CFTs, en particulier ici toutes celles décrivant des condensations de tachyon. Le plus gros succès de cette approche est l'obtention du potentiel symétrique et universel du tachyon en théorie des supercordes, qui révèle la présence de plusieurs minimums globaux -- d'où les solitons. \\
\end{itemize}

En outre au moins trois méthodes de théorie des champs peuvent \^etre distingu\'ees~: \\

\begin{itemize}\itemsep4pt
\item Les \emph{théories effectives du modèle sigma} qui étudient le groupe de renormalisation et font correspondre aux équations de flot des équations du mouvement~;
\item Les \emph{théories des champs de cordes cubiques} (SFT) qui utilisent une troncation cohérente du spectre de corde pour déterminer une action de basse énergie mais qui ne fournissent en général pas de formulation exacte~;
\item La \emph{théorie des champs de cordes ouvertes} (OSFT ou BSFT) qui utilise le formalisme de Belavin-Vilenkin (BV) pour exprimer une action décrivant l'espace des théories et dans laquelle les CFT ont un statut évidemment spécial. \\
\end{itemize} 

La méthode du modèle sigma et celle de OSFT ont naturellement une relation très intriquée, car elles sont toutes deux reliées au groupe de renormalisation de la théorie de surface. Dans ces différents contextes, plusieurs actions pour le tachyon de cordes ouvertes, éventuellement couplées aux autres champs non-massifs, ont été obtenues en théorie bosonique comme en théorie de supercordes. En général, elles diffèrent significativement les unes des autres, mais il faut bien mesurer qu'elles ne décrivent pas exactement la même chose. Le lecteur pourra lire par exemple cet article de Garousi~\cite{Garousi:2002wq} où une comparaison est faite. Ainsi l'action d'OSFT est très off-shell, tandis que l'action du modèle sigma est perturbativement off-shell, autour d'une solution particulière -- voir par exemple la discussion dans l'article de Kutasov et Niarchos~\cite{Kutasov:2003er}. \\

A pr\'esent, discutons brièvement du système brane-antibrane séparé et parallèle. Ce système est très intéressant pour au moins deux raisons. Premièrement, c'est le modèle à succès de physique branaire dynamique mettant en jeu divers champs non-massifs, éventuellement dynamiques -- par exemple la distance de séparation -- et des tachyons de cordes ouvertes. En outre, c'est un système non BPS, \cad qui brise spontanément la supersymétrie, qui a le bon goût de s'annihiler lorsque les deux branes sont coïncidentes. Il s'agit donc d’un candidat idéal pour chercher des modèles d'inflation réalistes -- par exemple les modèles KKLT~\cite{Kachru:2003aw} ou encore ce modèle d'Alexander de création de vortex~\cite{Alexander:2001ks}. Ce sont terminologiquement des modèles \emph{d'inflation branaire}. Deuxièmement, en dehors de la coïncidence, quand la séparation est de l'ordre de la distance de corde, la description de ce système est un peu approximative et très heuristique. C'est pourtant à cet ordre qu'apparaissent les tachyons et donc, dés cet instant, qu'il faut étudier la dynamique de leur condensation. Or le processus de condensation du tachyon à séparation non nulle est encore m\'econnue. D'ailleurs, existe-t-il bien un vide stable où le tachyon pourrait condenser ? \\

Le système brane-antibrane séparé se décrit en trois phases~: \\

\begin{itemize}\itemsep4pt
\item La première, à grande distance, est dominée dans sa dynamique par le potentiel coulombien attractif induit par l'amplitude à une boucle des cordes ouvertes entre les deux branes. Cependant, lorsque la distance est de l'ordre de l'échelle des cordes, l'amplitude de cordes ouvertes à l'ordre des arbres gagne en intensité. Or le fondamental du spectre des cordes ouvertes interbranaire admet une masse proportionnelle à la distance. Tant que la séparation excède la valeur critique $r_c = \pi \sqrt{2\alpha'}$ cette particule est massive donc stable\footnote{Dans la première phase, proche de la distance critique, le tachyon peut déjà condenser localement par effet tunnel -- nucléation et formation de bulle~\cite{Callan:1997kz}. Cependant, la constante de temps pour ce genre de phénomène est très grande par rapport au temps qu'il faut au système pour être attiré à l'intérieur de la phase tachyonique ; ce n'est donc pas un phénomène dominant.}. Elle induit à l'ordre d'une boucle un potentiel effectif pour le champ de distance, mais techniquement déjà inclus dans l'amplitude, à une boucle précédente. 
\item La deuxième phase concerne la valeur pr\'ecise de séparation $r_c$. Dans cette phase le fondamental interbranaire est non-massif. Tous les champs fondamentaux de l'ensemble des secteurs de cordes ouvertes hébergées sur les branes sont alors non-massifs. Il s'agit, comme nous le verrons, d'un point critique dans la CFT.
\item La troisième phase est la plus intéressante, car le fondamental interbranaire devient tachyonique. A l'ordre des arbres il est donc amené à condenser. Or l'ordre à une boucle n'est ici plus pertinent pour la raison suivante : le potentiel est instable donc on ne peut plus parler de correction quantique à une boucle. En effet, cela n'a de sens que si l'on peut parler de calcul perturbatif et ce n'est plus le cas ici\footnote{Pour cette raison aussi, l'amplitude à une boucle des cordes ouvertes cesse d'être un calcul physiquement pertinent. Et d'ailleurs, il diverge à cause du tachyon}. Notons en outre, qu'à l'ordre des arbres, le champ de distance est un module, donc, au moins à tachyon nul, il n'a pas de potentiel. \\
\end{itemize}

Au sein de la phase tachyonique, nous avons deux approches possibles pour décrire la condensation et le devenir du système~: la CFT ou les théories effectives. En CFT, nous avons accès à au moins un type de solution, le tachyon roulant, mais dont les calculs sont forts complexes et pour l'instant pas entièrement résolus. Il est possible qu'il existe aussi une solution de vortex mais nous ne la connaissons pas. En théorie effective cependant, il existe une action à l'ordre des arbres proposée par Garousi. N\'eanmoins, nous montrerons qu'elle n'est pas satisfaisante pour étudier la condensation temporelle du tachyon. \\

\section{Actions effectives Tachyon-DBI et domaine de validit\'e}
\label{sec:act_eff}

L'action de Garousi sur le système brane-antibrane est obtenue à partir de l'action de Sen de la brane non BPS de dimension maximale en type IIA. Nous présenterons donc en premier lieu cette action, puis dans la section~\refcc{sec:act_gar} nous verrons comment la relier à l'action de Garousi.

\subsection{Action de Sen abélienne TDBI}
\label{sec:act_sen}

L'action trouvée~\cite{Sen:1999md} par Sen répond à un certain nombre de critères imposés par la théorie de supercordes -- en bref, supersymétrie et forme DBI. Elle a été reformulée ensuite par Garousi~\cite{Garousi:2000tr} qui en a testé l'expression finale en comparant les éléments de matrice-S calculés autour du vide perturbatif de la théorie des cordes, à ceux calculés perturbativement à partir de l'action effective. Cette dernière est explicitement~:

\begin{align}\label{eq:Sen}
S_{sen} = - T_9 \int \di^{10}\sigma ~ V(T) \, \sqrt{-\det\parent{G_{ab} + B_{ab} + 2\pi \alpha' F_{ab} + \partial_a T \partial_b T}}
\end{align}

Discutons un moment du domaine de validité de cette action. D'après Sen~\cite{Sen:1999md} son domaine de validité est $T \gg 1$ et $\module{\partial_i T \partial_j T} \gg 1$ avec $(i,j)$ des indices spatiaux. Il suppose aussi que toute dérivée d'ordre supérieure ou égale à 2 est négligeable. La raison pour ces contraintes est que cette expression est validée telle que ses équations du mouvement admettent des solutions de ressaut (kink) et que les fluctuations des champs autour de ces solutions sont décrites par une action de type DBI. De cette manière on est en mesure d'identifier les solitons à des branes de dimension inférieure. 

En outre, nous disions que Garousi avait testé cette action en comparant des éléments de matrice-S autour du vide $T=0$. L'universalité de cette approche est discutable. En effet, un tachyon dans ce vide est au sommet de son potentiel, si bien que toute perturbation doit croître jusqu'à rapidement devenir non-perturbative~; en l'occurrence le roulement est inévitable. Néan\-moins, il reste possible d'obtenir des perturbations qui ne grossissent pas -- \cad d'énergie réelle -- en tronquant l'impulsion en dessous $k^2 =|m^2|$ avec $m^2$ la masse carrée (négative) du tachyon. Le long de ces perturbations, l'approche de Garousi est sans doute valable. Ces perturbations sont de genre espace et sont donc connectées non-perturbativement aux solutions de ressaut. Par conséquent, nous en déduisons que l'action de Sen-Garousi est au moins valide le long des tachyons de genre espace et autour de telles solutions. 

En revanche, on ne peut pas en dire autant pour les solutions roulantes -- d'énergie imaginaire -- si bien qu'il reste possible que cette action ne soit pas valide le long de celles-ci. En effet, la continuation des éléments de matrice-S des énergies réelles aux énergies imaginaires est ambiguë : faut-il toujours imposer la conservation de l'énergie? Cette question est évidemment reliée à la condition perturbative des éléments de matrice-S, condition qui est rapidement violée le long d'un tachyon roulant. 

Nous pouvons cependant rapporter le travail de Kutasov et Niarchos~\cite{Kutasov:2003er} dans cette direction. En effet, ils identifient dans la limite perturbative $x^0 \to -\infty$ les fonctions de corrélations de tachyons 

\begin{align}
\corr{T^+_{\vec k_1} T^+_{\vec k_2} \ldots T^- _{\vec p_1}T^-_{\vec p_2}\ldots}_{T^+}
\end{align}  

pour $\vec k_i, \vec p_j \sim 0 $ avec $T^\pm_{\vec k} = e^{\pm i\vec k. \vec x \pm \sqrt{1-k^2} x^0}$ dans le fond $T^+ e^{x^0/\sqrt 2}$ qui est exactement marginal, aux éléments de matrice-S correspondants, calculés dans la théorie des champs. Dans chaque cas ceux-ci doivent s'annuler. Dans le premier, par marginalité exacte des opérateurs de vertex, et dans le deuxième, parce que les champs correspondants sont les solutions exactes des équations classiques du mouvement. Cela implique et justifie que l'action effective à rechercher le long de la solution $T^+ e^{x^0/\sqrt 2}$ dans la limite $x^0 \to -\infty$ doit admettre $T^+e^{x^0/\sqrt 2} + T^- e^{-x^0/\sqrt 2}$ comme solution, et s'identifie, on-shell, à la fonction de partition sur le disque $S_{on} = Z[\psi^0 e^{X^0/\sqrt 2}]$. L'action effective tachyonique qu'ils obtiennent est la suivante~:

\begin{align}
S_T = T_p \int \di^{p+1} \sigma V(T) \sqrt{1 + \partial_\mu T \partial^\mu T}
\end{align}

en supposant encore que les dérivées secondes et supérieures sont négligeables. Cette action est au moins valable autour de tachyons quasi-homogènes. Ajoutons qu'elle est en outre valable tout le long de la solution $T^+ e^{x^0/\sqrt 2}$ \cad non-perturbativement. \\

A propos de cette action, le point suivant est important et éclaire sur les conditions de validité des actions effectives~: Kutasov et Niarchos étudient l'action le long d'une solution de tachyon roulant $T \propto e^{x^0/\sqrt 2}$ mais pas le long d'une solution générale $T \propto T^+ e^{x^0/\sqrt 2} + T^- e^{-x^0/\sqrt 2}$ et cela pour une raison bien précise: alors que la première solution interpole entre le vide perturbatif instable et \apriori le vide stable de cordes fermées, la solution générale interpole quant à elle entre deux vides stables de cordes fermées. Ainsi, le long de cette dernière il n'existe pas d'état asymptotique de corde ouverte. Par conséquent, des éléments de matrice-S d'interaction de champs de cordes ouvertes n'ont aucun sens. Or, puisque l'action est obtenue en comparant les éléments de matrice-S de la théorie des cordes \emph{autour d'une CFT} à ceux d'une théorie des champs \emph{autour d'une solution} une action sur des champs perturbatifs de cordes ouvertes le long de cette solution n'a aucun sens non plus. 

L'action obtenue autour de $T \propto e^{x^0/\sqrt 2}$ n'est donc au moins valide qu'autour de cette solution et non universellement. D'ailleurs, elle est effectivement montr\'ee invalide autour de la solution générale. 

En appliquant ce raisonnement à l'action de Sen, puis à celles obtenues par Garousi ci-dessous, il faut donc bien comprendre qu'elles ne sont pas nécessairement universelles et comme nous le disions, sont probablement surtout valables autour de fonds indépendants du temps.  

\subsection{Action de Garousi non abélienne TDBI}
\label{sec:act_gar}

Dans le but d'étendre cette construction à des systèmes de branes non BPS plus complexes -- mais aussi ultimement à des systèmes de branes et d'antibrane -- Garousi~\cite{Garousi:2000tr,Garousi:2004rd,Garousi:2007fn} propose simplement de non-abélianiser cette action, en transformant les champs en matrice d'un groupe de jauge de type $U(N)$ et les dérivées en dérivées covariantes sur ce même groupe de jauge. Il propose aussi d'introduire une trace symétrique sur le groupe de jauge $\Str$ dont il montre qu'elle se démarque sensiblement de la trace simple, pour reproduire les éléments de matrice-S, calculés en théorie des cordes. Cette trace consiste à symétriser son argument dans le groupe de jauge avant d'être appliquée. Pour deux branes non-BPS de dimension maximale en type IIA, nous aurions donc~:

\begin{align}
S = - T_9 \int \di^{10}\sigma ~ \Str V(T) \, \sqrt{\det\parent{G_{ab} + B_{ab} + 2\pi \alpha' F_{ab} + D_a T D_b T}}
\end{align}

avec tous les champs développés en secteurs le long du groupe $U(2)$ donc sur des matrices -- de Pauli $\sigma^{0,1,2,3}$ avec $\sigma^0=id$ en l'occurrence, voir figure~\refe{fig:secteur} -- et nous avons inclus la dérivée covariante ainsi que le tenseur de Maxwell non abélien~:

\begin{align}
F_{ab} &= \partial_a A_b - \partial_b A_a - i [A_a,A_b] \nonumber \\
D_a T &= \partial_a T - i [A_a,T]
\end{align}

\begin{wrapfigure}{l}{.5\linewidth}
\centering
\includegraphics[scale=.5]{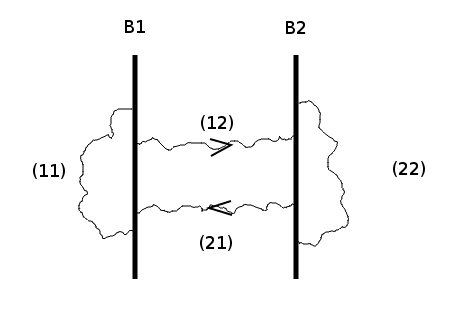}
\caption{\label{fig:secteur} \footnotesize{Les secteurs $(11)$ et $(22)$ s'organisent en facteurs de CP $\sigma^0$ et $\sigma^3$ tandis que les secteurs interbranaires $(12)$ et $(21)$ s'organisent en $\sigma^1$ et $\sigma^2$.}}
\end{wrapfigure}

Les actions des branes de dimensions inférieures sont obtenues par T-dualité le long des dimensions que l'on a décidé d'extraire du volume des branes. La construction est similaire à celle donnée par Myers dans~\cite{Myers:1999ps}. L'exactitude de cette méthode est discutable, en particulier en mettant en balance les contributions des fermions. Puisque l'action de Sen est au moins valide pour des tachyons de genre espace, cette contrainte doit s'appliquer également ici.\\

L'expression générale de l'action obtenue par T-dualité pour le système de plusieurs branes non-BPS localisées spatialement est compliquée et n'illuminera pas la discussion. On pourra la retrouver ici~\cite{Garousi:2004rd}. Une action à l'ordre des arbres pour des branes très écartées les unes des autres, typiquement pour $\ell \gg \alpha'$ n'a plus de sens à cause de la contribution du diagramme à une boucle dans le potentiel du champ de distance~ ; donc son domaine d'application concerne les courtes distances en plus des tachyons de genre espace.  \\

L'action du système brane-antibrane est ensuite obtenue~\cite{Garousi:2004rd} par application d'une projection sur les facteurs de Chan-Paton $\sigma^1$ et $\sigma^2$. En effet, le tachyon dans le système de deux branes non-BPS existe dans les 4 secteurs $U(2)$ ce qui n'est pas le cas du tachyon dans le système brane-antibrane où il n'existe que dans les secteurs $\sigma^{1,2}$. Or les opérateurs de vertex correspondant à ces tachyons étant identiques dans les deux systèmes, on peut effectivement s'attendre à ce que la physique soit rigoureusement identique à ce niveau, donc aussi l'action effective pour ces champs. Ce qui justifie une simple projection. Dans le même temps, à cause du mécanisme de higgs $U(2) \to U(1)\times U(1)$ provoqué par la séparation spatiale, seuls les secteurs $\sigma^{0,3}$ des champs de jauge et des scalaires transverses sont non-massifs. Ainsi, en annulant les secteurs non pertinents dans l'action non-abélienne de deux branes non-BPS nous devrions obtenir l'action non-abélienne d'une brane et d'une anti-brane (séparées). L'action obtenue est de la forme~\cite{Garousi:2004rd}~:

\begin{align}
S_{D\bar D} = - \int \di^{p+1} \sigma~ \parent{{\mathcal V}^{(1)}(\module{\tau},\ell)e^{-\Phi(X^{(1)})}\sqrt{-\det {\bf A}^{(1)}} + {\mathcal V}^{(2)}(\module{\tau},\ell)e^{-\Phi(X^{(2)})}\sqrt{-\det {\bf A}^{(2)}}}
\end{align}

avec $\ell^i = X^{(2) \, i}-X^{(1) \, i}$. Cette quantité a le sens de distance si la métrique est constante. Le champ $\tau$ est le tachyon complexe du système $D-\bar D$. L'indice $(\alpha)$ réfère à chaque brane. Etonnamment, les deux actions sont séparées et non regroupées en une seule. Le potentiel est explicitement :

\begin{align}
{\mathcal V}^{(\alpha)}(\module{\tau},\ell) &= \frac{\sqrt{\det Q^{(\alpha)}}}{\cosh \frac{\module{\tau}}{\sqrt{2}}} \nonumber \\ 
\det Q^{(\alpha)} &= 1+\frac{\module{\tau}^2}{4\pi^2\alpha'} \, \ell^i \ell^j \, g_{ij}(X^{(\alpha)})
\end{align}

avec $g_{ij}$ la métrique transverse aux branes. Supposons maintenant que $\ell\sim \sqrt{\alpha'}$. Alors dans l'ensemble de l'espace transverse qui sépare les branes, le fond  -- métrique et dilaton par exemple – peut \^etre suppos\'e constant dans les directions \emph{transverses} et ne plus dépendre que des coordonnées longitudinales avec $a=0\ldots p$. 

La matrice ${\bf A}^{(\alpha)}$ a une expression compliquée, que l'on retrouvera dans~\cite{Garousi:2004rd} mais donnons cependant son expression dans un fond trivial $g_{\mu\nu}=\eta_{\mu\nu}$, $B_{\mu\nu}=0$ et $\Phi=0$~:

\begin{align}
{\bf A}^{(\alpha)}_{ab} = \eta_{ab} + \partial_a X^{(\alpha)}_i \partial_b X^{(\alpha)}_j \, \eta^{ij} + 2\pi \alpha' F_{ab}^{(\alpha)} + \alpha' \, \frac{D_{a}\tau D_{b}\tau^* + D_{b}\tau D_{a}\tau^*}{2\det Q^{(\alpha)}} 
\end{align}

L'action se simplifie significativement le long de champs homogènes, \cad ne dépendant que du temps. C'est une situation à regarder attentivement, car plus aisée à étudier et directement reliée à la condensation homogène du tachyon du roulant qui nous intéresse spécifiquement. Notons que cette situation est exactement celle d'un système $D0-\bar D 0$ en type IIA. Si en outre les champs de jauge sont suppos\'es gelés et que les branes sont exactement parallèles alors nous aurons simplement~:

\begin{align}\label{eq:garousi_base}
S_{D\bar D} = - 2 \int \di^{p+1} \sigma~ {\mathcal V}(\module{\tau},\ell) \sqrt{1 - \frac{1}{1+\frac{\module{\tau}^2 \ell^2}{4\pi^2 \alpha'}} \parent{\frac{\dot \ell^2}{4} + \alpha' \module{\dot \tau}^2}}
\end{align}

Cette action sera la base de l'étude qui va suivre. Nous allons montrer qu'il n'existe pas de solution de tachyon roulant à distance constante et nous étudierons la résolution numérique de ses équations du mouvement. Avant cela étudions rapidement son développement à l'ordre quadratique~:

\begin{align}\label{eq:action_gar_quad}
S_{D\bar D} \sim - 2 \int \di^{p+1} \sigma~ \parent{1 - \frac{\dot\ell^2}{8} - \frac{\module{\dot \tau}^2}{2} + \frac{\module{\tau}^2 }{2}\parent{\frac{\ell^2}{4\pi^2 \alpha'} - \frac{1}{2}}}
\end{align}

A cet ordre l'action est donc trivialement celle d'un champ $\ell$ non massif et d'un champ complexe $\tau$ de masse $\alpha' m^2 = \ell^2/4\pi^2 \alpha' - 1/2$ ce qui \'etait attendu d'après l'étude du spectre perturbatif des cordes ouvertes autour du vide tachyonique $T=0$. La valeur de distance critique sera not\'e par la suite $\ell_c = \pi\sqrt{2\alpha'}$ et est telle que le tachyon est non-massif.

\section{R\'esolution num\'erique et r\'esolution exacte}
\label{sec:res}

Nous verrons dans cette section comment il est possible de résoudre numériquement l'action de Garousi. Dans la section~\ref{sec:eom_res} nous montrerons qu'elle décrit un système très chaotique ne rentrant pas significativement dans une phase de condensation. Dans la section~\refcc{sec:anal_conf} nous décrirons la solution exacte de condensation, nomm\'ee \emph{tachyon roulant} et qui constitue le sujet de cette thèse. Cette solution ne s'inscrit pas dans la théorie des champs proposée par Garousi. Nous finirons dans la section~\refcc{sec:plan} par décrire le plan de thèse et son objectif.

\subsection{Equations du mouvement de l'action de Garousi et résolution numérique}
\label{sec:eom_res}

Nous pouvons réécrire l'action~\refe{eq:garousi_base} en réinjectant le terme potentiel à l'intérieur de la racine sous une forme plus condensée. Nous utiliserons $\alpha'=1$~:

\begin{align}
{\mathcal L}_{gar} = \frac{1}{\cosh \frac{\module{T}}{\sqrt 2}} \sqrt{1 + \frac{\module{\tau}^2 \ell^2}{4\pi^2} - \frac{\dot \ell^2}{4} - \module{\dot \tau}^2}
\end{align}

Il est plus commode de passer dans le formalisme hamiltonien pour dériver les équations du mouvement et les résoudre numériquement. Nous notons $V(T)=1/\cosh T/\sqrt 2$ le potentiel tachyonique et $E$ l'énergie que nous définirons plus bas. On voit facilement dans le lagrangien ci-dessus que la phase du tachyon complexe n'a pas de potentiel puisque ce dernier ne dépend que du module. Par conséquent nous pouvons regarder des solutions à phase fixée\footnote{Laisser la phase libre peut être intéressant pour étudier la formation de vortex, mais dans notre cas nous cherchons simplement à étudier la condensation en module.}~; ce que nous ferons en notant $\module\tau=T$. Les équations de Hamilton sont~:

\eqali{
\dot T &= \frac{\Pi_{T}}{E} \parent{1+ \frac{T^2 \ell^2}{4\pi^2}} \nonumber \\
\dot \ell &= 4\frac{\Pi_{\ell}}{E} \parent{1+\frac{T^2 \ell^2}{4\pi^2}} \nonumber \\
\dot \Pi_{T} &= - \,  \frac{T \ell^2}{4\pi^2} \, \frac{E}{1+\frac{T^2 \ell^2}{4\pi^2}} + V^2 \frac{\tanh \frac{T}{\sqrt 2}}{E \sqrt 2 } \parent{1+\frac{T^2 \ell^2}{4\pi^2}} \nonumber \\
\dot \Pi_{\ell} &= -  \,  \frac{T^2 \ell}{4\pi^2}\,\frac{ E }{1+\frac{T^2 \ell^2}{4\pi^2}}
}

avec l'énergie

\begin{align}
E = \sqrt{\Pi_T^2 + 4 \Pi_{\ell}^2 + V^2}\sqrt{1+\frac{T^2 \ell^2}{4\pi^2}}
\end{align}

Tant que nous ne couplons pas le système à la gravité, l'énergie est conservée donc nous aurons $E$ constant. Nous voyons très facilement qu'en imposant $\dot \ell =0$ nous devons avoir $\Pi_{\ell}=0$. Or d'après la dernière ligne cela implique immédiatement $E=0$ (ou $\ell=0$ mais ce dernier cas n'est pas intéressant) si on suppose que nous avons bien un tachyon roulant, \cad non nul. Du coup ça n'est pas possible car il faut $E\neq 0$.

On en déduit qu'il n'existe pas de solution de tachyon roulant à distance constante dans l'action de Garousi. Nous montrons cependant dans la section~\refcc{sec:tach_roul_CFT} que cette solution existe pour tout $\ell<\ell_c$ ce qui s'oppose donc à l'expression de cette action et soulève un problème important, à mettre en balance avec l'argumentation précédente sur le domaine de validité de l'action de Garousi.  \\

La résolution numérique n'est pas triviale pour ce système car il est fortement couplé et non-linéaire. Nous avons néanmoins testé avec \textsc{Mathematica} de nombreuses configurations initiales, en particulier autour de $\dot T \sim 0$ et $T\sim 0$ avec $\dot \ell =0$ et $\ell \sim \ell_c$. Le résultat générique de ces résolutions est qu'à énergie conservée\footnote{En couplant à la gravité on s'attend à ce que résultat change puisqu'alors l'énergie n'est pas conservée et on doit au contraire obtenir qu'un flux d'énergie est extrait du système sous forme de graviton, suivant la discussion de Lambert \emph{et al.}~\cite{Lambert:2003zr}. C'est d'ailleurs bien ce qu'on observe en rajoutant ce degré de liberté dans les résolutions numériques mais nous n'en discuterons pas ici.} le système initie une condensation puis souvent décondense jusqu'à devenir très instable, comme on peut le voir sur la figure~\refe{fig:simu1}~; plusieurs de ces phases peuvent se produire d'affilée. Lors de la condensation, les branes oscillent autour de $\ell=0$ mais les oscillations ne sont pas sensiblement amorties au cours du temps\footnote{En ajoutant la gravité, la friction amortie significativement ces oscillations et le système semble tout à fait tendre vers une situation d'équilibre à distance nulle.} et le système ne semble donc pas atteindre un équilibre stable. 

\begin{figure}[h]\centering
\includegraphics[scale=.8]{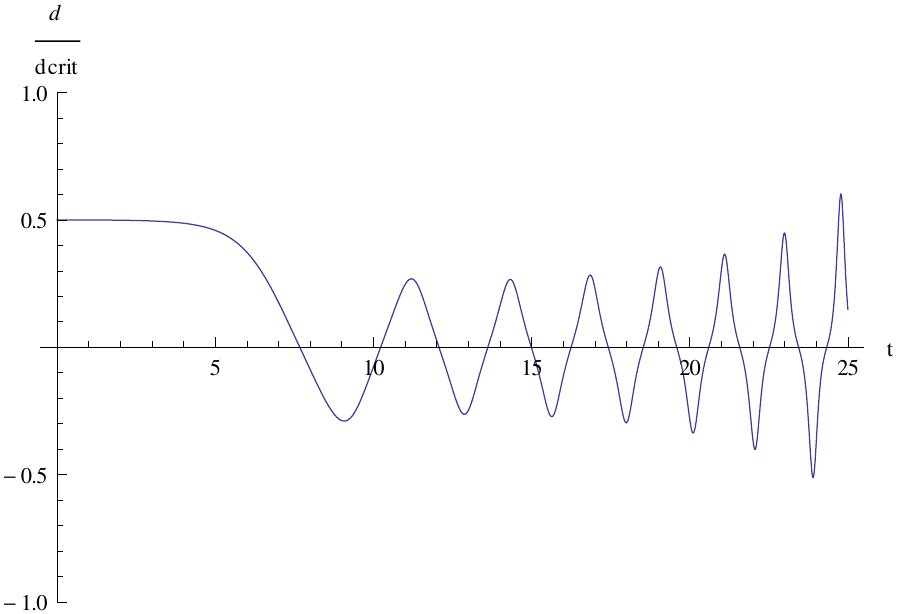} 
\includegraphics[scale=.8]{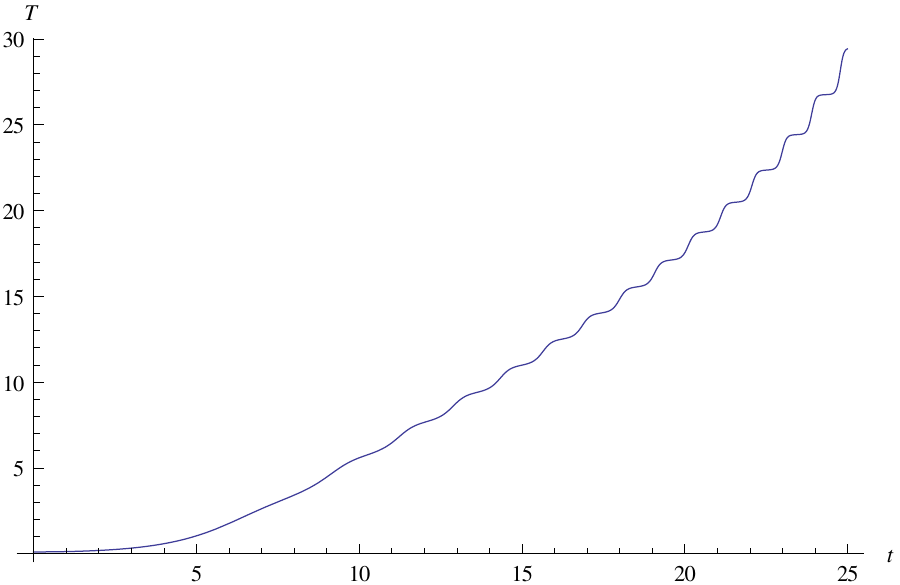}
\caption{\label{fig:simu1} \footnotesize{Résolution numérique typique des équations hamiltoniennes avec des conditions initiales $T(0)=0.1$, $\ell(0)=\ell_c/2$, $\Pi_T =0$ et $\Pi_\ell(0)=0$. Donc pour des valeurs initiales faibles du tachyon et de sa dérivée $\dot T(0)=0$. A gauche l'évolution temporelle du champ de distance et à droite celle du tachyon. Nous notons clairement une phase de condensation avec la distance tendant vers $\ell=0$ mais oscillant avec une amplification progressive amenant finalement à une instabilité du système. Par la suite la distance explose et le tachyon décondense, comme dans la figure~\refe{fig:simu2}.}} 
\end{figure}

\begin{figure}[h!]\centering
\includegraphics[scale=.8]{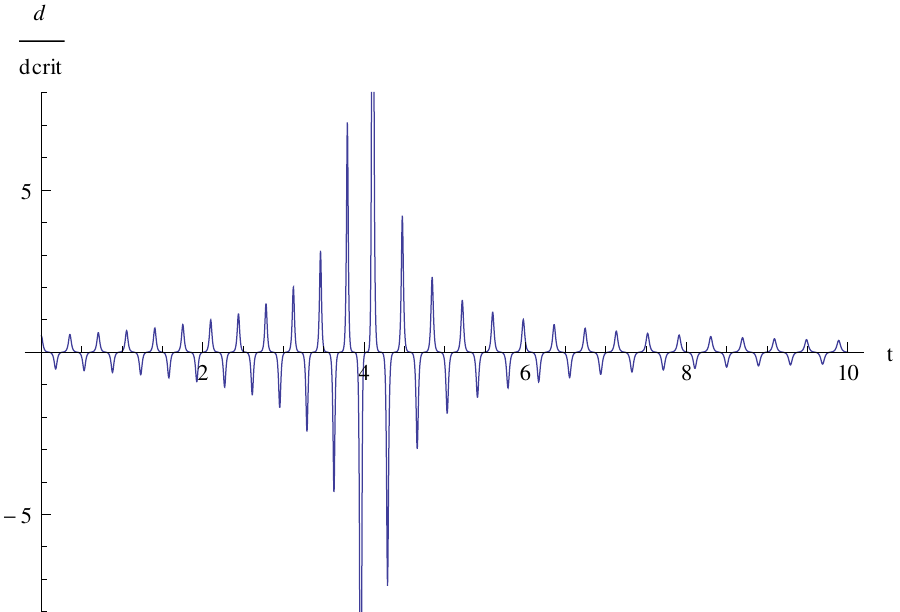} 
\includegraphics[scale=.8]{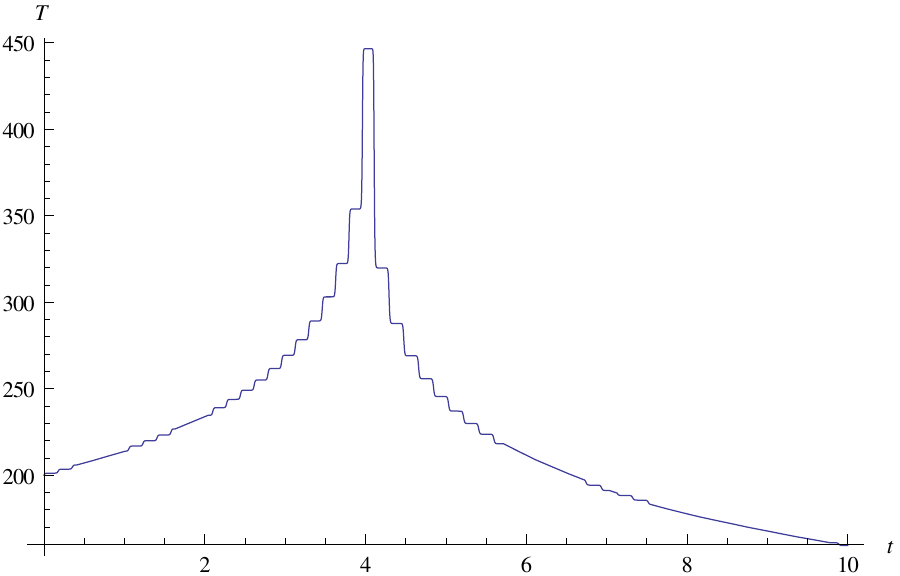}
\caption{\label{fig:simu2} \footnotesize{Résolution numérique typique des équations hamiltoniennes avec des conditions initiales $T(0)=200$, $\ell(0)=\ell_c/2$, $\Pi_T(0) =1$ et $\Pi_\ell(0)=0$. Donc pour des valeurs grandes du tachyon et de sa dérivée $\dot T(0) \sim \ell(0)T(0)/2\pi \gg 1$. A gauche l'évolution temporelle du champ de distance et à droite celle du tachyon. Nous notons clairement une phase de condensation puis de décondensation. La distance dépasse nettement sa valeur critique, donc assez vite le système est en dehors du champ de validité de l'action.}} 
\end{figure}

Il parait clair que les échanges énergétiques entre le tachyon et le champ de distance ne font pas dans un sens plus que dans l'autre si bien que le système ne stabilise pas dans une solution de tachyon roulant à séparation nulle mais oscille bien d'une phase de condensation à une phase de décondensation. Il ne parait pas évident qu'il puisse exister des conditions initiales qui améliorent ce comportement. On peut par exemple tester à des valeurs de distance initiale plus faibles ou à des valeurs de $\dot T \gg 1$, mais cela ne donne pas de comportement plus contrôlé, comme on peut le voir sur la figure~\refe{fig:simu2}. \\ 

Le fait que la résolution numérique ne propose pas de mode de condensation clair en espace plat alors qu'il s'agit d'un comportement attendu et fortement souhaité nous a poussé à étudier le système d'une façon plus analytique. Bien que nous y perdions l'aspect off-shell, nous avons opté pour mener l'étude dans le cadre des théories conformes sur la surface de corde, où la condensation de tachyon est généralement bien décrite par exemple dans le cas brane-antibrane coïncidente ou celui de la brane non BPS. En outre, il y a le fameux exemple de Kutasov et Niarchos démontrant qu'on peut quand même -- au moins dans certains cas particuliers -- déterminer une action effective off-shell à partir de résultats on-shell, avec l'avantage de se placer directement dans un fond condensant.  

\subsection{Etude de la théorie conforme de bord}
\label{sec:anal_conf}

Dans le syst\`eme brane-antibrane séparé, Bagchi et Sen~\cite{Bagchi:2008et} montraient que le tachyon roulant à distance constante était une CFT donc une solution des équations du mouvement, mais uniquement pour $\ell<\ell_c/\sqrt 2$. Le modèle sigma du tachyon roulant entre une brane et une antibrane parallèles et séparées d'une distance $r$ le long d'une direction que nous nommerons $X$ consiste en une déformation de l'action de surface -- sur le disque ou le demi plan complexe -- par un terme inséré sur le bord selon $S = S_{bulk} + \delta S$ avec~:

\begin{align}
S_{bulk} & = \frac{1}{2\pi \alpha'} \int_{\Sigma} \di^2 z ~ \parent{\partial X^\mu \bar \partial X_\mu + \frac{\alpha'}{2} \psi^\mu \bar \partial \psi_\mu + \frac{\alpha'}{2} \widetilde \psi^\mu \bar \partial \widetilde\psi_\mu} \nonumber \\
\delta S &= \parent{\begin{array}{cc}0 & 1 \\ 0 & 0 \end{array}} \otimes\oint_{\partial\Sigma} \lambda^+ \, \psi^+ e^{ir \wt X + \omega X^0} +  \parent{\begin{array}{cc}0 & 0 \\ 1 & 0 \end{array}} \otimes \oint_{\partial\Sigma} \lambda^- \, \psi^- e^{-ir \wt X + \omega X^0} 
\end{align}

avec $\psi^\pm = \pm ir \wt \psi + \omega \psi^0$ et $\lambda^\pm \in \mathbb C$. Le champ $\wt X$ est le T-dual Neumann de $X$ la direction Dirichlet transverse. Il s'agit d'une CFT $c=2$ par découplage -- dans les OPE -- des autres champs fondamentaux $X^{a,i}$ longitudinaux $(a)$ et transverses $(i)$. Au premier ordre la fonction bêta des tachyons est~:

\begin{align}
\beta_\pm = (\frac{1}{2}-r^2-\omega^2)\lambda^\pm
\end{align}

qui impose donc $\Delta^\pm = \omega^2+r^2 = 1$. En étudiant les OPE à N-points des tachyons, on trouve~\cite{Bagchi:2008et} qu'au-delà d'une certaine valeur de distance ($\ell \geq \ell_c/\sqrt 2$) des opérateurs marginaux, \cad de dimension $\Delta=1$ peuvent être produits. C'est ce qu'on appelle des \emph{résonances}. Les coefficients typiquement associés aux résonances divergent logarithmiquement dans l'échelle ultra-violette et brisent l'invariance conforme de la théorie. Ce point est assez gênant puisqu'il implique que le tachyon roulant serait non-marginal\footnote{La marginalité implique en général que la théorie est conforme, mais pas toujours -- voir section~\refcc{sec:mod_sig}.} dans le domaine de distance $\ell_c/\sqrt 2<\ell < \ell_c$ et marginal pour $\ell<\ell_c/\sqrt 2$. 

Toutefois, il nous a paru peu probable que le tachyon roulant ne fut pas une CFT dans l'intégralité du domaine sous-critique puisque rien ne semble pouvoir expliquer \emph{physiquement} un tel comportement du système. Physiquement la seule distance critique est $\ell_c$. Dans le système bosonique analogue, comme nous le verrons, ces mêmes résonances apparaissent, sauf que dans ce contexte, elles sont physiquement identifiables à un couplage du tachyon interbranaire $\sigma^{1,2}$ aux tachyons des secteurs $\sigma^{0,3}$. Evidemment dans le système brane-antibrane ces derniers sont rigoureusement absents par projection GSO. \\

En suivant la méthode de Gaberdiel \emph{et al.}~\cite{Gaberdiel:2008fn} que nous présentons dans la section~\refcc{sec:mod_sig}, nous avons montré~\cite{Israel:2011ut} que le tachyon roulant à distance constante était bien une CFT dans toute la phase tachyonique. Compte-tenu de ce que nous disions, il s'agissait donc de prouver que le tachyon roulant était exactement marginal et qu'en l’occurrence il ne produisait finalement pas de divergences logarithmiques dans la fonction de partition sur le disque.

Nous avons calculé les expressions exactes des OPE à l'ordre 2 et 4 dans le tachyon pour des valeurs de distance supérieures à $\ell_c/\sqrt 2$. Puis nous avons extrait analytiquement l'ensemble des divergences produites par intégration en appliquant une régularisation dite de \emph{point-splitting}\footnote{Elle consiste à tronquer la limite UV de toutes les fonctions de Green.}. Nous avons finalement pu constater que toutes les divergences logarithmiques s'annulaient ensemble comme attendu. Ce "miracle" est en outre clairement identifié comme une conséquence de la supersymétrie de surface. Pour cette raison et parce qu'il est trop ardu de calculer les OPE des tachyons aux ordres supérieurs\footnote{Mais nous avons pu vérifier numériquement à l'ordre 6} nous en avons déduit que le tachyon roulant était exactement marginal sur l'ensemble du domaine tachyonique.

\section{Fonction de partition, groupe de renormalisation et action effective quadratique}
\label{sec:fonction_partition_intro}

La preuve de la marginalité exacte du tachyon roulant, nous a permis de prolonger analytiquement le calcul de la fonction de partition à toute distance $\ell<\ell_c$. Malheureusement, nous n'avons pas pu calculer la fonction de partition analytiquement à tout ordre supérieur à 2. La raison étant que l'intégrale est multiple, ordonnée et engage un intégrande composé de nombreuses fonctions sinusoïdales où les variables d'intégrations sont fortement couplées, ressemblant fortement à celle d'un gaz de coulomb sur un cercle~\cite{Jokela:2007yc}, mais suffisamment différent pour être incalculable, même numériquement avec \textsc{Mathematica}. 

Cependant, nous avons pu calculer exactement l'ordre quadratique mais aussi les 5 premiers ordres à la distance $\ell_c/\sqrt 2$ et ce en développant une méthode diagrammatique de réduction des intégrales en super-espace. En exprimant la théorie du tachyon roulant directement en super-espace, nous faisons apparaître un terme de contact. Un traitement convenable de ce terme nous a montré qu'il jouait le rôle de contreterme -- un peu comme dans~\cite{Green:1987qu} -- pour supprimer la plupart des divergences de puissance dans les amplitudes\footnote{Nous avons cependant constaté qu'il n'était pas suffisant pour supprimer toutes les divergences de puissance et nous en avons déduit qu'il fallait ajouter des termes de contact d'ordre supérieur.}. En outre, ce terme de contact assure la continuité de la fonction de partition, et probablement des amplitudes en général, au moins en $\ell_c/\sqrt 2$ car en cette distance il n'est pas divergent mais fini. L'intérêt de la méthode diagrammatique était justement de traiter convenablement et avec facilité les contributions de ce terme de contact, directement en super-espace. La première motivation de ce calcul était de reconnaître dans les premiers termes du développement d'une fonction connue et dont nous pourrions déterminer une expression analytique continuable sur les autres valeurs de distance. Malheureusement, nous verrons qu'aucune fonction particulière ne peut être devinée à partir de cette expression.  \\

Toutefois, à l'aide du calcul à l'ordre quadratique et de la méthode de Kutasov et Niarchos, nous avons pu déterminer une expression exacte de l'action effective quadratique. En suivant leur raisonnement, cette action s'avère au moins valide le long du tachyon roulant~: 

\begin{align}\label{eq:act_kut_roul}
S = 2 T_p \int \di^{p+1}\sigma \, \Bigg[1 + \frac{1}{2}\partial_a r \partial^a r + \frac{1}{2 \sqrt{1-2r^2}} \partial_a T\partial^a T^*   - \frac{1}{4} \sqrt{1-2r^2} \module{T}^2 + \ldots \Bigg] 
\end{align}

En outre, par l'étude du modèle sigma des champs de tachyon et de distance le long de la th\'eorie conforme du tachyon roulant \`a distance constante, nous avons pu exprimer des fonctions b\^eta universelles à l'ordre quadratique dans les champs, \cad indépendantes du schéma de renormalisation. Nous les avons ainsi identifié, conformément à la discussion de la section~\refcc{sec:mod_sig} aux équations de mouvement quadratique de ces champs.

Par comparaison avec celles dérivées de l'action~\refe{eq:act_kut_roul} nous avons obtenu un bon accord, à une redéfinition des champs et des fonctions b\^eta près. Ceci indique au moins \`a cet ordre que l'action~\refe{eq:act_kut_roul} est compatible avec la physique interne des cordes interbranaires du syst\`eme brane-antibrane. Le calcul des ordres supérieurs est évidemment souhaité, mais il faudra pour cela développer de nouvelles méthodes, car toutes celles à disposition semblent inadéquates.  \\ 

Nous verrons cependant en conclusion qu'il existe un modèle intégrable correspondant au modèle de tachyon interbranaire \emph{constant}. Il s'agit du modèle Kondo~\cite{Fendley:1995kj,LeClair:1997sd} qui est en fait fortement relié au modèle de sine-Gordon de bord. Nous serons \'egalement en mesure de conjecturer la nature du vide de condensation du tachyon interbranaire \`a distance constante. Nous identifions ce vide \`a de la mati\`ere branaire remplissant l'espace d\'elimit\'e par la brane et l'antibrane.

\section{Plan de th\`ese}
\label{sec:plan}

	Dans cette partie introductive, la théorie des cordes et les connexions aux théories conformes seront présentées ainsi que les différents outils et concepts utilisés dans cette thèse. Les objets -- cordes fermées, ouvertes, branes -- de la théories de cordes et des supercordes seront présentés dans le chapitre~\refcc{chap:TDC_CFT} ainsi que les théories conformes et superconformes puis les théories conformes de bord (BCFT). Ces dernières seront reliées à la définition des états de bords, ce qui permettra d’introduire les branes et anti-branes, ainsi que les systèmes compos\'es \`a partir de ces objets, particuli\`erement le syst\`eme brane-antibrane. Le concept d'actions effectives du modèle sigma sera discuté dans le chapitre~\refcc{chap:TEC_sigma}, et les calculs du groupe de renormalisation abondamment utilisés au cours de cette thèse y seront également abordés. Enfin, dans le chapitre~\refcc{chap:cond_tach}, le problème de la condensation de tachyon en théorie des cordes sera pr\'esent\'e en détail et des exemples concrets de résolution seront proposés.

Les calculs concernant les objectifs de la thèse seront développés dans la partie~\refcc{part:tach_roul} ; et le modèle bosonique compos\'e de 2 branes parallèles et séparées analysé dans le chapitre~\refcc{chap:cond_bos}. L'\'etude sera ax\'ee sur le problème de la marginalité exacte du tachyon roulant interbranaire. Nous pr\'esenterons en dernier lieu l'analyse off-shell du mod\`ele sigma développé autour de la solution de tachyon roulant. En particulier, nous discuterons de la relation des fonctions b\^eta du groupe de renormalisation aux \'equations du mouvement des th\'eories des champs effectives. 

Dans le chapitre~\refcc{chap:tach_cond_susy}, le système compos\'e d'une brane et d'une antibrane parallèles et séparées sera analysé. Nous commencerons par montrer la marginalité exacte du tachyon roulant interbranaire \`a distance constante qui est le r\'esultat principal de cette th\`ese. Puis, comme dans le cas bosonique, le groupe de renormalisation du mod\`ele sigma off-shell autour de la solution roulante sera étudié. La fonction de partition en $r=r_c/\sqrt 2$ sera calculée et exprimée pour tout $r$. Enfin, l’action effective quadratique sera déterminée par la méthode de Kutasov et Niarchos, puis comparée aux équations de mouvement obtenues dans le cadre du groupe de renormalisation, puis aux actions effectives obtenues dans le passé par Sen dans le cas coïncident et proposées en dehors de la coïncidence par Garousi. 

Pour conclure la thèse (~\refcc{part:conclu}), les résultats seront rassemblés puis nous \'elargirons le champ de vision aux perspectives de continuation de ce travail de th\`ese. En particulier, nous discuterons de la possible identification du vide de condensation \apriori atteint asymptotiquement par le tachyon roulant, en \'etudiant un tachyon interbranaire constant dans un espace compact et en utilisant une T-dualit\'e. Nous discuterons \'egalement de la relation du modèle sigma du tachyon interbranaire constant et \`a distance constante avec le modèle Kondo et nous verrons en quoi cela peut mener \`a des r\'esultat int\'eressants. 

En~\hyperref[part:annexe]{Annexe}, l'article \emph{Rolling tachyon for separated brane-antibrane systems} publié dans \emph{Physical Review D} a été ajouté.

\chapter{G\'en\'eralit\'es~: Th\'eorie des cordes et th\'eories des champs conformes de surface}
\label{chap:TDC_CFT}

Dans ce chapitre, nous allons introduire en détail la théorie des cordes fermées bosoniques puis celle des supercordes et enfin nous présenterons les concepts liés aux théories de cordes ouvertes. Nous commen\c cerons par discuter dans la section~\refcc{sec:pres_trans} des transformations conformes et de leur relation à la théorie des cordes \emph{critique}. Puis dans la section~\refcc{sec:CFT_bos} nous présenterons la théorie bosonique par la définition de l'action et de l'amplitude Polyakov. On introduira les concepts d'opérateurs de vertex, d'OPE et d'états. Dans la section~\refcc{sec:SCFT} nous introduirons les théories de supercordes en commen\c cant par présenter l'action de surface de corde supersymétrique. Nous discuterons des symétries, des opérateurs de vertex, des états et du superespace. Dans la section~\refcc{sec:BCFT} nous aborderons la question des surfaces avec bord et des cordes ouvertes. Nous parlerons des conditions de bord et leur relation aux branes et au concept d'état de bord. Nous finirons par présenter les branes BPS en supercordes et les systémes non BPS, tels que brane-antibrane et brane non BPS.

\section{Transformations conformes et th\'eorie des cordes critique}
\label{sec:pres_trans}

Dans le cadre de la théorie des cordes, ces transformations et la théorie conforme qui y est associée seront maintenant introduites. Les cordes se propagent dans l'espace-temps en décrivant des surfaces. Si elles sont fermées, elles décrivent des surfaces tubulaires et si elles sont ouvertes, des nappes avec bords. Les interactions entre cordes sont obtenues en collant ces tubes ou ces nappes ensemble sous des formes typiques présentées dans la figure~\refe{fig:surface}. Les cordes décrivent donc en interagissant des variétés riemanniennes à 2 dimensions, \cad des surfaces lisses avec des bords, des poignées ou d'autres formes plus alambiquées telles que le ruban de Moebius ou la bouteille de Klein – voir~\cite{Nakahara:1990th} pour des d\'efinitions pr\'ecises.

Ces surfaces sont plongées dans un \emph{espace-cible}, en l'occurrence un espace-temps à $d+1$ dimensions\footnote{Cette valeur est pour l'instant arbitraire.}. La surface est localement, \cad sur chaque carte constituant l'atlas de la variété, décrite par un jeu de coordonnées $(x,y)\in {\mathbb R}^2$. Cette variété à 2 dimension est nommée $\Sigma$. Elle est munie d'une métrique intrinsèque, éventuellement définie globalement, exprimée dans ce jeu de coordonnées. Les notations suivantes sont utilisées pour la suite : $X^\mu$,  les coordonnées de l'espace-cible avec $\mu = 0 \ldots d$, et $M$ l’espace-cible dans lequel la surface est insérée. Du point de vue de l'espace-temps, $X^\mu$ est un vecteur. Mais, du point de vue de la surface chaque coordonnée est équivalente à une fonction sur les coordonnées $(x,y)$ d'une carte, nommée \emph{fonction d'insertion} et notée $\phi$.

\begin{figure}[h]
\centering
\begin{subfigure}[b]{0.5\textwidth}
\centering
\includegraphics[scale=.5]{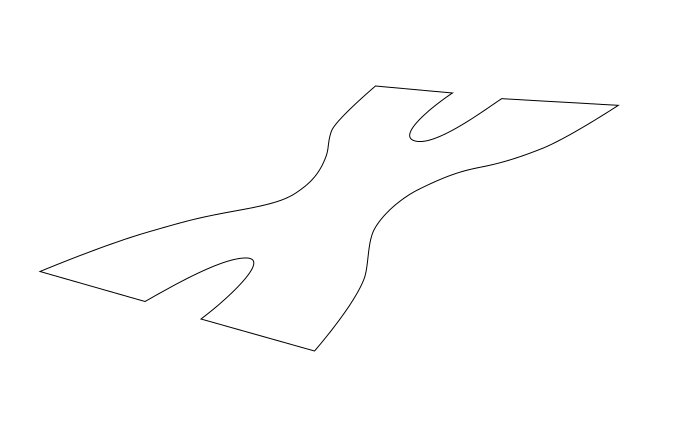}
\caption{Nappe de corde ouverte.}
\end{subfigure}
\begin{subfigure}[b]{0.4\textwidth}
\centering
\includegraphics[scale=.4]{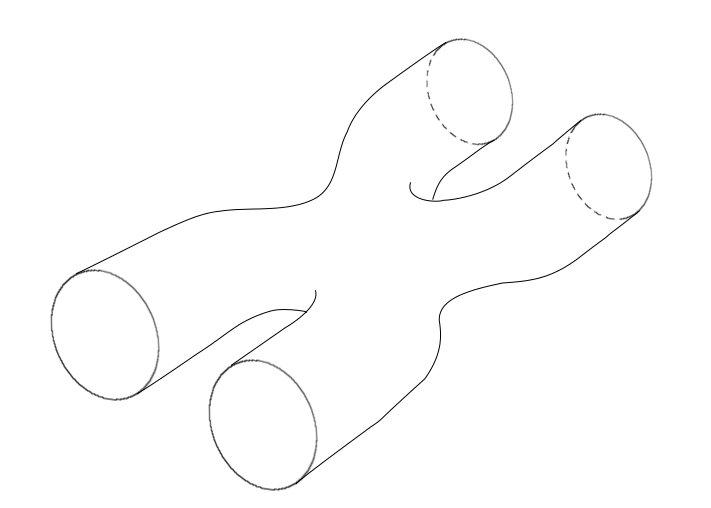}
\caption{Surface tubulaire de corde ferm\'ee.}
\end{subfigure}
\caption{\label{fig:surface} \footnotesize{Surfaces typiques décrites par des cordes en interaction. Elles représentant ici des diagrammes à l'ordre des arbres.}}
\end{figure}

\subsection{Difféomorphismes et transformations de Weyl}
Dans ce paragraphe,  $\phi$ sera défini et les difféomorphismes seront introduits. Soit $X$ une direction de l'espace-cible. L'insertion de cette surface dans l'espace-cible $M$ le long de cette direction est telle que pour tout point $p \in \Sigma$, $X(p)\in M$ est bijective et $X(p) = \phi \circ \sigma(p)$ avec $\sigma=(x,y)$ et $\phi$ une fonction bijective. Pour $\sigma \neq \sigma'$,

\eqna{\label{eq:diff}
X(p) = \phi \circ \sigma(p) = \phi' \circ \sigma'(p)
}

ce qui définit bien $\phi$ comme une fonction : $\phi'(\sigma')=\phi(\sigma)$. Par conséquent, pour tout système de coordonnées $\sigma$ et $\sigma'$ de $\Sigma$ reliés ensembles par une transformation régulière $C^\infty$, c'est-à-dire par un difféomorphisme, il existe $\phi$ et $\phi'$ tels que~\refe{eq:diff} est vérifiée. Ainsi, l'insertion est définie de façon continue tout au long du groupe des difféomorphismes de la surface $\Sigma$. Cette propriété implique qu'il n'y a pas -- qu'il ne doit pas y avoir -- de système de coordonnées privilégié sur toute surface. Au pire, il peut exister des classes de coordonnées disjointes. 

Il existe d'autres transformations de coordonnées qui ne sont pas des difféomorphismes mais qui peuvent tout de même constituer des symétries vis-à-vis de l'identité~\refe{eq:diff}. Le volume invariant d'une variété est d\'efini par l'intégrale~:

\eqna{\label{eq:volume}
V = \int_{\Sigma} \di^n x ~ \sqrt{\det g}
} 

avec $n$ la dimension et $g$ la métrique intrinsèque. Pour changer ce volume, il faut soit changer de métrique, soit changer de système de coordonnées. Puisque le volume est invariant par difféomorphisme, il est exclu que le changement de coordonnées souhaité soit lui-m\^eme un diff\'eomorphisme. Les \emph{transformations de Weyl} sont alors introduites. Il est commode de les définir par une transformation de la métrique, s'exprimant comme :

\eqna{
g_{ab} \; dx^a\otimes dx^b \to \Omega(x)^2 ~ g_{ab} \; dx^a \otimes dx^b
} 

Soit~\refe{eq:volume} donne :

\eqna{\label{eq:volume_weyl}
V \to V' = \int_{\Sigma} \di^n x ~\Omega(x)^{n} ~ \sqrt{ \det g} 
}

Par diff\'eomorphisme\footnote{Ces diff\'eomorphismes sont pr\'ecis\'ement les transformations conformes que nous introduisons plus bas.} et en conservant la forme de la m\'etrique, cette transformation est implémentée sous la forme d'un changement de système de coordonnées $x \to x'$ dont la m\'etrique est $g(x')$. Encore une fois, l'identité~\refe{eq:diff} s'appliquant, il toujours possible de trouver des fonctions d'insertion $\phi$ et $\phi'$ telles que $X(p)$ est conservée par transformation de Weyl. \\ 

Ainsi, une certaine insertion $X(\Sigma)$ peut-être conservée invariante par difféomorphismes et transformations de Weyl, quelque soit la variété. \\

Il existe une certaine catégorie de difféomorphismes, en fait ceux que nous venons d'utiliser, dites \emph{transformations conformes}, tels que pour $x=x(\widetilde x)$~:

\eqna{
g_{ab}(x) \; dx^a\otimes dx^b = \Omega(\widetilde x)^2 ~ g_{ab}(\widetilde x) \; d\widetilde x^a\otimes d\widetilde x^b
}

Or, en appliquant ensuite une transformation de Weyl, on obtient :

\eqna{
g_{ab}(x) \; dx^a\otimes dx^b \to g_{ab}(\widetilde x) \; d\widetilde x^a\otimes d\widetilde x^b
}

Donc, à une transformation de Weyl près, les transformations conformes sont des difféo\-morphismes qui conservent la métrique intrinsèque. Ceci implique qu'elles conservent aussi les angles. En fait, ce sont les transformations régulières les plus générales vérifiant cette propriété. \\

On verra que l'expression de l'action de théorie des cordes sur la surface de propagation, de par son invariance de Weyl, est directement invariante conforme ; c'est une propriété essentielle. 

\subsection{Invariance de Weyl et théorie des cordes critique}

Il vient d’être montré que les surfaces de cordes insérées dans l'espace-cible peuvent être paramétrisées librement, pourvu que la fonction d'insertion $\phi$ soit choisie convenablement. Or, cette insertion devrait être totalement indépendante de l'étalon de "volume" intrinsèque, c'est-à-dire le volume invariant~\refe{eq:volume} calculé à partir de la métrique intrinsèque, puisque physiquement la corde est définie par sa surface du point de vue de l'espace-cible. Par conséquent, la théorie décrivant la corde et définie sur la surface devrait être invariante de Weyl.

Toutefois, cette condition n'est pas indispensable\footnote{Elle n’est d'ailleurs pas impos\'ee dans des modèles équivalents en physique statistique, par exemple.}. Malgr\'e tout, en théorie des cordes, l'invariance de Weyl semble être une condition naturelle dans la mesure où il n'existe pas d'arguments physiques forts, dans l'espace-cible, pour ajouter un degré de liberté interne,  tel que cet étalon de volume intrinsèque, à la surface ; au contraire, des particules "physiques" ne pourront être introduites -- sur la couche de masse -- qu'à l'aide de cette contrainte. Quelques éclaircissements seront apportés par la suite dans la comparaison des actions de Nambu-Goto et de Polyakov et lorsque les amplitudes de Polyakov seront discutées. \\

Cette dernière condition si elle est remplie, définie la théorie des cordes \emph{critique}. L'invariance de cette théorie par difféomorphismes et transformation de Weyl en fait alors une \emph{théorie conforme}. \\

\subsection{Choix de jauge et brisures de symétrie}
L'existence de ces degrés de liberté de reparamétrisations n'est finalement qu'un artefact math\'e\-matique et non une propriété de l'objet inséré du point de vue de l'espace-cible ; ceux-ci doivent donc être fixées. Le terme de \emph{choix de jauge} est communément introduit car une telle redondance s'appelle une \emph{symétrie de jauge}.

Ces symétries peuvent apparaître \emph{localement} brisées et ce, classiquement ou quantiquement. Ceci a bien souvent des conséquences désastreuses pour la théorie -- anomalies, perte d'unitarité \etc. On cherchera donc presque toujours à réparer ces brisures, ou à empêcher leur apparition. Dans le cas de la symétrie de Weyl et en théorie des cordes \emph{critique}, ceci est reli\'e à la contrainte absolue d'invariance conforme des amplitudes. 

Les symétries de jauge pourront aussi apparaître \emph{globalement} brisées\footnote{J'entends par là que topologiquement, dans la globalité de la variété, il peut exister des classes de configurations géométriques inéquivalentes par $\text{diff}\times \text{Weyl}$.}. Cela implique en général que les propriétés intrinsèques doivent être organisées en classes d'\emph{équivalence} fixées en amont par des propriétés extrinsèques. Donc, à l'inverse des brisures locales, celles-ci ne sont pas pathologiques. En effet, ces brisures globales dépendent entre autre de la topologie de la surface, dont la nature est directement associée à des propriétés extrinsèques, physiquement pertinentes. Le tore, par exemple, possède un continuum de classes d'équivalence de métriques représentées par un nombre complexe, le \emph{module} qui correspond grossièrement à la forme du tore et à son vrillement. De manière générale, le terme \emph{module} est utilisé pour désigner un représentant de classe d'équivalence lorsque ce dernier est un paramètre continu -- par opposition à discret. En th\'eorie des cordes, la g\'eom\'etrie du tore correspond au calcul \`a une boucle des amplitudes du vide de cordes ferm\'ees. Dans ce cadre, le module est physiquement d\'eterminant.

Les transformations conformes ont été introduites ainsi que les difféomorphismes et les transformations de Weyl. L'introduction des transformations conformes en théorie des cordes va être analysée dans la suite à travers la descriptions des actions de Nambu-Goto et de Polyakov. Enfin, les amplitudes de Polyakov et les problèmes liés à l'invariance conforme seront décrits.

\section{Th\'eorie des champs conforme des cordes bosoniques}
\label{sec:CFT_bos}

Nous allons dans cette section introduire la plupart des outils de théorie des champs conforme appliqués directement à la théorie des cordes. Il s'agit donc spécifiquement d'introduire une CFT à \emph{2 dimensions}. Dans ce type de théorie, en cette dimension particulière, les transformations conformes sont générées par une infinité de générateurs. Ce sont donc des théories intégrables, \cad totalement résolubles et en particulier les calculs d'amplitudes (d'interactions entre cordes) peuvent être accomplis exactement. Nous montrerons d’abord comment la théorie des cordes peut être exprimée en termes de CFT. Puis nous analyserons les diverses symétries vérifiées par la théorie des champs sur la surface de cordes. Elles permettent de déterminer le spectre de masse des cordes ainsi que les expressions des états correspondants à chaque valeur de masse, \cad les particules décrites par les cordes. Nous finirons par montrer comment les produits d'opérateurs (OPE) contraints par la CFT permettent de calculer, ou pour le moins d'exprimer, exactement les amplitudes d'interaction. Pour plus de détail concernant les théories conformes, le lecteur pourra consulter la bible des théories conformes est l'ouvrage de Di Francesco et al.~\cite{DiFrancesco:1997nk}. Pour les aspects basiques de théorie conformes dans le contexte de la théorie des cordes, les ouvrages de Polchinski~\cite{Polchinski:1998rq} et de Kiritsis~\cite{Kiritsis:2007zz} sont des références. Des concepts plus avancées sont traités dans le second tome de Polchinski~\cite{Polchinski:1998rr} et dans la revue de Friedan, Martinec et Shenker~\cite{Friedan:1985ge}.

\subsection{Les théories conformes en théorie des cordes}

Avant tout, nous décrirons la dynamique classique des surfaces de cordes via la définition de leur action. Nous introduirons l'action de Nambu-Goto puis celle de Polyakov dont nous identifierons la théorie des champs qu'elle décrit à une CFT. Nous discuterons ensuite des amplitudes de Polyakov dans ce cadre. Nous verrons en particulier les nombreuses symétries et redondances dont il faut tenir compte. Nous introduirons alors les champs fantômes de Fadeev-Popov et les diverses contraintes vérifiées par les amplitudes en théories des cordes. Nous présenterons enfin les opérateurs de vertex, qui correspondent à des "particules" au sens d'une théorie quantique des champs, \cad des états (ou configurations) asymptotiques.

\subsubsection{Action de Polyakov}\label{sec:polyakov}
L'action des cordes, c'est-à-dire la quantité qu'une corde se propageant cherche à extrémiser, est naturellement donnée par la surface qu'elle trace lors de sa propagation dans l'espace-cible. Nous parlons ainsi de \emph{feuille d'univers}, par analogie à la particule ponctuelle traçant quant à elle une \emph{ligne d'univers}. D'après la section précédente, nous savons que l'aire invariante d'une surface insérée dans une variété plus grande est calculée à partir de la métrique induite par l'insertion $X$. Soit donc, en notant par abus de notation $X(p)=X \circ \sigma (p)$ : 

\eqna{
S_{NG}[X] \propto \int_{\Sigma} \di^2\sigma \, \sqrt{\det\parent{\partial_a X^\mu \partial_b X^\nu G_{\mu\nu}(X)}}
}

avec $G_{\mu\nu}(X)$ la métrique de l'espace-cible\footnote{Il faudrait aussi ajouter les champs anti-symétriques $B_{\mu\nu}$ et dilaton $\Phi$ mais nous reportons leur introduction à la section~\refcc{sec:mod_sig} qui traite du modèle sigma. Nous les négligerons pour l'instant.}. Cette dernière dépend ici explicitement du champ $X$ dans une démarche généraliste. La formule ci-dessus est l'expression de l'\emph{action de Nambu-Goto}, à partir de laquelle la trajectoire de la surface est extrémisée. Elle exprime ainsi une \emph{théorie des champs} pour la fonction d'insertion $X$. \\

En fait, ce n'est pas l'action la plus adéquate pour calculer des quantités physiques. Dans ce but, il est préférable d’utiliser l'\emph{action de Polyakov}~\cite{Polyakov:1981rd,Polyakov:1981re}. Celle-ci consiste à découpler la métrique induite $\gamma$ de la métrique intrinsèque $g$, de telle sorte que par extrémisation de cette action nous identifions $g=\gamma$ puis déduisons $S_{p}[\gamma,X]=S_{NG}[X]$. Ainsi, l'action de Polyakov décrit une théorie des champs à la fois pour la métrique intrinsèque et pour la fonction d'insertion de la feuille d'univers. Son expression est donnée par~: 

\eqna{\label{eq:act_pol}
S_p[g,X] = \frac{1}{4 \pi \alpha'} \int_{\Sigma} \di^2\sigma \sqrt{g} \, g^{ab} \partial_a X^\mu \partial_b X^\nu G_{\mu\nu}(X)
}

Le facteur de Regge $\alpha'= \ell_s^2$ avec $\ell_s$ la longueur de corde, est introduit de sorte que l'action soit sans dimension. En effet, $[X]=[\ell_s]$ et $[G,g]=1$. Nous distinguons dans cette action un terme cinétique pour le champ $X$ et un couplage à la métrique intrinsèque, donc à des degrés de libertés auxiliaires -- puisqu'ils n'ont pas de terme cinétique -- internes à la surface. \emph{Classiquement}\footnote{Ce n'est pas toujours évident au niveau quantique, \cad dans les calculs d'amplitudes. La métrique de fond $G_{\mu\nu}$ dans sa dépendance dans le champ $X$, doit toujours définir une théorie invariante conforme, tant au niveau classique qu'au niveau quantique.} cette action est invariante par difféomorphismes et par transformations de Weyl -- en supposant que $G_{\mu\nu}$ est elle-même invariante.

Il est important de constater que l'invariance de Weyl de $\sqrt g \; g^{ab}$ permet de déduire que la théorie décrite par cette action est effectivement conformalement invariante classiquement. A titre de contre-exemple, d'après la formule~\refe{eq:volume_weyl}, sur une variété de dimension supérieure à 2, le facteur $\sqrt g \; g^{ab}$ n'est pas invariant de Weyl et par conséquent la théorie immédiatement non-conforme. \\

L'action de Polyakov~\refe{eq:act_pol} décrit une \emph{théorie des champs conforme}, notée dans la suite CFT. \\

\subsubsection{Amplitudes de Polyakov}
A partir de cette action, nous définissons l'\emph{amplitude de Polyakov} : l'intégrale de chemin sur les champs $X$ et $g$. Cette intégrale est sommée naturellement sur toutes les géométries de surface, compactes, non équivalentes, reliant un certain nombre d'états asymptotiques de cordes. Ces états sont représentés par des \emph{opérateurs de vertex}. Ce sont des fonctionnelles de champs de feuille d'univers et de leurs dérivées $V[X,\partial^{n}X]$, à chacune desquelles est associé un champ d'espace-cible. Pour être bien précis, il faut introduire les modes d'oscillations des cordes, ce que nous ferons dans la section~\refcc{sub-sec:decomp_modes}. Un ensemble de modes d'oscillation -- un accord en quelque sorte -- est produit par une source "couplée" à la corde. Cette source peut être un champ d'espace-cible ou bien, plus complexe, une brane\footnote{Voir section~\refcc{sec:BCFT}.}. Il existe donc un certain nombre "d’extrémités" à la feuille d'univers où un ensemble de modes est sourcé par des champs ou des branes.

Ainsi, l'amplitude de Polyakov calcule une fonction de corrélation entre des états asymptotiques, chacun correspondant à un champ d'espace-temps. Le résultat est exprimé dans l'espace de Fourier des impulsions, donc une amplitude de Polyakov est en général un élément de matrice-S. Il faut noter que les états sont nécessairement connectés entre eux par une feuille d'univers, ce qui implique la préexistence d'une corde ; autrement dit, il n'y a pas de processus de création de corde. Nous utilisons donc, du point de vue de l'espace-cible, un formalisme de \emph{première quantification}. \\

Compte-tenu de ce qui a été précisé plus haut, pour un ensemble d'opérateurs de vertex $V_{\alpha}(k_{\alpha}^\mu)$, l'amplitude à N-points sur une feuille d'univers $\Sigma$ compacte, de nombre d'Euler $\chi$, doit s'écrire : 

\eqna{\label{eq:ampl_pol}
\corr{\prod_{\alpha=1}^N V_{\alpha} (k_{\alpha}^\mu) }_{\Sigma} = \int [d\omega] \int \frac{{\mathcal D}({}^{\omega}g)}{V(\text{Diff}\times\text{Weyl})} \int {\mathcal D}X \, e^{-S_p[{}^\omega g,X]} \prod_{\alpha=1}^N V_{\alpha} (k_{\alpha}^\mu) 
}

et l'amplitude de Polyakov complète sommée sur toutes les géométries compactes~:

\eqna{
\corr{\prod_{\alpha=1}^N V_{\alpha} (k_{\alpha}^\mu) } = \sum_{\Sigma \text{ compactes}} g_s^{-\chi}  \corr{\prod_{\alpha=1}^N V_{\alpha} (k_{\alpha}^\mu) }_{\Sigma}
}

Nous avons plusieurs remarques à formuler quant à cette expression et aux propriétés des éléments qu'on y a introduit~: \\

\begin{itemize}\itemsep4pt
\item[\emph{i)}] Un facteur $g_s$ a été ajouté dans la formule ci-dessus, un couplage de corde, et ce n'est pas arbitraire. En vérité, nous aurions d\^u ajouter un terme supplémentaire à l'action de Polyakov, vérifiant également les invariances par difféomorphismes et Weyl ; un terme purement topologique~: 

\eqn{
\lambda \, \int_{\Sigma} \frac{\di^2 \sigma}{2\pi} \sqrt g \, {\mathcal R} = \lambda \, \chi(\Sigma)
}

D'après cette formule, nous identifions naturellement $g_s = e^{\lambda}$. Le nombre d'Euler dépend du nombre de poignées (génus), de bords, et de cross-caps\footnote{Pour la définition voir Polchinski~\cite{Polchinski:1998rq}}~\cite{Nakahara:1990th} à la surface selon $\chi= 2-2g-b-c$. Ainsi le nombre d'Euler de la sphère est $\chi=2$, celui du tore $\chi=0$ et du disque $\chi=1$. Compte-tenu du classement des topologies selon les puissances dans le couplage, pour les cordes fermées, nous identifions la sphère à l'amplitude à l'ordre des arbres et le tore à l'amplitude à une boucle, \etc. Le disque serait quant à lui l'ordre des arbres pour les amplitudes de cordes ouvertes et l'anneau l'amplitude à une boucle. 

\item[\emph{ii)}]  L'intégrande doit être invariant par difféomorphisme et transformation de Weyl, d'après les arguments avancés dans la section précédente. Remarquons tout de même que du fait des intégrations sur toutes valeurs de $X$ et $g$, le résultat est évidemment indépendant du système de coordonnées et de la métrique intrinsèques ; ce qui n'interdit donc pas de s'intéresser à des intégrandes non invariants dans des problèmes de physique statistique par exemple. 

Ainsi, il faut vérifier que l'ensemble de l'intégrande et de la mesure sont bien invariants sous l'action du groupe $\text{Diff}\times\text{Weyl}$. Alors que l'invariance au niveau classique ne concerne que les variations de l'action, l'invariance au niveau quantique quant à elle tient compte des variations de l'ensemble des objets contenus dans l'amplitude, dépendant de paramètres intrinsèques. 

Polyakov montre~\cite{Polyakov:1981rd,Nakahara:1990th} que, dans Minkowski\footnote{Mais le résultat final sur le nombre de dimension est valable pour tout $G_{\mu\nu}$.}, par transformation de Weyl $g \to e^{2 \phi} g$, la mesure -- oublions un instant les modules -- se transforme selon : 

\eqna{\label{eq:anomalie}
{\mathcal D}X {\mathcal D} g \to e^{\frac{d+1-26}{24\pi^2} \int \di^2 z \sqrt g \parent{g^{ab}\partial_a \phi \partial_b\phi + {\mathcal R}\phi} } {\mathcal D}X {\mathcal D} g
}

La théorie des cordes critique impose donc que l'espace-temps cible soit composé de 25 dimensions d'espace et 1 de temps, soit 26 dimensions. Mentionnons qu'en théorie \emph{critique} des supercordes, la théorie des cordes supersymétrique, Polyakov montre~\cite{Polyakov:1981re} de façon équivalente qu'il faut $d+1=10$. Nous verrons au point v) ce qui concerne les opérateurs de vertex. 

\item[\emph{iii)}]  L'intégrale sur $\omega$ représente l'intégration sur les modules. Il peut y en avoir plusieurs, réels ou complexes. 

\item[\emph{iv)}] A module fixé, il reste à intégrer le long de la classe de métriques équivalentes par difféomor\-phisme et transformations de Weyl. Etant donné que la théorie est choisie invariante suivant ces transformations, l'intégration le long de cette classe est redondante et implique un surcomptage. Il convient donc de choisir une métrique de référence –- appelée métrique \emph{fiducielle} et notée ${}^{\omega}\widehat g$ -- puis d'intégrer sur les orbites du groupe $\text{Diff} \times \text{Weyl}$, et enfin de diviser par le volume de ce groupe. 

Afin de fixer la jauge nous utilisons la méthode de Fadeev-Popov, qui donne d'emblée la bonne mesure d'intégration. Cette dernière est exprimée en introduisant des champs abstraits, \cad des artefacts mathématiques, fermioniques et anti-commutants, nommés \emph{champs fantômes} et dont la mesure d'intégrale de chemin s'ajoute aux précédentes sous la forme~:

\begin{align}
\int {\mathcal D}b {\mathcal D}c~e^{-S_{gh}[b,c]}
\end{align}

L'action associée s'exprime selon~: 

\eqna{\label{eq:act_gh}
S_{gh}[b,c] = \frac{1}{2\pi} \int \di^2\sigma \, \sqrt{{}^\omega \widehat g} \; b_{ab} \; {}^\omega \widehat \nabla^a c^b
} 

L'ancienne symétrie de jauge géométrique qui vient d'être fixée se retrouve naturellement sous une autre forme dans l'action complète, mais devient maintenant une contrainte pure sur la théorie plutôt qu'une redondance. Il s'agit de la symétrie BRST. Les fantômes $(b_{ab},c^a)$ sont des champs conformes de dimension $(2,-1)$. Nous verrons cela plus en détail dans la section~\refcc{sub-sec:decomp_modes}.

\item[\emph{v)}]  Les opérateurs de vertex doivent être insérés sur la feuille d'univers de manière à représenter des états asymptotiques de cordes -- ou particules. Idéalement, ils seraient construits en perçant la surface sous la forme d'un trou circulaire puis en étirant la surface jusqu'à les envoyer à l'infini temporel~\cite{Cohen:1986pv}. En réalité, ce n'est pas la peine de procéder ainsi, grâce à l'invariance conforme. En effet, par cette invariance, nous pouvons changer localement de système de coordonnées tout en conservant la métrique fiducielle. De cette manière, n'importe quel point, à n'importe quelle coordonnée, peut devenir infiniment lointain de tous les autres. Ainsi, un trou circulaire de taille finie et envoyé à l'infini est conforme à une perforation ponctuelle, un \emph{poinçon}.  

Par conséquent, un état asymptotique doit correspondre à un opérateur de vertex inséré sur la surface sous la forme d'un poinçon, et ce en un point quelconque de cette surface. Puisqu'il n'y a pas de point d'insertion privilégié sur la feuille d'univers, il faut donc intégrer l'opérateur de vertex sur toute cette surface. L'expression générale d'un opérateur de vertex est alors : 

\eqn{
V_\alpha (k^\mu_\alpha) = \int_\Sigma \di^2 \sigma \, \sqrt{{}^\omega \widehat g} \; {\mathcal V}_\alpha(k^\mu_\alpha,\sigma) 
}

La condition d'invariance de Weyl de l'amplitude impose que cet opérateur soit lui-même invariant conforme. Pour cette raison l'opérateur ${\mathcal V}_\alpha(k^\mu_\alpha,\sigma)$ doit être un champ \emph{primaire} de la théorie conforme, une propriété introduite dans la section~\refcc{sub-sec:decomp_modes}. Mentionnons que dans Minkowski cette contrainte impose $(k^0)^2 - (k^i)^2 = m^2$, \cad la condition de couche de masse (!) pour la particule représentée par l'état asymptotique.

Avec la jauge fixée, tout opérateur de vertex est une fonctionnelle des champs $X$ mais \emph{aussi} des champs fantômes \apriori. 

\item[\emph{vi)}]  Il existe des difféomorphismes qui génèrent des transformations conformes, \cad qui peuvent être annulés par une transformation de Weyl. Cette symétrie est importante puisqu'il s'agit précisément de la symétrie conforme développée dans la section suivante. En fixant la jauge\footnote{Voir~\cite{Alvarez:1982zi,Polchinski:1998rq} pour plus de détails}, \cad en fixant la métrique, les transformations $\text{Diff}\times\text{Weyl}$ ne sont pas complètement fixées. Il existe une symétrie résiduelle~: la symétrie conforme. Autrement dit, $V(g) < V(\text{Diff}\times\text{Weyl})$. 

Du fait du surcomptage causé par cette symétrie résiduelle, l'amplitude finale -- à jauge fixée – doit \^etre divisée par un facteur de volume sur le groupe d'invariance conforme, nommé CKG -- groupe de Killing conforme. Ses éléments se nomment CKV -- vecteur de Killing conformes. Ce facteur sera noté $\Omega(CKG)$. Une autre option, est de fixer la position d'autant d'opérateurs de vertex qu'il y a de CKV. Sur la sphère on compte 3 CKV complexes dans le groupe de M\oe bius $SL(2,{\mathbb C})$, donc il suffit de fixer la position de 3 opérateurs de vertex. Sur le disque, on compte 3 CKV réels dans le groupe $SL(2,{\mathbb R})$. Il existe une autre méthode pour se débarrasser de ce surcomptage, utile lorsqu'il n'existe aucun opérateur à fixer – dans le cas du calcul de la fonction de partition par exemple. Elle consiste à extraire du calcul le nombre infini associé au surcomptage, en utilisant le formalisme de renormalisation -- voir entre autres~\cite{Tseytlin:1987ww,Andreev:1988bz,Fradkin:1985ys,Fradkin:1984pq}.

\item[\emph{vii)}]  Le théorème de Riemann-Roch permet de donner le nombre de modules $\mu$ et de CKV $\kappa$ en fonction de la topologie de la surface : 

\eqna{\label{eq:riem_roch}
\chi>0 ~: & \kappa = 3 \chi & \mu=0 \nonumber \\
\chi<0 ~: & \kappa = 0 & \mu= 3 \chi 
}
\end{itemize}

Pour conclure et compte-tenu de toutes les remarques précédentes, l'expression de l'amplitude sur une variété $\Sigma$ à jauge fixée est :

\eqna{\label{eq:ampl_fix}
\corr{\prod_{\alpha=1}^N V_{\alpha} (k_{\alpha}^\mu) }_{\Sigma} & =& \int \frac{d^\mu\omega}{\Omega(CKV)} \int {\mathcal D}X \, e^{-S_p - S_{gh}} \prod_{\alpha=1}^N V_{\alpha} (k_{\alpha}^\mu) \nonumber \\
&=&  \int d^\mu\omega \int {\mathcal D}X \, e^{-S_p - S_{gh}} \prod_{\alpha=1}^\kappa c\wt c{\mathcal V}_{\alpha} (k_{\alpha}^\mu,\sigma_\alpha)  \prod_{\beta=\kappa+1}^N V_{\beta} (k_{\beta}^\mu)
}

Chaque opérateur de vertex fixé doit être multipliée dans le bulk par la combinaison de champs fant\^omes $c\wt c$. Nous verrons pour quelle raison à la section~\refcc{sub-sec:decomp_modes}. Mentionnons simplement qu'ils remplacent la mesure $\di^2\sigma$ du point de vue des transformations conformes. \\

Nous avons introduit les théories conformes en théorie des cordes. En particulier, nous avons défini l'action de Polyakov, les amplitudes et les opérateurs de vertex, des quantités devant vérifier des conditions d'invariance de Weyl et de difféomorphismes donc d'invariance conforme. Nous allons maintenant définir concrètement ces opérateurs dans le contexte de la CFT.

\subsection{Les symétries dans l'action et l'amplitude de Polyakov}

Cette étude se base principalement sur les ouvrages de Polchinski et Kiritsis~\cite{Polchinski:1998rq,Polchinski:1998rr,Kiritsis:2007zz}. Le point de départ est l'action de Polyakov, \refe{eq:act_pol} et~\refe{eq:act_gh},  définie sur un espace-cible plat, \cad $G_{\mu\nu}(X)=\eta_{\mu\nu}$, avec la convention $\eta=\text{diag}(-1,1,1,1)$. Nous nous intéressons à la physique locale sur la surface, en se plaçant sur un voisinage $U$ de la variété $\Sigma$ : 

\eqn{
S_p[g,X] = \frac{1}{4 \pi \alpha'} \int_{U} \di^2\sigma \; \sqrt{\widehat g} \, \widehat g^{ab} \; \partial_a X^\mu \partial_b X^\nu \; \eta_{\mu\nu} + \frac{1}{2\pi} \int_U \di^2\sigma \, \sqrt{\widehat g} \; b_{ab} \; \widehat \nabla^a c^b
}

Localement, toute métrique sur une surface est conformalement plate, à un difféomorphisme près –- voir~\cite{Nakahara:1990th}. Il n'existe donc pas de module dans la description locale d'une surface. Ceci implique que, par invariance de Weyl, nous pouvons utiliser la métrique fiducielle suivante $\widehat g_{ab}=\delta_{ab}$ et paramétriser la surface par des coordonnées euclidiennes, à condition que la fonction $X$ vérifie bien quant à elle des contraintes de causalité. En outre, il n'existe pas de condition de réalité sur le système de coordonnées, que nous pouvons choisir dans ${\mathbb C}$ à partir du moment où le voisinage $U$ -- un ouvert –- est une surface de Riemann. Ainsi opterons-nous pour un système de coordonnées complexe, dans lequel la métrique est hermitienne et s'exprime comme $ds^2 = dz \otimes d\bar z$. Ce système est particulièrement pratique pour traiter la plupart des calculs. Dans ces coordonnées, l'action s'écrit comme : 

\eqn{\label{eq:act_plan}
S_p[g,X] = \frac{1}{2 \pi \alpha'} \int_{\mathbb C} \di^2 z \; \partial X^\mu \bar \partial X_\mu  + \frac{1}{2\pi} \int_{\mathbb C} \di^2 z \; \parent{b_{zz} \bar \partial c^z + b_{\bar z\bar z} \partial c^{\bar z}}
}

où nous notons simplement $\partial = \partial/\partial z$ et $\bar\partial = \partial/\partial \bar z$. La normalisation est toujours fixée suivant les conventions de Polchinski~\cite{Polchinski:1998rq}. Cette action est symétrique par transformations conformes et transformations de Poincaré sur les champs $X^\mu$, \cad les translations et les boost de Lorentz. \\ 

Certaines feuilles d'univers $\Sigma$ peuvent être décrites globalement par au moins un système de coordonnées fixé et en particulier si la surface est orientable, sur le plan complexe -- il s'agit alors d'une surface de Riemann. C'est le cas de la sphère $S^2$, du tore $T^2$ (et de tous les tores $T^n$ à plus forte raison) ou du disque $D^2$. 

Dans le cas de la sphère tous les points sont équivalents, mais ce n'est pas le cas du disque où les points de l'intérieur -- le \emph{bulk} – se distinguent des points du bord. Ainsi, l’étude de la physique des cordes, sur une géométrie de surface telle que la sphère, pourra \^etre faite au voisinage de n'importe quel point. En revanche dans le cas du disque nous discernons deux physiques distinctes, celle de l'intérieur -- vraisemblablement semblable à celle d'un voisinage de sphère -- et celle du bord -- de tout voisinage du bord pour être plus précis. \\

Cette distinction amène à définir deux théories des cordes sur les surfaces avec bord : les cordes fermées, celles du bulk, et les cordes ouvertes, celles du bord. Nous différencions ainsi les opérateurs de vertex du bulk et ceux du bord, qui ne sont pas associés aux même états asymptotiques. L'identification précédente est naturelle dans le sens où une corde fermée est un objet n'ayant pas de bord, tandis qu'une corde ouverte est un objet à deux bords, \cad chaque extrémité.

En ce sens, l'action~\refe{eq:act_plan} décrit la physique d'un voisinage du bulk, sans bord, paramétrisée sur le plan complexe entier ${\mathbb C}$, et correspond par conséquent à une théorie des cordes fermées. En revanche, pour décrire la physique des cordes ouvertes, il faut se placer sur un demi-plan complexe avec bord et on choisit en général le demi-plan supérieur $H_+ = \{z, \; \im \, z \geq 0 \}$. A noter qu'il faut éventuellement implémenter des conditions de bord sous la forme de termes de bord dans l'action~: 

\eqn{\label{eq:act_demi_plan}
S_p[g,X] = \frac{1}{2 \pi \alpha'} \int_{H_+} \di^2 z \; \partial X^\mu \bar \partial X_\mu + \frac{1}{2\pi} \int_{H_+} \di^2 z \; \parent{b_{zz} \bar \partial c^z + b_{\bar z\bar z} \partial c^{\bar z}} + \oint_{\partial H_+ = {\mathbb R}} \di z \; {\mathcal O}[X,b,c]
}

avec ${\mathcal O}$ un ensemble d'opérateurs de vertex \apriori primaires. Insistons cependant sur le fait que la théorie est ci-dessus définie sur des voisinages et non sur la variété entière. Par conséquent, nous ne pouvons pas calculer, dans cette description, une amplitude sur la sphère ou sur le disque~; il faudrait ajouter certaines conditions de périodicité, de symétries – relié au problème des CKV --  et de régularisation infra-rouge (IR). 

\subsubsection{Les symétries de l'action de cordes fermées}

Nous comptons plusieurs symétries dans l'action~\refe{eq:act_pol}, où la jauge n'est pas encore fixée. Les symétries \emph{internes} rencontrées précédemment : les difféomorphismes -- dont les transformations conformes -- et les transformations de Weyl. Mais aussi des symétries \emph{externes} -- dans le sens externes à la surface~: celles-ci concernent des transformations des champs eux-mêmes, entre eux, mais qui peuvent dépendre des coordonnées. Parmi elles se trouvent les translations $X^\mu \to X^\mu + a^\mu$ et les transformations spéciales -- boost -- de Lorentz $X^\mu \to \Lambda^\nu_\mu X^\nu$. L'ensemble de ces transformations forme le \emph{groupe de Poincaré}\footnote{Il peut exister d'autres symétries externes, auxquelles on se réfère en CFT sous l'appelation d'\emph{algèbre de courant}.} $(\Lambda,a)$. \\ 

Il serait plus judicieux de partir de l'action~\refe{eq:act_plan} dont la jauge est fixée. En effet, nous ne serions plus confrontés aux contraintes d'invariance de Weyl et de difféomorphismes directement, mais uniquement aux contraintes d'invariance par transformations conformes. Ainsi, dans la suite, tant que possible nous partirons de l'action~\refe{eq:act_plan}. \\

Avant de lister l'ensemble des symétries et des courants de Noether correspondants, donnons les équations du mouvement des différents champs, calculées à partir de l'action~\refe{eq:act_plan}. Sa variation lagrangienne donne les équations d'Euler-Lagrange suivantes : 

\eqna{\label{eq:eom}
\partial \bar\partial X^\mu = 0 \nonumber \\
\bar \partial c^z = \partial c^{\bar z} = 0 \nonumber \\ 
\bar \partial b_{zz} = \partial b_{\bar z\bar z} = 0
}

Nous en déduisons que $\partial X$ et $\bar \partial X$ sont respectivement holomorphe et anti-holomorphe. En ce qui concerne les fantômes, les conclusions sont similaires. Par conséquent, nous utiliserons la notation suivante~: 

\begin{align}
c^z = c(z) \qquad &,\qquad c^{\bar z} = \widetilde c(\bar z) \nonumber \\ 
b_{zz} = b(z) \qquad &,\qquad b_{\bar z \bar z} = \widetilde b(\bar z)
\end{align}

\subsubsection{Courants de Noether des champs $X$}

Les courants de Noether associés à ces symétries sont construits comme habituellement en théorie des champs –- voir~\cite{Itzykson:1980rh}. Le courant des symétries internes est associé aux transformations de la métrique, donc est représenté par le tenseur énergie-impulsion. Pour une transformation infinitésimale $\delta \sigma^a = \epsilon v^a(\sigma)$, avec $\sigma=(z,\bar z)$, on a : 

\eqna{\label{eq:tenseur-EP}
&& j_z = j(z) = i v^z T_{zz} + i v^{\bar z} T_{\bar z z} \nonumber \\
&& j_{\bar z} = \widetilde \jmath(z)=  i v^z T_{z \bar z } + i v^{\bar z} T_{\bar z \bar z} 
} 

Le champ $X$ se transforme selon $\delta X^\mu = - \epsilon v^a \partial_a X^\mu$ donc le tenseur énergie-impulsion vérifie~:

\eqna{\label{eq:tens_x}
&& T_{z \bar z} = T_{\bar z z} = 0 \nonumber \\
&& T_{zz} = T(z) = - \frac{1}{\alpha'} :\partial X^{\mu} \partial X_{\mu}: \nonumber \\ 
&& T_{\bar z \bar z} = \widetilde T(\bar z) = - \frac{1}{\alpha'} :\bar \partial X^{\mu} \bar \partial X_{\mu}:
}

Nous avons introduit la notation d'ordre normal $:\star:$ pour signifier que le produit intérieur est régulier au sens des OPE -- développement de produits d'opérateurs -- défini à la section~\refcc{sub-sec:decomp_modes}. Remarquons que $T^a_{\hphantom{a} a} = 2 T_{z \bar z}=0$, \cad que le courant des transformations de Weyl est nul. En outre, la conservation du courant impose que $\bar \partial (v^z T_{zz}) = \partial (v^{\bar z} T_{\bar z \bar z})=0$, de telle sorte que $v^z$ et $T_{zz}$ sont holomorphes, tandis que $v^{\bar z}$ et $T_{\bar z \bar z}$ sont anti-holomorphes, ce qui est vérifié le long des équations du mouvement~; d'où la notation utilisée $T(z)$ et $\widetilde T(\bar z)$. \\

Les transformations conformes sont portées par le vecteur tangent $v^a(z,\bar z)$ tel que $\epsilon v^z(z) = \alpha + \beta z + \gamma z^2$ avec des coefficients complexes infinitésimaux, regroupant dans l'ordre les translations ($\alpha$), les rotations et changements d'échelle ($\beta$) et les transformations conformes spéciales ($\gamma$). La transformation exacte s'écrit : 

\eqn{
z' = \frac{az+b}{cz+d} \qquad \text{avec } \parent{\begin{array}{cc}a & b \\ c & d \end{array}} \in SL(2,{\mathbb C})
}

De même $v^{\bar z}(\bar z) = \alpha + \beta \bar z + \gamma \bar z^2 $.  Les transformations conformes sont telles que les variables \mbox{(anti-)}holomorphes se transforment en variables \mbox{(anti-)}holo\-morphes $z\to f(z)$ et $\bar z \to \widetilde f (\bar z)$. Or puisque la métrique dans la jauge unitaire est $ds^2 = dz\otimes d\bar z$ elle devient $f(z)\widetilde f(\bar z) dz\otimes d\bar z$ ce qui est effectivement la transformation recherchée.  \\

Les transformations externes, \cad des champs entre eux, sont générées principalement -- mais il peut exister d'autres groupes de transformations, \cf les \emph{algèbres de courant} -- par le groupe de Poincaré dont la forme infinitésimale appliquée sur $X$ et sur la métrique d'espace-cible est : 

\eqna{
&& \delta X^\mu = \epsilon \; a^\mu(\sigma) + \epsilon \; \omega^\mu_{\hphantom{\mu} \nu} (\sigma) X^\nu \nonumber \\
&& \delta G_{\mu \nu} = \epsilon \; \omega_\mu^{\hphantom{\mu}\rho} G_{\rho\nu}
}

avec $\omega_{\mu\nu}=-\omega_{\nu\mu}$. Les courants associés respectivement à la translation et aux boost de Lorentz sont : 

\eqna{\label{eq:trans}
J^\mu_z = J^\mu (z) = \frac{i}{\alpha'} \partial X &\qquad & J^\mu_{\bar z}= \widetilde J^\mu  (\bar z)  = \frac{i}{\alpha'} \bar \partial X  \nonumber \\ 
j^{\mu\nu}_z (z,\bar z) = \frac{i}{\alpha'} : X^{[\mu} \partial X^{\nu]}: & \qquad &
  j^{\mu\nu}_{\bar z} (z,\bar z) = \frac{i}{\alpha'} :X^{[\mu} \bar \partial X^{\nu]}:
}

avec la convention $x^{[a} y^{b]} = (x^ay^b-x^by^a)/2$. Ces courants sont conservés classiquement le long des équations du mouvement. Notons cependant que le courant des transformations de Lorentz n'est pas un champ primaire à cause du facteur $X^\mu$.  \\

\subsubsection{Courants de Noether des champs fantômes}

N'oublions pas que le tenseur énergie-impulsion des champs fantômes doit être ajouté. Afin que l'action~\refe{eq:act_gh} soit invariante conforme, il faut les transformations des fantômes suivantes pour $\delta \sigma^a = -\epsilon \, v^a$ :

\eqna{
&& \delta b_{zz} = -\epsilon \; v^z \partial b + \epsilon \; (\partial v^z) \lambda b \nonumber \\ 
&& \delta c_{z} = -\epsilon \; v^{z} \partial c + \epsilon \; (\partial v^z) (1-\lambda) c
}

avec $\lambda=2$. Des formules équivalentes existent pour $\widetilde b$ et $\widetilde c$ en remplaçant simplement $z$ par $\bar z$. En variant l'action, nous calculons aisément le tenseur énergie-impulsion correspondant : 

\eqna{\label{eq:tens_gh}
&& T^g = :(\partial b)c:- 2 \partial(:bc:) \nonumber \\ 
&& \widetilde T^g = :(\partial \widetilde b) \widetilde c:- 2 \bar \partial(:\widetilde b\widetilde c:)
}

De même, ce tenseur est bien conservé le long des équations du mouvement, pour lesquelles $T^g$ et $\widetilde T^g$ sont holomorphes et anti-holomorphes. Le tenseur énergie-impulsion complet  associé à l'action~\refe{eq:act_plan} est ainsi :

\eqna{\label{eq:tens_x_gh}
T(z) =T^X + T^g \nonumber \\ 
\widetilde T(\bar z) =\widetilde T^X + \widetilde T^g
}

En tant que générateur des transformations conformes, le tenseur énergie-impulsion à un statut très important en CFT. C'est à partir de son action sur les fonctionnelles de champs que les opérateurs de vertex et les états asymptotiques sont définis. \\

Les fantômes vérifient aussi une symétrie externes, dénommée symétrie de nombre fantomatique -- \emph{ghost number} -- telle que $\delta b =-i\epsilon \, b$ et $\delta c = i\epsilon \, c$ et dont les courants sont~: 

\eqna{
&& j_g = - :bc: \nonumber \\ 
&& \widetilde \jmath_g = - :\widetilde b \widetilde c:
}

Ce courant est conservé le long des équations du mouvement, pour lesquelles $j_g$ et $\widetilde \jmath_g$ sont respectivement holomorphe et anti-holomorphe. \\

Si ces courants sont définis conservé classiquement, remarquons qu'ils ne le sont pas toujours au niveau \emph{quantique}. C'est potentiellement problématique car cela implique une brisure quantique de symétrie, soit des anomalies et éventuellement une rupture d'unitarité. 

\subsubsection{A propos de la symétrie BRST}
Pour approfondir l'étude de la théorie à jauge fixée, il faudrait s'intéresser au groupe de transformations conformes tenant compte des fantômes et qui se traduit sous la forme d'une symétrie externe, nommée BRST. Celle-ci mélange les fantômes aux champs $X$, suivant la transformation infinitésimale~: 

\eqna{\label{eq:BRST}
& \delta_B X^\mu = i \epsilon \parent{c \partial + \widetilde c \bar \partial} X^\mu \nonumber \\ 
& \delta_B b = i \epsilon \parent{T^X + T^g} \qquad \delta_B \widetilde b = i \epsilon \parent{\widetilde T^X + \widetilde T^g}   \nonumber \\ 
& \delta_B c = i\epsilon c \partial c \qquad \delta_B \widetilde c = i\epsilon \widetilde c \bar \partial \widetilde c
}

Puisque $CFT \subset BRST $, la théorie des champs vérifiant cette symétrie est une CFT sur laquelle \emph{a priori} un certain nombre de contraintes supplémentaires sont posées, en particulier sur l'expression des états asymptotiques de la théorie -- et des opérateurs de vertex -- se traduisant par des contraintes physiques. Le courant associé à cette symétrie est : 

\eqna{
&& j_B = cT^m + \frac{1}{2}:cT^g: + \frac{3}{2} \partial^2 c \nonumber \\ 
&& \widetilde \jmath_B = \widetilde c\widetilde T^m + \frac{1}{2}:\widetilde c\widetilde T^g: + \frac{3}{2} \bar \partial^2 \widetilde c
}

Ici le tenseur énergie-impulsion est $T^m$ où $m$ signifie "matière", au lieu de $T^X$, afin de généraliser l'expression à toute théorie des champs non fantômes -- comme par exemple les fermions en supercordes. Ce courant est conservé le long des équations du mouvement, pour lesquelles $j_B$ et $\widetilde \jmath_B$ sont respectivement holomorphe et anti-holomorphe.

L'expression ci-dessus montre que la transformation BRST est en fait une transformation conforme avec $v^z(z)=c(z)$ et respectivement $v^{\bar z}(\bar z)=\widetilde c(\bar z)$.

\subsection{Dimensions conformes, champs primaires et états}
\label{sub-sec:decomp_modes}

Comment un champ se transforme-t-il par transformations conformes ? Pour répondre à cette question autant définir un champ par ses lois de transformations. Ainsi nous pourrions étendre aux champs la notion de tenseur de sorte qu'ils soient covariants par transformation conforme. Cette propriété permettrait de définir aisément des quantités physiques, \cad invariantes conformes. Les \emph{champs primaires}\footnote{Un \emph{champ} est, de manière générale, une fonctionnelle des champs fondamentaux de la théorie de surface, ici$ X$ et ses dérivées successives} sont définis par la loi de transformation suivante –- voir en particulier l'ouvrage de Di Francesco~\cite{DiFrancesco:1997nk}~:

\eqna{\label{eq:transf_conf_primaire}
{\mathcal O}'(z',\bar z') = \parent{\frac{\partial z'}{\partial z}}^{-h}  \parent{\frac{\partial \bar z'}{\partial \bar z}}^{-\bar h} {\mathcal O}(z,\bar z)
}

Le champ ${\mathcal O}$ ainsi défini est dit \emph{champ primaire de poids} $\parent{h,\bar h}$, où ces quantités sont des nombres \emph{a priori} réels. La dimension conforme du champ désigne $\Delta=h + \bar h$ et paramétrise l'évolution d'un champ selon un changement d'échelle rigide $(z,\bar z) \to (\zeta z,\zeta \bar z)$. Cette quantité est à rapprocher de la notion de \emph{dimension} pour un champ, rencontrée dans le cadre du formalisme de renormalisation en théorie quantique des champs. Nous définissons aussi le spin $S=h-\bar h$ qui caractérise l’évolution d’un champ selon une rotation rigide $(z,\bar z) \to (e^{i\phi} z,e^{-i\phi} \bar z)$. D'après la formule~\refe{eq:transf_conf_primaire}, les transformations holomorphes et anti-holomorphes sont tout à fait séparées. C'est un point crucial des transformations conformes dans la jauge unitaire sur le plan complexe.  

\subsubsection{Correspondance état-opérateurs~: espace de Hilbert}

Nous pourrions parler de ces champs en terme d'\emph{opérateurs tensoriels}. Ils devraient alors être définis en relation à un espace de Hilbert. En effet, certains champs primaires sont exprimables sous la forme de tenseurs conventionnels, c'est-à-dire définis sur la base de l'espace tangent $T_p \Sigma$ en un point $p$ de la surface. Il s'agit des champs primaires de poids entiers notés $\parent{n,\bar n}$. D’autres en revanche, les champs primaires de poids non entiers $(h,\bar h) \notin {\mathbb R}^2$, ne peuvent pas être définis sur une telle base. 

Il faut donc introduire une base généralisée $\{ \ket{h,\alpha_{(i)},z}\otimes \ket{\bar h,\widetilde \alpha_{(i)},\bar z} \}$ en tout point $(z,\bar z)$ sur un espace de Hilbert des champs primaires

\eqna{
{\mathcal H}=\underset{(h,\bar h) \in {\mathbb R}^2}{\bigoplus} {\mathcal H}_h \otimes {\mathcal H}_{\bar h}
}

Les paramètres $(\alpha_{(i)},\wt \alpha_{(i)})$ forment un ensemble de propriétés discriminant des ket de même poids et résultant d'un classement en représentations de symétries supplémentaires – les algèbres de courant. Autrement dit, chaque espace de Hilbert ${\mathcal H}_h$ est lui-même décomposé dans une base hilbertienne ou généralisée $\{\ket{h,\alpha_{(i)},z}\}_{\alpha_{(i)}}$. Le poids $h$ est en général calculé en fonction des $\alpha_{(i)}$. Ces espaces de Hilbert constituent les espaces de \emph{plus haut poids} en CFT -- voir la présentation claire de~\cite{Recknagel:1998ih}. 

Les bras $\bra{h,\alpha_{(i)},z}$ sont également introduits tels que~:

\begin{align}
\braket{h,\alpha_{(i)},z}{h',\alpha'_{(i)},z'} = \delta_{h,h'} \, \delta^{(\ell)}(\alpha_{(i)}-\alpha'_{(i)}) \, \delta(z-z')
\end{align}

Le bra contient une distribution delta de dirac $\bra z = \bra {\delta_z}$. Chaque ket et bra subit les transformations conformes selon : 

\eqna{
&& \ket{h,\alpha_{(i)},z'}' = \parent{\frac{\partial z'}{\partial z}}^{-h}  \ket{h,\alpha_{(i)},z} \nonumber \\ 
&& \bra{h,\alpha_{(i)},z'}' = \parent{\frac{\partial z'}{\partial z}}^{h-1}  \bra{h,\alpha_{(i)},z}
}

avec $\ell$ le nombre total de propriétés $\alpha_{(i)}$. La transformation du bra provient de la distribution delta de Dirac. Et respectivement pour les états correspondant aux opérateurs anti-holomorphes. Ainsi tout champ primaire -- local -- de poids $(h,\bar h)$ s'exprime de manière générale sous la forme d'un tenseur invariant selon : 

\eqna{
{\mathcal O}^{(h,\bar h)}(z,\bar z)  = \sum_{\alpha_{(i)}}{\mathcal O}^{(h,\bar h)}_{\alpha_{(i)},\widetilde \alpha_{(j)}} (z,\bar z) \bra{h,\alpha_{(1)},\alpha_{(2)},\ldots,z}\otimes \bra{\bar h,\widetilde \alpha_{(1)},\widetilde \alpha_{(2)},\ldots,\bar z} 
}

De sorte qu'un opérateur local est défini par l'action d'un tenseur ${\mathcal O}^{(h,\bar h)} = \int \di^2 z \; {\mathcal O}^{(h,\bar h)}(z,\bar z) $ sur un ket de base : 

\eqna{\label{eq:corr_op_et}
{\mathcal O}^{(h,\bar h)} \Big(| h,\alpha_{(i)},z \left.\right> , |\bar h,\bar \alpha_{(i)},\bar z \left.\right>\Big) = {\mathcal O}^{(h,\bar h)}_{\alpha_{(i)},\bar\alpha_{(i)}}(z,\bar z)
}

Les vecteurs $| h,\alpha_{(i)},z \left.\right>$ sont appelés des \emph{états} et la relation ci-dessus~\refe{eq:corr_op_et} peut être entendue comme une correspondance \emph{états - opérateurs}. Puisque l'état est piqué en un point particulier $z(p)$ de la surface, l'opérateur obtenu est bien défini en un seul point et correspond donc à un état asymptotique tel que cela a été défini auparavant~\refcc{sec:polyakov}. 

La correspondance \emph{états - opérateurs} n'est pas triviale à déterminer exactement. Il faut introduire les notions d'OPE puis d'action des courants sur les opérateurs et d'algèbre de courant. \\

Il existe une classe d'opérateurs locaux plus grande que celles des opérateurs primaires, dont la transformation par changement d'échelle rigide $z \to \zeta z$ est donnée par : 

\eqna{\label{eq:transf-op}
{\mathcal A}'(\zeta z',\bar \zeta \bar z') = \zeta^{-h} \bar \zeta^{-\bar h} {\mathcal A}(z,\bar z)
} 

Nous appelons aussi $(h,\bar h)$ les poids de l'opérateur ${\mathcal A}$. Les champs primaires constituent une classe de cet ensemble d'opérateurs. Les opérateurs non primaires, sont conventionnellement nommés opérateurs \emph{descendants}. 

Une correspondance état-opérateur peut être définie à travers un développement équivalent à celui effectué précédemment. Ces opérateurs descendants sont covariants le long d'un sous-groupe des transformations conformes, qui ne concerne que les changements d'échelles rigides. Soit donc ${}^d {\mathcal H}$ l'espace de Hilbert défini par : 

\eqna{
 {}^d{\mathcal H}=\underset{(h,\bar h) \in {\mathbb R}^2}{\bigoplus} {}^d{\mathcal H}_h \otimes {}^d{\mathcal H}_{\bar h} 
}

et tel que ${\mathcal H} \subset {}^d{\mathcal H}$. Ainsi ${}^d{\mathcal H}$ est généré par ${\mathcal H}$ en appliquant les algèbres de courant et de Virasoro sur les kets de plus haut poids. La base $\{\ket{h,\alpha_{(i)},z}\otimes\ket{\bar h,\widetilde \alpha_{(i)},\bar z}\}_{\alpha_{(i)},\widetilde \alpha_{(j)}}$ et les opérateurs descendants sont également définis via : 

\eqna{
{\mathcal A}^{(h,\bar h)} \Big(| h,\alpha_{(i)},z \left.\right> , |\bar h,\widetilde\alpha_{(i)},\bar z \left.\right>\Big) = {\mathcal A}^{(h,\bar h)}_{\alpha_{(i)},\widetilde \alpha_{(i)}}(z,\bar z)
}

Notons que ${\mathcal A}^{(h,\bar h)} $ n'est pas à proprement parler un tenseur, mais un objet covariant dans les changements d'échelles rigides. 

\subsubsection{OPE et courants}

Nous définissons le produit intérieur entre n'importe quels opérateurs -- donc pas nécessairement primaires -- par l'OPE :

\eqna{
{\mathcal A}_i (z){\mathcal A}_j(w)= \sum_k C_{ij}^{\hphantom{ij}k}(z,w) \; {\mathcal A}_k(w)
}

En théorie quantique des champs, ce produit est défini dans la limite où les insertions sont très proches. Il n'est généralement pas convergent. Dans une CFT à deux dimensions, toute OPE est convergente avec un rayon de convergence égal à la distance entre les opérateurs. Suivant des contraintes de transformations propres aux théories conformes, nous pouvons déterminer la forme générale des OPE : 

\eqna{\label{eq:OPE}
{\mathcal A}_i (z_i,\bar z_i){\mathcal A}_j(z_j,\bar z_j)= \sum_k \frac{C_{ij}^{\hphantom{ij}k}}{z_{ij}^{h_i+h_j-h_k} \bar z_{ij}^{\bar h_i+\bar h_j-\bar h_k}} \; {\mathcal A}_k(z_j,\bar z_j)
}

avec la notation $z_{ij}=z_i-z_j$ et $C_{ij}^{\hphantom{ij}k}$ un coefficient \emph{a priori} complexe. Insistons sur le fait que ces opérateurs ne sont pas nécessairement primaires et qu’en général seule une très restreinte minorité d’opérateurs est primaire. Particulièrement intéressante est l'OPE du tenseur énergie-impulsion avec un opérateur quelconque -- y compris lui-même. En effet, la transformation d'un opérateur le long d'un groupe généré par un courant est donné schématiquement par l'intégrale suivante~: 

\eqna{
\delta{\mathcal A}(z,\bar z) = \oint_{C_z} \di w \; \delta v(w) \, j(w) {\mathcal A}(z) + c.c.
}  

L'intégrale est définie le long d'un contour fermé autour de la position de l'opérateur. Le courant $j(w)$ est exprimé en terme des champs de la théorie et est donc lui-même un opérateur. Pour des raisons de cohérence, afin que les symétries soient conservées aussi au niveau quantique, le courant doit toujours \^etre un champ primaire. En utilisant~\refe{eq:OPE}, la variation infinitésimale $\delta {\mathcal A}$ s'exprime sous la forme d'un développement sur des opérateurs primaires et descendants. \\

Selon~\refe{eq:tenseur-EP} la transformation conforme d'un opérateur est donnée par~:  

\eqna{
\delta{\mathcal A}(z,\bar z) = \oint_{C_z} \di w \; \delta v^z(w) \, T(w) {\mathcal A}(z,\bar z) + c.c.
}

Puisque $T$ est de poids $(2,0)$ et d'après la transformation~\refe{eq:transf-op} du champ ${\mathcal A}$, nous devons avoir~:

\eqna{
T(w) {\mathcal A}^{(h,\bar h)}(z,\bar z) = \ldots + \frac{h}{(w-z)^{2}} {\mathcal A}^{(h,\bar h)}(z,\bar z) + \frac{1}{(w-z)} \partial {\mathcal A}^{(h,\bar h)}(z) + \ldots
}

Les termes cachés ici ne sont pas connus \emph{a priori} et dépendent de l'opérateur en question. En présence d'un champ primaire ${\mathcal O}$ le développement se réduit à~: 

\eqna{
T(w) {\mathcal O}^{(h,\bar h)}(z,\bar z) = \frac{h}{(w-z)^{2}} {\mathcal O}^{(h,\bar h)}(z,\bar z) + \frac{1}{(w-z)} \partial {\mathcal O}^{(h,\bar h)}(z) + \ldots
}

En l'occurrence, puisque $T$ doit être un champ primaire, il faut : 

\eqna{
T(w) T(z) = \frac{2}{(w-z)^{2}} T(z) + \frac{1}{(w-z)} \partial T(z) + \ldots
}

Néanmoins, de manière générale ce n'est pas ce qui est obtenu En effet, dans le calcul de l'amplitude~\refe{eq:anomalie} il apparaissait une anomalie proportionnelle à $d+1-26$. Cette anomalie se retrouve naturellement dans la transformation conforme de $T$ : 

\eqna{
T(w) T(z) = \frac{c/2}{(w-z)^4} + \frac{2}{(w-z)^{2}} T(z) + \frac{1}{w-z} \, \partial T(z) + \ldots
}

La constante $c$ s'appelle la \emph{charge centrale} et caractérise l'anomalie. Dans le contexte du calcul~\refe{eq:anomalie} nous trouvons $c=d+1-26$. Cette valeur est obtenue après avoir fixé la jauge, \cad en prenant comme point de départ l'amplitude~\refe{eq:ampl_fix}, soit le tenseur energie-impulsion~\refe{eq:tens_x_gh}. \\

Afin d'effectuer ces calculs explicitement, il faut connaître les OPE de base, \cad celles des champs fondamentaux de la théorie $X$, $b$ et $c$, ainsi que l'\emph{ordre normal}. Les expressions des OPE sont obtenues à partir des équations \emph{quantiques} du mouvement, \cad des équations d'opérateurs s'appliquant à l'intérieur d'un corrélateur. 

L'ordre normal d'une fonctionnelle $:{\mathcal F}[X,b,c]:$ est défini tel que l'ensemble est un opérateur dont le corrélateur est régulier dans les coordonnées de la feuille d'univers, soit $\corr{:{\mathcal F}[X,b,c]:} = \text{reg}$. De manière équivalente, l'ordre normal est défini par soustraction des singularités de la fonctionnelle : 

\eqali{
:{\mathcal F}[X,b,c]: = {\mathcal F}[X,b,c] \; - \; \text{sing.}
}

Dans le cas où la fonctionnelle est le produit intérieur de deux fonctionnelles régulières -- supposons-les holomorphes pour simplifier -- ${\mathcal F}(z)$ et ${\mathcal G}(w)$, nous obtenons une définition complémentaire de l'OPE en utilisant l'ordre normal~:

\eqali{
{\mathcal F}(z){\mathcal G}(w) & = \text{sing.} \; + \; :{\mathcal F}(z){\mathcal G}(w): \nonumber \\
& =  \text{sing.} \; + \;  \sum_{n\in {\mathbb N}} \frac{(z-w)^n}{n!} \, :(\partial^n {\mathcal F}){\mathcal G}(w):
}

L'application d'un simple développement de Taylor sur ${\mathcal F}$ permet de passer de la première ligne à la deuxième. Maintenant, à partir de l'action~\refe{eq:act_plan} nous pouvons calculer les fonctions de corrélation suivantes~:

\eqna{
\corr{X^\mu(z,\bar z)X^\nu(w,\bar w)} &=& - \alpha' \, \frac{\eta^{\mu \nu}}{2}  \ln \module{z-w}^2 \nonumber \\
\corr{b(z)c(w)} &=& \frac{1}{z-w} \nonumber \\ 
\corr{b(z)b(w)} &=& \corr{c(z)c(w)} = 0 
}

Rappelons ici que les champs $b$ et $c$ anticommutent. Par conséquent $b(z)c(w)=c(z)b(w)$. Sachant que sur une variété sans bord, $\corr{:X(z,\bar z)X(w,\bar w):}=\corr{X(z,\bar z)}^2 =0$ par imparité de l'intégration sur $X$, le même argument s'appliquant sur les champs fantômes, nous dérivons les OPE suivantes~:

\eqna{
X^\mu(z,\bar z) X^\nu(w,\bar w) &=& - \alpha' \, \frac{\eta^{\mu \nu}}{2} \ln \module{z-w}^2 \nonumber \\ && + \, \sum_{n\in {\mathbb N}} \parent{\frac{(z-w)^n}{n!} \,:(\partial^n X^\mu )X^\nu(w,\bar w): + \frac{(\bar z - \bar w)^n}{n!} \,:(\bar \partial^n X )X(w,\bar w):} \nonumber \\
b(z)c(w) &=& \frac{1}{z-w} \, + \, \sum_{n\in {\mathbb N}} \frac{(z-w)^n}{n!} \,:(\partial^n b )c(w):  \nonumber \\
}  

En utilisant ces développements et à partir des formules~\refe{eq:tens_x_gh},~\refe{eq:tens_x} et~\refe{eq:tens_gh}, nous obtenons ~:

\eqna{\label{eq:transf_champ}
&& T(w) \partial X^\mu(z) = \frac{1}{(w-z)^{2}} \partial X^\mu(z) + \frac{1}{w-z} \, \partial^2 X^\mu(z) + \text{reg.} \nonumber \\
&& T(w) b(z) = \frac{2}{(w-z)^{2}} b(z) + \frac{1}{w-z} \, \partial b(z) + \text{reg.} \nonumber \\
&& T(w) c(z) = \frac{-1}{(w-z)^{2}} b(z) + \frac{1}{w-z} \, \partial c(z) + \text{reg.} \nonumber \\
&& T(w) T(z) = \frac{(d+1-26)/2}{(w-z)^4} + \frac{2}{(w-z)^{2}} T(z) + \frac{1}{w-z} \, \partial T(z) + \text{reg.}
}

Ainsi, $\partial X^\mu$, $b$ et $c$ sont-ils des champs primaires de poids respectifs $(1,0)$, $(2,0)$ et $(0,-1)$. Dans la dernière formule nous retrouvons bien le développement attendu. Les transformations conformes ne sont générées correctement au niveau quantique, \emph{à jauge fixée}, qu'à la condition que l'anomalie s'annule, \cad quand la charge centrale s'annule, donc si et seulement si $d+1=26$. C'est une définition complémentaire de la théorie \emph{critique} des cordes, mais encore une fois reliée à la contrainte d'invariance de Weyl que nous souhaitons imposer intuitivement à la théorie. \\

La vérification de la conservation des courants au niveau quantique est immédiate : 

\eqna{
&& T(z) j^\mu (w) = \frac{1}{(z-w)^2} j^\mu(w) + \frac{1}{z-w} \partial j^\mu(w) + \text{reg} \nonumber \\
&& T(z) j^{\mu \nu} (w,\bar w) =\frac{1}{(z-w)^2} j^{\mu\nu}(w) + \frac{1}{z-w} \partial j^{\mu\nu}(w) + \text{reg} \nonumber \\
&& T(z) j^g(w) = \frac{-3}{(z-w)^3} + \frac{1}{(z-w)^2} j^{g}(w) + \frac{1}{z-w} \partial j^{g}(w) + \text{reg} \nonumber \\
&& T(z) j_B(w) = \frac{d+1-26}{2(z-w)^4} + \frac{1}{(z-w)^2} j_B(w) + \frac{1}{z-w} \partial j_B(w) + \text{reg}
}

Les deux premiers courants sont des tenseurs -- donc conservés -- tandis que les deux derniers ne le sont pas. Le courant de nombre fantomatique et le courant de BRST admettent chacun une anomalie. Dans le cas de $j^g$, l'anomalie est reliée à l'existence de \emph{modes zéro} des champs fantômes et implique le théorème Riemann-Roch~\refe{eq:riem_roch} – voir~\cite{Friedan:1985ge}. L'anomalie de $j_B$ est quant à elle reliée à la dimension de l'espace-cible et impose toujours $d+1=26$. C'est une nouvelle perspective sur la définition de la théorie critique des cordes. \\

\subsubsection{Op\'erateurs de vertex int\'egr\'es, non-int\'egr\'es et \'etat du vide}

Les opérateurs de vertex doivent \^etre invariants conformes mais aussi et surtout invariants BRST. Cela garantie l'invariance conforme \emph{quantique} de toute amplitude. En raison de l'existence dans certaines géométries de vecteurs de Killing conformes (CKV), il faut aussi imposer, dans les amplitudes, à un certain nombre d'opérateurs de vertex d'être fixés en des coordonnées arbitraires. Il existe donc une dichotomie d'opérateurs~: les op\'erateurs intégrés et les op\'erateurs fixés. 

Les opérateurs intégrés sont de la forme~:

\eqna{
V = \int \di^2 z \, {\mathcal A}^{(h,\bar h)}(z,\bar z) 
} 

L'invariance BRST sera dans ce cas totalement équivalente à l'invariance conforme et impose simplement que $(h,\bar h)=(1,1)$ et que ${\mathcal A}$ soit un champ primaire. C’est-à-dire dans l'écriture utilisée précédemment~:

\eqna{
V = \int \di^2 z \, {\mathcal O}^{(1,1)}(z,\bar z)
} 

Imposer au champ d'être primaire contraint fortement les paramètres dont il dépend ; en particulier, l'impulsion et la polarisation. \\

Les opérateurs non intégrés, fix\'es, doivent être, afin de v\'erifier l'invariance BRST, de la forme : 

\eqna{\label{eq:op_fix}
V= :c \, \widetilde c \, {\mathcal O}^{(1,1)}(z,\bar z) :
}

Il s'agit bien d'un opérateur de poids $(0,0)$ qui est donc bien invariant conforme et invariant BRST. Il s'agit de la forme la plus naturelle pour définir un état physique asymptotique, \cad piqué en un point quelconque de la surface, via la correspondance~\refe{eq:corr_op_et}. \\

L'état du vide est naturellement défini dans ce contexte comme correspondant au champ $c(z) \, \widetilde c(\bar z)$. Cela permet de définir convenablement les amplitudes du vide, à un point et à deux points. Dans les notations précédentes, le ket du vide $\ket{\Omega} = \ket{0}\otimes\ket{\downarrow}$ et son homonyme anti-holomorphe, s'expriment par : 

\eqna{\label{eq:vide_phys}
\Omega^{(-1,-1)}\parent{\ket{\Omega,z},\ket{\widetilde \Omega,\bar z}} = c(z) \, \widetilde c(\bar z)
}

Il faut à présent étudier plus en détail la correspondance état-opérateur que nous avons introduit à plusieurs reprises. En particulier, nous allons présenter les opérateurs de création, d'annihilation et de symétrie.

\subsubsection{Décomposition des champs de cordes fermées et des courants}

Rappelons les équations de mouvement~\refe{eq:eom} de l'action~\refe{eq:act_plan} : 

\eqna{\label{eq:eom_bis}
\partial \bar\partial X^\mu = 0 \nonumber \\
\partial \widetilde c = \bar \partial c = 0 \nonumber \\ 
\partial \widetilde b = \bar \partial b = 0
}

Puisque $\partial X$, $b$, $c$ et $\bar \partial X$, $\widetilde b$, $\widetilde c$, sont respectivement holomorphe et anti-holomorphe. Ils se décomposent en séries de Laurent~:

\eqna{\label{eq:laurent}
&& \partial X^\mu = -i \sqrt{\frac{\alpha'}{2}} \sum_{n \in {\mathbb N}} \alpha^\mu_n \, z^{-n-1} \qquad \bar \partial X^\mu = -i \sqrt{\frac{\alpha'}{2}} \sum_{n \in {\mathbb N}} \widetilde \alpha^\mu_n \, \bar z^{-n-1} \nonumber \\
&& c = \sum_{n \in {\mathbb N}} c_n \, z^{-n+1} \qquad \widetilde c = \sum_{n \in {\mathbb N}} \widetilde c_n \, \bar z^{-n+1} \nonumber \\
&& b = \sum_{n \in {\mathbb N}} b_n \, z^{-n-2}  \qquad \widetilde b = \sum_{n \in {\mathbb N}} \widetilde b_n \, \bar z^{-n-2}
} 

Les conventions sont toujours celles de Polchinski~\cite{Polchinski:1998rq}. Dans ces conventions, les transformations conformes de chacun des champs sont correctement implémentées. Nous pouvons promouvoir les équations~\refe{eq:eom_bis} et~\refe{eq:laurent} au niveau quantique, c'est-à-dire à l'intérieur d'un corrélateur, ce qui implique qu'elles sont aussi valables en décrivant les champs $X$ et les fantômes par des opérateurs agissant dans un certain espace de Hilbert. Par conséquent, nous pouvons considérer que les coefficients $\alpha^\mu_n$, $c_n$ et $b_n$ sont des opérateurs. \\

Ces opérateurs peuvent \^etre définis en relation inverse à partir de l'expression des champs par~:

\eqna{\label{eq:correspondance}
&& \alpha^\mu_n = \sqrt\frac{2}{\alpha'} \oint \frac{\di z}{2\pi} z^n \partial X^\mu(z) \nonumber \\ 
&& \widetilde \alpha^\mu_n = -\sqrt\frac{2}{\alpha'} \oint \frac{\di \bar z}{2\pi} \bar z^n \bar \partial X^\mu(\bar z) \nonumber \\ 
&& c_n = -i \oint \frac{\di z}{2\pi} z^{n-2} c^z(z) \nonumber \\ 
&& b_n = -i \oint \frac{\di z}{2\pi} z^{n+1} b_{zz}(z)
}

et de façon  similaire pour $\widetilde c$ et $\widetilde b$. D'après les deux premières expressions, par unicité de valeur de $X$ dans l'espace-cible, nous trouvons $\alpha^\mu_0 = \widetilde \alpha^\mu_0$. En effet, le long d'un chemin fermé $\gamma$,

\eqn{
\sqrt\frac{2}{\alpha'}\oint_\gamma \frac{dX^\mu}{2\pi} = 0 = \oint_{\gamma(z,\bar z)} \parent{ \frac{\di z}{2\pi} \partial X^\mu + \frac{\di \bar z}{2\pi} \bar \partial X^\mu } = \alpha^\mu_0 - \widetilde \alpha^\mu_0
}

Le mode $\alpha_0^\mu$ est l'opérateur d'impulsion \emph{de la corde}, noté $\alpha^\mu_0 = \sqrt\frac{2}{\alpha'} p^\mu$. \\ 

La décomposition complète de $X$ peut alors être déduite de~\refe{eq:laurent} sous la forme~: 

\eqn{\label{eq:dev_x}
X^\mu(z,\bar z) = x^\mu_0 - i \frac{\alpha'}{2} p^\mu \ln \module{z}^2 - i \sqrt{\frac{\alpha'}{2}} \sum_{n \in {\mathbb N}^*} \parent{ \alpha_n \, z^{-n} + \widetilde \alpha_n \, \bar z^{-n}}
}

L'opérateur position $x^\mu_0$ est associé \emph{classiquement} à une position de référence de la corde en une coordonnée de référence donnée sur la surface.

Ainsi par transformation conforme, la surface de corde complète -- la sphère ici – est décrite sous la forme d'un cylindre infiniment long ; ce qui correspond bien à l'intuition d'une corde fermée se propageant. Les coordonnées de la surface changent de la façon suivante : $z \to e^{- i t - i\sigma}$ et $\bar z \to e^{- i t + i\sigma}$ avec $\sigma \in [0,2\pi]$ et $t \in {\mathbb R}$. Ces deux coordonnées décrivent bien un cylindre. En terme de ces coordonnées,~\refe{eq:dev_x} devient :

\eqn{
X^\mu(t,\sigma) = x^\mu_0 - \alpha' \, p^\mu \; t - i \sqrt{\frac{\alpha'}{2}} \sum_{n \in {\mathbb N}^*} e^{i n t} \parent{ \alpha_n \,  e^{i n \sigma} + \widetilde \alpha_n\,  e^{- i n \sigma}}
}

D'après cette formule $p^\mu$ est bien assimilé à l'impulsion de la corde. Les coefficients $\alpha^\mu_n$ correspondent quant à eux à des modes de vibration de la corde. Ils sont de deux types~: les modes gauches $\alpha_n$ et les modes droits $\widetilde \alpha_n$, puisque les ventres et les noeuds des oscillations se déplacent respectivement le long des trajectoires $t+\sigma = \text{const.}$ et $t-\sigma = \text{const}$. Conventionnellement, les quantités droites et gauches sont respectivement notées avec et sans tilde. De plus, le terme \emph{mode} peut désigner l'entier $n$ ou l'opérateur $\alpha_n$. 

Cette discussion s'applique aussi aux modes des fantômes $c_n$, $\widetilde c_n$, $b_n$ et $\widetilde b_n$. \\

La méthode pour décomposer les courants de symétries et en particulier le tenseur énergie-impulsion est semblable à celle que nous venons de présenter. Nous ne nous intéressons ici qu'au tenseur énergie-impulsion et au courant des translations d'espace-cible~\refe{eq:trans}~: 

\eqna{
T(z) = \sum_n L_n z^{-n-2} \nonumber \\
J^\mu(z) = i \sum_{n} \alpha^\mu_n z^{-n-1}
}

Parmi les opérateurs $L_n$, trois ont un statut particulier, $L_{-1}$, $L_0$ et $L_1$. Ils génèrent les transformations conformes complètes, \cad le long de $v(z)=\alpha + \beta z + \gamma z^2$. Ils vérifient une algèbre $SL(2,{\mathbb C})$. Des formules équivalentes sont obtenues pour les courants anti-holomorphes. Outre ces opérateurs particuliers, remarquons l'existence d'une infinité de générateurs $L_n$ de translations sur la surface. Cela fait de cette CFT une théorie intégrable, \cad totalement résoluble. Cette propriété est assurée par l'holomorphicité du tenseur énergie-impulsion. Cette contrainte n'est en général pas vérifiée dans une variété de dimension $d\neq 2$ parce que la conservation de l'énergie-impulsion -- mais aussi des autres courants de symétrie -- ne se réduit pas en d'aussi simples contraintes d'(anti-)holomorphicité. Les algèbres sont calculées à l'aide des OPE suivantes $T J$, $TT$ et $JJ$ : 

\eqna{
&& \oint_{C_0} \di w \oint_{C_w} \di z  \; z^{m+1}w^{n+1} \, T(z) T(w) =  [L_m,L_n] = (m-n)L_{m+n} + \frac{c}{12}(m^3-m)\delta_{m+n} \nonumber \\
&& \oint_{C_0} \di w \oint_{C_w} \di z  \; z^{m}w^{n} \,J^\mu(z) J^\nu(w) = [\alpha^\mu_m,\alpha^\nu_n] = m \delta_{m+n} \eta^{\mu\nu} \nonumber \\ 
&& \oint_{C_0} \di w \oint_{C_w} \di z  \; z^{m+1}w^{n} \,T(z) J^\mu(w) = [L_m,\alpha^\mu_n] = -n \alpha^\mu_{m+n}
}

En première ligne, est identifiée l'algèbre de Virasoro ${\mathfrak{vir}}$ dont $L_{-1}$, $L_0$ et $L_1$ constituent bien une sous-algèbre ${\mathfrak{sl}}(2,{\mathbb C})$. Le générateur $L_0$ est particulier car il est associé aux changements d'échelle rigides. Il mesure donc le poids $h$ d'un opérateur descendant ou primaire -- respectivement $\widetilde L_0$ mesure $\bar h$ -- dans sa décomposition de Laurent suivant son action d'algèbre. L'opérateur $\alpha^\mu_0$ est associé à l'impulsion de la corde, il mesure donc $p^\mu$. Le commutateur $[L_0,\alpha_0^\mu]=0$ indique que les vecteurs propres de $L_0$ se décomposent le long des vecteurs propres de $\alpha_0^\mu$. Le calcul de l' OPE $JX$ montre que l'opérateur position $x^\mu$ vérifie le commutateur habituel en mécanique quantique $[x^\mu,p^\nu]=i \eta^{\mu\nu}$. 

L'algèbre des opérateurs $\alpha_n^\mu$ est une algèbre d'opérateurs de création et d'annihilation, ce qui est clair en les redéfinissant par $\widehat \alpha_n^{\mu}=\alpha_n^{\mu}/\sqrt n$. Les seuls commutateurs non nuls sont alors~:

\eqna{
[\widehat \alpha_m^{\mu},\widehat \alpha_{-m}^{\nu}] = \eta^{\mu\nu}
} 

Par convention, les $\alpha_{n>0}$ sont choisis en tant qu'opérateurs d'annihilation et les $\alpha_{-n}=(\alpha_{n>0})^\dagger$ en tant qu'opérateurs de création. Par convention, $p^\mu$ est créateur et $x^\mu$ annihilateur. Partant d'un vide $\ket{0}\otimes\ket{\bar 0}$ par application récursive des opérateurs de création et d'annihilation, un espace de Fock est décrit. Ce vide est défini par les actions : 

\eqna{
\alpha_{n\geq 0}\ket{0}=0 \qquad \widetilde \alpha_{n \geq 0}\ket{\bar 0}=0 
}

Les vecteurs de cet espace sont caractérisés par les valeurs propres de $\alpha_0$ et de $L_0$, par conséquent notés $\ket{h,\vec k}$. D'après la signature minkowskienne de la métrique, l'espace de Fock n'est pas bien défini à cause des états de normes négatives. Cependant, ces états découplent par invariance BRST. Par conséquent, si on ne s'intéresse qu'aux états physiques, il est suffisant de ne s'attarder que sur ces états sur la couche de masse et dans la jauge du cône de lumière, \cad $X^0$ et $X^1$ fixés donc gelés. Cependant, ce n'est pas une jauge manifestement invariante de Lorentz, donc en général, $X^0$ et $X^1$ sont laissés libres et la contrainte de couche de masse est imposée implicitement. \\

La même algèbre -- anti-commutante –- est développée pour les champs fantômes~: 

\eqna{
&& \{ c_n,b_m \} = \delta_{n+m}  \nonumber \\ 
&& [L_0,c_n] = -n c_n \qquad [L_0,b_n]=-n b_n
}

Cette définition donne lieue à l'existence de deux familles de créateurs et d'annihilateurs. La première famille d'annihilateurs est définie par  $c_{n\geq0}$ avec $b_{-n}=(c_{n\geq0})^{\dagger}$ les créateurs correspondants. La deuxième famille est générée par la famille $b_{n\geq0}$ et $c_{-n}=(b_{n\geq0})^{\dagger}$ respectivement annihilateurs et créateurs. Il faut introduire deux vides $\ket{\downarrow} = c_1\ket{0}$ et $\ket{\uparrow}=c_0 c_1\ket{0}$ tels que : 

\eqna{
c_{n>0} \ket{\downarrow}=0 \qquad b_{n>0} \ket{\uparrow}=0 \nonumber \\ 
c_0 \ket{\downarrow} = \ket{\uparrow} \qquad  b_0 \ket{\uparrow} = \ket{\downarrow} \nonumber \\ 
\Rightarrow \quad c_{n\geq 0} \ket{\uparrow} = 0 \qquad b_{n\geq 0} \ket{\downarrow} = 0 
}

Tout ce matériel peut être utilisé pour exprimer le tenseur énergie-impulsion et donc $L_n$, d'après les formules~\refe{eq:trans}. Cependant, il faut définir un ordre normal d'opérateur $\normal{~}$ équivalent à l'ordre normal des champs~$:~:$. Il doit être tel que si l'on évalue un courant dans le vide, le résultat est défini pour tout $z$ et compatible avec la définition du champ primaire du vide correspondant~\refe{eq:vide_phys}. Pour cela, il faut définir $\normal{~}$ tel que dans l'expression résultante, tous les annihilateurs sont à droite et les créateurs à gauche, à une constante additive près.  \\

Le vide fondamental et son adjoint sont choisi conformément à~\refe{eq:vide_phys}, ce qui revient à se placer dans la \emph{jauge de Siegel} $b_0\ket{phys}=0$, \cad : 

\eqna{\label{eq:siegel_gauge}
&& \ket{\Omega}=\ket{0}\otimes\ket{\downarrow}=c_1 \ket{0}\otimes\ket{0}_{gh} \nonumber \\
&& \bra{\Omega}=\bra{0}\otimes\bra{\downarrow}=\bra{0}\otimes \bra{0}_{gh} c_{-1}c_0
}

Et nous obtenons alors pour expression des modes du tenseur énergie-impulsion, la formule suivante~: 

\eqna{
L_m = \frac{1}{2} \sum_{n \in {\mathbb N}} \normal{\alpha^\mu_{m-n} \alpha_{\mu \, n}} + \sum_{n \in {\mathbb N}} (n+m) \normal{b_{m-n}c_n} + \delta_m
}

La constante additive $\delta_m$ doit permettre de retrouver l'identité $\bra{\Omega}L_0\ket{\Omega} = -1$ par comparaison à $\corr{T(z)}=-z^{-2}$. Et on trouve $\delta_m = -\delta_{m,0}$. Enfin, remarquons l'identité suivante~: 

\eqna{
&& L_0 \, \alpha^\mu_{-n}\alpha^\nu_{-m} \ket{\Omega} = (n+m) \alpha^\mu_{-n}\alpha^\nu_{-m} \ket{\Omega} \nonumber \\
&& L_0 \, c_{-n}b_{-m} \ket{\Omega} = (n+m)  c_{-n}b_{-m} \ket{\Omega}
}

En théorie bosonique, tout vecteur propre peut donc être caractérisé par 3 nombres :le poids, l'impulsion et le mode $N$. On écrit $\ket{h,\vec k,N,\downarrow} \otimes \ket{\bar h,\vec k, \widetilde N,\downarrow}$ tel que~:

\begin{align}\label{eq:valeurs_propres} 
&\alpha^\mu_0 \ket{h,\vec k,N,\downarrow} = \ket{h,\vec k,N,\downarrow} k^\mu \nonumber \\
& L_0\ket{h,\vec k,N,\downarrow} = \ket{h,\vec k,N,\downarrow} h \qquad \text{avec } h = \alpha' \frac{k^2}{4} + N -1
\end{align} 

et similairement sur les kets du secteur droit. Il faut en outre imposer au tenseur énergie-impulsion les contraintes de Virasoro. Elles demandent l’annulation du tenseur énergie-impulsion sur la surface de corde et sont une conséquence des équations du mouvement de la métrique de la surface~\cite{Kiritsis:2007zz}. En effet, sur une surface de Riemann, le tenseur d'Einstein-Hilbert vérifie trivialement $G_{ab}=0$ puisque $R_{ab}= R g_{ab}/2$. Alors, $T_{ab} =0$ et cela se traduit en langage opératoriel par~:

\begin{align}
L_{m\geq 0} \ket{phys}=0
\end{align}
 
sur tout état physique $\ket{phys}$. En effet, puisque $\corr T = \bra{phys}\sum L_{-n} z^{n-2}\ket{phys}$ il suffit d'imposer cette contrainte en utilisant $L_n = L^\dagger_{-n}$ par hermiticité du tenseur énergie-impulsion. Les contraintes de Virasoro deviennent donc une condition de \emph{physicité} des états de l'espace de Hilbert. Si l'on retourne à la formule~\refe{eq:valeurs_propres}, cela impose entre autres~:
 
\begin{align}
& h = \alpha' \frac{k^2}{4} + N -1 = 0 \nonumber \\ 
& \bar h = \alpha' \frac{k^2}{4} + \widetilde N -1 = 0
\end{align}

Il en résulte une condition d'identification des niveaux d'excitation $N=\widetilde N$ appelée en anglais \emph{level-matching} et la formule de masse~:

\begin{align}
m^2 = - k^2 = \frac{4}{\alpha'} (N-1)
\end{align}

Le fondamental $N=0$ est clairement tachyonique, c'est une caractéristique de la théorie des cordes bosoniques, qui admet des tachyons de cordes fermées et ouvertes. L'ensemble d'états suivant $N=1$ est non-massif. Ces derniers sont très intéressants car ils admettent parmi eux un champ tensoriel de spin 2~: le graviton. Les modes suivants sont tous massifs. Puisque la limite naturelle de théorie des cordes est $\alpha'= \ell_s^2 \to 0$, les gaps de masse carré $\Delta m^2 = 4/\alpha'$ sont quasi-infinis, de sorte qu'en oubliant le fondamental tachyonique, les cordes massives découplent et ne restent que les cordes non-massives. Ces dernières constituent ainsi la partie du spectre de corde physiquement intéressante.

\subsubsection{Correspondance états-opérateur de vertex~: définition explicite}

La correspondance entre les états et les opérateurs est obtenue en utilisant~\refe{eq:correspondance}. Dans l'hypothèse où les champs peuvent être définis sans ambiguïté, $\partial X(z)$ par exemple, holomorphes réguliers en leur point d'insertion $z$, alors il est possible de relier les opérateurs de création à des opérateurs de vertex par développement en série de Taylor. Ceci permet de définir explicitement la correspondance état-opérateurs. Ainsi, à position fixée elle est donnée par~:

\begin{align}
\ket{k^\mu,\Omega,z} \quad\longrightarrow & \quad :c(z) e^{i k_\mu X^\mu}: \nonumber \\
\alpha^\mu_{-n}\ket{\Omega,z} \quad\longrightarrow &  \quad :c(z) \partial^{n} X(z): \nonumber \\
c_{-n} \ket{\Omega,z} \quad\longrightarrow &  \quad :c(z) \partial^{n+1} c(z): \nonumber \\
b_{-n} \ket{\Omega,z} \quad\longrightarrow &  \quad :c(z) \partial^{n-2} b(z):
\end{align}

La première identité est obtenue par comparaison entre l'action du courant de translation sur l'opérateur de vertex et l'action de $\alpha_0$ sur l'état. Seule la partie holomorphe est présentée ici. Évidemment, les formules équivalentes existent pour la partie non-holomorphe. 

De manière générale, ces opérateurs sont des \emph{descendants}. Seule une minorité d'entre eux sont des opérateurs \emph{primaires}. Ceux-ci correspondent aux \emph{états de plus haut poids} vérifiant par définition $L_{n\geq 0}\ket{\psi}=0$ et contenus dans l'espace de Hilbert ${\mathcal H}$.

Tout état du spectre -- et son opérateur de vertex associé -- est obtenue par combinaison de l'ensemble des modes créateurs sur le ket d'impulsion $k^\mu$. Par la correspondance ainsi donnée, il suffit ensuite de remplacer chaque mode par son opérateur.

\subsection{Spectre physique et contrainte BRST, états et opérateurs de vertex}

La formule de masse précédente permet de décomposer l'ensemble des états de l'espace de Hilbert le long d'un spectre de masse. A chaque niveau de masse, formellement identifié par $N$, il existe un certain nombre d'états physiques, la condition étant que  $L_{m\geq 0} \ket{phys}=0$. En toute rigueur, puisque la jauge a été fixée en introduisant des champs fantômes, il faut plutôt imposer la condition d'invariance BRST $Q_B \ket{phys}=0$. La charge BRST $Q_B$ est définie par l'intégrale de contour~:

\begin{align}
Q_B = \frac{1}{2\pi i} \oint \parent{\di z~j_B(z) -  \di \bar z~\wt \jmath_B(\bar z)}
\end{align}

dont le contour encercle l'opérateur sur lequel la charge agit. La contrainte BRST est formellement similaire à la contrainte d'invariance conforme puisque $CFT \subset BRST$. Mais de nouvelles contraintes liées à la cohomologie BRST ${\mathcal H}_{\text{fermé}}/{\mathcal H}_{exact}$ sont imposées. La contrainte BRST constitue la condition de fermeture et construit l'espace de Hilbert des états fermés BRST ${\mathcal H}_{\text{fermé}}$. Les états exacts BRST sont donnés par $Q_B \ket{\chi}$ avec $\ket{\chi}$ un état quelconque non physique, car pour $c=0$ la nilpotence $Q_B^2=0$ est vérifiée. La nilpotence implique en outre que $Q_B \ket{\chi}$ est un état \emph{nul}. L'espace de Hilbert des états exacts BRST  ${\mathcal H}_{exact}$ est obtenu ainsi. 

La condition de \emph{jauge de Siegel}, brièvement introduite et utilisée précédemment dans la formule~\refe{eq:siegel_gauge}, est obtenue en choisissant l'espace de Hilbert appartenant à la cohomologie BRST. Il s’agit en réalité d’un choix arbitraire du vide fantôme $\ket{\downarrow}$ équivalent à l'inclusion des champs $c \wt c(z)$ dans tout opérateur de vertex à position fixée. Ainsi, le vide $\ket{\uparrow}$ aurait \apriori pu être choisi en tant que vide fondamental. L'étude de la cohomologie BRST montre cependant que $\ket{\uparrow}=Q_B \ket{\downarrow}$ et que $\ket{\uparrow}$ n'appara\^it donc pas comme un bon état de départ pour construire l'espace de Hilbert de la cohomologie. \\

Les deux premiers niveaux sont alors~: \\

\begin{itemize}\itemsep4pt
\item Le tachyon fondamental $\ket{k,0,\downarrow}\otimes \wt{\ket{k,0,\downarrow}} $ qui vérifie simplement $\alpha' m^2=-4$. L'opérateur de vertex correspondant est obtenue par la correspondance introduite précédemment et qui donne à position fixée~: 
\begin{align}
{\mathcal V}_T = c\wt c \, e^{i k_\mu X^\mu}(z)
\end{align}

et intégré simplement~:
\begin{align}
V_T = \int \di^2 z \, e^{i k_\mu X^\mu}(z)
\end{align}

\item Le premier état excité $\ket{k,1,\downarrow}\otimes \wt{\ket{k,1,\downarrow}} $ de masse $m^2=0$. Les deux secteurs droits et gauches sont équivalents. Le détail du secteur gauche est le suivant~:
\begin{align}
\ket{k,1,\downarrow} = \zeta_\mu \, \alpha_{-1}^\mu \ket{k,0,\downarrow}
\end{align}

dont la polarisation $\zeta_\mu$ vérifie la relation $\zeta_\mu k^\mu =0$ mais n'est déterminée qu'à une transformation de jauge près $\zeta_\mu \sim \zeta_\mu + \sqrt{2\alpha'}\, \beta k_\mu$ avec $\beta$ un nombre. Pour l'état complet nous devons plutôt introduire la polarisation $\zeta_{\mu\nu}$ mais le nombre de contraintes est alors double. En tout cela fait $24\times 24$ degrés de liberté.

 L'opérateur de vertex correspondant à cet état est simplement~:
\begin{align}
{\mathcal V}_\zeta = c\wt c \, \zeta_{\mu\nu} \partial X^\mu \bar \partial \wt X^\nu e^{i k_\mu X^\mu}(z)
\end{align}

Par la décomposition $\zeta_{\mu\nu} = h_{(\mu\nu)} + b_{[\mu\nu]} + \eta_{\mu\nu}\Phi$ cet opérateur décrit le \emph{graviton}, tenseur symétrique sans trace, le champ de \emph{Kalb-Ramond} anti-symétrique, et le \emph{dilaton} $\Phi$ scalaire et correspondant ici à la trace.\\
\end{itemize}

\subsection{Amplitudes et OPE}

Cette section et ce chapitre seront terminés en étudiant les amplitudes et leurs relations aux OPE définis précédemment. Une fois la correspondance connue entre les états et les opérateurs de vertex, on peut exprimer les amplitudes à N-points sur la surface de corde, \cad les éléments de matrice-S on-shell. 
Puisqu'une amplitude s'écrit linéairement, il n’est pas possible d’exprimer une amplitude générale à plus de 2 points uniquement à partir d'états asymptotiques de type $\ket{\psi,z}$. Il est donc nécessaire d'exprimer les états sous forme de fonctions d'opérateurs agissant sur les états asymptotiques du vide ou plus généralement sur des états physiques. Les opérateurs de vertex jouent ce rôle. En outre, il y a correspondance entre les opérateurs qui agissent sur l'espace de Hilbert et les champs de la théorie conforme. Cette correspondance est établie par la formule~\refe{eq:corr_op_et} et s'exprime pour les amplitudes, ici à l'ordre des arbres, par~:

\begin{multline}\label{eq:amplitude}
\int \prod_i \di^2 z_i \, \bra{\psi,\infty} T[\hat V(z_1,\bar z_1) \hat V(z_2,\bar z_2)] \ldots \ket{\psi',0} \\ = \int \prod_i \di^2 z_i \, \corr{T[V'_\psi(\infty) V(z_1,\bar z_1) V(z_2,\bar z_2) \ldots V_{\psi'}(0)]}
\end{multline}

avec $T$ l'opérateur d'ordre en temps conforme, \cad qui ordonne dans le plan complexe $|z_1|>|z_2|>\ldots$. Les points $z\to 0$ et $z\to\infty$ dans le plan complexe sont les équivalents conformes des infinis temporels asymptotiques -- sur la sphère, ils correspondent aux pôles sud et nord. Les opérateurs sont des poinçons de la surface aux points d'insertion. On note les opérateurs avec un chapeau. A gauche, les opérateurs $\hat V$ sont des fonctions des opérateurs fondamentaux $(\alpha_n,c_n,b_n,\ldots)$. A droite, les champs $V$ sont des fonctionnelles des champs fondamentaux $(X^\mu,b,c,\ldots)$. Attention l'opérateur $V'_\psi(\infty)$ est défini de manière particulière car, techniquement, il faut subdiviser la sphère en deux hémisphères collés à l'équateur avec certaines conditions de collage conformes~\cite{Polchinski:1998rq,Polchinski:1998rr}. Donc $V'$ est le transformé conforme de $V$ suivant la transformation correspondante à la fonction de transition à l'équateur, qui est $z\to u = 1/z$. \\

L'amplitude sans insertion s'appelle \emph{fonction de partition}, à l'ordre des arbres ici et est notée $Z$. L'amplitude dépend évidemment de la théorie de surface, \cad de l'action de surface $S$ ou de l'hamiltonien $H$ suivant la description choisie et de la géométrie -- par exemple le disque ou la sphère, mais aussi le cylindre ou le tore. En ce sens, la fonction de partition est une donnée importante car elle caractérise la théorie quantique, nue. C'est en particulier le cas de la fonction de partition cylindrique. Nous y reviendrons dans le cadre des supercordes, dans la section~\refcc{sub-sec:GSO}.

Du point de vue des opérateurs, la définition de la théorie, \cad de l'action de surface, modifie les relations de commutations. Du point de vue des champs, elle modifie les formules d'OPE. En général, une amplitude est complètement déterminée par l'expression des OPE. En fait en théorie conforme, la connaissance des OPE à 2 points~\refe{eq:OPE}, \cad les valeurs des fonctions à 3 points, suffit à résoudre entièrement la théorie, donc à calculer toutes les amplitudes. Par exemple, à partir de l'OPE des champs $X^\mu$ en espace plat et dans le plan complexe, l'OPE de N opérateurs $:e^{i k_\mu X^\mu}(z_i,\bar z_i):$ est exactement déterminée~:

\begin{align}\label{eq:OPE_tach_general}
\prod_{i=1}^N :e^{i k^{(i)}_\mu X^\mu}(z_i,\bar z_i): = \prod_{1\leq i<j}^N \module{z_i-z_j}^{\alpha' \frac{k^{(i)}\cdot k^{(j)}}{2}} ~ :\prod_{i=1}^N e^{i k^{(i)}_\mu X^\mu}(z_i,\bar z_i):
\end{align}

Dans ce cas la fonction de corrélation de ce produit donne l'intégrale~:

\begin{align}
\int \prod_i^N \di^2 z_i~\prod_{i=1}^{N-1} \Theta(\module{z_i}-\module{z_{i+1}})~ \prod_{1\leq i<j}^N \module{z_i-z_j}^{\alpha' \frac{k^{(i)} \cdot k^{(j)}}{2}} ~ \corr{ :\prod_{i=1}^N e^{i k^{(i)}_\mu X^\mu}(z_i,\bar z_i):}
\end{align}

La valeur du corrélateur régulier est simplement une fonction delta de Dirac à D dimensions $\delta^{(D)}(\sum_i k^{(i)})$. La fonction $\Theta(x)$ est la fonction th\^eta de Heaviside, valant $1$ pour tout $x>0$ et $0$ sinon. Par la suite, pour obtenir exactement l'amplitude~\refe{eq:amplitude} sur la sphère, il faut d'abord fixer la position de trois opérateurs de sorte de fixer la jauge $SL(2,\mathbb C)$ du groupe des CKV sur la sphère, typiquement sont choisies les coordonnées $(0,1,\infty)$. Puis, il faut appliquer la transformation conforme $z \to 1/z$ à l'opérateur le plus à gauche qui reçoit la position $\infty$. Cela s'exprime finalement\footnote{On renomme les coordonnées $z_i$ de sorte que sont fixées $(z_{1},z_{2},z_{3})=(0,1,\infty)$.} par~:

\begin{align}
\delta^{(D)}(\sum_{i=1}^N k^{(i)}) ~ \int \prod_{i=4}^{N} \di^2 z_i~\prod_{i=4}^{N-1} \Theta(\module{z_i}-\module{z_{i+1}})~\prod_{4\leq i<j}^{N} \module{z_i-z_j}^{\alpha' \frac{k^{(i)} \cdot k^{(j)}}{2}} \prod_{i=4}^{N}\module{1-z_i}^{\alpha' \frac{k^{(2)} \cdot k^{(j)}}{2}}
\end{align}

Les intégrales finales s'expriment en générale à l'aide de fonctions $\Gamma$ et constituent des amplitudes de Veneziano. La formule présentée ci-dessus est la fonction de corrélation de N tachyons de cordes fermées en théorie bosonique. La formule d'OPE~\refe{eq:OPE_tach_general} sera réutilisée dans les calculs de la fonction de partition du système brane-antibrane séparé. \\

Nous passerons maintenant à la définition de la théorie de supercordes et des théories superconformes.

\section{Supercordes et th\'eorie superconforme}
\label{sec:SCFT}

En théorie bosonique, le spectre de particules ne contient que des bosons et ne peut donc pas correspondre à une physique réaliste qui contient, comme nous le savons bien par expérience, des particules fermioniques. En fait, il est possible de construire des spineurs d'espace-cible~ \cite{Polyakov:1981re, Polchinski:1998rr, Kiritsis:2007zz, Friedan:1985ge, Green:1987sp} en théorie des cordes. Nous allons nous concentrer sur le formalisme des cordes RNS, qui introduit les fermions dans la surface de corde. Mentionnons cependant qu'il existe aussi un formalisme dans lequel on introduit les fermions directement dans l'espace-cible~ : les cordes de Green-Schwarz. Nous n'aborderons pas leur sujet ici car nous n'en ferons pas usage, mais je renvois aux ouvrages~ \cite{Green:1987sp,Green:1987mn}. Dans le formalisme RNS, il faut ajouter des degrés internes fermioniques $(\psi_a^\mu,\wt \psi_b^\mu)$ en nombre \apriori arbitraire sur la surface de corde, \cad avec $a=1 \ldots N_L$ et $b=1\ldots N_R$. Avec les bosons $(X_L^\mu,X^\mu_R)$ ils forment sur la surface de cordes une théorie supersymétrique $N=(N_L,N_R)$. En réalité, on devrait plutôt parler de théorie de supergravité à 2 dimensions, car les transformations de supersymétrie sont définies localement sur la surface.

Nous obtenons ainsi la théorie de \emph{supercordes} qui est une théorie de particules bosoniques et fermioniques vérifiant éventuellement une ou plusieurs supersymétries d'espace-cible. On distingue 5 théories de supercordes de dimension $10$~ : types IIA et IIB qui sont des supergravités de type II donc ${\mathcal N}=2$~ ; type I, une supergravité de type I donc ${\mathcal N}=1$~ ; et enfin deux théories hétérotiques définies sur les groupes $SO(32)$ ou $E_8\times E_8$. Ces deux dernières sont obtenues en imposant une supersymétrie de surface $N =(0,2)$ et les 3 premières sont construites de sorte qu'elles vérifient une supersymétrie de surface $N=(1,1)$. Nous n'étudierons dans la suite que les types IIA et IIB car il s'agit du cadre de mon travail de thèse. Les cordes de type I sont non-orientées et couplent naturellement à des cordes ouvertes non orientées, tandis que les types II sont orientées et \apriori des théories de cordes fermées exclusivement -- mais ça n'est pas tout à fait le cas comme nous le verrons.

\subsection{Action et symétrie superconforme}

Pour construire les théories de type II en formalisme RNS, il faut introduire l'action suivante sur le plan complexe directement, selon les conventions de Polchinski~\cite{Polchinski:1998rr}~:

\begin{align}
S_{II} = \frac{1}{2\pi \alpha'} \int_{\mathbb C} \di^2 z ~ \parent{\partial X^\mu \bar \partial X_\mu + \frac{\alpha'}{2} \psi^\mu \bar \partial \psi_\mu + \frac{\alpha'}{2} \widetilde \psi^\mu \bar \partial \widetilde\psi_\mu}
\end{align} 

Il s'agit d'une théorie conforme, en l'occurrence libre puisqu'il n'y a pas de couplages, de charge centrale holomorphe $c=3D/2$ et antiholomorphe $\wt c = 3D/2$. Les bosons sont chacun de charge $(c,\wt c)=(1,1)$ et les fermions holomorphes (ou gauches) $\psi^\mu$ sont de charge $(1/2,0)$ tandis que les fermions antiholomorphes (ou droits) $\wt \psi$ de charge $(0,1/2)$. Pour simplifier, plaçons nous maintenant dans minkowski $G_{\mu\nu}=\eta_{\mu\nu}$. Nous reviendrons plus tard sur des configurations plus générales dans la discussion sur le modèle sigma.

\subsubsection{Champs de matière}

Le tenseur énergie-impulsion total est ais\'ement obtenu~ :

\begin{align}\label{eq:tenseur_conforme}
T(z) = - \frac{1}{\alpha'} \partial X^\mu \partial X_\mu - \frac{1}{2} \psi^\mu \partial \psi_\mu
\end{align}

Les fermions vérifient les équations du mouvement suivantes~ :

\begin{align}
\partial \wt \psi^\mu = 0 \nonumber \\ 
\bar \partial \psi^\mu = 0
\end{align}

De sorte que $\psi$ est holomorphe et $\wt \psi$ anti-holomorphe. Une caractéristique essentielle de cette théorie est la supersymétrie de surface. Elle s'exprime par les transformations suivantes~ :

\begin{align}
&\delta X^\mu(z,\bar z) = \sqrt{\frac{\alpha'}{2}} \parent{-\eta(z) \psi^\mu(z)-\eta(z)^* \wt \psi^\mu(\bar z)} \nonumber \\ 
&\delta \psi^\mu(z) = \sqrt{\frac{2}{\alpha'}} \, \eta(z) \partial X^\mu(z)  \nonumber \\ 
&\delta \wt\psi^\mu(\bar z) = \sqrt{\frac{2}{\alpha'}} \, \eta(z)^* \bar \partial X^\mu(\bar z)  
\end{align}

avec $\eta(z)$ un paramètre fermionique donc anti-commutant. Elles se nomment \emph{transformations superconformes} et sont générées par le courant de supersymétrie~ :

\begin{align}\label{eq:tenseur_superconforme}
& j_\eta(z) = \eta(z) T_F(z) \qquad \text{et}\qquad  \wt \jmath_\eta(\bar z) = \eta(z)^* \wt T_F(\bar z) \nonumber \\
& T_F = i \sqrt\frac{2}{\alpha'} \psi^\mu \partial X_\mu \qquad \text{et}\qquad \wt T_F = i \sqrt\frac{2}{\alpha'} \wt \psi^\mu \bar \partial X_\mu 
\end{align}

L'ensemble fermé constitué des transformations conformes et superconformes forment l'\emph{algèbre superconforme} sachant que le commutateur de deux courants superconformes donne un courant conforme. \\

Les OPE fondamentales des champs sont dérivées à partir de l'action et sont~ :

\begin{align}\label{eq:X_psi}
X^\mu(z,\bar z) X^\nu(w,\bar w) &\sim - \frac{\alpha'}{2} \eta^{\mu\nu}\ln \module{z-w}^2 \nonumber \\ 
\psi^\mu(z)\psi^\nu(w) & \sim \frac{\eta^{\mu\nu}}{z-w} \nonumber \\ 
\wt\psi^\mu(\bar z)\wt\psi^\nu(\bar w) & \sim \frac{\eta^{\mu\nu}}{\bar z-\bar w}
\end{align}

Desquels l'algèbre superconforme suivante est d\'eduite~ :

\begin{align}
T(z)T(w) &\sim \frac{3(d+1)/4}{(z-w)^4} + \frac{2}{(z-w)^2} T(w) + \frac{1}{z-w} \partial T(w) \nonumber \\ 
T(z)T_F(w) &\sim \frac{3/2}{(z-w)^2} T_F(w) + \frac{1}{z-w} \partial T_F(w) \nonumber \\ 
T_F(z)T_F(w) &\sim \frac{d+1}{(z-w)^3} + \frac{2}{z-w} T(w)
\end{align}

Nous identifions donc $c=3(d+1)/2$. Une théorie des champs vérifiant cette symétrie superconforme est appelée SCFT, \cad \emph{super-CFT}. Ici le nombre de supersymétries de surface en fonction des secteurs gauche et droit est $N=(1,1)$. Nous pourrions cependant ajouter plus de supersymétries, de sorte que $T_F \to T_F^a$ avec $a=1\ldots N$. Nous ne calculerons pas explicitement les transformations conformes de tous les champs. Mentionnons simplement que $\psi$ est de poids $(1/2,0)$ et $\wt\psi$ de poids $(0,1/2)$ tandis qu'ils participent à la charge centrale de $c=\wt c =1/2$. Pour les champs $X$ rien ne change par rapport à la section précédente. 

\subsubsection{Fantomes et super-fantomes}

Nous ne devons pas oublier qu'il faut ajouter des fantômes une fois que la jauge de la symétrie SCFT est fixée par la méthode de Fadeev-Popov. Ici en revanche, nous devrons ajouter une nouvelle paire de fantômes $(\beta,\gamma)$ bosoniques et commutants, aux fantômes déjà introduits dans la théorie bosonique. Pour ces champs, nous construisons le tenseur énergie-impulsion et le générateur super-conforme~ :

\begin{align}\label{eq:tenseurs_ghost}
T^{gh} &= T_{bc} - \frac{\gamma \partial \beta}{2} - \frac{3}{2}  \beta\partial \gamma \nonumber \\
T_F^{gh} &= -i \parent{c \partial \beta - \frac{\gamma b}{2} + \frac{3}{2} \partial c \beta} 
\end{align}

avec $T_{bc}$ le tenseur énergie-impulsion des fantômes $(b,c)$. Nous n'expliciterons pas ici les transformations conformes de tous ces champs, mais nous mentionnerons simplement que $(\beta,\gamma)$ sont holomorphes de poids $(3/2,-1/2)$. Leur contribution totale à la charge centrale est $c=11$ tandis que les fantômes $(b,c)$ contribuent de $c=-26$. \\ 

Comme nous l'avons fait précédemment, nous pouvons maintenant construire un courant BRST. Il est ici donné par la combinaison~ :

\begin{align}
j_B = \gamma T_F^m + c T^m + \frac{1}{2} \parent{c T^{gh} + \gamma T_F^{gh}} 
\end{align}

et son équivalent anti-holomorphe. Ultimement c'est ce qu'il faudra appliquer aux opérateurs de vertex une fois la jauge fixée dans l'amplitude pour en faire des opérateurs physiques, \cad invariants superconformes.

\subsubsection{Supercordes critiques et amplitude de Polyakov}

La limite de théorie critique est atteinte quand l'amplitude est exactement invariante conforme, en particulier au niveau quantique. L'anomalie étant proportionnelle à la charge centrale, il faut imposer $c_{tot} =0$, soit ici~ :

\begin{align}
c_{tot} &=\frac{3(d+1)}{2} +11 -26 \nonumber \\ &= \frac{3(d +1 - 10)}{2} = 0
\end{align}

Ceci impose $d+1=10$ comme nous l'avancions dans la section bosonique. L'amplitude de Polyakov en théorie critique à jauge fixée est finalement~ :

\eqali{\label{eq:ampl_pol2}
\corr{\prod_{\alpha=1}^N V_{\alpha} (k_{\alpha}^\mu) }_{\Sigma} = \int\frac{[d\omega]}{\Omega(SCKV)}  \int {\mathcal D}b{\mathcal D}c{\mathcal D}\beta{\mathcal D}\gamma \int {\mathcal D}X {\mathcal D}\psi{\mathcal D}\wt\psi \, e^{-S_p[{}^\omega g]-S_{gh}} \prod_{\alpha=1}^N V_{\alpha} (k_{\alpha}^\mu) 
}

en divisant par le groupe des \emph{super}-CKV\footnote{La g\'en\'eralisation supersym\'etrique des vecteurs de Killing conformes (CKV) introduit dans la section~\refcc{sec:CFT_bos}.}. Il est équivalent de fixer les positions d'autant d'opérateurs que de SCKV mais la procédure exacte nécessite d'introduire la bosonisation et les \emph{picture} que nous ne verrons pas ici mais que la littérature décrit déjà abondamment -- voir par exemple~\cite{Friedan:1985ge, Polchinski:1998rr}. Les opérateurs de vertex ont des formes plutôt complexes s'ils sont décomposés sur les champs $X$ et $\psi$ mais nous les introduirons plutôt dans le formalisme du super-espace où ils sont plus aisés à définir.

\subsection{Modes et états asymptotiques}

\subsubsection{Secteurs R et NS et séries de Laurent}

Afin d'introduire les modes comme en théorie bosonique, nous pouvons simplement développer les champs holomorphes et anti-holomorphes en série de Laurent. Ce ne sera cependant pas aussi direct ici car nous devons distinguer deux types de secteurs. En effet, puisque toute observable est bosonique, l'expression est nécessairement paire dans les fermions. Ainsi deux types de conditions périodiques sur le plan complexe peuvent \^etre impos\'ees~ :

\begin{align}
\text{Neveu-Schwarz (NS)}~: \qquad &\psi^\mu (e^{i 2\pi }z) = \psi^\mu (z)  \nonumber \\ 
\text{Ramond (R)}~: \qquad &\psi^\mu (e^{i 2\pi }z) = -\psi^\mu (z)
\end{align}

Dans la description du cylindre, le secteur NS est anti-périodique par rotation complète autour de la direction compacte et le secteur R périodique. Le long de ces secteurs les séries de Laurent sont~ :

\begin{align}
\text{NS}~: \qquad &\psi^\mu (z) = \sum_{r \in \mathbb Z +\frac{1}{2}}\psi^\mu_{r} \, z^{-r-1/2} \nonumber \\ 
\text{R}~: \qquad &\psi^\mu (z)= \sum_{r \in \mathbb Z } \psi^\mu_{r} \, z^{-r-1/2}
\end{align}

Et de même dans le secteur anti-holomorphe en remplaçant $\psi$ par $\wt \psi$ et $z$ par $\bar z$. Il y a une coupure dans le secteur R ce qui est important pour les relations d'anti-commutation des opérateurs. Sans rentrer dans les d\'etails elles sont donn\'ees par~ : 

\begin{align}
\acomm{\psi^\mu_r}{\psi^\nu_s} =\{\wt\psi^\mu_r,\wt\psi^\nu_s\} = \eta^{\mu\nu}\delta_{r,-s}
\end{align}

La correspondance entre les modes et les opérateurs est permise par les identités intégrales suivantes~ :

\begin{align}\label{eq:correspondance_susy}
\psi_r &= \frac{1}{2\pi i}\int \di z \, z^{r-1/2} \psi(z) \nonumber \\ 
\wt \psi_r &= \frac{1}{2\pi i} \int \di z \, z^{r-1/2} \wt \psi(z)
\end{align}

Dans la limite où chaque champ peut \^etre d\'efini correctement en un point donné en extrayant les divergences des séries de Laurent par une généralisation de l'ordre normal, les modes correspondent précisément à des opérateurs de vertex réguliers en leur point d'insertion. Par exemple pour un champ NS~ :

\begin{align}
\psi_{-1/2 - r} \longrightarrow \frac{1}{r!} \partial^r \psi(0)
\end{align} 

Pour ce qui est des champs R, il n'existe pas de relation aussi simple à cause de la coupure dans l'intégrande ($r$ est entier). Il faut en fait définir des opérateurs de spin, que nous n'introduirons pas ici, mais dont nous mentionnons simplement l'existence. Les décrire nécessiterait d'introduire la bosonisation et ce sont des considérations qui sont en dehors du sujet de la présente thèse. \\

Les champs fantômes superconformes $(\beta,\gamma)$ étant obtenus par supersymétrie à partir des fantômes $(b,c)$ et le générateur de supersymétrie \'etant un spineur de surface vérifiant lui-même les conditions NS ou R, nous avons que le développement en série de Laurent de ces fantômes est~ :

\begin{align}
&\beta(z)=\sum_{r \in \mathbb Z + \nu} \beta_r \, z^{-r-3/2} \nonumber \\ 
& \gamma(z)=\sum_{r \in \mathbb Z + \nu} \gamma_r \, z^{-r+1/2}
\end{align}

avec $\nu=0$ dans secteur NS et $\nu=1/2$ dans R. Ils vérifient le commutateur $\comm{\gamma_r}{\beta_s}=\delta_{r,-s}$. En terme de tous ces opérateurs, les expressions des opérateurs du tenseur énergie-impulsion et du tenseur super-conforme se développent suivant~ :

\begin{align}
& T_F(z) = \sum_{r \in \mathbb Z +\nu} G_r \, z^{-r-3/2} \nonumber \\ 
& T(z) = \sum_n L_n \, z^{-n-2}
\end{align}

dont l'algèbre complète est donnée pour $c=0$ par~ :

\begin{align}
\comm{L_m}{L_n} &= (m-n)L_{m+n} \nonumber \\ 
\acomm{G_r}{G_s} &= 2 L_{r+s} \nonumber \\ 
\comm{L_m}{G_r} &= \frac{m+2r}{2} G_{m+r}
\end{align}

A partir des expressions ~\refe{eq:tenseur_conforme}, ~\refe{eq:tenseur_superconforme} et ~\refe{eq:tenseurs_ghost} nous obtenons~ :

\begin{align}
&L_m = L_m^{mat} + L_m^{gh} + a \, \delta_{m,0} \nonumber \\ 
& \quad L_m^{mat} = \frac{1}{2} \sum_{n \in \mathbb Z}\normal{\alpha_{m-n}^\mu \alpha_{\mu,n}} + \frac{1}{4}\sum_{r \in \mathbb Z + \nu}(2r-m)\normal{\psi^\mu_{m-r} \psi_{\mu,r}} \nonumber \\ 
& \quad L_m^{gh} = \sum_{n \in \mathbb Z}(m+n)\normal{b_{m-n}c_n} + \sum_{r \in \mathbb Z + \nu} \frac{1}{2}(m-2r)\normal{\beta_{m-r}\gamma_r} \\
&G_r = G_r^{mat} + G_r^{gh}  \nonumber \\ 
&\quad G_r^{mat} = \sum_{n \in \mathbb Z} \alpha_n^\mu \psi_{\mu,r} \nonumber \\ 
&\quad G_r^{gh} = -\sum_{n \in \mathbb Z}\parent{\frac{1}{2}(2r+n)\beta_{r-n}c_n + 2 b_n\gamma_{r-n}}
\end{align}

avec $a=-1/2$ dans le secteur NS et $a=0$ dans secteur R. 

\subsubsection{Constructions des états physiques}

Les états physiques construits dans la théorie \`a jauge fix\'ee doivent être tels qu'ils sont invariants BRST. Ils sont obtenus par application des divers modes que nous venons d'introduire sur le vide NS ou R et doivent correspondre bijectivement à des opérateurs de vertex éventuellement primaires ou descendants de manière plus générale, par la correspondance état-opérateur que nous avons déjà vu dans la théorie bosonique. La contrainte d'invariance BRST s'exprime par la formule de fermeture~ :

\begin{align}
Q_B \ket{phys} = 0 
\end{align}

et puisque $Q_B^2 =0$ quand $c=0$ nous savons que tout état physique doit appartenir à la cohomologie BRST $\sim {\mathcal H}_\text{fermé}/{\mathcal H}_{\text{exact}}$ où un état exact est simplement $\ket{phys'} = Q_B \ket\chi$. Les vides NS et R sont tels que~ :

\begin{align}
&L_0 \ket{0}_{NS} = -\, \frac{1}{2}\ket{0}_{NS} \qquad L_0 \ket{0}_{R} = 0 \nonumber \\
&\psi^{NS}_{r>0} \ket{0}_{NS} = 0 \qquad \psi^R_{r>0} \ket{0}_R = 0 \qquad \psi^R_{0} \ket{0}_R =  \ket{0'}_R
\end{align}

Les modes zéros du secteur R $\psi_0^\mu$ jouent le rôle de matrices $\gamma^\mu$ en 10 dimensions et forment les générateurs d'une algèbre de Clifford $\acomm{\psi^\mu_0}{\psi^\nu_0} \propto \eta^{\mu\nu}$. Les vides $\ket{0}_R$ en sont une représentation spinorielle de dimension 32 et notée ${\bf 32}$ réductible en deux représentations de Weyl $\bf 16 + \bf 16'$ suivant la valeur propre de l'opérateur de chiralité, respectivement $+1$ pour l'une et $-1$ pour l'autre. Cet opérateur de chiralit\'e est ici noté $(-1)^F$ dans le secteur gauche et $(-1)^{\overline F}$ dans le secteur droit. L'opérateur $F$ nommé \emph{nombre fermionique} anti-commute avec tout fermion de surface de corde, y compris les fermions fantomes et ceux du secteur NS. 

Le vide NS contient une contribution de fantômes telle qu'il vérifie~ :
\begin{align}
(-1)^F\ket{0}_{NS} = -\ket{0}_{NS}
\end{align}

La construction des états et des opérateurs est assez similaire au cas bosonique, donc nous ne développerons pas la méthode. Il est plus intéressant de comprendre comment apparaissent les différentes théories de type II, par exemple, en introduisant la fonction de partition et les projections GSO. \\

En effet, par les valeurs propres de l'opérateur de chiralité et par la décomposition RNS sur chaque secteur droit et gauche, quatre secteurs se distinguent pour les cordes fermées ainsi que quatre jeux de valeurs propres de chiralité~ :

\begin{align}\label{eq:comb_sect}
\NS\pm \NS\pm \qquad \NS\pm \R\pm \qquad \R\pm \NS\pm \qquad \R\pm \R\pm
\end{align}

Les secteurs NS-NS et R-R sont bosoniques -- un bi-spineur est un boson tenseur -- et les secteurs NS-R et R-NS spinoriels donc fermioniques. Le spectre développé dans ces secteurs peut être réduit en classes supersymétriques et non-supersymétriques, comme nous allons maintenant le voir en introduisant la projection GSO.

\subsubsection{super-espace et opérateurs de vertex}

Mais avant cela, étudions brièvement le super-espace et comment y définir une action de surface de corde et des opérateurs de vertex. Le super-espace $N=(1,1)$ est défini par le couple de coordonnées $(z,\bar z,\theta,\bar \theta)$ avec $(\theta,\bar \theta)$ des variables de Grassmann anti-commutantes. Nous y définissons un superchamp $\Phi$ par développement de Taylor~ :

\begin{align}
\Phi(z,\bar z)=\phi(z,\bar z) + \theta \psi(z,\bar z) + \bar \theta \wt \psi(z,\bar z) + \theta \bar \theta F(z,\bar z)
\end{align}

où les champs $\phi$, $\psi$, $\wt \psi$ et $F$ sont superpartenaires par action de supersymétrie. En général, $F$ est champ auxiliaire et doit s'annuler sur son équation du mouvement. Ainsi, nous introduisons les champs bosoniques et fermioniques superpartenaires dans un seul et même champ $\mathbb X$ tel que~ :
\begin{align}
\mathbb X(z,\bar z) =  X(z,\bar z) + i \sqrt{\frac{\alpha'}{2}} \theta \psi(z) + i \sqrt{\frac{\alpha'}{2}} \bar \theta \wt \psi(\bar z) + \theta \bar \theta F(z,\bar z)
\end{align}

En introduisant la super-dérivée $D =\partial_\theta + \theta\partial$ et la super-intégrale $\int \di^2 z \di^2 \theta$ qui s'appliquent selon les conventions habituelles, l'action correspondante, toujours dans Minkowski, peut s'\'ecrire~ :

\begin{multline}
S_{super} = \frac{1}{2\pi\alpha'}\int \di^2 z \di^2 \theta ~ D\mathbb X^\mu \bar D\mathbb X_\mu \\ 
=  \frac{1}{2\pi\alpha'}\int \di^2 z ~ \parent{\partial X^\mu \bar \partial  X_\mu + \frac{\alpha'}{2} \psi^\mu \bar \partial \psi_\mu + \frac{\alpha'}{2} \wt\psi^\mu \partial \wt \psi_\mu + \frac{\alpha'}{2} F^2}
\end{multline} 

Nous avons décomposé les champs dans la seconde ligne et appliqué la dérivée et l'intégrale. Le champ auxiliaire s'intègre trivialement et disparaît simplement sans conséquence pour redonner l'action originelle. En général, la convention $\alpha'=2$ est choisie, de sorte que le facteur $\sqrt{2/\alpha'}$ disparaît, ce qui est souvent plus commode à utiliser. Cependant nous utiliserons plutôt dans la suite $\alpha'=1$ qui est une convention plus habituelle et commune avec la théorie bosonique. \\

Les OPE du champ $\mathbb X$ sont directement dans le super-plan complexe~ :
 
\begin{align}
\mathbb X^\mu(z,\theta) \mathbb X^\nu(w,\theta') = - \frac{\eta^{\mu\nu}}{2} \ln\module{z-w - \theta \theta'}^2
\end{align} 

En développant les membres de gauche et droite on retrouve les formules explicites des OPE des champs $X$ et $\psi$ données en début de section dans la formule~\refe{eq:X_psi}. \\

Dans ce formalisme, les expressions des divers opérateurs de vertex \emph{intégrés}, correspondant aux champs du secteur NS, sont ais\'ement obtenues. Dans le secteur R, la construction est plus délicate car il faut introduire des champs de spin, ce que nous ne ferons pas ici car nous n'en aurons pas spécifiquement besoin, mais nous recommandons le lecteur vers~\cite{Polchinski:1998rr}. \\

\begin{itemize}\itemsep4pt
\item Le tachyon NS est exprimé par l'opérateur de vertex intégré dans le super-espace~ :
\begin{align}
{\mathcal V}_T = \int \di^2z \di^2\theta~e^{ik_\mu {\mathbb X}^\mu}
\end{align}

Il doit vérifier la condition de masse $\alpha' m^2 = - k^2 = -2$ qui apparaît naturellement en demandant que l'opérateur complet intégré soit invariant conforme. Le poids de l'opérateur exponentiel est $(h,\bar h)=(k^2,k^2)/4$ et celui de $\theta$ et $\bar\theta$ respectivement $(-1/2,0)$ et $(0,-1/2)$. 
\item Le premier état excité NS est donné par l'opérateur de vertex suivant~ :

\begin{align}
{\mathcal V}_\zeta = \int \di^2z \di^2\theta~\zeta_{\mu\nu}\,D\mathbb X^\mu \bar D \mathbb X^\nu \, e^{ik_\mu {\mathbb X}^\mu}
\end{align}

de masse $m^2=0$. On l'identifie naturellement au graviton, Kalb-Ramond et dilaton comme dans la théorie bosonique, sachant que la polarisation tensorielle $\zeta_{\mu\nu}$ vérifie également des conditions d'orthogonalité avec l'impulsion $k^\mu \zeta_{\mu\nu}=0$ ce qu'on montre en étudiant en détail l'invariance BRST des états eux-mêmes. \\
\end{itemize}

En développant ces opérateurs et en intégrant sur les coordonnées grassmannienne, une formule complètement décomposée sur les champs $X$ et $\psi$, mais aussi $F$ en général, est obtenue. Leur formule explicite est relativement complexe et longue, d'où l'intérêt du formalisme de super-espace.

\subsection{Construction des théories de type IIA et IIB~: Fonction de partition et projection GSO}
\label{sub-sec:GSO}

Une projection permet de r\'eduire le spectre en deux classes\footnote{Il ne s'agit cependant pas d'un partitionnement.} cohérentes IIA et IIB\footnote{Par d'autres projections, on trouve aussi IIA' et IIB' identiques à IIA et IIB, mais aussi OA et OB qui ne sont pas supersymétriques.}. Dans chacune de ces classes, la supersymétrie doit permettre de supprimer l'énergie du vide. Elles sont données explicitement par~ :

\begin{align}\label{eq:comb_II}
&\IIA~:~\NS+ \NS+ \qquad  \R+ \NS+ \qquad \NS+ \R- \qquad \R+ \R- \nonumber \\ 
&\IIB~:~\NS+ \NS+ \qquad  \R+ \NS+ \qquad \NS+ \R+ \qquad \R+ \R+
\end{align}

En théorie des cordes fermées, l'énergie du vide est calculée au premier ordre à partir de la fonction de partition à une boucle dans le vide par la formule $F= k T \ln Z$, avec $T$ la température et $k$ la constante de Boltzmann, et $Z=Z_{S^2} + Z_{T^2}$. La fonction de partition est normée de telle sorte que $Z_{S^2}=1$ et par conséquent $F=k T Z_{T^2}$. L'énergie du vide est ainsi reliée au premier ordre au diagramme de corde toroïdal. 

Comme nous l'avions vu dans la section~\refcc{sec:pres_trans}, il existe plusieurs classes de tores représentées par un nombre complexe, le module $\tau=\tau_1+i\tau_2$ dont les valeurs sont prises sur la variété du groupe $SL(2,\mathbb C)$. En outre, puisque le tore est une boucle, après dérivation d'un spectre complet explorant un espace de Hilbert ${\mathcal H}_{tot}$ nous pouvons exprimer la fonction de partition comme la trace de l'opérateur de translation sur la surface de corde paramétrée par le module. 

\begin{align}
Z_{T^2} = \tr_{{\mathcal H}_{tot}} e^{2\pi i \tau_1 P - 2\pi \tau_2 H}
\end{align}

Les opérateurs de translation $H=L_0+\overline{L_0}$ et $P=L_0-\overline{L_0}$ respectivement \emph{hamiltonien} et \emph{spin} sont dérivés du tenseur énergie-impulsion sur la sphère, car ce sont des opérateurs locaux qui ne dépendent pas de la topologie mais uniquement de la théorie des champs locale. Pour calculer le diagramme du tore, il faut cependant se replacer, par transformation conforme, dans la description cylindrique de la corde, à la différence de la description faite sur le plan complexe. Par conséquent il faut transformer $H$ et $P$ qui ne sont en g\'en\'eral pas des opérateurs primaires, sauf si $(c,\overline c)=0$. Dans le cas contraire, la transformation exacte donne $H\to H - \frac{c+\overline c}{24}$ et $P \to P - \frac{c-\overline c}{24}$, soit~ :

\begin{align}\label{eq:fonc_part_tore}
Z_T(\tau) &= \tr e^{2\pi i \tau_1 P -2\pi \tau_2 H } \nonumber \\ 
&= (q\widetilde q)^{-c/24} \tr q^{L_0} \widetilde q^{\overline{L_0}}
\end{align}

La contrainte $c=\overline c$ est imposer pour l'absence d'anomalie gravitationnelle. Nous avons introduit la notation $q=e^{2\pi i \tau}$ et $\widetilde q =e^{-2\pi i \overline\tau}$.

Maintenant, par application de $H$ et $P$ dans un fond trivial, les secteurs fondamentaux $X^\mu$ ($c=1$) mais aussi $\psi^\mu$ ($c=1/2$), avec $\mu = 0 \ldots d$, ainsi que tous les champs fantômes, ne se mélangent pas entre eux. C'est une conséquence de la séparation du tenseur énergie-impulsion en chacune de ces contributions et de l'absence de mélange par OPE. Il faut comprendre que chaque secteur $\mu$ est en fait une représentation irréductible de l'algèbre de Virasoro. En revanche, ils peuvent se mélanger entre eux par des transformations externes, par exemple groupe de Poincaré ou supersymétrie. A l'inverse chaque secteur peut vérifier une symétrie interne supplémentaire le long de laquelle la représentation de Virasoro est réductible, nous en avons vu un exemple dans l'étude du tachyon par Sen, ce qu'on appelle les algèbres de courant. 

D'après sa forme, nous pouvons factoriser la fonction de partition~\refe{eq:fonc_part_tore} sur autant de secteurs existants. Nous comptons $d$ secteurs $\mu$ et 1 secteur de champs fantômes supersymétriques $\{c,b,\beta,\gamma\}$, soit~ :

\begin{multline}
Z_T(\tau) = (q\widetilde q)^{(10-d)/16} \parent{\tr_{X} \, q^{L_0} \widetilde q^{\overline{L_0}}}^d \times \parent{\tr_{\psi} \,(-1)^{{\bf F}} q^{L_0} \widetilde q^{\overline{L_0}}}^d \\ \times \parent{\tr_{b,c}  \, (-1)^{{\bf F}} q^{L_0} \widetilde q^{\overline{L_0}}} \times \parent{\tr_{\beta,\gamma} \, q^{L_0} \widetilde q^{\overline{L_0}}}
\end{multline}

Nous avons inclus l'opérateur de nombre fermionique d'espace-cible $\bf F$, nécessaire pour rétablir la périodicité imposée par la géométrie du tore. La contribution des fantômes compense exactement deux facteurs bosoniques et fermioniques, de sorte que $d \to d-2$. Remarquons en outre que la charge centrale est proportionnelle à $d-10$ et par conséquent ne s'annule qu'à condition que l'espace ait 10 dimensions. 

Afin qu'elle soit bien d\'efinie, cette fonction doit en outre vérifier l'invariance par transformation modulaire $\tau\to\tau+1$ et $\tau\to-1/\tau$ qui conserve le tore. Or toutes les combinaisons~\refe{eq:comb_sect} ne vérifient pas cette invariance, seulement celles que nous avons cité, dont surtout~\refe{eq:comb_II}, le font. La \emph{projection GSO} est introduite pour les théories IIA et IIB telles que dans la première les secteurs R droits sont de chiralité $-1$ et tous les autres $+1$ ; dans la deuxième tous les secteurs doivent \^etre de chiralité $+1$. Du point de vue de la trace sur les secteurs $\psi$ et $\overline \psi$, nous avons donc~ :

\begin{align}
\IIA~:~ \tr_{\psi,\overline \psi} ~ \hat{P}_{GSO}\,(-1)^{{\bf F}} q^{L_0} \widetilde q^{\overline{L_0}} =& \parent{\tr_{NS} \,\frac{1+(-1)^F}{2} q^{L_0}  - \tr_{R} \,\frac{1+(-1)^F}{2} q^{L_0} } \nonumber \\ &\qquad\times \parent{\tr_{NS} \,\frac{1+(-1)^{\overline F}}{2} \widetilde q^{\overline{L_0}}  - \tr_{R} \,\frac{1-(-1)^{\overline F}}{2} \widetilde q^{\overline{L_0}}}  \nonumber \\ 
\IIB~:~ \tr_{\psi,\overline \psi} ~ \hat{P}'_{GSO}\,(-1)^{{\bf F}} q^{L_0} \widetilde q^{\overline{L_0}} =& \parent{\tr_{NS} \,\frac{1+(-1)^F}{2} q^{L_0} - \tr_{R} \,\frac{1+(-1)^F}{2} q^{L_0}} \nonumber \\ & \qquad \times \parent{\tr_{NS} \,\frac{1+(-1)^{\overline F}}{2}  \widetilde q^{\overline{L_0}} - \tr_{R} \,\frac{1+(-1)^{\overline F}}{2} \widetilde q^{\overline{L_0}}} 
\end{align}

Ces théories vérifient une supersymétrie d'espace-cible. Cela se voit entre autres par la suppression de la fonction de partition dans Minkowski. En effet, elle s'exprime en fonction de l'identité \emph{abstruse} de Jacobi~ :

\begin{align}
\vartheta_3^{\hphantom 3 4} - \vartheta_2^{\hphantom 3 4}-\vartheta_4^{\hphantom 3 4} = 0 \quad \Longrightarrow \quad Z_{T^2}=0
\end{align}

Les fonctions $\vartheta_i$ sont les fonctions th\^eta de Jacobi~\cite{Polchinski:1998rq}. Par conséquent dans ces théories, l'énergie du vide, exprimée plus tôt en fonction de $Z_{T^2}$ s'annule elle-même, ce qui est une propriété essentielle de la supersymétrie. Dans une supergravité d'espace-cible définie sur un espace de Minkowski ceci est bien cohérent avec une constante cosmologique nulle. \\

Le spectre de masse associé à ces théories, grâce à la projection GSO, vérifie une supersymétrie d'espace-cible ${\mathcal N} =2$ et admet donc deux gravitinos. En outre, à la différence des théories bosoniques, il est dépourvu de tachyon. Cela est dû à la supersymétrie qui impose d'avoir à chaque niveau d'excitation autant de fermions que de bosons. En effet, il n'existe pas de partenaire supersymétrique au tachyon, car du point de vue de la supersymétrie de la surface de corde, il est le fondamental. Ainsi, du point de vue de l'espace-cible, il briserait la supersymétrie. \\ 

Nous n'irons pas plus loin dans la description des théories de supercordes de type II, car il faut maintenant décrire le passage des cordes fermées aux cordes ouvertes, ce qui nécessite d'introduire les conditions aux bords conformes qui définissent des BCFT, les opérateurs de vertex du bord et les branes.

\section{Surfaces avec bord : cordes ouvertes, branes et th\'eories conformes de bord}
\label{sec:BCFT}

Les théories des cordes ouvertes s'\'etudient sur des surfaces d'univers avec bord. On s'intéresse par conséquent à la physique sur un voisinage $U$ avec bord $\partial U \neq\varnothing$. Ce voisinage peut \^etre d\'ecrit par le demi-plan supérieur $U=H_+$. L'action correspondante est définie par~:

\begin{align}
S_p[X,\psi] = \frac{1}{2 \pi \alpha'} \int_{H_+} \di^2 z \; \parent{\partial X^\mu \bar \partial X_\mu + \frac{1}{2} \psi^\mu \bar \partial \psi_\mu + \frac{1}{2} \wt\psi^\mu \partial \wt \psi_\mu}+ \oint_{\partial H_+ = {\mathbb R}} \di z \; {\mathcal O}[X,\psi]
\end{align}

où sur le bord nous avons inclus des conditions de bord paramétrisées par un opérateur de vertex ${\mathcal O}[X,\psi]$ ce qui définit en général une théorie \emph{non libre}, \cad que ces opérateurs tiennent lieu de termes d'interaction -- nous verrons cela un peu plus en détail lorsque nous introduirons le concept de \emph{modèle sigma}. Cependant, il n'est pas nécessaire d'ajouter un opérateur sur le bord pour imposer une condition au bord, de sorte que la théorie peut néanmoins être libre. Par exemple, les conditions de Neumann et Dirichlet -- qui définissent des théories de cordes ouvertes libres -- imposent qu'en $z=\bar z$~:

\begin{align}\label{eq:condition_ND}
N~:~\partial X &= \bar \partial X \qquad \text{et} \qquad \psi = \zeta \wt \psi \nonumber \\ 
D~:~\partial X &= -\bar \partial X \qquad \text{et} \qquad \psi = -\zeta \wt \psi 
\end{align}

avec $\zeta=\pm 1$ la structure de spin que l'on introduit plus rigoureusement dans les conditions de collage du super-espace sur le bord, \cad $\theta=\zeta\bar\theta$.  \\

Les opérateurs de vertex correspondant aux états asymptotiques de cordes ouvertes sont naturellement définis par des poinçons sur le bord, \cad que tel état de corde ouverte $\ket{\psi,z}$ est défini en un point $z$ du bord. Il correspond à l'opérateur de vertex ${\mathcal V}_\psi (z)$ mais selon une relation pas forcément triviale -- \cf problème des fantômes et des champs fermioniques en supercordes.  

De même que dans le cas des cordes fermées, l'ajout de déformations à l'action peut être interprétées comme autant d'excitations sourcées dans la surface de cordes -- et sur le bord -- par un fond externe, de l'espace-cible. Le fait que le couplage de l'espace-cible à la corde se fasse le long du bord suggère qu'il existe des sources externes macroscopiques connectées directement au bord, \cad formant l'espace-cible du bord de surface. C'est ce qu'on verra par le suite sous le nom de \emph{brane}.

\subsection{Conditions de bord générales et branes}

De manière générale~\cite{Gaberdiel:2001zq,Onogi:1988qk}, les conditions de bord s'appliquent sur les différents courants de symétries de sorte que la symétrie est conservée ou non. Par exemple, les conditions de Neumann conservent la symétrie de translation, tandis que les conditions de Dirichlet la brisent. On interprète que les conditions de Neumann laissent libres les extrémités des cordes ouvertes tandis que les conditions de Dirichlet les fixent sur des hyperplans, constituant des défauts topologiques plongés dans l'espace-cible. Bien que ces hyperplans semblent à première vue complètement gelés, en vérité on peut montrer qu'ils sont des objets dynamiques, car on peut imposer des conditions aux bords plus générales avec des degrés de liberté dynamiques. Ces hyperplans sont nommés \emph{D-branes} en général\footnote{Il existe d'autres types d'hyperplan, par exemple les \emph{orientifolds},}. \\ 

La symétrie conforme doit \^etre conservée par les conditions aux bords. C'est une contrainte forte mais qui est naturelle afin que la théorie soit invariante conforme y compris le long du bord, \cad corresponde à une situation physique. De fait, il faudra toujours avoir sur le bord, l'identité du tenseur énergie-impulsion suivante~:

\begin{align}\label{eq:condition_T}
T(z) = \wt T(z)
\end{align} 

La formule de collage pour $T_F$ est semblable. La formule générale pour les conditions au bord des courants que l'on notera $J^a$ est~:

\begin{align}\label{eq:condition_generale}
J^{(a)}(z) - \Omega(\wt J^{(a)} (z)) =0
\end{align}

avec $\Omega$ un automorphisme~\cite{Recknagel:1998ih,Recknagel:1998ut} de l'algèbre de courant. Cet automorphisme doit être tel que le tenseur énergie-impulsion vérifie toujours~\refe{eq:condition_T}, en utilisant la construction de Sugawara du tenseur en fonction des courants~:

\begin{align}
T(z) = \sum_{a,I} :J_I^{(a)} (z)J_I^{(a)}(z):
\end{align}

sachant que chaque courant $J^{(a)}$ est éventuellement décomposé le long d'une algèbre dont les générateurs sont indexés par le nombre $I$. L'exemple le plus fameux est sûrement $\wh{su}(2)_1$ au rayon auto-dual dans une théorie compactifiée sur un tore. Le tenseur énergie-impulsion étant un cas particulier de courant, la comparaison de la formule~\refe{eq:condition_generale} avec la formule~\refe{eq:condition_T} montre que dans son cas l'automorphisme est trivialement $\Omega=id$ l'identité.

Les conditions Neumann et Dirichlet le long des champs scalaires $X$ correspondent à des contraintes sur le courant des translations $J^\mu = \partial X^\mu $. De telle sorte que sont identifi\'es respectivement $\Omega_N= id$ et $\Omega_D = -id$. En terme des modes d'oscillations cela se traduit par l'identification~:

\begin{align}\label{eq:conditions_demi-plan1}
N~: \qquad \alpha_n &= \wt \alpha_n \nonumber \\ 
D~: \qquad \alpha_n &= -\wt \alpha_n
\end{align}

Ainsi que sur les fermions de surface en supercorde d'après~\refe{eq:condition_ND}~:

\begin{align}\label{eq:conditions_demi-plan2}
N~: \qquad \psi_r &= \zeta \wt \psi_r \nonumber \\ 
D~: \qquad \psi_r &= - \zeta \wt \psi_r
\end{align}

En tenant compte de ces conditions au bord, nous pouvons ensuite définir les états de la même manière qu'en corde fermée en appliquant les modes sur l'état du vide, ou d'impulsion $k^\mu$. En supercordes, l'état de vide est toujours NS ou R sachant que ces conditions s'appliquent maintenant comme des conditions de bord relatives, \cad entre un bord et l'autre sachant qu'une corde ouverte est décrite dans l'espace-cible plutôt sous la forme d'une nappe. A la différence des cordes fermées et du fait des conditions de collages, nous n'avons plus qu'un seul vide. Nous verrons en détail un peu plus tard le cas des supercordes ouvertes. 

Notons tout de suite que les conditions de collage sur les modes diffèrent significativement lorsque l'on se place sur le disque plutôt que sur le demi-plan complexe. Dans l'expression sur le disque, nous avons sur le cercle unité les conditions de collage suivantes des modes de cordes ouvertes~\cite{Cardy:1989ir,Ishibashi:1988kg}~:

\begin{align}
&N~: \qquad \alpha_n = -\wt \alpha_{-n}  \quad , \quad \psi_r = i\zeta \wt \psi_{-r}  \nonumber \\
&D~: \qquad \alpha_n = \wt \alpha_{-n}  \quad , \quad  \psi_r = - i\zeta \wt \psi_{-r}
\end{align}

Notons la différence de signe -- et l'ajout d'un facteur $i$ -- du fait de la transformation conforme~\cite{Cardy:1989ir} du plan vers le disque qui laisse apparaître un facteur $(-1)^S$ avec $S$ le spin de $\partial X$ ou de $\psi$ ici, respectivement $1$ et $1/2$. En se plaçant sur le disque unité l'expression de la théorie -- sans insertion sur le bord -- est meilleure pour décrire l'amplitude -- à l'ordre des arbres -- correspondante en terme de cordes fermées. En effet, l'origine du disque correspond en général à $t\to -\infty$ avec $t$ le "temps" sur la surface de corde, donc au temps asymptotique pour une corde fermée. De sorte que le bord $S^1$ constitue ainsi une configuration de corde fermée à temps fini, mais qui n'est du coup pas un état asymptotique. La notion d'\emph{état de bord} doit \^etre introduite. Nous l'aborderons dans cette section~; mentionnons simplement que les conditions de bord se présentent sous la forme de contraintes de construction de cet état de bord~\cite{Ishibashi:1988kg,Gaberdiel:2001zq,Gaberdiel:2002my}.

\subsection{Quantification des cordes ouvertes, opérateurs de vertex et états de bord}

La quantification des cordes ouvertes se fait de manière quasiment identique au cas des cordes fermées puisque le développement en mode d'oscillation et les relations aux opérateurs sont des constructions locales. Simplement, les conditions aux bords imposent les contraintes supplémentaires que nous avons vu entre les modes droits et les modes gauches. Typiquement, le nombre de degrés de liberté est divisé par deux -- en l'occurence on verra que le nombre de supersymétrie d'espace-cible n'est qu'au mieux à moitié conservé. En outre, les fonctions de corrélations seront l\'eg\`erement modifiées en fonction des conditions de bord. 

De cette manière lorsque l'on impose les conditions de Neumann~\refe{eq:conditions_demi-plan1} et~\refe{eq:conditions_demi-plan2}, les modes droits sont exactement égaux aux modes gauches le long des directions concernées par ces conditions $\alpha^\mu_n = \wt \alpha^\mu_{n}$. De sorte que le spectre de masse des cordes ouvertes Neumann est simplement $m^2 = -k^2 = N-a$ avec $a=1$ en bosonique, $a=1/2$ pour NS et $a=0$ pour R. La dépendance dans l'impulsion est différente du cas de corde fermée. Pour obtenir cette formule, il faut étudier les OPE de bord et les appliquer au produit $T(z)e^{ik_\mu X^\mu}(0)$ sur le bord afin d'obtenir le poids conforme de l'opérateur d'impulsion. Par la correspondance on obtient ainsi la valeur propre de $L_0 \ket{k^\mu,\Omega}$. 

Si on impose des conditions de Dirichlet, la relation est différente $\alpha^\mu_n = -\wt \alpha^\mu_{n}$ mais le spectre de masse est invariant, à part que le nombre de degrés de liberté d'impulsion change. En effet, rappelons que nous avions $p^\mu \propto \alpha_0^\mu + \wt \alpha_0^\mu$ de sorte que le long d'une coordonnée Dirichlet, $p^\mu = 0$ donc le générateur de translation spatiale est brisé et la coordonnée fixée sur le bord. Pour les cordes ouvertes on ne définit donc d'impulsion que le long des coordonnées Neumann. 

\subsubsection{Correspondance état-opérateur sur le bord}

De manière absolument identique que dans la partie précédente, nous pouvons aussi définir des opérateurs de vertex correspondant aux états construits par application des modes créateurs sur le vide. En théorie bosonique, la relation entre $\alpha_n$ et ${\mathcal V}_n (X)$ est simplement donnée par~\refe{eq:correspondance}. En théorie supersymétrique, nous aurons~\refe{eq:correspondance_susy}. Les opérateurs de vertex que nous obtenons par ce biais sont naturellement définis sur le -- ou les -- bord(s) de la surface. Par exemple, sur la géométrie du disque, ils sont construits sur le cercle $S^1$ et sur la géométrie conforme du demi-plan complexe, sur l'axe réel $\mathbb R$. En effet, une insertion dans l'intérieur de la surface peut toujours \^etre contenue dans un voisinage ne contenant aucun bord, donc est défini par une théorie de corde fermée exclusivement. Du coup, seul un objet du bord peut relever de la physique du bord, soit des cordes ouvertes. Cela amène naturellement à définir l'opérateur de vertex comme intégré le long du bord de la surface exclusivement~:

\begin{align}
V_o[X^\mu,\psi^\mu,\text{\etc}] = \oint_{\partial\Sigma} {\mathcal V}[X^\mu,\psi^\mu,\text{\etc}]
\end{align}

En définissant une théorie des cordes fermées, \cad une physique de l'intérieur, compatible avec les conditions au bord imposées, 
nous pouvons définir l'opérateur de bord comme la projection de l'opérateur de bulk sur le bord. 

\subsubsection{OPE sur surface avec bord et ordre normal de bord}

Cependant, il y a une petite subtilité, car lorsqu'un opérateur par exemple sur le demi-plan complexe supérieur, au point $z$ s'approche du bord, son point d'insertion $z$ s'approche de son image conjuguée complexe $\bar z$. Or les fonctions de Green des champs fondamentaux sont elles-mêmes modifiées pour satisfaire aux conditions de bord. On a plus précisément les OPE sur $H_+$ avec des conditions de Neumann (Dirichlet)\footnote{En utilisant $\alpha'=1$.}~:

\begin{align}\label{eq:Green_H}
X^\mu (z,\bar z) X^\nu(w,\bar w) &= - \frac{\eta^{\mu\nu}}{2} \ln \module{z-w}^2 \pm \frac{\eta^{\mu\nu}}{2} \ln \module{z-\bar w}^2 \nonumber \\ 
\psi^\mu(z) \wt \psi^\nu(\bar w)& = \pm \zeta \, \frac{\eta^{\mu\nu}}{z-\bar w} \nonumber \\ 
\psi^\mu(z) \psi^\nu(w) & = \frac{1}{z-w} \nonumber \\ 
\wt \psi^\mu(\bar z) \wt \psi^\nu(\bar w) & = \frac{1}{\bar z- \bar w}
\end{align}

avec $(+)$ pour Neumann et $(-)$ pour Dirichlet. On voit donc l'ajout d'un terme supplémentaire dans l'OPE des scalaires. Nous pouvons exprimer les fonctions de Green aussi sur le disque par transformation conforme. On trouve~:

\begin{align}\label{eq:Green_D}
X^\mu (z,\bar z) X^\nu(w,\bar w) &= - \frac{\eta^{\mu\nu}}{2} \ln \module{z-w}^2 \pm \frac{\eta^{\mu\nu}}{2} \ln \module{1-z\bar w}^2 \nonumber \\ 
\psi^\mu(z) \wt \psi^\nu(\bar w) &= \zeta \, \frac{2 \eta^{\mu\nu} \sqrt{z \bar w}}{1-z\bar w} \nonumber \\ 
\psi^\mu(z) \psi^\nu(w) & = \frac{\sqrt{zw}}{z-w} \nonumber \\ 
\wt \psi^\mu(\bar z) \wt \psi^\nu(\bar w) & = \frac{\sqrt{\bar z \bar w}}{\bar z- \bar w}
\end{align}

avec sur le disque unité $z=\rho e^{i\phi}$ pour $\rho \in [0,1]$ et $\phi \in [0,2\pi[$. Nous voyons donc par ces fonctions de Green, qu'un opérateur de vertex peut interagir avec lui-même, comme si le demi-plan complexe était doublé et qu'une image de l'opérateur de vertex existait en $z'=\bar z$. C'est ce qui est appel\'e communément l'\emph{astuce de doublement} -- en anglais \emph{doubling-trick} -- car tous les opérateurs anti-holomorphes peuvent \^etre d\'efinis comme étant les continuations analytiques, dans le demi-plan complexe inférieur, des opérateurs holomorphes de $H_+$. De sorte qu'en général on aura l'OPE du bulk vers le bord~:

\begin{align}
{\mathcal V}(z,\bar z) = \frac{\bnormal{{\mathcal V}_b(z)}}{\module{z-\bar z}^{h+\bar h-h_b}} +\ldots
\end{align}

où l'\emph{ordre normal du bord} est d\'efini par les symboles étoilés $\bnormal{\cdot}$ de telle sorte que l'opérateur qui y est contenu est régulier en tout point du bord. Nous avons rajouté des points de suspension pour signifier que techniquement un développement de Taylor en opérateurs descendants pourrait \^etre fait le long du bord, en fonction de $2i y = z-\bar z$. 

\subsubsection{OPE sur le bord}

Nous pouvons définir les champs fondamentaux sur le bord directement et calculer leurs OPE en fonction des variables de bord. Simplement nous aurons pour $z\to \bar z$ sur demi-plan $H_+$~:

\begin{align}
X^\mu(z,\bar z) &\longrightarrow X^\mu(z) = 2 X^\mu_L (z) ~\text{ (Neumann) ou }~ x_0^\mu ~\text{ (Dirichlet)}\nonumber \\ 
\frac{1}{\sqrt 2}\parent{\psi(z) \pm\zeta \wt \psi(\bar z)} &\longrightarrow \Psi(z) = \frac{(1 + \zeta^2)}{\sqrt 2} \, \psi(z) = \sqrt 2 \psi(z)
\end{align}

avec $(+)$ pour Neumann et $(-)$ pour Dirichlet. Nous avons utilisé la condition de collage sur la variable de Grassmann en super-espace~:

\begin{align}
\theta = \zeta \bar \theta \quad \Longrightarrow \quad \mathbb X(z) = X(z) + \frac{i}{\sqrt 2 } \theta \parent{\psi(z) +\zeta \wt \psi(z)}
\end{align}

et $\psi = \pm \zeta \wt\psi$ puis nous avons choisi d'inclure le facteur $1/\sqrt 2$ à l'intérieur de la définition du champ fermionique, de sorte que sur le bord~:

\begin{align}\label{eq:convention_psi}
\mathbb X(z) = X(z) + i \theta \Psi(z)
\end{align}

qui est une convention commune dans la littérature. Nous avons divisé le champ scalaire en deux contributions, holomorphe et anti-holomorphe, respectivement $X_L$ et $X_R$. Etant donn\'e que la condition de collage ne concerne que les modes d'oscillations, seul le mode zéro $x_0^\mu$ subsiste avec des conditions de Dirichlet. Mais en présence de conditions Neumann $X_L=X_R$. Alors nous avons les OPE sur le bord~:

\begin{align}
X^\mu(z) X^\nu(w) &= -2\eta^{\mu\nu} \ln \module{z-w} + \ldots \nonumber \\ 
\Psi(z) \Psi(w) & = \frac{2}{z-w}
\end{align}

Nous utiliserons beaucoup ces formules dans la suite. Elles peuvent directement exprim\'ees en super-espace sous la forme~:

\begin{align}
\mathbb X^\mu(z) \mathbb X^\nu(w) = - 2 \eta^{\mu\nu} \ln \module{z-w-\theta_z \theta_w}
\end{align}

Ainsi toute OPE sur le bord entre opérateurs descendants dépendant des champs fondamentaux s'exprime par la formule~:

\begin{align}
{\mathcal A}_{h_1}(z_1) {\mathcal A}_{h_2}(z_2) = \sum_k \frac{C_{12}^{\hphantom{12}k} }{\module{z_1-z_2}^{h_1+h_2 - h_k}}  {\mathcal A}_{h_k}(z_2) 
\end{align}

Ici $h$ est la somme des poids holomorphe et anti-holomorphe. 

\subsubsection{Etats de bord}

En discutant des conditions de bord, nous avons à maintes reprises introduit le concept d'état de bord. Nous disions entre autre qu'en se plaçant dans une description adéquate pour décrire telle amplitude dont la géométrie est celle d'un bord sans insertion, le formalisme des cordes fermées pouvait \^etre utilis\'e afin d'effectuer le calcul. Nous avions choisi le disque unité dont le bord est le cercle $S^1$. Le bord ressemble à une insertion sur la sphère mais particulière puisqu'elle peut \^etre atteinte en un temps fini, à la différence des inclusions habituelles d'état asymptotique -- qu'il faut un temps infini pour atteindre, par définition. Or, du point de vue d'une évolution d'un état de corde fermée initialement piqué à l'origine, le bord constitue aussi une configuration de corde fermée, \cad un état. En outre, par transformation conforme, l'origine peut\^etre envoy\'ee à l'infini temporel futur tout en conservant le cercle unité. Ainsi, la configuration au bord peut-elle assimil\'ee \'egalement \`a une condition initiale pour une corde ferm\'ee. Les configurations de cordes fermées devant être cohérentes avec les conditions aux bord imposées, nous construisons donc un état de bord $\ket B$ tel qu'il vérifie~\cite{Gaberdiel:2001zq,Gaberdiel:2002my,Cardy:1989ir,Ishibashi:1988kg}~:

\begin{align}
\parent{L_n - \wt L_{-n}} \ket B = 0 \nonumber \\
\parent{J^a_n - (-1)^s \Omega \circ \wt J^a_{-n}} \ket B = 0
\end{align}

avec $J^a$ le courant holomorphe qui se décompose suivant une série de Laurent $J^a(z) = \sum_n J^a_n z^{-n-h}$ et $s=h-\bar h$ le spin -- dans cet exemple, $\bar h=0$ donc $s=h$. En toute généralité nous avons laissé l'action d'un automorphisme $\Omega$ sur le courant anti-holomorphe. La plupart du temps il est quasi-trivial, \cad $\Omega=\pm id$ mais nous pouvons trouver des exemples plus larges, tels que les branes conformes~\cite{Gaberdiel:2001zq}. \\ 

Sachant qu'un état de bord peut \^etre interpr\'et\'e comme une configuration initiale ou finale de corde fermée, toute amplitude sur le disque -- sans insertion au bord -- doit pouvoir s'exprimer comme~:

\begin{align}
A_B[\ldots] = \bra{0}\ldots\ket{B} = \corr{\ldots}_{B,D^2}
\end{align}

avec un nombre arbitraire d'insertion dans l'intérieur. Nous voyons que l'amplitude doit refléter la dualité corde ouverte/corde fermée du diagramme, de telle sorte qu'en terme de corde fermée $\ket B$ est une source et en terme des cordes ouvertes $B$ est une condition aux bords, éventuellement représentée par une déformation de bord dans l'action~\cite{Gaberdiel:2002my} de type $\oint_{S^1} {\mathcal O}_B$.  \\

\begin{wrapfigure}{r}{.45\linewidth}
\centering
\includegraphics[scale=.5]{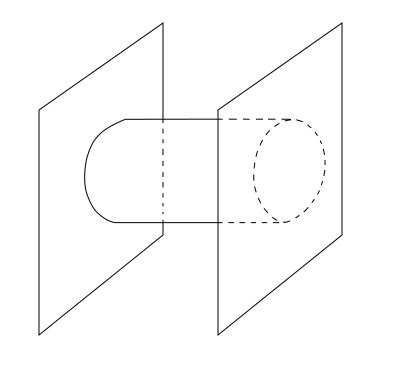}
\caption{\label{fig:diag_echange} \small{Diagramme cylindrique~ : \'echange de corde ferm\'ee entre deux branes ou fonction de partition \`a une boucle des cordes ouvertes.}}
\vspace{.5cm}
\end{wrapfigure}

Des diagrammes à plus d'un bord sont imaginables. Par exemple avec deux bords, le diagramme cylindrique d'échanges de cordes fermées entre deux branes repr\'esent\'e sur la figure~\refe{fig:diag_echange} est dual à la fonction de partition à l'ordre d'une boucle $Z_{\alpha \beta}$ des cordes ouvertes avec les deux conditions aux bords $\alpha$ et $\beta$. En introduisant les états de bords $\ket \alpha$ et $\ket \beta$ nous aurions l'amplitude~\cite{Cardy:1989ir} fonction du module $\tau$~:

\begin{align}
Z_{\alpha\beta}(\tau) = \bra{\alpha} e^{-\frac{\pi H_c}{\tau} } \ket{\beta}
\end{align}  

La relation de cette amplitude à la fonction de partition des cordes ouvertes permet d'introduire les formules de Verlinde~\cite{Verlinde:1988sn,Cardy:1989ir} et les conditions de Cardy~\cite{Gaberdiel:2002my,Cardy:1989ir} pour construire des théories conformes de bord.

Nous proposions d'identifier l'état de bord à une brane, mais nous pouvons tout aussi bien identifier l'état de bord à plusieurs branes, \cad une superposition d'états de bord, pas forcément coïncidantes. C'est ce que nous faisons explicitement lorsque nous construisons les branes BPS en théorie des supercordes.  L'interprétation en terme de cordes ouvertes est  cependant plus délicat à cause de l'apparition de secteurs de cordes ouvertes tendues entre différentes branes. En outre, à l'ordre des arbres il est difficile de vraiment faire sens de la superposition linéaire des états de bords, à cause justement des secteurs interbranaires qui ne trouvent pas d'équivalent direct en terme d'état de bord trivial, tout cela est relié au caractère non-abélien de la théorie effective sous-jacente\footnote{Nous comprenons qu'à cause des secteurs interbranaires le système ne peut pas \^etre r\'eduit à un ensemble de branes isolées et indépendantes, à moins que l'on se place dans une limite où ce secteur est trop massif et découple.}. Nous d\'evelopperons ce point plus bas, en introduisant les facteurs de Chan-Paton. Mais tout de suite, explicitons brièvement l'expression de l'état de bord d'une seule brane.

\subsubsection{Etat de bord d'une brane bosonique}

En théorie bosonique, l'état de bord correspondant à des conditions Dirichlet le long des directions $i=p+1,\ldots 26$ et Neumann le long de $a=0 \ldots p$ est celui représentant le spectre de cordes fermées sourcées par une $Dp$-brane. Il vérifie les contraintes~:

\begin{align}
\parent{L_n - \wt L_{-n}} \ket{Dp} &= 0 \nonumber \\
\parent{\alpha^a_n + \wt \alpha^a_{-n}} \ket{Dp} &= 0  \nonumber \\
\parent{\alpha^i_n - \wt \alpha^i_{-n}} \ket{Dp} &= 0  \nonumber \\ 
x^i \ket{Dp} &= a^i  \ket{Dp}
\end{align} 

avec $a^i$ la position de la brane. L'état de bord vérifiant ces conditions est, à une normalisation près~:

\begin{align}
\ket{Dp,\bf a} \propto \int \prod_{i=p+1}^{26} \di k^i \, e^{i k^i a^i} \wh{\ket{Dp,\bf k}}
\end{align}

Le ket $\wh{\ket{Dp,\bf k}}$ est un état cohérent dont l'expression est donnée par~:

\begin{align}
\wh{\ket{Dp,\bf k}} = \exp\croch{\sum_{n>0} \parent{-\sum_{i} \alpha^i_{-n} \wt \alpha^i_{-n} + \sum_{a>1} \alpha^a_{-n} \wt \alpha^a_{-n}} } \ket{0,\bf k}
\end{align}

dans la jauge du cône de lumière, \cad pour $X^0$ et $X^1$ jaugés, qui permet de s'abstraire des questions de fantômes. L'état fondamental est ici simplement un vecteur propre d'impulsion $k^i$. Nous verrons un peu plus loin la construction d'une brane en supercordes. 

\subsection{Branes et facteurs de Chan-Paton}

Lorsque plusieurs branes sont ins\'er\'ees dans l'espace-cible et que l'on veut décrire la théorie des cordes ouvertes, il faut s'inquiéter de classer les différents \emph{secteurs} de ces cordes en fonction des points d'attache de leurs extrémités. En effet, il existe une différence très nette entre les cordes tendues d'une brane à une autre et celles attachées à une seule et même brane, puisque topologiquement les unes ne sont pas continuellement d\'eformables en les autres – voir figure~\refe{fig:brane_sep},

\begin{wrapfigure}{l}{.5\linewidth}
\centering
\includegraphics[scale=.5]{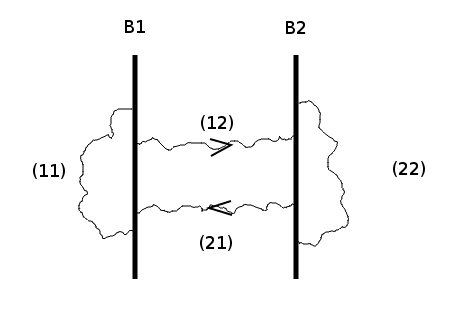}
\caption{\label{fig:brane_sep} \small{Diff\'erents secteurs de cordes ouvertes dans le syst\`eme compos\'e de deux branes.}}
\end{wrapfigure}

Du point de vue de la théorie des champs sous-jacente et déduite du spectre de cordes ouvertes, le fait que plusieurs secteurs existent implique la différentiation dans l'action correspondante des champs associés à chaque secteur. Prenons deux branes coïncidentes. En calculant le spectre on trouve aisément que les champs sont tous identiques pour tous les secteurs. Sur chacun par exemple en théorie bosonique, nous avons un tachyon puis un boson vecteur de jauge non massif. Du point de vue de l'action effective il doit appara\^itre une forme de symétrie puisque ces secteurs sont finalement indiscernables. 

\subsubsection{Branes coïncidentes, groupe et facteurs de CP}

En fonction du nombre de branes coïncidentes, le nombre de secteurs augmente et le degré de symétrie également. On montre que le groupe le long duquel les secteurs forment une représentation est $U(N)$ avec $N$ le nombre de branes coïncidentes. Les facteurs de Chan-Paton (CP) sont les représentants matriciels de ce groupe~\cite{Paton1969516} et peuvent être reliés à des charges attachées aux extrémités -- des quarks~\cite{Marcus:1986cm}. Par exemple, en présence de deux branes, le groupe est $U(2)$ et les facteurs de CP sont les matrices de Pauli $\sigma^{1,2,3}$ plus l'identité $I=\sigma^0$. Les champs d'espace-cible sont ainsi développés le long du groupe $U(N)$ et puisque nous disions que les champs correspondent à des couplages des termes de bord, alors ces derniers se décomposent également le long de ce groupe~:

\begin{align}
\delta S = \sum_n^{\frac{N(N-1)}{2}} \Lambda^n \otimes \oint_{\mathbb R} \phi_n(z) 
\end{align}

avec $\Lambda^n \in U(N)$ un facteur de CP. Cela permet de généraliser les conditions de bord et de développer le formalisme du modèle sigma non linéaire à des champs non-abéliens. Dans cette description, il est plus d\'elicat de parler de formules de collages des courants mais aussi d' états de bords ~ : il faudrait les généraliser au cas non-abélien\footnote{ Cela existe probablement dans la littérature mais je n'en ai pas pris connaissance.}.

\subsubsection{Séparation non nulle entre branes et brisure spontanée de symétrie}

Lorsque les branes sont s\'epar\'ees, cette symétrie est spontanément brisée ; par exemple en s\'eparant une seule brane du système la sym\'etrie $U(N)$ est bris\'ee en $U(1)\times U(N-1)$. Ce processus ressemble beaucoup au mécanisme de Higgs et pour être plus précis il s'agit exactement du même mécanisme, quoiqu'ici l'interprétation est géométrique, car le secteur de cordes ouvertes interbranaire acquiert un terme de masse supplémentaire proportionnel à la distance de séparation.  

Pour modéliser cette séparation sur le bord il faut inclure une \emph{ligne de Wilson} sur la coordonnée conjuguée à la direction de séparation. Si $X$ est cette direction -- la variable conjuguée à l'impulsion -- alors $\wt X$ est la coordonnée duale -- la variable conjuguée à l'extension et la position spatiale de la corde. En théorie des cordes ouvertes, lorsque $X$ vérifie une condition de Dirichlet, $\wt X$ vérifie une condition de Neumann. Les deux sont reliées par T-dualité le long de coordonnées compactes. \\

Soient deux branes bosoniques. Les facteurs de CP $\sigma^0$ et $\sigma^3$ correspondent aux cordes ouvertes dont les extrémités sont attachées à une seule et même brane. On peut montrer que la déformation~:

\begin{align}
\sigma^3 \otimes \oint \frac{r}{2} \partial \wt X
\end{align}  

modifie le spectre des cordes ouvertes des secteurs $\sigma^1$ et $\sigma^2$ en $m^2=r^2+N-1$ et correspond exactement à séparer les branes d'une distance $r$. Les cordes ouvertes des secteurs $\sigma^0$ et $\sigma^3$ conservent quant à elles la formule $m^2=N-1$. On voit explicitement dans ces formules que le spectre de masse brise effectivement la symétrie, que l'on récupère dans la limite $r\to 0$. La déformation supersymétrique correspondante devrait être simplement~:

\begin{align}
\sigma^3 \otimes \oint \frac{r}{2} D \wt{\mathbb X}
\end{align}

Pour un spectre de masse modifié en $m^2=r^2+N-1/2$ dans le secteur NS et $m^2=r^2+N$ dans le secteur R. En fait, ça n'est pas tout à fait correct car techniquement il faudrait introduire des degrés de liberté supersymétriques sur le bord -- des fermions de bord -- pour remplacer les facteurs de CP. On ne les introduira que bien spécifiquement dans le chapitre des supercordes.

\subsection{Cordes ouvertes en théories IIA et IIB}

Les théories IIA et IIB concernent \apriori des cordes fermées, mais n'excluent pas toutefois la présence de cordes ouvertes et des interactions avec celles-ci. La théorie des champs supersymétrique décrite par ce spectre est une supergravité de type IIA ou IIB et nous savons qu'il y existe des solutions solitoniques brisant une fraction de la supersymétrie, des états BPS,\footnote{Bogomol'nyi-Prasad-Sommerfield.} nommées $p$-branes. Or nous en déduisons qu'il doit exister dans le spectre des champs de jauge -- des formes différentielles -- de diverses dimensions et qui se couplent naturellement à ces objets multidimensionnels. Or d'après l'étude du spectre de cordes fermées, c'est bien ce que l'on trouve.

En outre, comme nous l'avons vu, les diagrammes d'interaction entre des p-branes, via un échange de cordes fermées, sont des cylindres. Or nous pouvons décrire ceux-ci en terme d'échange de cordes fermées se propageant d'une brane à l'autre, mais nous pourrions aussi les décrire en terme d'une boucle de cordes ouvertes -- en admettant leur existence et qu'elles s'avèrent cohérentes -- dont les extrémités seraient attachées à chaque p-branes. A \emph{priori} rien n'empêche cette identification si elle fonctionne~; donc à une transformation modulaire près, l'expression du diagramme dans ces deux descriptions doit être identique. Du coup, une théorie de cordes ouvertes cohérente avec les th\'eories de cordes fermées de type IIA ou IIB, devrait pouvoir \^etre exprim\'ee, telle que les cordes ouvertes ont les extrémités sont attachées exclusivement à des p-branes. Par similarité au modèle bosonique, ces branes sont nomm\'ees  $Dp$-branes, du fait des conditions typiquement Dirichlet des cordes ouvertes. Au passage, si le spectre de cordes ouvertes duales est supersymétrique, alors en espace plat, $Z_{\text{anneau}}$ la fonction de partition à l'ordre d'une boucle dans les cordes ouvertes, doit s'annuler similairement au cas des cordes fermées dans le vide. \\ 

Les conditions de cohérence entre la description de cordes ouvertes et celle de cordes fermées, ainsi que leur interaction mutuelle, montrent que les D-branes brisent explicitement la supersymétrie et exactement une moitié. 

\subsubsection{Spectre de cordes ouvertes et état BPS}

Pour les cordes ouvertes, il existe aussi des secteurs NS et R et il faut aussi introduire une projection GSO. Il est nécessaire d'avoir sur le bord $T=\widetilde T$ afin que les difféomorphismes y soient bien définis~ : il n'y a qu'une seule variable sur chaque bord. Ainsi, par cohérence, les générateurs de supersymétrie sur la surface doivent vérifier une identité semblable mais plus générale $G=\Omega(\widetilde G)$ avec $\Omega$ un automorphisme. On en déduit que les cordes ouvertes ne vérifient qu'une seule supersymétrie de surface, soit $N=1$. Le spectre résultant ne peut quant à lui que suivre cette contrainte, puisqu'il n'y a plus qu'un seul secteur indépendant qui est soit NS soit R -- dans le secteur R, nous avons un vide fondamental spinoriel simplement $\bf 16$ soit les degrés de liberté d'un seul spineur en dimension 10. Le générateur de supersymétrie d'espace-cible est donc unique le long des cordes ouvertes et ${\mathcal N}=1$. 

Or les cordes ouvertes se couplant à des cordes fermées, le spectre résultant de ces dernières ne peut que vérifier également une supersymétrie ${\mathcal N}=1$. Nous obtenons donc que les D-branes brisent spontanément la supersymétrie ${\mathcal N}=2$ en ${\mathcal N}=1$ et que ces objets sont par conséquent des états $1/2$-BPS. Autrement dit, dans le bulk la théorie est \apriori maximalement supersymétrique mais le spectre des cordes émises et reçues par la brane doit en briser la moitié.  

\subsubsection{Branes chargées et anti-branes}

Nous devons maintenant introduire les anti-branes que nous présenterons de trois manières différentes mais complémentaires. Tout d'abord notons que les D-branes sont des objets chargés, du fait entre autre qu'elles sont couplées aux champs différentiels de jauge R-R des théories IIA et IIB. Ce couplage est effectivement décrit par un terme de type Chern-Simons\footnote{Techniquement, il comprend aussi des termes de couplages aux champs $F$ et $B$, mais nous les négligerons pour l'instant.}. Pour une $Dp$-brane de charge $Q_p$ nous aurons~: 

\begin{align}
S_{CS} = Q_p \int_{p+1} C_{(p+1)}
\end{align}

avec $C_{(p+1)}$ la n-forme différentielle correspondante au champ de jauge autorisée en type IIA (p pair) et en type IIB (p impair). Cela est tel que l'intégrale de flux de ce champ autour de la brane, \cad le long d'une sphère $S^{d-p-2}$ est~:

\begin{align}
Q_p \propto \int_{S^{d-p-2}} \star dC_{p+1}
\end{align}

Nous avons utilisé le dual de Hodge $\star$ comme habituellement. En outre, la charge n'est pas ici une donnée continue, mais elle est quantifiée. La raison est simple~: le couplage de la brane aux champs de jauge suggère que les équations de mouvement de ces derniers soient de type Maxwell et Bianchi. Or la dualité de Poincaré des formes différentielles implique qu'un champ donné peut agir comme champ électrique pour une brane ou comme champ magnétique pour une autre -- duale à la première du coup. L'étude du monopole de Dirac permet ensuite de d\'eduire que, les branes \'etant chargées électriquement et magnétiquement, il faut quantifier les charges. Sachant que rien n'impose la positivité absolue des charges, il doit donc exister autant d'objets de charges négatives que d'objets de charges positives. Dans l'absolu, il peut aussi exister des objets fondamentaux non-chargés et on verra que ceux-ci sont non BPS, \cad qu'ils brisent totalement la supersymétrie d'espace-cible -- ils seront importants pour nous car ils peuvent admettre des tachyons dans leur spectre. \\

Par conséquent, nous avons des branes ($Q>0$), des anti-branes ($Q<0$) et des branes non BPS ($Q=0$). \\

\subsubsection{Supersymétrie et anti-brane}

Une manière complémentaire de les introduire est d'étudier la brisure de supersymétrie induite par les conditions aux bords des cordes ouvertes. L'étude de l'effet de la T-dualité sur les fermions montre que la chiralité du secteur droit est inversée, de sorte que nous avons effectivement que $\IIB \leftrightarrow \IIA$ par T-dualité. Il en résulte l'introduction d'un opérateur de transformation de parité $\beta^\perp$ dans l'espace-cible agissant sur les spineurs~\cite{Polchinski:1998rr,Bachas:1998rg}. Le symbole $\perp$ signifie que la parité n'agit que sur les coordonnées transverses à la brane. S'il s'agit d'une $Dp$-brane les conventions d'indices $a=0\ldots p$ et $i = p+1\ldots 10 \in \, \perp$ sont choisies. Cet opérateur est explicitement~:

\begin{align}
\beta_p^\perp = \prod_{m \in \perp} \beta^m \qquad \text{avec } \quad \beta^m = \Gamma \Gamma^m 
\end{align}

Nous avons utilisé les matrices $\Gamma^\mu$ généralisées à toute dimension. La matrice $\Gamma$ est la généralisation de $\gamma_5$. La supersymétrie non brisée correspond aux générateurs~:

\begin{align}
(Q_{p})_\alpha = Q_\alpha + \parent{\beta^\perp_p \widetilde Q}_\alpha
\end{align}

La condition aux bords peut en réalité être plus générale, car on pourrait aussi bien appliquer une transformation d'espace-cible $\Omega$ sur les spineurs donc sur $Q_p$, $Q$ et $\widetilde Q$ de telle sorte que cela peut se réécrire en terme d'une transformation de $\beta^\perp_p$~:

\begin{align}
(Q_{p})_\alpha = Q_\alpha + \parent{\Omega^{-1}(\beta^\perp_p) \widetilde Q}_\alpha
\end{align}

La transformation la plus simple est une rotation -- la plus générale correspond à une transformation de Poincaré -- dans l'espace transverse à la brane. Alors~\cite{Polchinski:1998rr,Bachas:1998rg}~:

\begin{align}
(Q_{p})_\alpha = Q_\alpha + \parent{\rho^{-1}\beta^\perp_p\rho \, \widetilde Q}_\alpha
\end{align}

Or la formule reliant la supersymétrie à la charge R-R, est~:

\begin{align}
\acomm{Q_\alpha}{\overline{\widetilde Q}_\beta} = -2 \sum_{p} Q^{R}_{M_1\ldots M_p} \epsilon^{M_1\ldots M_p} \parent{\beta^\perp \Gamma^0}_{\alpha\beta}
\end{align}

Et par conséquent, nous aurons via la transformation $\Omega$~:

\begin{align}
\acomm{Q_\alpha}{\overline{\widetilde Q}_\beta} = -2 \sum_{p}Q^{R}_{M_1\ldots M_p} \epsilon^{M_1\ldots M_p} \parent{ \Omega( \beta^\perp) \Gamma^0}_{\alpha\beta}
\end{align}

En l’occurrence pour ce qui nous intéresse, si $\Omega=-1$, autrement dit si $\rho$ est une rotation de $\pi$ dans une direction transverse donnée, nous avons immédiatement que $\Omega(Q^R)=-Q^R$. Le cas de rotation quelconque est moins trivial mais est intéressant car en présence d'une autre brane fixée, l'angle d'intersection est un paramètre tel que pour des valeurs génériques la supersymétrie est totalement brisée~\cite{Polchinski:1998rr,Epple:2003xt,Jones:2003ew}, sauf dans certaines configurations~\cite{SheikhJabbari:1997cv} qui peuvent être $1/4$, $1/8$, $3/16$ ou $1/16$ BPS. \\

Nous en concluons qu'une anti-brane est une brane tournée de $180^\circ$ dans l'espace transverse. Remarquons maintenant que les supersymétries conservées par l'anti-brane sont complè\-tement orthogonales à celles conservées par la brane correspondante, puisqu'il s'agit précisément des supersymétries brisées dans son cas. \\ 

\subsubsection{Etats de bord et anti-brane}

Enfin, brane et antibrane peuvent \^etre introduites en utilisant le formalisme des états de bords~\cite{Gaberdiel:2000jr,Gaberdiel:2001zq,Gaberdiel:2002my}. Un état de bord est défini de telle sorte qu'il vérifie les conditions de bord subies par les différentes algèbres sur la surface de corde. Soit $\ket{Bp,a^\mu,\zeta}$ l'état de la brane localisée en $a^\mu$ avec $\zeta=\pm 1$ la structure de spin -- condition de collage entre modes droit et gauche des spineurs sur le bord. En suivant la construction de Garberdiel~\cite{Gaberdiel:2000jr} avec ses conventions, cet état doit vérifier dans la jauge du cône de lumière ($\mu=0,1$) qui sont des coordonnées définies Dirichlet et sont donc transverses, sur le bord du disque~:

\begin{align}
&p^b \ket{Bp,a^\mu,\eta} = 0  & b = 2\ldots p+2 \nonumber \\
&(\alpha^b_n + \widetilde \alpha^b_{-n}) \ket{Bp,a^\mu,\eta} = 0  & b = 2\ldots p+2 \nonumber \\ 
&(\alpha^\mu_n - \widetilde \alpha^\mu_{-n}) \ket{Bp,a^\mu,\eta} = 0  &\mu = p+3 \ldots 9 \nonumber \\ 
&(x^\mu - a^\mu) \ket{Bp,a^\mu,\eta} = 0  &\mu = 0,1,p+3 \ldots 9 \nonumber \\ 
&(\psi^b_r +i\eta \widetilde \psi^b_{-r}) \ket{Bp,a^\mu,\eta} = 0  & b = 2 \ldots p+2 \nonumber \\ 
&(\psi^\mu_r - i\eta \widetilde \psi^\mu_{-r}) \ket{Bp,a^\mu,\eta} = 0  &\mu = p+3 \ldots 9
\end{align}

et quelque soit le secteur (NS ou R). Notons que cet état ne peut que décrire un D-instanton, à cause des conditions Dirichlet le long des coordonnées du cône de lumière. Cependant, Gaberdiel explique qu'il est possible de revenir à une définition de brane standard par double rotation de Wick~: une sur $X^0$ et une autre sur une coordonnée spatiale longitudinale à la brane. Cette manière de faire permet de s'abstraire des problèmes de normes négatives de long de la direction temporelle. 

L'état de bord précédent s'exprime par transformée de Fourier à une normalisation près en fonction de l'état cohérent $ \ket{Bp,k^\mu,\eta}$~:

\begin{align}\label{eq:def_bound}
\ket{Bp,a^\mu,\eta} \propto \int \prod_{\mu=0,1,p+3,\ldots 9} \di k^\mu \, e^{i k_\mu a^\mu} \ket{Bp,k^\mu,\eta}
\end{align}

Cet état cohérent est maintenant donné par~:

\begin{multline}
\ket{Bp,k^\mu,\eta} = \exp \Bigg(\sum_{n>0} \croch{-\frac{1}{n}\sum_{a=2}^{p+2}\alpha^a_{-n} \widetilde\alpha^a_{-n} + \frac{1}{n}\sum_{\mu=p+3}^{9}\alpha^\mu_{-n} \widetilde\alpha^\mu_{-n} }  \\ + i\eta \sum_{r>0} \croch{-\sum_{a=2}^{p+2}\psi^a_{-n} \widetilde\psi^a_{-n} + \sum_{\mu=p+3}^{9}\psi^\mu_{-n} \widetilde\psi^\mu_{-n}}\Bigg)\ket{Bp,k^\mu,\eta}^{(0)}
\end{multline}

avec $n$ entier ou demi-entier suivant le secteur (R ou NS) et $\ket{Bp,k^\mu,\eta}^{(0)}$ l'état fondamental de moment $k^a=0$ et $k^\mu \neq 0$ et tel que dans le secteur R-R~:

\begin{align}
&(\psi^a_0 + i\eta \widetilde \psi^a_{0})\ket{Bp,k^\mu,\eta}_{RR}^{(0)} = 0 & a = 2 \ldots p+2 \nonumber \\ 
&(\psi^\mu_0 - i\eta \widetilde \psi^\mu_{0})\ket{Bp,k^\mu,\eta}_{RR}^{(0)} = 0  &\mu = p+3 \ldots 9
\end{align}

Dans le secteur NS-NS, l'état de bord vérifie du fait de la définition fantomatique du vide NS-NS, \cad $(-1)^F \ket{0}_{NS} = - \ket{0}_{NS}$~:

\begin{align}
&(-1)^F \ket{Bp,a^\mu,\eta}_{NSNS} = - \ket{Bp,a^\mu,-\eta}_{NSNS} \nonumber \\ 
&(-1)^{\widetilde F} \ket{Bp,a^\mu,\eta}_{NSNS} = - \ket{Bp,a^\mu,-\eta}_{NSNS} 
\end{align}

De sorte que seule la combinaison $\ket{Bp,a^\mu}_{NSNS} = \ket{Bp,a^\mu,+}_{NSNS} - \ket{Bp,a^\mu,-}_{NSNS}$ est invariante GSO en théorie IIA ou IIB. Je rappelle qu'il faut une chiralité (+) dans le secteur NS, d'après~\refe{eq:comb_II}. Dans le secteur R-R on trouvera cependant~:

\begin{align}
&(-1)^F \ket{Bp,a^\mu,\eta}_{RR} = \ket{Bp,a^\mu,-\eta}_{RR} \nonumber \\ 
&(-1)^{\widetilde F} \ket{Bp,a^\mu,\eta}_{RR} = (-1)^{p+1} \ket{Bp,a^\mu,-\eta}_{RR} 
\end{align}

Ce qui donne encore une seule combinaison invariante GSO $\ket{Bp,a^\mu}_{RR} = \ket{Bp,a^\mu,+}_{RR} + \ket{Bp,a^\mu,-}_{RR}$ pour $p$ pair en type IIA et $p$ impair en type IIB. L'état de bord complet de la brane est la combinaison linéaire des deux secteurs. La dualité corde ouverte/corde fermée de la fonction de partition cylindrique impose que les normalisations relative et absolue soient spécifiquement fixées. Ainsi nous avons deux états distincts~:

\begin{align}\label{eq:brane}
&\ket{Dp,a^\mu} = \ket{Bp,a^\mu}_{NSNS} + \ket{Bp,a^\mu}_{RR}  \nonumber \\
&\ket{\overline Dp,a^\mu} = \ket{Bp,a^\mu}_{NSNS} - \ket{Bp,a^\mu}_{RR}
\end{align}

Parmi lesquels nous distinguons la brane $(+)$ et l'anti-brane\footnote{La brane tournée d'un angle quelconque a un état de bord plus compliqué à calculer~\cite{deAlwis:1999cg}.} $(-)$. Il parait donc que par rotation de $\pi$, seul le secteur R-R est modifié, ce qui est naturel puisqu'il est spinoriel à la différence du secteur NS-NS. 

\subsection{Systèmes de branes BPS et non BPS}

Nous pouvons construire toute sorte de systèmes de branes, coïncidentes~\cite{Myers:1999ps}, séparées, secantes avec angles~\cite{Polchinski:1998rr,Epple:2003xt,Jones:2003ew,Hashimoto:1997gm}, non secantes avec angles~\cite{Hashimoto:1997gm}, boostées~\cite{Dorn:2005vg,Durin:2005ts,Hashimoto:1997gm} \etc. Nous étudierons rapidement un simple système de deux branes identiques parallèles puis le système d'une brane parallèle à une anti-brane. 

\subsubsection{Système de deux branes parallèles}

La fonction de partition cylindrique calculée par échange de cordes fermées entre ces deux branes s'exprime simplement par~:

\begin{align}
\int_0^\infty \di \ell \bra{Dp,a^\mu} e^{-\ell H_c} \ket{Dp,b^\mu}
\end{align}

avec $H_c$ l'hamiltonien de cordes fermées dont les détails se trouvent dans~\cite{Gaberdiel:2000jr}. Avec les renormalisations correctes, elle vaut simplement après transformation modulaire $\ell = 1/2t$ la fonction de partition annulaire des cordes ouvertes projetées GSO~:

\begin{align}
\int_0^\infty \frac{\di t}{2t} \parent{\tr_{NS}\frac{1+(-1)^F}{2} e^{-2 t H_o} -  \tr_{R}\frac{1+(-1)^F}{2} e^{-2 t H_o}}
\end{align}

$H_o$ est l'hamiltonien de cordes ouvertes dont les détails peuvent \^etre trouv\'es \'egalement dans~\cite{Gaberdiel:2000jr}. La projection GSO ci-dessus est bien celle qui permet de retrouver une supersymétrie d'espace-cible ${\mathcal N}=1$, car l'intégrande est nul -- encore gr\^ace à une identité de Jacobi. Ainsi, le spectre de cordes ouvertes est libre de tachyon~: c'est un système stable. Tant que les deux branes sont strictement parallèles, elles conservent les mêmes générateurs de supersymétrie et par conséquent le système a le même degré de supersymétrie qu'une brane seule, \cad qu'il est $\nicefrac{1}{2}$ BPS.

\subsubsection{Système brane-anti-brane}

A l'inverse, la projection GSO précédente n’apparaît plus pour une fonction de partition calculée entre une brane et une anti-brane parallèles, ce qui brise donc la supersymétrie et laisse éventuellement apparaître un tachyon~:

\begin{align}
\int_0^\infty \di \ell \bra{\overline Dp,a^\mu} e^{-\ell H_c} \ket{Dp,b^\mu} &\propto \int_0^\infty \frac{\di t}{2t} \parent{\tr_{NS}\frac{1-(-1)^F}{2} e^{-2 t H_o} -  \tr_{R}\frac{1-(-1)^F}{2} e^{-2 t H_o}} 
\end{align}

Le problème est que la projection sur le secteur NS autorise le fondamental de cordes ouvertes à exister dans le secteur interbranaire, puisqu'il est de chiralité négative de part sa nature fantomatique. Soit $\vec {y_1}$ et $\vec {y_2}$ les positions spatiales des deux branes. Nous identifions naturellement $\norm{\vec y}$ avec la distance les séparant. Par exemple, pour une D-particule et une anti D-particule, nous aurons~\cite{Banks:1995ch} explicitement~:

\begin{align}
Z &\propto \int_0^\infty \frac{\di t}{2t} (2\pi t)^{-3/2} e^{-\frac{t}{2} \frac{\vec y^2}{4\pi^2 \alpha'}} \frac{\eta(it)^{-9}}{\theta_1(0|it)}\sum_{\alpha=2,3,4}^{\sing{-,+,+}} e_\alpha \theta_\alpha(0|it)^4 \nonumber \\ 
&\propto \int_0^\infty \frac{\di t}{2t} (2\pi t)^{-3/2} e^{-\frac{t}{2} (\frac{\vec y^2}{4\pi^2 \alpha'}-1/2)} f(t)
\end{align}

avec $e_2=-e_3=-e_4=-1$. La fonction $f(t)$ est telle que $f(t\to 0)\to 0$ et $f(t \to \infty)\to 1$. Cette intégrale est clairement divergente pour $\norm{\vec y}< \pi\sqrt 2$. Cette divergence est le fait d'un tachyon~ : ce dernier appara\^it explicitement en calculant le spectre de cordes ouvertes tendues entre les 2 branes et en appliquant la projection $P_{D\bar D}=\frac{1-(-1)^F}{2}$ sur les deux secteurs NS et R. En utilisant la même technique que dans le cas bosonique, \cad en étudiant la contribution de distance dans le spectre de corde tendue s'enroulant une fois le long d'une direction compacte, et en appliquant la projection, nous trouvons que le spectre des cordes interbranaires est effectivement, en fonction des secteurs~:

\begin{align}
\NS - \qquad & m^2 = N + \frac{\norm{\vec y}^2}{4\pi^2}  - \frac{1}{2} \nonumber \\ 
\R -  \qquad & m^2 = N  + \frac{\norm{\vec y}^2}{4\pi^2}
\end{align}

Le fondamental NS est bien tachyonique tant que $\norm{\vec y}<\pi \sqrt 2$. Par conséquent, nous identifions une distance critique $\ell_c=\pi\sqrt 2$ différente de celle obtenue dans le cas bosonique -- nous obtenions alors $\ell_c=2\pi$ -- en laquelle le bi-fondamental est non-massif.

\subsubsection{Brane non BPS}

A partir du système brane-antibrane que nous venons d'introduire et qui brise totalement la supersymétrie d'espace-cible, nous pouvons construire une brane unique conservant cette propriété. La méthode que Sen~\cite{Sen:1998ii,Sen:1999mg} utilise pour la construire, consiste à appliquer un certain orbifold sur le système initial. Il obtient ainsi une brane non chargée, puisque le système initial ne l'est pas et non BPS. Sen l'a naturellement appelé \emph{brane non-BPS}. Cette dernière a une dimension complémentaire\footnote{L'orbifold conserve la dimension de la brane mais transforme IIA en IIB et inversement.} des branes BPS, \cad qu'elle a $p$ impair (pair) en type IIA (IIB), en cohérence avec sa neutralité par comparaison aux champs de jauge disponibles dans ces théories-ci.   \\

Rapidement, nous allons résumer la construction que propose Sen. Partons d'une brane et d'une anti-brane coïncidentes en type IIA (par exemple), \cad de dimension $p$ pair. Appliquons l'orbifold $(-1)^{{\bf F}_L}$ qui inverse la parité des spineurs d'espace-cible le long des modes gauches des cordes fermées. Cet orbifold projette, sous la forme $1+(-1)^{{\bf F}_L}$ par exemple dans le calcul d'une fonction de partition toroïdale, la théorie de type IIA sur celle de type de IIB. En outre, l'effet sur l'état de bord de la brane de cet opérateur est de modifier le signe devant l'état $\ket{Bp}_{RR}$ dans~\refe{eq:brane} et par conséquent $(-1)^{{\bf F}_L}$ transforme une brane en une antibrane.  Ainsi, seul le système de ces deux objets est invariant par cet orbifold.

Maintenant, de même qu'en théorie bosonique, nous introduisons des facteurs de Chan-Paton pour représenter les divers secteurs de cordes ouvertes sur le système brane-antibrane. Puisque nous avons 2 branes, le groupe de jauge est $U(2)$ et nous utiliserons encore les matrices de Pauli $\sigma^{0,1,2,3}$. Or nous avons dit que $(-1)^{{\bf F}_L}$ échange la brane et l'antibrane ; par conséquent si $\Lambda$ est un facteur de CP quelconque, la transformation est représentée par le facteur\footnote{Ce pourrait aussi être $\sigma^2$ mais c'est exactement identique.} $\sigma^1$ et $\Lambda \to (\sigma^1)^{-1} \Lambda \sigma^1$. De sorte que les seuls facteurs invariants par cette orbifold sont $\sigma^1$ et $\sigma^0$ dont on en déduit que ces deux secteurs sont indépendants, puisque $\comm{\sigma^0}{\sigma^1}=0$. Par conséquent, la théorie des champs résultante est abélienne. \\

L'objet que nous obtenons est donc élémentaire, puisque $\sigma^3$ étant le secteur associé à la position relative de la brane et de l'antibrane a disparu. En outre, il est de dimension anormale par rapport aux branes BPS naturelles. Enfin, comme il est issu du système brane-antibrane, il brise explicitement la supersymétrie et admet éventuellement un -- et un seul -- tachyon, celui du secteur $\sigma^1$, dans son spectre de cordes ouvertes. Dans ce dernier secteur, le spectre de masse est~:

\begin{align}
\NS \quad \alpha' m^2 &= N -\, \frac{1}{2} \nonumber \\
\R \quad \alpha' m^2 & = N 
\end{align}

Typiquement, la brane non BPS est tachyonique donc instable, mais il existe des techniques permettant de stabiliser une telle brane, \cad faire disparaître le tachyon. La brane non-BPS a été très étudiée du point de vue de la condensation du tachyon, qui s'avère mieux contrôlée que dans le cas bosonique, entre autre parce que le potentiel effectif est symétrique et possède des minima stables. L'action effective a été bien contrainte et sa forme est maintenant presque canonique.

\chapter{G\'en\'eralit\'es~: th\'eories effectives et mod\`ele sigma}
\label{chap:TEC_sigma}
 
Nous commen\c cerons ce chapitre en pr\'esentant les actions effectives et en pr\'ecisant les termes. Nous parlerons du potentiel effectif et de l'introduction du tachyon dans ce cadre. Puis dans la section~\refcc{sec:mod_sig} nous verrons un cas particulier de construction d'action effective en th\'eorie des cordes qui est le mod\`ele sigma. Puisque cela est li\'e au groupe de renormalisation nous pr\'esenterons dans le m\^eme temps les divers sch\'emas de renormalisation et les calculs des \'equations de fl\^ot -- fonctions b\^eta -- que nous utiliserons par la suite. 

\section{Th\'eories effectives~: actions, potentiels et tachyon}
\label{sec:th_eff}

En théorie quantique des champs~\cite{Itzykson:1980rh,Peskin:1995ev}, l'objet de base est l'action \emph{fondamentale} sur les champs fondamentaux -- quantiques -- de la théorie, notée généralement $S[\phi]$. La dynamique complète est encodée dans son expression, qui comprend des termes cinétiques et des termes potentiels. Ces champs sont développés linéairement autour d'une valeur "classique" résolvant les équations classiques du mouvement, qui en première approximation sont dérivées de l'action fondamentale. En effet, le traitement quantique perturbatif\footnote{Ce n'est pas le cas dans le traitement non-perturbatif, qui ne nécessite pas un tel développement.} ne peut que concerner des perturbations des champs, des valeurs "microscopiques" en opposition à des valeurs "macroscopiques". 

Cependant, la prise en compte d'effets quantiques conduit en g\'en\'eral à modifier, parfois drastiquement, les valeurs des champs classiques, qui du coup ne vérifient plus les équations fondamentales du mouvement. En particulier cela se produit en tenant compte des termes divergents et en les soustrayant proprement du calcul -- \cad renormalisation. Cela amène entre autre à modifier la masse physique en une masse nue. Ainsi le champ fondamental doit vérifier les équations du mouvement en fonction de la masse nue, et le champ "classique" -- \cad conforme aux observations -- quant à lui doit vérifier les équations du mouvement en fonction de la masse physique. Il existe donc une claire dichotomie entre ce qui est défini fondamentalement et ce qui apparaît (semi)classiquement, \cad effectivement. \\

Nous sommes donc amen\'es \`a introduire le concept d'\emph{action effective}, dont il existe trois définitions, suffisamment différentes pour les distinguer. 

\subsection{Actions effectives}
 
Commençons par ce que l'on pourrait nommer l'action "semi-classique" \emph{effective}, généra\-le\-ment notée $\Gamma[\varphi]$. Celle-ci tient compte -- idéalement -- de tous les effets quantiques et dont les équations du mouvement ont pour solution la valeur "classique" du champ autour de laquelle la théorie quantique est définie. Cette action~\emph{effective} a donc pour objet un champ classique "off-shell" ; elle est composée de termes cinétiques et potentiels. En particulier, on définit le \emph{potentiel effectif} qui, associé à un terme cinétique standard -- quadratique dans la dérivée première du champ -- doit être minimisé. La valeur du champ classique qui minimise le potentiel effectif est le champ classique "on-shell". Il correspond à la valeur observable du champ dans le vide $\varphi_{cl} = \corr \varphi$ ; par extension, on l'appelle donc "vide" de la théorie, car il caractérise l'état macroscopique du vide. \\

Or, il existe une deuxi\`eme définition d'\emph{action effective}. La théorie des champs fondamentale, en dehors des problèmes de divergences, doit être régularisée pour des raisons associées aux mesures effectuées en laboratoire. En effet, il peut être souhaitable de ne connaître la théorie que jusqu'à une certaine échelle d'énergie -- qu'on nommera \emph{cut-off} -- par exemple l'énergie atteinte lors d'une collision dans un accélérateur de particules. Or pour des raisons quantiques, toutes les énergies supérieures à cette échelle sont accessibles au système physique -- dans des délais temporels infiniment courts en raison de la relation d'incertitude d'Heisenberg -- et vont donc \apriori contribuer au processus. Cependant, on peut tenir compte de l'intégralité de ces effets quantiques d'un coup en les regroupant -- par intégration -- dans un ensemble de termes dans une action \emph{effective} à l'échelle d'énergie souhaitée. Il s'agit de la construction de l'\emph{action effective wilsonienne} (voir par exemple \cite{Bilal:2007ne}) et notée $\Gamma_\mu[\phi]$, avec $\mu$ le cut-off. \\

Dans la limite où certaines interactions sont hors de portée d'un système pour des raisons énergétiques, il existe aussi une procédure consistant à intégrer tous les champs ne pouvant être produit qu'à travers ces interactions et donc à négliger tous les effets quantiques de haute énergie. L'action obtenue pour le champ étudié est donc une approximation de basse énergie de l'action wilsonienne. Par exemple, en théorie des particules, l'échelle d'énergie est souvent très basse par rapport à $M_{GUT}=10^{15} \, \text{GeV}$, qui est la limite à laquelle le modèle standard dans sa définition fondamentale cesse d'être pertinent ; donc l'expression de l'action du modèle standard n'est \apriori qu'une approximation d'une théorie plus fondamentale.

Ainsi, oune \emph{action effective de basse énergie} est \'egalement introduit. Elle est not\'ee g\'enérale\-ment $S_{eff}[\phi]$ et telle que~:

\begin{align}
\Gamma_\mu[\phi] \stackrel{E \ll \mu}{\sim} S_{eff}[\phi] + o(\frac{E}{\mu})
\end{align}

Insistons bien sur le fait qu'aucune des deux n'est la même action effective que la première définie au-dessus, car ici les champs y apparaissant sont toujours quantiques -- on doit toujours les intégrer dans l'intégrale de chemin -- et non classiques\footnote{Du point de vue des composantes des champs de haute énergie il s'agit bien de la même définition, mais cette action résultante n'est pas explicitement invariante de Lorenz, contrairement à la première.}.  C'est une distinction importante car on peut avoir à faire à ces deux types d'action effective en théorie des cordes et donc aussi dans l'étude de cette thèse, en particulier lorsque l'on discutera des traitements off-shell de la condensation de tachyon. \\

Dans la suite nous appellerons donc \emph{action effective} l'action $\Gamma$ définie selon la première définition. La seconde $\Gamma_\mu$ sera nommée \emph{action effective wilsonienne}. Enfin la troisième, $S_{eff}$ désignera l'\emph{action effective de basse énergie}. 

\subsection{Minimisation du potentiel effectif}

Nous avons vu que le potentiel effectif apparaissant dans l'expression de l'action effective devait être minimisé. C'est un processus classique naturel que l'on retrouve dans diverses applications mécaniques, dans le sens où l'on peut assimiler ce potentiel à une donnée énergétique qu'il convient de minimiser. En l’occurrence, ce comportement apparaît immédiatement dans la résolution des équations du mouvement. Le potentiel effectif possède un ou plusieurs minima locaux et un ou plusieurs minima globaux. 

Autour d'un minimum, le potentiel est toujours convexe ; ainsi en première approximation, il y est quadratique et est donc celui d'un oscillateur harmonique -- pour les perturbations quantiques du champ autour de sa valeur classique "constante" correspondante. Le potentiel quadratique y est caractérisé par une constante de couplage que l'on nomme naturellement \emph{masse carrée} et telle que $m^2 \geq 0$. Autrement dit, l'action fondamentale autour de ce vide pour les perturbations $\phi=\delta\varphi$ est à l'ordre quadratique~:

\eqali{\label{eq:action_effective_masse}
S[\phi] \propto \int \di^n x ~\parent{ \frac{1}{2} \partial_\mu \phi \partial^\mu \phi - \frac{m^2}{2} \phi^2}
}

En négligeant dans un premier temps les effets quantiques \emph{tunnel} -- non perturbatifs -- de décroissance d'un minimum local vers un minimum global, on peut considérer que le "vide" associé à ce minimum local est stable. Par opposition, toute configuration classique initialement "constante" en dehors de ce minimum sera instable car appelée à devenir dynamique, dans la direction d'un minimum local. 

Il existe donc comme dans toute théorie des champs des solutions statiques et des solutions dynamiques, pouvant chacune avoir éventuellement des dépendances temporelles ou spatiales non triviales. Pour ce qui est des dépendances spatiales, on parle de \emph{soliton} et pour la dépendance temporelle, on peut parler d'\emph{instanton}. Mais insistons encore sur le point suivant~: chacune de ces solutions classiques représente un vide du point de vue de la théorie des perturbations quantiques autour de la valeur classique du champ.

Ainsi, une théorie quantique des champs  est d\'efinie dans un vide dépendant soit de l'espace, soit du temps, soit des deux, soit d'aucun. Cependant, la résolution d'une théorie quantique le long d'un vide dont les dépendances spatio-temporelles sont non triviales est peu aisée et en général seule la théorie quantique autour des points de la trajectoire du champ où le vide est constant est d\'ecrite. Les configurations pour lesquelles asymptotiquement -- à ses extrémités -- la trajectoire tend vers un -- ou des -- vide(s) constant(s), sont particuli\`erement int\'eressantes, car le plus souvent topologiquement non-triviales.

\subsection{Maxima locaux du potentiel et champs tachyoniques}

\subsubsection{Maxima locaux et tachyons}

Les maxima locaux sont des points très particuliers du potentiel effectif, car les configurations constantes en ceux-ci résolvent aussi les équations du mouvement. Or il apparaît immé\-dia\-te\-ment que ces solutions ne peuvent pas être stables, puisque toute perturbation microscopique entraîne une chute exponentielle du champ classique le long de la pente du potentiel, d'une part ou de l'autre du maximum.

Ainsi, le vide n'est pas quantiquement stable et on ne peut donner de sens au terme de "perturbation", ce qui proscrit toute étude perturbative autour d'un tel vide. 
Cependant, on peut être intéressé par l'étude des trajectoires partant naturellement de ce vide et classer en particulier toutes celles qui rejoignent asymptotiquement un vide stable. C'est exactement le type de solution qui va nous intéresser dans notre étude du tachyon.

En effet, ces maxima sont immédiatement associés à des "perturbations" tachyoniques dans la théorie quantique correspondante pour la raison suivante. A l'instar de la théorie quantique effective autour d'un minimum du potentiel, on peut développer le potentiel effectif à l'ordre quadratique, \cad sous la forme~\refe{eq:action_effective_masse}, et obtenir autour du maximum un oscillateur harmonique. L'équation de mouvement du champ est alors simplement $(-\square + m^2) \phi =0$ avec $\square = - \eta^{\mu\nu}\partial_\mu \partial_\nu$ et $m^2<0$. On identifie donc que la constante de couplage correspondante est une masse carrée négative. De sorte que la perturbation est un \emph{tachyon}. C'est par exemple la situation que l'on rencontre dans le mécanisme de Higgs. \\

\subsubsection{Tachyons, perturbations et condensation}

La notion de \emph{perturbation} correspond ici \`a la d\'efinition suivante~ : valeur de champ suffisamment faible pendant un intervalle de temps fini ou infinitésimal, telle que tout terme d'interaction de la théorie -- fondamentale ou effective – peut \^etre trait\'e comme vertex d'interaction à développer autour de la théorie libre du champ. Sachant que le vide où apparaît une perturbation tachyonique est instable, l'intervalle de temps sera d'autant plus court que la masse carrée sera négative. Par la relation d'incertitude d'Heisenberg, nous avons à peu de choses près $\Delta t \sim 1/\module{m}$. Cette relation s'obtient aussi en remarquant que puisque la masse carrée est négative, la masse en elle-même -- \cad l'énergie au repos -- est imaginaire pure. Alors la dépendance temporelle habituelle des modes perturbatifs de la théorie libre qui va comme $e^{i E t}$ donne pour $E= -i \module m$ l'évolution temporelle de la "perturbation" de tachyon $e^{\module m t}$ ce dont on déduit la constante de temps donnée plus haut. 

A des temps plus élevés que $1/\module m$ nous ne pouvons plus faire sens du terme "perturbation" et donc fatalement du terme "particule". Ceci explique donc que la notion de tachyon en tant que particule n'est donc pas particulièrement bien définie dans le temps. En revanche, si on applique la relation de masse $m^2=E^2-p^2$ en évitant le domaine d'énergie imaginaire, il faut alors imposer que le vecteur énergie-impulsion est de genre espace, soit que l'impulsion est $p^2 > \module m^2$. Dans ce contexte, ce qu'on pourrait appeler \emph{particule tachyonique} viole significativement la causalité sur des distances de l'ordre de $1/\module p < 1/\module m$.

Parce que les modes d'impulsion faible sont inévitablement produits dans le vide instable, le destin du champ tachyonique est donc de rouler le long du potentiel. Pourtant l'issue de ce phénomène est incertaine et dépend fortement de la forme du potentiel effectif. S'il existe des vides stables dans la direction du roulement, alors il est possible que le champ finisse par atteindre au moins un de ces vides et s'y condense en relâchant de l'énergie. \emph{In fine} nous pouvons espérer obtenir un système exactement défini dans le vide stable. C'est précisément ce qu'on cherche à produire dans le mécanisme de Higgs en imposant au champ un potentiel de type \emph{chapeau mexicain}~\cite{Binetruy:2006ad} qui admet un ensemble de vides stables décrit par $U(1)$. La phase dynamique de condensation n'est en g\'en\'eral pas d\'ecrite, car la th\'eorie est habituellement \'etudi\'ee directement autour du vide stable. Cependant la question de cette évolution se pose dans les modèles cosmologiques, entre autre à cause des créations de \emph{défauts topologiques} -- \eg cordes cosmiques. 

En effet, l'exploration du potentiel effectif suggère l'existence d'autres modes de condensation~: par exemple ces condensations purement spatiales introduisant des défauts topologiques -- \emph{solitons} -- interpolant entre plusieurs vides stables distincts\footnote{Dans le cas du Higgs, il s'agit de solutions de vortex car l'ensemble des vides stables est continu. L'interpolation est donc angulaire.}. Ces défauts seraient comme "posés" au maximum instable du potentiel, lieu d'échanges de tachyons anti-causaux et donc d'épaisseur $\sim 1/\module m$.  Des condensations hybrides, que l'on nomme \emph{inhomogènes}, sont aussi imaginables, telles qu'elles combinent un roulement du champ et une séparation spatiale dans plusieurs vides distincts. Notons que cela est soumis à des conditions sur la conservation de l'énergie car la création de défauts topologiques -- des formes de murs de domaines ou de cordes cosmiques dans le cas présent -- n'est pas gratuite énergétiquement, comme on le sait bien dans les études des métaux ferromagnétiques à propos des murs de domaines.

\section{Mod\`ele sigma non lin\'eaire et groupe de renormalisation}
\label{sec:mod_sig}

	Nous allons brièvement présenter ce qu'on entend par modèle sigma et en quoi cela est en relation avec la théorie des cordes. En particulier, nous mettrons cela en contact avec le groupe de renormalisation et les fonctions bêta que nous introduirons dans la section suivante. 

\subsection{Modèle sigma, équations de flots et équations du mouvement}

Le modèle sigma est une théorie des champs couplée,dépendant d'un certain nombre de paramètres -- dont par exemple la métrique sur l'espace des champs qui apparaît sous la forme $\kappa_{ij}(\phi) \partial_a \phi^i \partial^a \phi^j$. Le principe est de classer l'ensemble des théories des champs renormalisables et d'éventuellement décrire la topologie de l'espace des théories. Les différentes théories des champs sont caractérisées par des flots de renormalisation des couplages dont les équations de flots sont données par les \emph{fonctions bêta} du groupe de renormalisation. Parmi ces théories, sont importantes celles qui sont des points fixes du groupe de renormalisation~\cite{Banks:1987qs,Harvey:2000na}-par définition- invariantes d'échelle. Les CFT font partie de ces types de théories sauf qu'elles sont invariantes d'échelle localement et ont donc, comme nous avons pu le voir, un statut très particulier. 

En théorie des cordes, on peut naturellement définir un modèle sigma sur la surface de corde. En effet, l'action de Polyakov sur le plan complexe étant donnée par~:

\begin{align}
S_{p} = \frac{1}{2\pi \alpha'}\int \di^2 z \, G_{\mu\nu}(X) \partial X^\mu \bar \partial X^\nu
\end{align} 

nous voyons que la métrique sur les champs $X$ est identifiée à la métrique d'espace-cible. Nous disions plus tôt que les théories des cordes devaient être définies invariantes conformes, \cad si on veut calculer des éléments de matrice-S on-shell. Autrement dit, l'existence de particules réelles est soumise à condition dans sa description interne et en particulier, nous voyons que le fond géométrique de l'espace-cible sera lui-même contraint, au moins localement. En effet, pour définir une CFT, une condition nécessaire est l'invariance d'échelle globale. Par conséquent, l'équation de flots de $G_{\mu\nu}$ doit être schématiquement telle que la métrique en est bien un point fixe~:

\begin{align}
\beta_{G} = \frac{\di G}{\di \ln \mu} = 0
\end{align}

avec $\mu$ le facteur d'échelle de renormalisation. Il semble, dans une certaine mesure, que cela est très similaire à la définition d'une équation du mouvement, quoiqu'une telle équation est dérivée d'un principe de moindre action~; ce qui fait que la relation entre l'équation du mouvement et les fonctions bêta du groupe de renormalisation n'est pas nécessairement triviale~\cite{Tseytlin:1986ws,Callan:1985ia,Tseytlin:1986zz}. 

En remarquant que $G_{\mu\nu}\partial X^\mu \bar \partial X^\nu$ est très similaire à l'opérateur de vertex d'un graviton, nous comprenons que tout champ contenu dans le spectre doit pouvoir apparaître également dans l'action sur la surface. En fait, d'après la forme et parce l'action est exponentiée, on comprend que la contribution de la métrique est équivalente à la définition d'un état cohérent de graviton. Par conséquent, il est naturel d'introduire les autres champs de la même manière. Nous aurons ainsi de façon très générale, avec $\Phi$ le dilaton, $B_{\mu\nu}$ le Kalb-Ramond, $A_\mu$ les champs de jauge de cordes ouvertes et $T$ le tachyon de corde ouverte sur le bord, en théorie bosonique\footnote{En théorie bosonique on pourrait aussi rajouter le tachyon de corde fermée.} ou supersymétrique\footnote{L'inclusion des champs R-R ou des fermions est trop délicate nous n'en parlerons pas ici.} une action de surface suivante~:

\begin{multline}\label{eq:mod_sig_act}
S_{\text{surf}} = \frac{1}{2\pi\alpha'} \int \di^2 \sigma \, \sing{\parent{G_{\mu \nu}(X) \, \eta^{ab} + B_{\mu\nu}(X)\, \epsilon^{ab} } \partial_a X^\mu \partial_b X^\nu + \Phi(X) \, {\mathcal R}}  \\ + i \oint \di \sigma^a \, A_\mu(X) \, \partial_a X^\mu   + \oint \di \sigma \, T(X)
\end{multline} 

qui définit bien une théorie des champs couplée à 2 dimensions. On parle de chacune de ces contributions en terme de \emph{déformation}. Il faudrait techniquement ajouter aussi des champs massifs mais nous supposons qu'ils découplent dans la limite $\alpha' \to 0$. Pour les supercordes, l'action peut s'\'ecrire directement en super-espace, avec $\sigma^a \in H_+$ sous la forme~:

\begin{multline}\label{eq:mod_sig_act_super}
S_{\text{super}} = \frac{1}{2\pi\alpha'} \int \di^2 z \di^2 \theta \, \sing{\parent{G_{\mu \nu}(\mathbb X)  + B_{\mu\nu}(\mathbb X)} D \mathbb X^\mu \bar D \mathbb X^\nu + \Phi(\mathbb X) \, {\mathcal R}_{super}}  \\ + i \oint \di \sigma \di \theta \, A_\mu(\mathbb X) \, D \mathbb X^\mu   + \oint \di \sigma \di\theta \, T(\mathbb X)
\end{multline} 

que l'on peut décomposer après intégration sur les variables de Grassmann en fonction de $X^\mu$ et $\psi^\mu$ ainsi que leurs homologues anti-holomorphes. Simplement à cause du développement

\begin{align}
A(\mathbb X^\mu) = A(X) + \partial_\mu A(X)\parent{\theta \psi^\mu + \bar \theta \wt \psi^\mu + \theta\bar \theta \, F^\mu}  - \partial_\mu \partial_\nu A(X) \, \theta\bar \theta \,\psi^\mu \tilde \psi^\nu 
 \end{align} 

Les champs du modèle sigma ne sont donc plus simplement $G_{\mu \nu}$ \etc. mais aussi leurs dérivées premières et éventuellement secondes. \\ 
 
Maintenant, chacun des champs de~\refe{eq:mod_sig_act} dans son expression en fonction des champs fondamentaux $X^\mu$ devra schématiquement être solution des équations de flots données par~:

\begin{align}\label{eq:beta}
\beta_i = \frac{\di \varphi_i}{\di \ln \mu} = 0    \qquad \text{ pour tout $i$}
\end{align} 

Ce n'est pas l'unique contrainte car cela ne fait que définir une théorie invariante d'échelle globale. Afin d'exprimer une CFT, il faut aussi que les champs soient invariants d'échelle locale, par conséquent, on impose à toutes les déformations d'être des opérateurs primaires -- de la CFT libre -- \cad de poids $\Delta = (1,1)$ dans le bulk et simplement $\Delta = 1$ sur le bord~:

\begin{align}
V^i_{bulk} = \int \di^2 \sigma \sqrt{g}  \,\mu^i\, {\mathcal O}_{(1,1)} (z,\bar z) \qquad\text{et}\qquad V^i_{bord} = \oint \di y \, \sqrt{g_{yy}} \,\mu^i\, {\mathcal O}_{1} (y)
\end{align}

Parce que l’on s'attend à ce que les équations imposées par les fonctions bêta soient équi\-va\-len\-tes à des équations du mouvement dérivées d'une action -- en l’occurrence \emph{effective} -- on comprend que les champs lorsqu'ils définissent une CFT correspondent à des solutions classiques de ces équations. Ils forment donc, ce qu'on appelle des \emph{fonds} ou \emph{vides} dans le jargon des actions effectives, que nous avons définis dans la section précédente. \\ 

L'objectif de l'étude des modèles sigma en théorie des cordes est donc de trouver l'expression de cette action effective dont les équations du mouvement admettent des CFT pour solutions. L'interprétation physique de cette action effective est sujette à caution, car on ne l'assimile pas nécessairement à une action effective définie dans l'espace-cible. Par exemple, dans la définition de l'OSFT -- théorie des champs de cordes ouvertes -- de Witten~\cite{Witten:1992qy,Witten:1992cr,Marino:2001qc,Gerasimov:2000zp}, l'action effective obtenue est définie sur l'espace des théories des champs et non sur l'espace-cible. En revanche, dans la définition de Tseytlin \emph{et al.}~\cite{Andreev:1988bz,Tseytlin:1987ww,Andreev:1988cb,Tseytlin:2000mt,Tseytlin:1986ti} ils construisent à partir de la fonction de partition off-shell renormalisée et non-intégrée sur les modes zéro $x^\mu$ des champs $X^\mu$, une action off-shell définie sur l'espace-cible\footnote{Cependant, ils insistent sur certaines ambiguïtés associées à des redéfinitions des champs non fixées par le groupe de renormalisation. En particulier, l'existence de plusieurs schéma de renormalisation nourrit cette ambiguïté car il n'existe pas de schéma \emph{naturel} et la dépendance des fonctions bêta, dans le schéma, peut être forte.}. Nous nous placerons dans ce formalisme plutôt que dans celui de Witten qui est trop difficile à utiliser pour ce que nous avons à calculer. \\

Dans le formalisme de fonction de partition de Tseytlin \emph{et al.} l'action est donc construite à partir de la formule suivante~:

\begin{align}\label{eq:eg_act_part}
S[\mu_i] = Z_R[\mu_{i,R}] = \int \di^D x^\mu Z_R'[\mu_{i,R}(x)]
\end{align}

où $\mu_{i,R}$ sont les couplages $\mu_i$ renormalisés et $Z'$ la densité de fonction de partition. Les couplages dans ces théories doivent être définis \emph{relevants} de telle sorte qu'ils ont bien un point fixe UV -- ou ,dans la meilleure situation, interpolent entre deux points fixes UV et IR, le long du groupe de renormalisation~\cite{Harvey:2000na}. A l'inverse, les couplages \emph{irrelevants} n'ont pas nécessairement -- et en général -- de point fixe UV en $\mu^i \to \infty$ et par conséquent la théorie résultante n'est souvent pas renormalisable~\cite{Tseytlin:2000mt,Kubota:1988qn}. Par conséquent, on ne peut pas utiliser les couplages irrelevants comme des perturbations de champs d'espace-cible.

Dans~\refe{eq:eg_act_part} l'extraction de la mesure sur les modes zéro est naturelle à partir de la mesure de l'intégrale de chemin. En notant $X^\mu=x^\mu+\hat X^\mu$ que nous avons séparé en mode zéro + modes d'oscillateurs, elle s'exprime selon~:

\begin{align}
\int {\mathcal D}^DX^\mu = \int d^D x^\mu \int {\mathcal D}^D \hat X^\mu
\end{align}

De sorte que le lagrangien effectif d'espace-cible se définit naturellement par la relation~:

\begin{align}
{\mathcal L}(\mu_i) = Z_R'[\mu_{i,R}(x)]
\end{align}

Cette relation n'est pas tout à fait exacte car elle dépend de la théorie étudiée. En théorie bosonique et en présence de tachyons, la relation doit être modifiée~\cite{Witten:1992cr,Tseytlin:2000mt} en ${\mathcal L}(\mu_i) = (1+\beta_i \partial_i)Z[\mu_i]$. Nous avons enlevé l'indice R par commodité et nous avons noté $\partial_i=\partial/\partial \mu_i$. En revanche, elle est \apriori exacte en théorie supersymétrique. Dans tous les cas, lorsque l'on se place spécifiquement sur une CFT, \cad sur une solution des équations du mouvement, nous avons toujours~:

\begin{align}
{\mathcal L}_{on-shell}(\mu^{on}_i) = Z_R'[\mu^{on}_{i,R}(x)]\Bigg\vert_{\text{CFT}}
\end{align}

Ces formules ont été utilisées avec succès pour obtenir les actions~\cite{Tseytlin:1986ti,Metsaev:1987qp} Born-Infeld (BI) mais aussi obtenir les actions effectives de tachyon dans les systèmes brane-antibrane coïncident ou de brane nonBPS~\cite{Tseytlin:2000mt,Kutasov:2003er}. \\

La formule qui relie les équations du mouvement de l'action effective $S$ aux fonctions bêta des champs $\phi_i$ est de manière très générale donnée par~:

\begin{align}
\frac{\delta S}{\delta \phi_i} = \kappa_{ij}(\phi) \beta_j
\end{align}

avec $\kappa_{ij}(\phi)$ un coefficient tel que la covariance est rétablie. Sans ce coefficient, la relation est en générale fausse à cause de l'indépendance dans le schéma de renormalisation de l'équation du mouvement d'une part et la dépendance de la fonction bêta en ces schémas de l'autre. Cependant, si les fonctions bêta sont universelles alors nous pouvons \apriori choisir $\kappa_{ij}$ constant et indépendant des champs. Cela se produit lorsque des résonances apparaissent – voir section suivante. Dans ce cas, les divergences sont logarithmiques et les contributions aux fonctions bêta universelles, \cad invariantes par changement de schéma de renormalisation.  \\

Pour résumer, si l'on impose à l'action de n'être construite qu'à partir de champs primaires relevants, la condition~\refe{eq:beta} suffit à définir une CFT.

\subsection{Schémas de renormalisation et fonctions bêta}
\label{sec:RG}

Dans cette section nous allons introduire plus en détail le calcul des fonctions bêta en fonction du schéma de renormalisation choisi. Nous en distinguerons 2 en particulier~: le schéma de soustraction minimale et le schéma de Wilson. Le formalisme de renormalisation n'est en effet pas défini de manière unique. \\

Il existe plusieurs façons de renormaliser et cela commence par le type de régularisation utilisée. Nous parlerons ici des renormalisations des divergences UV uniquement. Nous trouvons la \emph{régularisation dimensionnelle} qui redéfinit la dimension de l'espace en $d-\varepsilon$ avec $\varepsilon$ infinitésimale~; elle est souvent utilisée en théorie des champs car elle ne brise pas les invariances de Poincaré ni de jauge -- sauf en théorie des cordes. Nous trouvons aussi la \emph{régularisation brutale UV} qui a le mérite d'être plus immédiate et consiste à poser une limite ultra-violette dans l'espace des impulsions~; le problème est qu'alors la symétrie de Poincaré n'est plus manifeste. Une méthode similaire s'applique directement dans l'espace des positions -- on parle alors de \emph{point-splitting} (en anglais). Elle consiste à décaler infinitésimalement les pôles dans les fonctions de Green autour du point de divergence. Cela revient à tronquer l'espace des positions divergentes en dessous d'un seuil infinitésimal $\varepsilon$. C'est cette dernière méthode que nous utiliserons sur la surface de corde, car elle ne brise pas l'invariance de Weyl à la différence de son homologue dimensionnelle. Il existe enfin la \emph{régularisation zeta}, qui consiste à identifier une fonction dans un domaine de paramètres sans divergence et de procéder à la continuation analytique dans le domaine divergent. Mais cela a le désavantage de ne pas prouver que cette dernière est autorisée.

Une fois que la régularisation est choisie, il reste à se placer dans un certain \emph{schéma de renormalisation}. Il s'agit d'une méthode permettant d'extraire les divergences dépendant du paramètre infinitésimal $\varepsilon$ dans un calcul d'amplitude. La méthode de \emph{soustraction minimale} consiste à retrancher les divergences en insérant un ensemble de contreterme dans la définition de l'action et en espérant -- il faut ultimement le vérifier -- que l'ajout du contreterme ne rajoute pas plus de divergences qu'il n'en enlève. La méthode de \emph{Wilson} consiste à laisser les couplages dépendre explicitement de $\varepsilon$ de telle sorte que les divergences s'annulent ordre par ordre dans l'amplitude. Dans chacun de ces cas, il existe une définition des fonctions bêta que nous allons maintenant aborder en détail.

\subsubsection{Schéma minimal de soustraction}

Ce schéma de renormalisation pose comme principe que toute divergence UV surgissant dans un calcul d'amplitude – par exemple une fonction de partition -- doit être soustraite par le biais d'un ensemble de contretermes ajouté à l'action fondamentale. Ceci amène à définir, comme à l'accoutumée en théorie des champs, des couplages physiques $\mu$ et des couplages nus $\mu_B$. En l’occurrence, dans le cas qui nous occupe, nous avons une théorie fondamentale définie sur une surface délimitée par un bord, par exemple le demi-plan supérieur $H_+$. Supposons que seuls les couplages de bord sont non-triviaux et sont appelés à être renormalisés. Nous définissons donc l'action complète renormalisée et \emph{déformée}\footnote{On nomme \emph{déformation}, et on la note génériquement $\delta S$, tout terme supplémentaire à l'action de la théorie fondamentale libre telle que $S=S_f + \delta S$.} selon~:

\begin{align}\label{eq:action_renorm}
S_{R} = S_{bulk} + \sum_a \mu^a_B \oint_{\mathbb R} \phi_a  = S_{bulk} + \sum_a \ell^{h_a-1} \mu^a \oint_{\mathbb R} \phi_a + S_{ct}
\end{align}

où les champs primaires $\phi^a$ forment un ensemble complet et fermé par OPE~:

\begin{align}\label{eq:OPE_base}
\phi_a(x)\phi_b(y) = \sum_c \frac{C_{ab}^{\hphantom{ab}c}}{(x-y)^{h_a+h_b-h_c}} \phi_c(y) 
\end{align}

et nous avons introduit l'échelle de renormalisation $\ell$. Très schématiquement, on retranche brutalement les divergences UV et on étudie le flot de renormalisation résultant des couplages physiques~; autrement dit, il faut comprendre que le cut-off UV n'est pas une échelle, mais simplement, une régularisation dont le résultat final ne dépend pas. 

Le choix est fait ici, d'introduire les cut-offs dans l'espace des positions et non dans celui des moments comme il est fait habituellement en théorie des champs pour la simple raison que la théorie libre est ici une CFT et que les corrélateurs sont connus exactement dans cet espace. La régularisation utilisée -- nommée \emph{point-splitting} en anglais -- consiste à empêcher les opérateurs de s'approcher à moins de $\varepsilon$ et de s'éloigner de plus de $L$. Cela revient à ajouter, dans toute OPE à 2 points, les fonctions thêta de Heaviside $\theta(|x-y|-\epsilon)\theta(L-|x-y|)$. Cette régularisation brise explicitement la symétrie de Poincaré sur la surface, mais puisque le résultat final ne doit pas dépendre des cut-offs, elle n'est juste plus manifeste dans le développement mais recouverte \emph{in fine}. \\ 

Les contretermes $S_{ct}$ doivent générer toutes les soustractions de divergences obtenues par OPE des déformations $\mu^a \oint_{\mathbb R} \phi_a$. En d'autres termes, par développement de $e^{-S_{ct}}$ toutes ces divergences doivent \^etre supprimées de telle sorte que l'amplitude à calculer converge, dans la limite $\ell \to \infty$.

Nous pouvons calculer sans difficulté ces contretermes au deuxième ordre dans les couplages. Supposons que nous calculions une amplitude dont les insertions ont des OPE régulières avec la déformation (OPE régulière). Cela revient à s'intéresser à la fonction de partition\footnote{La résolution des divergences de la fonction de partition doit résoudre automatiquement le problème des divergences apparaissant par OPE avec des insertions, parce que la fonction de partition génère toutes les amplitudes.} d'une théorie des champs réduite. On se s'intéresse donc ici qu'aux OPE internes au développement du facteur $e^{-\delta S}$. Du coup, nous avons~:

\begin{align}\label{eq:dev_exp}
e^{-\sum_a \ell^{h_a-1} \mu^a \oint \phi_a} = 1 - \sum_a \ell^{h_a-1} \mu^a \oint \phi_a + \frac{1}{2} \sum_{a,b} \ell^{h_a+h_b-2} \mu^a\mu^b \oint\oint_{(L,\varepsilon)} \phi_a \cdot \phi_b  + \ldots
\end{align}

On montre sans difficulté que le deuxième ordre s'écrit, en utilisant~\refe{eq:OPE_base}~:

\begin{multline}\label{eq:OPE_int}
\sum_{a,b} \ell^{h_a+h_b-2} \mu^a\mu^b \int\di\omega \int_{w+\epsilon}^{w+L} \phi_a(z) \phi_b(w) \\ = \sum_{a,b} \ell^{h_a+h_b-2} C_{ab}^{\hphantom{ab}c} \, \mu^a\mu^b \int\di\omega \, \phi_c(w) \, \int_{w+\epsilon}^{w+L} \frac{1}{(z-w)^{h_a+h_b-h_c}}
\end{multline}

A présent, trois situations se présentent. Soit $h_a+h_b-h_c<-1$ ,auquel cas, nous avons une divergence UV~; soit $h_a+h_b-h_c=-1$, auquel cas, la divergence est à la fois UV et IR -- ce qu'on nomme une résonance\footnote{Le nombre $h_i-1$ est la quantité qui apparaît en exposant du facteur d'échelle $\ell$ dans la définition de la déformation non renormalisée. On peut y voir la valeur propre de l'opérateur de dilatation $L_0$ sur la surface, similaire comme on le sait à un hamiltonien. Or, hamiltonien $\sim$ énergie $\sim$ fréquence, ce qui explique le jargon.} car $h_a-1 + h_b-1= h_c-1$~; et enfin, soit $h_a+h_b-h_c>-1$ auquel cas la divergence est IR et par conséquent nous ne nous en occuperons pas car il n'est pas nécessaire de renormaliser, si bien que la fonction bêta sera triviale. \\

Exprimons tout de suite la fonction bêta des couplages de façon générale en fonction du contreterme. On rappelle qu'elle est définie par la formule~:

\begin{align}
\beta^a = \ell \frac{\di \mu^a}{\di\ell}
\end{align}

Comme nous le présentions dans la formule~\refe{eq:action_renorm}, les couplages nus sont reliés aux couplages physiques par le biais des contretermes, soit schématiquement~:

\begin{align}\label{eq:coup_nu}
\mu^a_B = \ell^{h_a-1} \parent{\mu^a + \delta \mu^a_{ct}(\epsilon,\ell) }
\end{align}

Or ces couplages nus doivent être indépendants de $\ell$, car ils sont des paramètres fondamentaux, divergents certes, mais fixes de la théorie. On en déduit l'équation~:

\begin{align}
\ell \frac{\di \mu_B^a}{\di\ell} = 0 = (h_a-1) \ell^{h_a-1} \parent{\mu^a + \delta \mu^a_{ct}(\epsilon,\ell) } + \ell^{h_a-1} \parent{\beta^a + \ell \frac{\di \delta \mu^a_{ct}(\epsilon,\ell)}{\di\ell} }
\end{align}

Ce qui donne pour expression de la fonction bêta~:

\begin{align}\label{eq:beta_expl}
\beta^a = (1-h_a) \mu^a + \croch{(1-h_a)\delta \mu^a_{ct}(\epsilon,\ell) - \ell \frac{\di \delta \mu^a_{ct}(\epsilon,\ell)}{\di\ell}}
\end{align}

La partie crochetée de la formule ci-dessus peut-être calculée explicitement et sans difficulté au deuxième ordre du développement des déformations, mais comme nous l'avons vu, dépend des valeurs respectives des poids entrant en jeu dans le calcul des OPE. Nous allons maintenant calculer ces fonctions bêta en fonction du résultat de l'intégrale~\refe{eq:OPE_int} donc de la valeur de la combinaison $h_c-h_a-h_b$. \\

Le premier cas correspond à $h_c-h_a-h_b<-1$. Il est le plus simple et le plus rapide à traiter. Il suffit d'intégrer~\refe{eq:OPE_int}, ce qui donne~:

\begin{align}
\sum_{a,b} \ell^{h_a+h_b-2} C_{ab}^{\hphantom{ab}c} \, \mu^a\mu^b \, \frac{L^{1+h_c-h_a-h_b}-\varepsilon^{1+h_c-h_a-h_b}}{1+h_c-h_a-h_b} \int\di\omega \, \phi_c(w) 
\end{align}

Pour bien analyser les divergences, le plus pratique est encore de ne centrer l'étude que sur un couplage, disons donc $\mu^c$, et de choisir tous les autres couplages nus. Il est même plus correct de procéder ainsi, car tous les couplages peuvent être renormalisés et à tout ordre ; or c'est l'ensemble "couplage + contreterme" qui doit être développé. Réécrivons donc la formule inspirée de la précédente~:

\begin{align}
\sum_{a,b} C_{ab}^{\hphantom{ab}c} \, \mu_B^a\mu_B^b \, \frac{L^{1+h_c-h_a-h_b}-\varepsilon^{1+h_c-h_a-h_b}}{1+h_c-h_a-h_b} \int\di\omega \, \phi_c(w) 
\end{align} 

Il est à présent très clair que tant que $1+h_c-h_a-h_b<0$, seule la limite UV est divergente. Par conséquent, il faut ajouter à $\mu^c$ le contreterme~:

\begin{align}
\delta S_{ct} = - \sum_{a,b} C_{ab}^{\hphantom{ab}c} \, \mu_B^a\mu_B^b \, \frac{\varepsilon^{1+h_c-h_a-h_b}}{1+h_c-h_a-h_b} \int\di\omega \, \phi_c(w) 
\end{align}

D'où l'on extrait :

\begin{align}
\delta \mu^c = - \ell^{1-h_c} \sum_{a,b} C_{ab}^{\hphantom{ab}c} \, \mu_B^a\mu_B^b \, \frac{\varepsilon^{1+h_c-h_a-h_b}}{1+h_c-h_a-h_b} 
\end{align}

La fonction bêta est immédiatement déduite de l'expression~\refe{eq:beta_expl}. Par invariance des couplages nus, les deux termes dépendant de $\delta \mu^c$ se compensent pour donner simplement~:

\begin{align}
\beta^c = (1-h_c)\mu^c
\end{align}

Ainsi, en présence d'une divergence UV purement de type \emph{puissance}, le contreterme ne modifie pas le flot de renormalisation. \\

Le cas suivant $h_c-h_a-h_b=-1$ se nomme \emph{résonance}. L'intégrale~\refe{eq:OPE_int} est alors logarithmique :

\begin{align}
\int_\varepsilon^L \frac{\di z}{z} = \ln\frac{L}{\varepsilon}
\end{align} 

Nous avons à la fois une divergence UV et une divergence IR mais nous devons uniquement soustraire du calcul la divergence UV. Cependant, pour que l'argument du logarithme soit sans dimension, il faut y inclure l'échelle de renormalisation sous la forme $\ln \ell/\varepsilon$. Ainsi, le contreterme est facilement déduit des calculs précédent et doit être~:

\begin{align}
\delta S_{ct} = \sum_{a,b} C_{ab}^{\hphantom{ab}c} \, \mu_B^a\mu_B^b \, \ln\frac{\ell}{\varepsilon} \int\di\omega \, \phi_c(w) 
\end{align}

D'où nous extrayons :

\begin{align}
\delta \mu^c = \ell^{1-h_c} \sum_{a,b} C_{ab}^{\hphantom{ab}c} \, \mu_B^a\mu_B^b \, \ln\frac{\ell}{\varepsilon}
\end{align}

et la fonction bêta~:

\begin{align}
\beta^c = (1-h_c)\mu^c - \sum_{a,b} C_{ab}^{\hphantom{ab}c} \mu^a\mu^b  + o(\ln\frac{\ell}{\varepsilon})
\end{align}

Nous avons remplacé les couplages nus par leur expression~\refe{eq:coup_nu}. Donc en toute généralité il existe un terme dépendant explicitement de $\varepsilon$ et $\ell$. Notons, cependant, que souvent ces termes sont d'ordres plus élevés que quadratiques et sont donc négligeables en première approximation -- il se peut d'ailleurs qu'ils finissent par s'annuler en tenant compte de l'ensemble des contributions. \\ 

En ce qui concerne le dernier cas divergent IR, nous ferons simplement la remarque suivante~: si les divergences IR peuvent paraître importantes dans la théorie définie sur le plan complexe, elles ne le sont pas vraiment pour un vrai calcul d'amplitude en théorie des cordes\footnote{Le plan complexe est conforme à un \emph{voisinage} de la variété à 2 dimension et non l'intégralité de la variété. Pour la sphère par exemple, ce peut \^etre un hémisphère.}. En effet, la limite infrarouge est naturellement fixée par la géométrie complète de la surface, \cad soit la sphère soit le disque. Si l'on fait en sorte que la limite UV soit bien définie et que les flots de renormalisations soient nuls, alors le résultat ne dépend simplement ni de la taille de la sphère ni de celle du disque -- elle correspond naturellement à l'échelle infrarouge. Mais intéressons-nous maintenant au schéma de Wilson.

\subsubsection{Schéma wilsonien}

Dans ce schéma, nous nous intéressons directement au comportement ultra-violet de la théorie renormalisée. L'idée est de tronquer la théorie à une certaine échelle, ici $\varepsilon$ et d'exprimer une action effective à cette échelle, ce que l'on avait introduit par la notation $\Gamma_\varepsilon$, de telle sorte que le calcul résultant est fini dans l'UV. Cette action effective s'écrit dans notre cas comme~:

\begin{align}\label{eq:action_wilson}
\Gamma_\varepsilon = S_{bulk} + \sum_a \varepsilon^{h_a-1} \mu^a(\varepsilon) \oint \phi_a
\end{align}

Encore une fois nous ne nous intéressons qu'aux déformations de bord. Les couplages dépendent maintenant explicitement du cut-off UV. A la différence du schéma minimal, il n'y a pas de contreterme, ils sont déjà inclus dans $\mu^a(\varepsilon)$. L'expression de ces derniers est obtenue de telle sorte que la théorie ne dépende pas asymptotiquement ($\varepsilon\to 0$) de la régularisation UV, ce qui s'exprime par~: 

\begin{align}\label{eq:wilson_cont}
\varepsilon \partial_\varepsilon \parent{ e^{-\Gamma_\varepsilon}} \stackrel{\varepsilon \to 0}{\longrightarrow} 0 
\end{align}

et donne, d’emblée, la fonction bêta des couplages selon~:

\begin{align}
\beta^a = \varepsilon \partial_\varepsilon \mu^a
\end{align}

Nous constatons donc que dans ce schéma tout ce qui concerne les divergences infrarouges est laissé de c\^oté. Autrement dit, seul le comportement \emph{local} de la théorie est étudié. Chaque couplage s'exprime selon~:

\begin{align}
\mu^a(\varepsilon) = \varepsilon^{1-h_a} \sing{\mu^a_R + \delta\mu^a(\mu_R^b,\varepsilon , L)}
\end{align} 

Nous avons introduit un couplage renormalisé $\mu^a_R$ indépendant du cut-off et un "contreterme" dépendant des deux cut-offs ainsi que d'autres couplages renormalisés, en toute généralité. L'ensemble réinjecté dans l'action~\refe{eq:action_wilson} permet d'écrire~:

\begin{align}\label{eq:wils_sous}
\Gamma_\varepsilon(\mu^a) &= S_{bulk} + \sum_a \mu_R^a(\varepsilon) \oint \phi_a + \sum_a \delta \mu^a(\mu_R^b,\varepsilon,L) \oint \phi_a \nonumber \\ 
 & = S_{bulk} + \sum_a \mu_R^a \oint \croch{\phi_a}_R = \Gamma_R(\mu^a_R)
\end{align}

où nous introduisons des champs primaires renormalisés. Notons que la deuxième ligne est le point de départ du schéma minimal et qu'alors simplement $\croch{\phi_a}_R = \phi^B_a$. L'expression ci-dessus permet d'écrire toute amplitude comme~:

\begin{align}
{\mathcal A}_\varepsilon [\mu^a] = {\mathcal A}_R [\mu^a_R]
\end{align}

Reprenons le développement de la section précédente et adaptons-le au cas qui nous intéresse. Nous avons~\refe{eq:dev_exp}~:

\begin{align}\label{eq:dev_exp_wils}
e^{-\sum_a \ell^{h_a-1} \mu^a \oint \phi_a} = 1 - \sum_a \epsilon^{h_a-1} \mu^a \oint \phi_a + \frac{1}{2} \sum_{a,b} \epsilon^{h_a+h_b-2} \mu^a\mu^b \oint\oint_{(\ell,\varepsilon)} \phi_a \cdot \phi_b  + \ldots
\end{align}

Encore une fois, le second ordre après OPE devient :

\begin{align}
 \sum_{a,b} \varepsilon^{h_a+h_b-2} C_{ab}^{\hphantom{ab}c} \, \mu^a\mu^b \int\di\omega \, \phi_c(w) \, \int_{w+\epsilon}^{w+L} \frac{1}{(z-w)^{h_a+h_b-h_c}}
\end{align}

Pour $h_a+h_b-h_c \neq 1$, nous pouvons calculer l'intégrale génériquement :

\begin{align}
 \sum_{a,b}  C_{ab}^{\hphantom{ab}c} \, \mu^a\mu^b  \, \frac{\parent{\frac{L}{\varepsilon}}^{h_c+1-h_a-h_b} - 1}{h_c+1-h_a-h_b}~\times\varepsilon^{h_c-1} \int\di\omega \, \phi_c(w)
\end{align}

Compte-tenu de l'équation~\refe{eq:wilson_cont} et du développement~\refe{eq:dev_exp_wils}, nous devons résoudre~:

\begin{align}\label{eq:big_eq}
\varepsilon \partial_\varepsilon \parent{ \varepsilon^{h_c-1} \parent{ \mu^c - \sum_{a,b}  C_{ab}^{\hphantom{ab}c} \, \mu^a\mu^b  \, \frac{\parent{\frac{L}{\varepsilon}}^{h_c+1-h_a-h_b} - 1}{h_c+1-h_a-h_b}}} \to 0
\end{align}

Ce qui donne l'équation~:

\begin{multline}
\varepsilon^{h_c-1} \Bigg( (h_c-1) \mu^c + \beta^c  - \sum_{a,b}  C_{ab}^{\hphantom{ab}c} \parent{\beta^a \mu^b + \mu^a \beta^b + (h_c-1)\mu^a\mu^b}\frac{\parent{\frac{L}{\varepsilon}}^{h_c+1-h_a-h_b} - 1}{h_c+1-h_a-h_b} \\ + \sum_{a,b}  C_{ab}^{\hphantom{ab}c} \, \mu^a\mu^b  \, \parent{\frac{L}{\varepsilon}}^{h_c+1-h_a-h_b} \Bigg) \to 0
\end{multline}

A l'ordre quadratique rien sinon l'ordre linéaire dans les fonctions bêta des couplages $\mu^a$ et $\mu^b$ ne doit contribuer. Du reste, elles sont triviales compte-tenu de la formule ci-dessus. Supposons donc que nous avons pour tout $a$~:

\begin{align}
\beta^{a} = (1-h_{a})\mu^{a} + \delta\beta^{a}
\end{align}

avec $\delta\beta^a$ d'ordre 2. La formule précédente à l'ordre quadratique dans les couplages toujours, devient simplement~:

\begin{align}
\varepsilon^{h_c-1} \Bigg( \delta\beta^c + \sum_{a,b}  C_{ab}^{\hphantom{ab}c} \mu^a \mu^b \Bigg) \to 0
\end{align}

Dans le cas où $h_c+1-h_a-h_b<0$, on trouve~:

\begin{align}
\delta\beta^c = - \sum_{a,b}  C_{ab}^{\hphantom{ab}c} \mu^a \mu^b  \quad \Longrightarrow \quad \beta^c = (1-h_c)\mu^c - \sum_{a,b}  C_{ab}^{\hphantom{ab}c} \mu^a \mu^b
\end{align}

Dans le cas contraire, si $h_c+1-h_a-h_b>0$ le terme est non divergent et ne contribue simplement pas à la fonction bêta qui est donc trivialement~:

\begin{align}
\beta^c = (1-h_c) \mu^c 
\end{align}

Ceci implique que le groupe de renormalisation n'est pas sensible ici aux divergences IR. Le cas intermédiaire, résonant, $h_c+1-h_a-h_b = 0$ n'est pas difficile à résoudre en revenant à la formule~\refe{eq:big_eq} et en remplaçant~:

\begin{align}
\frac{\parent{\frac{L}{\varepsilon}}^{h_c+1-h_a-h_b} - 1}{h_c+1-h_a-h_b} \longleftrightarrow \ln \frac{L}{\varepsilon}
\end{align}

Alors nous obtenons encore~:

\begin{align}
\beta^c = (1-h_c)\mu^c - \sum_{a,b}  C_{ab}^{\hphantom{ab}c} \mu^a \mu^b
\end{align}

Nous pouvons condenser toutes ces formules sous la forme~:

\begin{align}
\beta^c = (1-h_c)\mu^c - \sum_{a,b}  C_{ab}^{\hphantom{ab}c} \mu^a \mu^b ~ \Theta(h_a+h_b-h_c-1)
\end{align}

avec $\Theta(x\geq 0)=1$ et $0$ sinon.

\subsubsection{Comparaison entre les deux schémas et remarques sur les résonances}

Dans ce schéma, les fonctions bêta sont donc différentes de celles du schéma de soustraction minimale. Cela tient au fait que, dans ce sch\'ema, le groupe de renormalisation analyse directement le comportement de la théorie dans l'UV. Les quantit\'es d\'eriv\'ees dans les deux sch\'emas sont cependant reli\'es par une simple redéfinition des champs, comme par exemple dans la formule~\refe{eq:wils_sous}. Le calcul détaillé de Gaberdiel \emph{et al.}~\cite{Gaberdiel:2008fn} est int\'eressant \`a cet \'egard. Ainsi, la différence entre les fonctions b\^eta de ces deux sch\'emas ne tient qu'à une redéfinition des champs. Ceci implique que l'interprétation des fonctions bêta en tant qu'équations du mouvement est délicate, et il faut comprendre que l'identification n'est possible qu'\emph{\`a une redéfinition des champs près}. 

Toutefois, à la résonance, les fonctions bêta des deux schémas sont exactement égales. C'est un résultat important impliquant que les résonances fournissent des contributions universelles aux fonctions bêta, \cad qui ne dépendent pas du schéma de renormalisation et sont donc invariantes par redéfinition des champs. Naturellement, nous sommes tentés d'interpréter ces fonctions bêta comme équations du mouvement. Nous verrons dans les sections~\refcc{sec:off_sep} et~\refcc{sec:renorm_susy} que cela est encore assez d\'elicat, du fait des ambigu\"it\'es de red\'efinition des fonctions b\^eta par des termes proportionnels \`a des fonctions b\^eta, \cad nuls lorsqu'elles sont identifi\'ees \`a des \'equations du mouvement, selon $\beta_i=0$.

\chapter{G\'en\'eralit\'es~: Condensation de tachyon de cordes ouvertes}
\label{chap:cond_tach}
 
Nous allons \`a pr\'esent discuter des théories pré-existantes de condensation de tachyon dans des systèmes de branes. Dans la section~\refcc{sec:cond_bos}, nous développerons les concepts et les outils dans le cadre de la théorie bosonique, o\`u nous verrons en d\'etail les diff\'erents modes de condensation de tachyon ainsi que les objets qu'ils produisent. Puis, dans la section~\refcc{sec:cond_susy} nous les adapterons \`a la question de la condensation de tachyon en th\'eorie des supercordes. 

Il convient d'étudier tout syst\`eme pr\'ealablement en th\'eorie bosonique, car l'intuition peut y prendre plus de place. En effet, la dynamique des fermions est souvent peu intuitive et peut faire perdre de vue la pertinence d'une étude. En outre, il est souvent pratique de tester préalablement des outils en théorie bosonique car plus simple à manipuler. Voici quelques références utiles et détaillées sur la condensation de tachyon~\cite{Sen:2004nf,Sen:2002nu,Sen:2000kd,Sen:2002in,Sen:1999xm,Kutasov:2003er}. \\

Dans la section suivante nous discuterons du potentiel effectif du tachyon et nous verrons comment introduire la condensation ainsi que les vides stables et instables. Nous montrerons que par condensation on peut atteindre un vide stable non tachyonique et nous étudierons sa nature. Nous verrons qu'il s'identifie à une théorie des cordes fermées~\cite{Sen:1999mh,Sen:2000kd,Bergman:2000xf}. Nous \'etablirons dans les sections~\refcc{sec:2.1.2} et~\refcc{sec:2.1.3} ensuite l'existence de solutions \`a d\'ependence spatiale ou temporelle permettant d'interpoler de vide stable \`a vide stable ou de vide instable \`a vide stable. Enfin dans la section~\refcc{sec:2.1.4} nous discuterons bri\`evement de la connexion des CFT de condensation aux mod\`eles int\'egrables.

\section{Tachyon, vide stable et potentiel effectif}
\label{sec:tach_vide}

Au sein de certaines théories des cordes, des excitations tachyoniques~\cite{Polchinski:1998rq,Polchinski:1998rr} sont identifiées dans le spectre de masse des cordes ouvertes ou fermées. Or ces théories décrivent des théories de cordes "classiques", \cad qui ne sont pas exprimées en raison de champs quantifiés, et pour lesquelles il n'existe pas \emph{a priori} de notion de champ de corde et de potentiel effectif. Cependant, compte-tenu de ce qui a été développé plus haut, ces théories des cordes doivent être identifiées à des maxima locaux dans des théories des champs effectives au sein desquelles, les cordes sont les quanta de perturbation des champs quantiques correspondants\footnote{On pourra lire sur les actions effectives en théorie des cordes~\cite{Fradkin:1985ys,Fradkin:1984pq}.}.

Ainsi, toute théorie des cordes ayant un ou des tachyons dans son spectre est-elle localisée dans le paysage d'une théorie des champs effective en un vide instable~\cite{Tseytlin:2001ah,Tseytlin:2000mt,Sen:2004nf}. Par conséquent, du point de vue d'une théorie des champs de cordes, il ne s'agit pas d'un vide pertinent à étudier. A l'inverse, s'inspirant de ce qui a été dit, il serait imaginable~\cite{Sen:2002nu,Sen:1999mh} d’imposer à la théorie des cordes de quitter son vide instable et d'évoluer du vide initialement "tachyonique" vers son vide stable\footnote{La littérature emploie souvent, par un évident abus de langage, que le "vide tachyonique" est le vide stable. Il ne convient pas d’utiliser cette appellation trompeuse quitte à s'éloigner du choix des auteurs. Dans cette thèse le "vide tachyonique" est défini par le vide \emph{instable}.}. Ce procédé est précisément ce que l'on entend par \emph{condensation du tachyon} en théorie des cordes. Il existe en fait des modes de condensation plus généraux que celui, dynamique, qu'on vient de présenter. On peut en construire dépendant de l'espace, d'autres du temps et certains des deux, \cad en général des modes de condensation \emph{localement dépendant} des coordonnées de l'espace-temps cible. Enfin on peut aussi imaginer simplement transporter le système \emph{à la main} dans son vide stable. Il s'agit alors de condensation \emph{statique homogène}.

Il existe cependant une condition importante à respecter~ : le mode de condensation doit être une solution des équations du mouvement de l'action effective. Comme nous l'avons vu dans le chapitre précédent, cette propriété s'exprime aussi dans la théorie de surface de corde, en termes de contrainte de marginalité exacte, \cad d'invariance conforme de la théorie.  

\subsubsection{Vide stable et condensation statique homogène}

L'étude du vide stable est de première importance. En effet, il s'agit d'identifier sa nature et sa composition en éléments fondamentaux -- branes, cordes. Nous pourrions alors déterminer la nature du système stable correspondant et par conséquent ce en quoi un système instable pourrait évoluer. Dans un premier temps, nous allons examiner un cas de condensation homogène et statique, \cad que nous placerons \emph{à la main} le tachyon constant en son -- ou un de ses -- vide stable. Puis, dans la section suivante, approfondir la question de la condensation locale.

Dans ce cas de solution constante, on s'attend~\cite{Sen:1999mh} intuitivement à ce que ce vide obtenu soit celui des cordes fermées à $26$ dimensions en théorie bosonique et $10$ dimension en théorie de supercordes. En effet, la symétrie de Poincaré dans le volume d'univers de la brane est conservée par cette solution et le vide est par définition stable. Or un vide de corde ouverte ne semble pas compatible avec ces propriétés, puisqu'en premier lieu le tachyon doit avoir disparu\footnote{Une façon de se débarrasser d'un tachyon sur une brane est d'ajouter de nouveaux champs. Par exemple un champ de jauge constant peut permettre de décaler le spectre de masse suffisamment pour compenser la soustraction à l'origine du tachyon. Cependant, c'est un procédé ad-hoc qui n'aide pas à comprendre le mécanisme "naturel" de condensation de tachyon puisqu'il s'en débarrasse.} et qu'en second lieu on impose à la solution d'être indépendante des coordonnées le long de la brane. Notons cependant que rien n'empêcherait \apriori l'existence d'une brisure spontanée de la symétrie de Poincaré ; mais il s'agirait alors d'un phénomène relevant d'une condensation locale. \\

Il est possible de prouver~\cite{Sen:2004nf,Sen:2000kd,Bergman:2000xf,Yi:1999hd} que ce vide est effectivement celui d'une théorie des cordes fermées, autant en théorie bosonique qu'en supercordes, \cad que la brane -- sur laquelle n'existe que des cordes ouvertes -- a la capacité d'imiter un vide de corde fermée de dimension maximale dans l'espace-temps complet. En effet, la théorie de corde fermée obtenue est bien uniforme dans l'espace à 26 ou 10 dimensions et non uniquement le long du volume de la brane. Ceci est justifié par l'argument que la tension de la brane est (quasi-)nulle dans ce vide, de telle sorte qu'elle s'y déforme -- par fluctuations puisque la brane est entité dynamique -- infiniment dans les dimensions transverses\footnote{Dans la limite de tension nulle, le syst\`eme devrait croiser la valeur de la tension de corde donc la brane devrait finalement se désintégrer en cordes.} jusqu'à se répandre dans l'espace entier. 

Plus concrètement, il se produit une matérialisation des flux électriques~\cite{Sen:2000kd} le long de la brane. Ils sont accompagnés d'un confinement~\cite{Bergman:2000xf,Yi:1999hd} leur donnant une forme linéaire ou bien circulaire. Enfin, leur tension est quantifiée en la valeur de la tension de corde ; ce qui implique qu'ils se matérialisent sous forme de cordes fondamentales. Par exemple, en connectant les deux extrémités \emph{séparées} d'une corde ouverte, ils reforment une corde fermée. Il semble que la brane finisse par se désintégrer de cette manière. \\

Il en découle un certain nombre de propriétés essentielles qui doivent être vérifiées par le potentiel effectif et les caractéristiques observables du système branaire instable, en particulier l'énergie, la pression et les diverses charges -- une brane étant couplée à divers champs de cordes fermées.  

\subsection{Contraintes sur le potentiel effectif}

Nous noterons le potentiel effectif $V(T)$. La propriété d'existence d'un vide de corde fermée implique que la brane soit autour ce vide de tension nulle~\cite{Sen:1999mh,Sen:2004nf}~; ce qui est équivalent à demander qu'elle soit totalement dissipée. Le long du potentiel, la tension de la p-brane -- la densité d'énergie de masse en somme -- est donnée par la combinaison~:

\begin{align}\label{eq:tension_brane}
{\mathcal T}_p = T_p \, V(T)
\end{align}

On suppose que le vide est atteint pour $T=T_0^{(i)}$ pour $i=1\ldots N$, si en toute généralité il existe $N$ vides stables\footnote{On suppose qu'ils sont tous des minima globaux. La question des minima locaux est plus difficile à adresser. En outre, l'ensemble de ces vides peut ne pas être dénombrable, ce qui se produit dans le cas $D-\bar D$, puisque le tachyon est complexe.}. Ainsi, l'hypothèse impose~:

\begin{align}
V(0)=1 ~~\text{et}~~ V(T_0^{(i)})=0
\end{align}

Notons dés à présent que ce résultat est compatible avec la plupart des potentiels effectifs obtenus en théorie des champs de cordes ouvertes (OSFT)~\cite{Kutasov:2000aq,Marino:2001qc}, à part en théorie bosonique~\cite{Gerasimov:2000zp,Witten:1992cr,Witten:1992qy} pour des problèmes d'analycité et d'asymétrie, mentionnés dans la note~\notref{note-analy} ci-dessous. En outre, quelques actions effectives de tachyon obtenues dans le cadre des théories conformes avec bord, confirment ce comportement~\cite{Kutasov:2003er}.

Le potentiel de tachyon a en outre été prouvé universel par Sen~\cite{Sen:1999xm}, \cad qu'il ne dépend pas des valeurs classiques des autres champs de fond comme par exemple la métrique ou le dilaton. Il s'en suit que l'on peut tout aussi bien étudier la condensation de tachyon en espace ouvert ou en espace compact, et ce en utilisant une même expression du potentiel. Par conséquent, on s'attend à ce que les solutions obtenues ne diffèrent pas significativement dans un cas par rapport à l'autre. Une représentation souvent utilisée, presque canonique aujourd'hui est~: 

\begin{align}\label{eq:pot_sym_can}
V(T) = \frac{1}{\cosh \alpha T}
\end{align}

avec $\alpha=1/2$ en théorie bosonique\footnote{\label{note-analy} Ce point est peu délicat et dépend de quel tachyon on parle. Par exemple, cette forme est correcte pour un tachyon interbranaire dans le cadre d'un système de branes bosoniques parallèles. Mais elle ne l'est \apriori plus pour un tachyon vivant sur une seule brane parce que le potentiel y est asymétrique -- voir figure~\refe{fig:pot_bos}. On pourrait faire une continuation analytique des supercordes vers la théorie bosonique, mais le lagrangien effectif obtenu n'est pas compatible avec le calcul de la fonction de partition~: la correspondance est non-analytique autour de $T=0$. On pourra par exemple comparer les études de Tseytlin~\cite{Tseytlin:2000mt} avec la discussion de Kutasov et Niarchos dans~\cite{Kutasov:2003er}.} et $\alpha=1/\sqrt 2$ en supercordes. Il est représenté sur la figure~\ref{fig:pot_sup}. 

\begin{figure}[h!]
\centering
\includegraphics[scale=1]{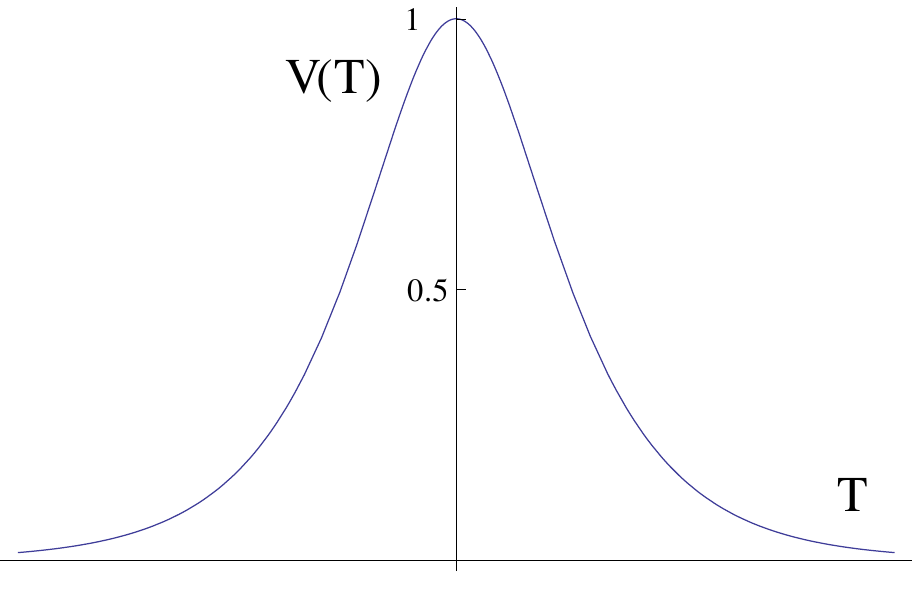}
\caption{\label{fig:pot_sup} Potentiel effectif canonique pour le champ de tachyon.}
\end{figure}

L'argument $T$ peut aussi être remplacé par un module complexe, par exemple dans le cas $D-\bar D$. Ainsi si le tachyon est réel, on dénombre généralement deux vides distincts $T_0^\pm\to \pm \infty$ qui vérifient mais brisent spontanément la symétrie $\mathbb Z_2$. Tandis que si le tachyon est complexe l'ensemble des vides n'est plus dénombrable mais vérifie et brise spontanément la symétrie $U(1)$. \\

Dans le cadre des supercordes, on trouve fréquemment dans la littérature les représentations suivantes~:

\begin{align}
V(T) \in \sing{ e^{- \beta T^2} \, , \, \frac{1}{1 + \gamma T^2} }
\end{align}

avec $\beta$ et $\gamma$ des constantes arbitraires ici -- elles peuvent être réabsorbées par redéfinition des champs dans le tachyon. On s'attend en général à ce que la théorie correspondante puisse être amenée, par redéfinition des champs, sous une forme telle que le potentiel effectif y est canonique.\\

En revanche, dans le cadre des cordes bosoniques~\cite{Gerasimov:2000zp,Witten:1992cr,Tseytlin:2000mt} pour un tachyon de corde ouverte dont les extrémités sont attachés à une même brane, on obtient par calcul direct dans le cadre de la BSFT,~:

\begin{align}
V(T) = e^{-T}\parent{1+T}
\end{align}

Comme nous pouvons le constater dans sa représentation graphique -- figure~\ref{fig:pot_bos} -- le potentiel n'est pas minimisable dans la région $T<0$ de telle sorte que toute condensation y est catastrophique, \cad perpétuelle et sans jamais atteindre de vide stable, ce qui ne peut évidemment pas être d'un grand intér\^et.\footnote{Les effets tunnel vers cette région ne devraient pourtant pas \^etre négligeables. Ils seraient même probablement dominants, mais n'oublions pas que la théorie bosonique est de toute fa\c con pathologique \`a cause du tachyon de corde ferm\'ee.}. On ne s'intéresse donc généralement qu'à la partie minimisable $T>0$ du potentiel. Dans ce cas, on peut se ramener à l'étude d'un tachyon dans un potentiel symétrique. \\

\begin{figure}[h!]
\centering
\includegraphics[scale=1]{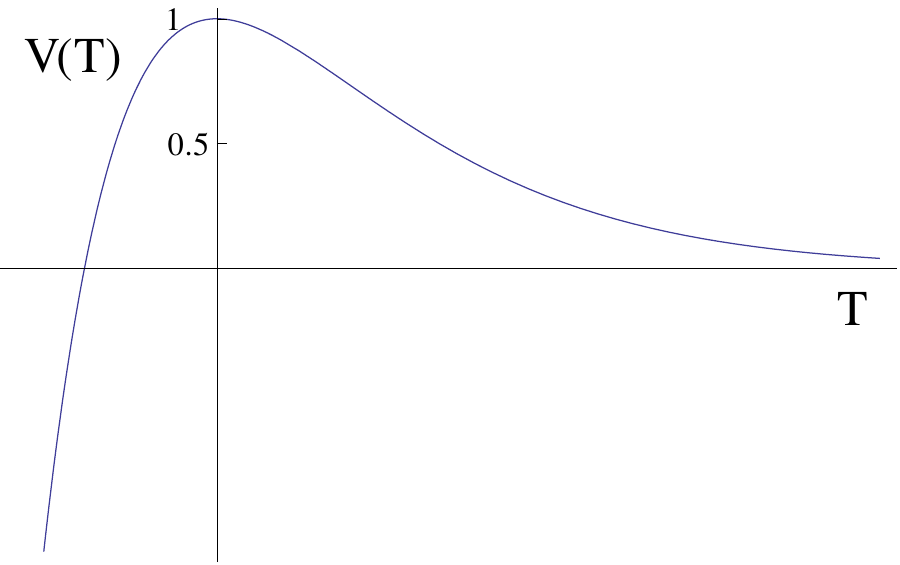}
\caption{\label{fig:pot_bos} Potentiel effectif pour le champ de tachyon obtenu explicitement en théorie des champs de corde bosonique.}
\end{figure}

\subsubsection{Contraintes sur les observables du système}

Nous disions que la nature du vide de corde fermée impose des contraintes sur les observables de la théorie branaire, par exemple l'énergie, la pression ou les diverses charges. 

Premièrement, l'énergie doit \^etre identiquement nulle dans l'espace-temps entier, restaurant ainsi la symétrie de Poincaré brisée spontanément par la brane. Or la densit\'e d'\'energie de la brane dans le vide est donn\'ee par sa tension~\refe{eq:tension_brane}. Le profil de la brane dans le vide tachyonique \'etant en outre une fonction delta de Dirac $\delta^{(D-p-1)}(x^\perp)$ -- on suppose que la brane est placée en $x^\perp =0$ qui par symétrie de translation permet toujours de conserver un point de vue général -- nous avons alors, du vide instable au vide stable, la transition suivante~:

\begin{align}
\varepsilon = T_p \, \delta^{(D-p-1)}(x^\perp) ~ \longrightarrow~ \varepsilon_0 = T_p V(T_0) = 0
\end{align}

où $\delta^{(D-1)}(0)$ est simplement le volume du vide à un instant $t$. La densité d'énergie $\varepsilon$ s'annule donc dans le vide stable \`a condition que le potentiel s'annule \'egalement. \\

Deuxièmement, la pression aussi est censée s’annuler~: on nomme \emph{pression} les composantes diagonales $T_{ii}$ du tenseur énergie-impul\-sion perpendiculaires à celle de l'énergie $\epsilon=T_{00}$. En effet, si l'on s'attend bien à obtenir un espace-temps vérifiant une symétrie de Poincaré, homogène, alors cet espace -- en l'absence de contrainte \emph{a priori} -- doit être plat. Rappelons que cela est justifié par l'argument que la tension effective de la brane tend vers zéro et que par conséquent elle fluctue sans coût énergétique en tout point de l'espace, de fa\c con totalement homogène --  autrement dit les fluctuations ne sont plus contrôlées. De la sorte, on peut justifier que la pression doit également tendre vers zéro et même être tout à fait nulle dans le vide de corde fermée. \\

Enfin, rappelons que l'on assimile une brane à une source de cordes fermées. Entre autre, nous avons vu qu'elle constitue une source d'énergie et de pression, et plus généralement d'énergie-impulsion~: $T_{\mu\nu} \neq 0$. Or par la relation d'Einstein $G_{\mu\nu} \propto T_{\mu\nu}$ elle constitue donc aussi une source de graviton. De même on sait qu'elle constitue une source pour le champ $B_{\mu\nu}$ et pour le dilaton $\Phi$. En supercorde, il faut en outre étudier son couplage aux champs Ramond-Ramond via les termes de \emph{Wess-Zumino}, ce que nous verrons un peu plus en détail dans la section~\refcc{sec:cond_susy}. L'expression conjecturée~\cite{Sen:1999md,Garousi:2000tr} de l'action effective de basse énergie -- à l'ordre des arbres -- sur une brane bosonique \footnote{Au moins, en ce qui concerne le tachyon interbranaire d'un système de deux branes bosoniques parall\`eles. Voir discussion dans la note~\notref{note-analy}. Mais cette forme est probablement valable de fa\c con générale.} ou une brane non BPS -- au terme WZ près -- est~: 

\begin{align}\label{eq:TDBI}
S_{eff} = -T_p \int d^{p+1} \sigma \, e^{-\Phi} \, V(T)\, \sqrt{-\det\parent{G_{ab} + B_{ab} + 2\pi \alpha' F_{ab} + \partial_a T \partial_b T}} 
\end{align}

On nomme cette action TDBI pour \emph{tachyon}-Dirac-Born-Infeld. Les champs $G_{ab}$ et $B_{ab}$ sont ici les "pullback", sur le volume d'univers de la brane, de la métrique et du champ de Kalb-Ramond d'espace-cible. Quant au champ $F_{ab}$, il s'agit du tenseur de Maxwell du champ de jauge $U(1)$ de corde ouverte, hébergé sur le volume de la brane. Dans le cas d'une simple brane bosonique ou non BPS, le tachyon n'est pas couplé à ce champ. Si ce dernier survit à la condensation, alors que les cordes ouvertes doivent avoir disparu, nous avons un problème. Cela se produit \'egalement dans le système $D-\bar D$. On pourra lire à ce propos les discussions de Sen~\cite{Sen:2000kd,Sen:1999md} ainsi que de Bergman et de Yi~\cite{Bergman:2000xf,Yi:1999hd}. Il est admis à présent qu'à cause du potentiel tachyonique ce champ électrique est confiné en tube de flux autour du vide stable, permettant ainsi, comme on l'a expliqué plus haut, la formation de cordes fermées.

En effet, le potentiel peut être réabsorbé dans le terme cinétique du champ de jauge, mais il réapparaît alors dans l'expression de la constante de couplage de jauge en $1/V$. Par conséquent, lorsque $V\to 0$, la constante de couplage tend vers l'infini et la théorie de jauge est fortement couplée. Un phénomène non-perturbatif analogue se produit sur des M-branes~\cite{Bergman:2000xf} et on trouve effectivement que cela mène au confinement du champ de jauge sous la forme de tube de flux.   \\

Maintenant, suite à la condensation, si nous sommes en présence d’un vide de corde fermée, alors les seules cordes pouvant apparaître proviendraient de fluctuations quantiques du vide. Or dans un vide stable, les valeurs des observables ne sont pas modifiées par les fluctuations quantiques, puisque cela serait paradoxal. Par conséquent, il faut trouver que toutes les sources de champs s'annulent en $T_0^{(i)}$. 

C'est ce que l'on obtient en faisant tendre $T\to T_0^{(i)}$ dans l'action~\refe{eq:TDBI} puisqu'alors $S_{eff}\to 0$. Cependant, il faut être plus rigoureux que cela pour converger sur une conclusion, puisque comme nous l’avons montré on peut toujours réabsorber le potentiel tachyonique dans les champs. Une étude rigoureuse des \emph{états de bords}~\cite{Sen:2002nu,Sen:1999mh,Sen:2004nf} permet de montrer que les sources de cordes fermées disparaissent effectivement. Plus précisément, l'annulation de l'état de bord dans le vide stable $\ket{B}\to 0$ est identifiée sans équivoque à la disparition complète de la brane représentée par cet état de bord. Enfin, l'absence de brane implique l'absence de corde ouverte. \\ 

Ainsi, nous venons de voir que le vide stable du tachyon est bien le vide des cordes fermées, bien que la théorie initiale soit celle d'une brane, \cad dans laquelle les excitations fondamentales sont des cordes ouvertes. Maintenant, nous pouvons étendre cette étude à des modes de condensation locaux dans l'espace-temps, \cad brisant spontanément la symétrie de Poincaré dans l'espace-cible donc associés à la formation de défauts topologiques.

\section{En système de branes bosonique}
\label{sec:cond_bos}

Nous sommes à présent amenés à décrire et à classifier l'ensemble des solutions de condensation dérivées de l'action effective TDBI~\refe{eq:TDBI}. Pour ce que nous souhaitons étudier, on peut cependant se restreindre à l'action effective tachyonique~:

\begin{align}
S = T_p \int \di^{p+1} \sigma~ V(T) \sqrt{1+\partial_a T\partial^a T}
\end{align}

De plus nous savons que cette action est valide universellement -- \cf chapitre~\refcc{chap:mot} -- à la différence de l'action TDBI~\refe{eq:TDBI}. Comme nous le disions ces solutions peuvent être de différents types. Nous venons de présenter le cas constant. Or ces solutions peuvent aussi être~: \emph{homogènes} si elles dépendent uniquement du temps et \emph{inhomogènes} si elles dépendent aussi de l'espace. Nous verrons aussi un cas particulier de solution dépendant du temps interpolant entre un vide instable et un vide stable.

\subsection{Solutions de condensation spatiale~: ressaut et rebond}
\label{sec:2.1.2}

Nous verrons tout d'abord le cas des solutions inhomogènes statique, qui correspondent donc à une désintégration spatiale d'une $p$-brane instable. On en distingue deux sortes~: les solutions interpolant entre deux vides stables distincts, auxquels on se réfère dans la littérature sous le nom de \emph{ressaut} ; et les solutions interpolant d'un vide stable vers lui-même, ce que l'on nomme \emph{rebond}. Dans chaque cas, en utilisant l'interprétation qu'un vide stable est un vide de corde fermée, on s'attend à ce que la solution représente une $(p-1)$-brane.

Dans le cas bosonique, ces deux solutions décrivent en réalité strictement la même chose par une sorte d'équivalence des théories conformes les décrivant sur la surface de la corde\footnote{Notons cependant que la solution de rebond a été beaucoup étudiée dans la littérature en OSFT~\cite{Moeller:2000jy,Harvey:2000tv,Kostelecky:1988ta}.}. En outre, en supercordes, il n'existe que des solutions de ressaut -- et vortex -- pour des questions de topologie. On s'attend donc à ce que le ressaut soit l'objet fondamental d'intérêt dans la condensation inhomogène. Par la suite on y attachera donc plus d'importance qu'au rebond. 

\subsubsection{Solution de ressaut}

Nous nous baserons sur les articles de Sen~\cite{Sen:1999mh,Sen:2002nu} principalement. Les détails pourront s'y trouver, nous ne ferons donc qu'un survol de son étude. Ses arguments vont comme suit. \\

Comme nous le disions, le potentiel de tachyon est universel. Il s'en suit que l'on peut tout aussi bien étudier la condensation de tachyon en espace ouvert ou en espace compact, et ce en utilisant une même expression du potentiel. Par conséquent, on s'attend à ce que les solutions obtenues ne diffèrent pas significativement dans un cas par rapport à l'autre. 

Ainsi, Sen propose d'étudier dans le cadre de la théorie conforme de bord la condensation de tachyon en espace compact de type cylindrique -- où la T-dualité est définie.  Cette étude est particulièrement simplifiée par l'apparition d'une symétrie\footnote{Plus spécifiquement une algèbre de courant} $SU(2)_L \times SU(2)_R$ cachée~\cite{Polchinski:1998rq}, à une certaine valeur de rayon de compactification $\wt R = \sqrt{\alpha'}$. Cette symétrie concerne uniquement le champ scalaire compact, et implique simplement que les modes de ce champ vérifient une symétrie supplémentaire permettant donc de les classer. 

La configuration que Sen propose d'étudier est la suivante. Supposons que l'espace n'est compactifié que le long d'une seule direction que nous noterons $X$ de rayon $R$ et que deux branes superposées s'enroulent autour de cette direction. Il existe 4 secteurs de cordes ouvertes se transformant selon la symétrie de jauge $U(2)$ et plus précisément dans la représentation adjointe de $U(2)$, composée des matrices de Pauli $\sigma^{0,1,2,3}$. Dans chacun de ces secteurs il existe un tachyon -- c'est un cas particulier en théorie bosonique. Toutefois, seul un secteur est réellement intéressant ici~: il s'agit du secteur interbranaire $\sigma^{1,2}$ car il est l'unique secteur à admettre un tachyon en théorie des supercordes. Rappelons qu'en théorie bosonique, l'existence du tachyon de corde fermée est pathologique~; ultimement, les travaux accomplis dans ce contexte sont destinés à être réutilisés en théorie des supercordes. Pour cette raison, Sen ne se focalise que sur la condensation de ce tachyon, et nous suivrons cet engagement.

Ensuite, nous l'avons dit plus haut, La condensation de tachyon en théorie bosonique est en général mal définie pour $T \leq 0$ à cause du puits de potentiel infini. Cependant, parce que nous étudions un tachyon du secteur interbranaire, nous avons que $V(T)$ est défini symétrique ${\mathbb Z}_2$ sous la forme~\refe{eq:pot_sym_can}. Par conséquent on s'attend à trouver des solutions topologiquement non triviales interpolant entre deux vides distincts.\\

En suivant la méthode de Sen, il est possible de ne construire qu'un seul soliton. Dans ce but, il impose au tachyon d'être anti-périodique en allumant une demi-unité de ligne de Wilson dans la direction compacte, \cad le long de $X$ dans le secteur minimalement couplé à $\sigma^{1,2}$~:

\begin{align}
\sigma^3 \otimes \frac{i}{2 R} \oint \partial X
\end{align}

Dans cette configuration, Sen montre que le mode tachyonique associé à $X$ et vérifiant cette anti-périodicité

\begin{align}
T(x) = \alpha \, \cos \frac{x}{2 R}
\end{align}

devient non massif en $R_c = \wt R/2$. Comme il est en outre possible d'obtenir une dimension compacte de rayon $R/2$ en imposant un \emph{twist} $h_X = e^{i p_X \, \pi R }$ sur une dimension compacte de rayon $R$, \cad en ne sélectionnant que les champs identifiés sous cette translation $X \to X + \pi R $. Nous pouvons donc exprimer la théorie compactifiée au rayon $R_c$ sous la forme d'une théorie au rayon $\wt R$. Or chose pratique, en $R=\wt R$ l'opérateur de vertex du tachyon 

\begin{align}\label{eq:tachyon}
{\mathcal V} = \sigma_1 \otimes \, \alpha \, \cos X \equiv \sigma_1 \otimes \, \frac{\alpha}{2} \, \partial  \phi
\end{align}

est identifié à une ligne de Wilson, en terme du courant $SU(2)$ $\partial\phi$, et est donc exactement marginal pour toute valeur de $\alpha$. Lorsque nous avons cette propriété, on dit que la théorie est une CFT et les couplages de cette CFT constituent les fonds classiques des champs correspondants dans l'espace-cible -- \cad les solutions des équations du mouvement de l'action effective. En dehors du rayon auto-duale $R=\wt R$ en revanche, Sen montre que la marginalité exacte n'est assurée qu'en deux valeurs $\alpha=0$ et $\alpha=1/2$ parce qu'autrement le tachyon développe un tadpole. 

Le fond $\alpha=0$ est clairement identifié au vide tachyonique instable. A l'inverse, le fond $\alpha=1/2$ est identifié avec une solution de ressaut. Le profil de surface de corde correspondant est sensiblement favorable\footnote{En fait, l'espace-cible à cause du facteur de CP voit plutôt $T(x)^2$. Dans ce point de vue, la solution ressemble plus à un rebond, mais cela prouve simplement que l'interprétation est un peu ambiguë. Voir plus loin.} à cette interprétation~(voir figure~\ref{fig:ressaut}).

\begin{figure}[h!]
\centering
\begin{minipage}{.45\linewidth}
\centering
\includegraphics[scale=0.85]{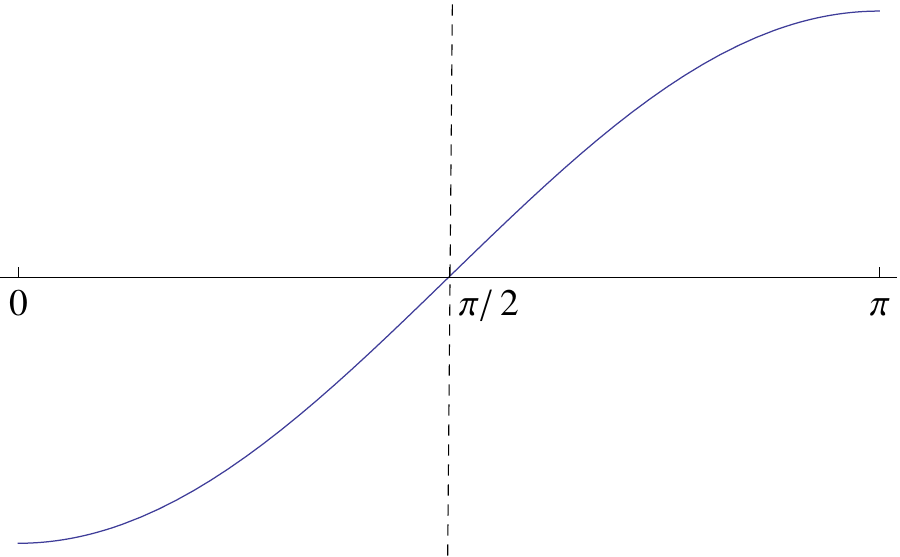}
\caption{\label{fig:ressaut} Solution de ressaut sur $x \in \{0 ; \pi \}$. Le soliton est localisé en $x=\pi/2$.}
\end{minipage}
\hfill
\begin{minipage}{.45\linewidth}
\centering
\includegraphics[scale=0.85]{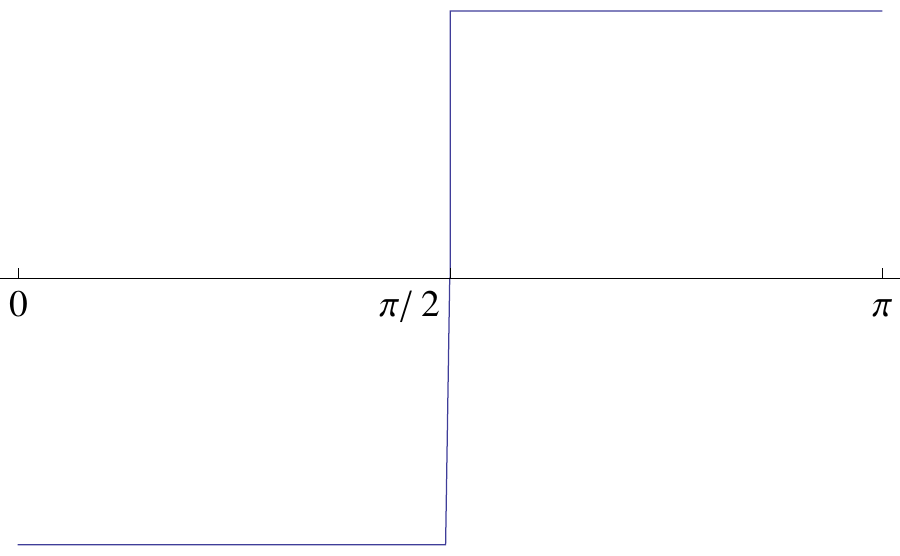}
\caption{\label{fig:heavi} Solution de ressaut dans l'espace-cible}
\end{minipage}
\end{figure}

Sen montre en effet que la CFT associée à ce ressaut est bien celle d'une brane localisée au point de l'espace où $T=0$, soit ici $x=\pi/2$. Les valeurs de $T$ correspondant au vide de corde fermée sont donc dans cette description $T=\pm 1/2$. Cependant, il faut noter qu'il s'agit là de la description du processus sur la feuille d'univers d'une corde, \cad que l'opérateur de vertex~\refe{eq:tachyon} correspondant au fond est ce que \emph{perçoit} la corde et non ce qui est effectivement dans l'espace-cible. C'est une distinction importante. En réalité, dans l'espace-cible, le profil de la solution ressemblerait à une fonction d'Heaviside~(voir figure~\ref{fig:heavi}). \\

En suivant la construction de Sen, nous avons donc obtenu la solution de ressaut dans un espace compact. On s'attend évidemment à ce que cette solution existe aussi dans la limite de décompactification, \ie pour $R\to \infty$. Comme nous avons vu, c'est en effet le cas tant que $\alpha=1/2$. Or Sen montre aussi que le fait d'augmenter le rayon n'a bien aucune incidence sur la nature de la solution, \cad qu'il s'agit toujours d'une $(p-1)$-brane. Cependant, la forme de la déformation sur le worldsheet change. Naïvement, on l'écrirait proportionnelle à $\cos X/2R$, mais alors elle ne serait plus marginale et il faudrait ajouter des perturbations supplémentaires pour conserver cette propriété. \\

\subsubsection{Solution de rebond}

Nous disions plus haut que l'on pouvait aussi décrire cette brane de codimension 1 sous la forme d'un rebond. Cela se fait naturellement en supprimant la ligne de Wilson et en rétablissant la périodicité. Au rayon auto-dual $R=\wt R=1$, nous construisons un tachyon tel que simplement~:

\begin{align}
T(x) = \sigma^1 \otimes \alpha \cos X \equiv \sigma^1 \otimes \alpha \partial \phi
\end{align}

Comme précédemment, ce tachyon est exactement marginal pour tout $\alpha$ au rayon auto-dual et uniquement en $\alpha=\{0,1/2\}$ en toute autre valeur du rayon. Par conséquent il constitue toujours une solution des équations du mouvement. Or on peut montrer qu'il décrit \emph{visiblement} en cette valeur une interpolation spatiale d'un vide stable vers lui-même, prenant la forme de deux $(p-1)$-brane localisées en $x=\{0,\pi R\}$. Cela est suggéré par la forme du tachyon\footnote{Encore une fois la donnée importante est le tachyon carré que l'on peut aisément se représenter à partir de la figure~\refe{fig:rebond}.} de la surface de corde conjugué au fait qu'on obtient explicitement que les solitons sont localisés en $x=\{0,\pi\}$. C'est bien ce qu'on appelle une solution de rebond (voir figure~\ref{fig:rebond}). 

\begin{figure}[h!]\centering
\includegraphics[scale=1]{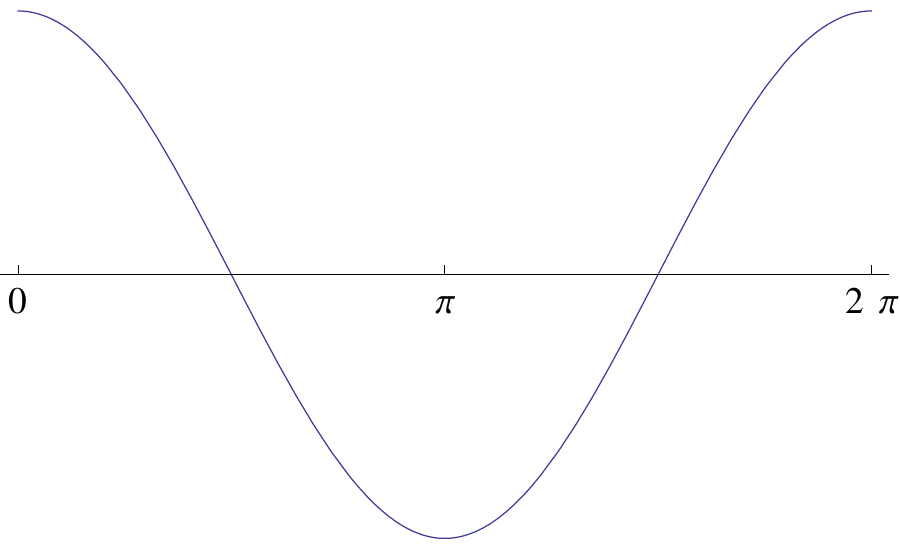}
\caption{\label{fig:rebond} Profil de rebond avec $\alpha=1/2$ centré sur la brane localisée en $x=\pi$.}
\end{figure}

Ce résultat est donnée explicitement dans~\cite{Sen:2002nu}. Sen y obtient une expression pour la "fonction d'onde" de la brane -- plus spécifiquement, il caractérise l'état de bord $\ket{B} = \ket{B}_{c=25} \otimes \ket{B}_{X} \otimes \ket{B}_{gh}$ dans la direction\footnote{Les autres directions ne sont pas relevantes ici car découplées.} $X$ par~:

\begin{align}
\ket{B}_X \propto \parent{ \sum_{n=-\infty}^{+\infty} \sin^{2n}(\alpha \pi) e^{i 2 n X(0)}} \ket{0}_c
\end{align}

avec $\ket{0}_c$ le vide invariant de corde fermée $SL(2,\mathbb C)$. Nous avons modifié un peu sa formule pour l'adapter à notre présentation et telle qu'ici le tachyon contient le facteur CP $\sigma^1$ qui ne sélectionne que les puissances paires dans la formule ci-dessus. Ainsi, la source des cordes ouvertes, \ie la fonction d'onde de la brane est proportionnelle à~:

\begin{align}\label{eq:profil_2}
f(x) = \sum_{n=-\infty}^{+\infty} \sin^{2n}(\alpha \pi) e^{i 2n x}
\end{align}

Or en $\alpha=1/2$, cela se resomme précisément sous la forme d'un peigne de Dirac~:

\begin{align}
f(x) = \pi \, \delta_{\pi} (x)  
\end{align} 

qui décrit donc deux branes de codimension $1$ localisées en $x=0$ et $x=\pi$ sur un espace compact de rayon $R=1$. En généralisant à tout $R$ ce mode de condensation doit donc faire correspondre une paire de branes coincidentes à un ensemble de branes séparées par une distance $\Delta = \pi R$. Notons que ce système est stable \emph{géométriquement} pour tout $R$ par symétrie du système. Au passage, la déformation $T(x) = \sigma^1\otimes (1/2) \cos X$ pourrait aussi décrire, si la direction $X$ était non compacte, un ensemble \emph{infini} de branes périodiquement espacées d'une distance $\Delta x = \pi$. Cet ensemble est également géométriquement stable parce qu'il est infini.

\subsubsection{Vide de corde fermée}

Maintenant, nous avons ici accès à une forme de preuve sur la nature du vide stable, car nous verrons que l'on peut contraindre par un court raisonnement les valeurs asymptotiques du potentiel effectif $V(T)$. Nous proposons de le montrer succinctement en suivant les arguments de Sen. \\

Notons $\pm T_0$ les valeurs dans l'espace-cible du tachyon au vide stable. D'un point de vue énergétique, on veut prouver que l'on a~:

\begin{align}\label{eq:cont_pot}
V(\pm T_0) = 0 ~~\text{et}~~ V(0) + 2 T_p = 0
\end{align}

avec $T_p$ la tension de chacune des branes. Or, si cela n'est pas vérifié alors l'énergie totale de la configuration est~:

\begin{align}
\int \di T \, V(T) \to \infty
\end{align}

A l'inverse avoir montré que le ressaut est une $p-1$-brane ayant une tension finie $T_{p-1}$, prouve qu'il faut~:

\begin{align}
\int \di T \, V(T) = T_{p-1} < \infty
\end{align}

de sorte que les contraintes~\refe{eq:cont_pot} sont immédiatement validées. Notons entre autre que la formule canonique du potentiel en théorie bosonique $2 T_p/\cosh (T/2\alpha')$ valable pour un tachyon dans le secteur $\sigma^1$ permet d'obtenir en intégrant le tachyon le long du demi-espace ouvert\footnote{Il faut tenir compte de l'orbifold $h_X$ qui identifie l'espace-cible à la moitié de l'espace total.} $x \in [0,\pi R[$~:

\begin{align}
\int_{-\infty}^{+\infty} \di T \, V(T) =  2\pi \alpha' \, T_p = T_{p-1}
\end{align}

Ainsi, les solutions de ressaut et de rebond (en $\alpha=1/2$) décrivent bien une brane de codimension 1 entourée de part et d'autre d'un vide de corde fermée. \\

A partir de la solution statique inhomogène, il est possible d'obtenir par continuation analytique -- on parle de \emph{rotation de Wick} dans ce cas -- une solution dépendante du temps et dont les propriétés découlent immédiatement de celles de la solution de ressaut. Il s'agit de la solution de tachyon \emph{roulant}, \cad dynamique.

\subsection{Solutions de condensation temporelle~: S-brane, solutions hybrides}
\label{sec:2.1.3}

Il existe trois sortes de solutions dépendant du temps, bien distinctes cette fois-ci~: les solutions asymptotiquement stables, auquel on se réfère en général sous le nom de S-brane \emph{complète} \footnote{\emph{S} signifie \emph{space-like}, \cad de genre espace.}~\cite{Gutperle:2002ai,Durin:2005ts}~; et les solutions interpolant entre le vide tachyonique instable et un vide stable, qui correspondent à ce que l'on appelle \emph{demi} S-brane. 

Il existe aussi des solutions inhomogènes ou hybrides, mélangeant une condensation spatiale et une condensation temporelle que nous aborderons brièvement en dernier lieu.

\subsubsection{S-brane complète }

La solution de S-brane complète est obtenue~\cite{Sen:2002nu} directement à partir de la solution de ressaut par continuation analytique d'un espace-cible euclidien vers un espace-cible minkowskien -- ce qu'on appelle une rotation de Wick. Appelons $X^0_E$ le temps euclidien et $X^0$ le temps minkowskien. La rotation de Wick effectue la continuation $X^0 = i X^0_E$ avec $X^0_E \in \mathbb R$. La validité de cette continuation implique que tout calcul effectué en temps euclidien est égal au calcul équivalent effectué en temps minkowskien. On étudiera ici un tachyon sur une seule brane. Son potentiel est asymétrique mais ce mode de condensation n'explore que le domaine $T>0$. \\

Le tachyon suivant décrit sur la feuille d'univers une solution de ressaut en temps euclidien~:

\begin{align}\label{eq:eucli}
T_E= \lambda \cos X^0_E
\end{align}

En admettant que la direction $X^0_E$ n'est pas compactifiée, cet opérateur de vertex est exactement marginal pour toute valeur de $\lambda$ et correspond donc à une solution des équations du mouvement. Par rotation de Wick, on obtient le tachyon dépendant explicitement du temps 

\begin{align}\label{eq:complete}
T = \lambda \cosh X^0
\end{align}

D'après le profil, ce profil décrit une solution de "rebond" temporel. Or un rebond devrait \^etre une brane de codimension $1$ dans le volume d'univers de la brane instable ; donc le profil~\refe{eq:complete} devrait repr\'esenter une brane de codimension $1$ et de genre espace, ce qu'on appelle une $S$-brane, \cad un hyperplan de genre espace le long duquel les cordes ouvertes devraient v\'erifier des conditions de Dirichlet. \\

Cette interprétation est fausse, nous allons maintenant voir pourquoi. Premièrement, la solution~\refe{eq:eucli} décrit un ensemble de rebonds situés périodiquement en $x=\pi [2\pi]$. C'est ce qu'on appelle des \emph{instantons}. Or ce rebond temporel que l'on décrit par~\refe{eq:complete} serait quand à lui localisé en $x^0=0$. Deuxièmement, comme nous l'avons vu, le rebond est effectivement une brane de codimension $1$ mais uniquement pour $\lambda=1/2$. Or, en cette valeur, le rebond temporel est en réalité une configuration stationnaire d'énergie nulle, \cad le vide de corde fermée $T=T_0$. En effet, l'énergie associée à~\refe{eq:complete} et la fonction d'onde de la brane instable sont données~\cite{Sen:2002nu} par les expressions~: 

\begin{align}\label{eq:ener_complete}
E &= \frac{T_p}{2} \parent{ \cos(2\pi \lambda) + 1} \leq T_p \nonumber \\
f(x^0) &= \frac{1}{1 + e^{x^0} \sin\lambda \pi } + \frac{1}{1+e^{-x^0}\sin\lambda \pi } - 1
\end{align}

Cette solution dite de S-brane \emph{complète} n'est donc jamais exactement une $S$-brane, mais plutôt une tentative \emph{ratée}, \cad que le système n'a pas l'énergie nécessaire pour reconstruire la brane instable. En outre, une S-brane est interprétée comme un objet non-perturbatif associé à un effet tunnel d'un vide (méta)stable vers un vide stable distinct. Or ici, le vide initial et le vide final sont exactement les mêmes, il n'y a donc pas d'effet tunnel. Par conséquent, parler de S-brane, en tout cas dans cet exemple-ci est délicat. 

La situation associée au tachyon interbranaire étudié précédemment serait en ce sens sûrement plus pertinente puisque son potentiel admet deux vides stables distincts. Dans ce cadre, avec le facteur CP $\sigma^1$ la d\'eformation~\refe{eq:eucli} est \'equivalente \`a une solution de rebonds p\'eriodiquement espac\'es et topologiquement non-triviaux. Elle d\'ecrit donc vraiment un ensemble d'instantons et la solution minkowskien correspondante~\refe{eq:complete} doit d\'ecrire un effet non-perturbatif et non-trivial. Dans la littérature cet effet est nomm\'e S-brane bien qu'il ne d\'ecrit pas sp\'ecifiquement un hyperplan localis\'e avec des conditions de Dirichlet\footnote{Dans la littérature pour faire la distinction, cet hyperplan de genre espace est nomm\'e DS-brane~\cite{Durin:2005ts} pour \emph{Dirichlet} S-brane.}.     \\

Notons enfin, qu'il y a une importante différence -- énergétique -- entre la condensation temporelle et la condensation spatiale. En effet, l'énergie est une donnée qui doit être temporellement conservée, tandis qu'elle n'a pas de telle contrainte spatialement. Ainsi la solution de ressaut est entourée de part et d'autre d'un vrai vide de corde fermée à 26 dimensions, \cad de densité d'énergie nulle.  Or ce ne peut être le cas de la solution temporelle que l'on vient de décrire. Si $E\neq 0$ l'énergie étant conservée, même en $T\to +\infty$ la théorie ne peut être celle d'un vide de corde fermée à 26 dimensions. Premièrement, si l'énergie dans le "vide" est non nulle c'est qu'il existe une source quelque part ; dans le cadre de la théorie des cordes ce ne peut être qu'une brane. Deuxièmement, la théorie est celle d'une condensation de tachyon sur une brane instable, on s'attend donc à ce que l'énergie soit stockée dans ce volume et non au dehors ; ce qu'on voit facilement puisque les conditions de Dirichlet sur les coordonnées transverses à la brane instable ne sont pas modifiées par le tachyon. 

Ensuite, le calcul de la pression~\cite{Sen:2002in} montre que~:

\begin{align}
T_{ij} \propto f(x_0) \delta_{ij}
\end{align}

Or d'après~\refe{eq:ener_complete} $f(x^0)$ s'annule pour $x^0 \to \pm\infty$, \cad pour $T_{WS} \to + \infty$.\footnote{J'indique $WS$ pour "worldsheet".} Compte-tenu qu'on ne connait pas la correspondance exacte entre une solution sur la surface d'univers et une solution d'espace-cible, il est \apriori difficile de justifier l'identification de $T_{WS} \to +\infty$ avec le vide stable $T_0 = + \infty$. Cependant, puisque la pression s'annule, le système tend à devenir stationnaire asymptotiquement. Il parait donc raisonnable de faire cette identification. En dehors du comportement asymptotique, on ne peut pas plus identifier l'expression de $T_{WS}$ à celle de la solution d'espace-cible. On le voit bien en $\lambda=1/2$ puisque $T_{WS}$ dépend du temps alors que le système physique est tout à fait stationnaire.

Enfin, nous avons vu que le vide asymptotique stable du tachyon devait correspondre à une théorie des cordes fermées. Bien qu'il soit maintenant clair que la symétrie de Poincaré sur l'ensemble de l'espace-cible n'est pas restaurée ici, les arguments que nous avions avancés restent valables et en particulier celui du confinement qui ne dépend que de la valeur asymptotique du potentiel tachyonique. Ainsi, il apparaît que le vide asymptotique est en fait composé d'un ensemble de cordes fermées d'énergie $E\neq 0$ confinées dans le plan de la brane instable.\\

D'après ces considérations, la solution de rebond temporel correspond donc physiquement à un gaz initial de cordes fermées non relativistes -- pression nulle -- conspirant à former une brane instable mais échouant et revenant à son état initial. On considère qu'il s'agit d'un système très peu physique car le degré d'ajustement est extrême.

\subsubsection{Demi S-brane}

Comme nous disions il existe une autre solution, dérivée en fait du cas précédent. Il s'agit du tachyon initialement (pour $x^0\to-\infty$) plac\'e au vide instable puis roulant asymptotiquement vers le vide stable (en $x^0\to +\infty$). C'est une solution physiquement plus pertinente que la précédente parce qu'elle d\'ecrit un processus naturel de d\'estabilisation. 

En extrapolant les résultats précédents, on s'attend à ce que l'énergie soit constante et conservée à $E=T_p$ et que la pression chute et s'annule en $x^0\to+\infty$. En outre, on s'attend aussi à ce que le contenu physique et sa répartition spatiale asymptotiques suivent le schéma précédent, \cad un agrégat de cordes fermées non relativistes confinées et se propageant dans le volume d'univers de la brane initiale. Compte-tenu de l'énergie à disposition qui va en $1/g_s$ on peut imaginer que les particules décrites par les cordes seront extrêmement massives.

On dérive les formules de l'énergie, de la fonction d'onde de la brane et de la pression à partir des formules précédentes. Rappelons que la S-brane correspond au tachyon~: 

\begin{align}\label{eq:complete2}
T_f(x^0) = \lambda \cosh x^0
\end{align} 

En revanche, la demi S-brane doit être décrite par la solution~:

\begin{align}
T_{h}(x^0) = \zeta e^{x^0}
\end{align}

Remarquons tout d'abord que la constante de couplage $\zeta$ n'est pas vraiment pertinente puisqu'on peut la réabsorber par translation temporelle. Maintenant, pour obtenir l'un à partir de l'autre, il faut appliquer à la solution de S-brane \emph{i)} une translation temporelle $x^0 \to x^0-\ln \lambda$, puis \emph{ii)} prendre $\lambda \to 0$ et enfin \emph{iii)} appliquer de nouveau une translation temporelle pour faire apparaître $\zeta$.  

Il pouvait apparaître à partir de~\refe{eq:complete2} que la seule solution telle que $E=T_p$ est la brane instable fixée au sommet du potentiel ; on voit donc qu'il n'en est rien et qu'il existe au moins une autre solution. Maintenant, en appliquant la transformation proposée sur $E$ et $f(x^0)$ on trouve~:

\begin{align}
E = T_p \quad \text{et} \quad f(x^0) = \frac{1}{1 + \pi \zeta e^{x^0}}
\end{align}

On a donc bien la formule souhaitée pour l'énergie, indépendante de $\zeta$. D'autre part, l'état de bord le long de la direction $X^0$, \cad $\ket{B_0} \propto f(x^0) \ket{0}_c$, s'annule bien en $x_0\to+\infty$ et tend vers $\ket{0}_c$ en $x^0 \to -\infty$. Similairement pour la pression, on trouve qu'elle chute et s'annule en $x^0 \to +\infty$ et tend vers une constante asymptotiquement dans le passé. \\ 

On comprend que cette voie de condensation est très importante pour les raisons suivantes. D'une part, on justifie qu'il s'agit d'une solution d'une grande pertinence physique, dans le sens où l'on ne s'interroge pas de savoir de quelle façon telle brane instable est apparue -- qui est une autre question -- mais de savoir comment elle va évoluer, et finalement se désintégrer et en quoi ; ce dont on peut répondre. D'autre part, la solution de demi S-brane est asymptotiquement libre dans le passé, contrairement à la S-brane complète, ce qui autorise à étudier les éléments de matrice-S (voir~\cite{Kutasov:2003er}).

\subsubsection{Condensation temporelle et production de particule}

Les questions associées à la production de cordes fermées (non duales à des cordes ouvertes) proposées dans le cas bosonique par Lambert \emph{et al.} dans~\cite{Lambert:2003zr} puis continuées par Karczmarek \emph{et al.} dans~\cite{Karczmarek:2003xm} sont aussi applicables ici, puisqu'il n'y a aucune raison de les négliger. On pourra aussi lire~\cite{Sen:2004zm,Kluson:2003rd,Kluson:2003qk}. L'intégrité de la brane en condensation temporelle n'est donc pas garantie et elle doit simplement s'évaporer au cours du temps sous la forme de ces cordes fermées non duales ultra-massives et non-relativistes. 

De sorte que la solution de tachyon roulant n'est physiquement valable que pour un temps relativement bref, de l'ordre de la constante de temps du processus, \cad $1/\module{m} \sim \alpha'$. En effet, le couplage aux cordes fermées implique que l'énergie n'est pas conservée dans le volume de la brane, contrairement à ce qu'on a pu supposer et démontrer en négligeant cette question. Notons cependant qu'il a été proposé -- voir par exemple~\cite{Sen:2004zm} -- que les cordes fermées produites correspondraient en fait à la matière tachyonique -- \cad seraient les cordes fermées duales produites par confinement -- de sorte que la théorie de corde ouverte du tachyon roulant serait finalement valable. A l'heure actuelle ça n'est toujours qu'une hypothèse.

\subsubsection{Solutions hybrides}

On peut imaginer construire des solutions dynamiques dont la condensation donne lieue à une production -- ou collision -- de ressaut ou de rebond. En l’occurrence, on sait que le tachyon suivant est marginal~:

\begin{align}
T(X^0,\vec X) = \lambda e^{\omega X^0} \cos ( \vec k \cdot \vec X)
\end{align}

pour $\omega^2 =1- \vec k ^2$. Cette déformation a été étudiée par Larsen \emph{et al.} dans~\cite{Larsen:2002wc} ainsi que par Sen dans~\cite{Sen:2002vv} qui montre que la marginalité exacte n'est atteinte qu'en $k=1/\sqrt 2$. Par rotation dans le volume de la brane, on peut réexprimer $ \vec k \cdot \vec X = \module k \cdot X_{\vec k}$ avec $X_{\vec k}$ la direction pointée par $\vec k$. On montre par étude de la CFT et des états de bords que cette solution représente effectivement dans un espace non compact, pour $x^0 \to \infty$ un ensemble d'objets de codimension 1 equi-espacées d'une distance $\Delta x = \sqrt 2 \pi$. Ces objets sont \apriori des $D(p-1)$ branes mais leur tension est légèrement supérieure à $T_{p-1}$ donc il a été proposé~\cite{Sen:2002vv} que l'énergie en surplus est portée par des tachyons condensant sur chacune de ces branes.

\subsection{Connexion aux théories conformes et modèles intégrables}
\label{sec:2.1.4}

Les solutions de ressaut et de demi S-branes décrivent comme nous l'avons vu des théories conformes. Celles-ci sont en fait bien connues sous les noms respectifs de \emph{modèle de bord de sine-Gordon} et de \emph{théorie de bord de Liouville}. Ces modèles ont été extensivement étudiés dans la littérature~\cite{Fateev:1997nn,Baseilhac:2002kf,Teschner:2003qk,Fateev:2000ik}, en particulier en temps que modèles de physique statistique à 2 dimensions. Sine-Gordon de manière générale\footnote{La théorie de bord en est une extension naturelle.} est connue pour être un modèle intégrable -- \cad totalement soluble -- et Liouville pour décrire une théorie des cordes \emph{non-critique}. 

Des outils mathématiques très puissants ont été développés pour résoudre ces théories – sachant qu'elles sont r\'esolvables –  et sont donc d'une grande aide dans l'étude de la condensation de tachyon. Il y eu des tentatives pour définir le tachyon roulant depuis une théorie de Liouville par rotation de Wick~\cite{Gutperle:2003xf,Gutperle:2002ai,Schomerus:2003vv}, ce qui défini la \emph{théorie de Liouville de genre temps} (TBL). En outre, il existe des extensions supersymétriques de ces modèles~\cite{Nepomechie:2001qr,Nepomechie:2001xk,Baseilhac:2003gc,Baseilhac:2002kf,Mattik:2005du} donc cela ne s'arrête pas au cas bosonique. En ce qui concerne le calcul des fonctions de corrélation, de la fonction de partition, des effets non-perturbatifs sur la surface de cordes, de la construction des états de bord, pour ces théories l'essentiel est déjà connu. Malheureusement, elles ne représentent pas l'ensemble des modèles de tachyon condensant qui semblent consister en des généralisations de ces modèles intégrables, comme par exemple pour ce qui nous intéressera le modèle Kondo -- bosonique~\cite{Baseilhac:2002kf,Oshikawa:1996dj,LeClair:1997sd,Fendley:1995kj} et supersymétrique.

\section{Condensation de tachyon en système non BPS instable}
\label{sec:cond_susy}

	Les systèmes de brane non-BPS et de brane-antibrane coïncidentes sont très bien connus et ont été étudiés en détail par nombre d'auteurs (Sen, Kutasov, Larsen, Garousi...)~\cite{Sen:2003tm,Sen:1998ii,Sen:1998sm,Kutasov:2000aq,Harvey:2000na,Larsen:2002wc,Kraus:2000nj,Garousi:2000tr,Garousi:2007fn}. La condensation de tachyon y est relativement bien comprise et dans une certaine mesure\footnote{Voir section~\refcc{chap:mot}.} nous avons une action effective dont la forme, et en particulier le potentiel, sont assez bien contraints. Comme nous disions dans la section~\refcc{chap:mot} les diverses solutions de condensation -- ressaut, antiressaut, vortex, S-brane -- entrent dans le schéma de relations de descente entre branes de la théorie K.   \\ 

Rappelons sa forme canonique~:

\begin{align}
V(T) = \frac{1}{\cosh \frac{T}{\sqrt 2}}
\end{align}

Sa représentation graphique est donnée dans la figure~\refe{fig:potentiel_super_sup}. Dans le système brane-antibrane, le tachyon $T$ de la formule ci-dessus est remplacé par le module complexe $\module{T}$. De sorte que le potentiel est indépendant de la phase du tachyon et est donc explicitement symétrique $U(1)$. Le tachyon d'espace-cible y condenserait en $\module{T} = \pm \infty$. Dans le cas de la brane non-BPS, le tachyon est réel et par conséquent le potentiel correspond exactement à ce qui est représenté sur la figure~\refe{fig:potentiel_super_sup} qui nous le voyons est symétrique $\mathbb Z_2$. La mise en valeur de ces symétries est importante pour construire des solitons topologiquement non triviaux donc stables, comme nous le verrons prochainement. 

\begin{figure}[h!]\centering
\includegraphics[scale=1]{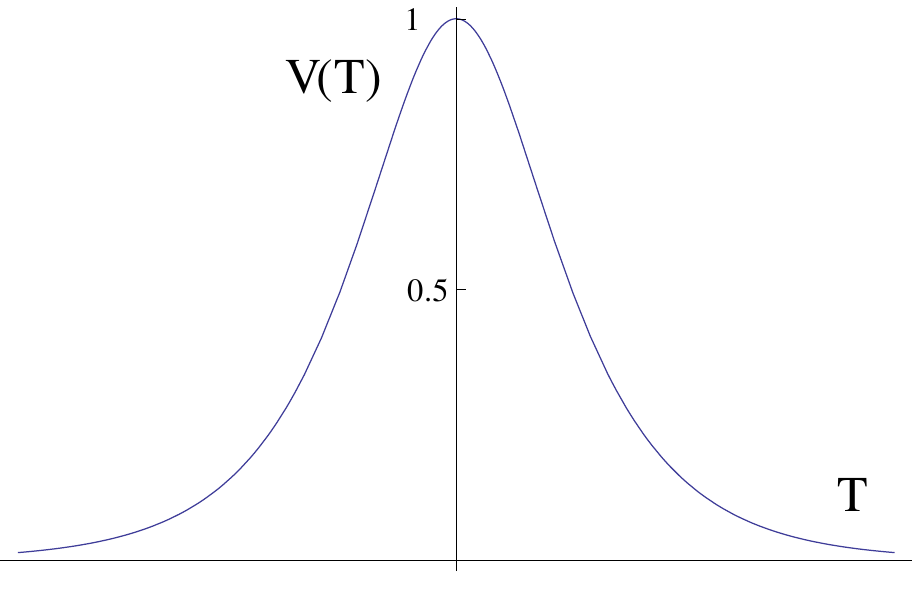}
\caption{\label{fig:potentiel_super_sup} Forme canonique du potentiel tachyonique en théorie supersymétrique.}
\end{figure}

Nous allons  brièvement rappeler la forme de l'action effective TDBI obtenue en supercordes que nous avions introduite plus tôt. Nous verrons d'abord le cas de la brane non-BPS puis celui des branes $D-\bar D$. Nous ajouterons la contribution des champs de Ramond-Ramond dans le terme de Wess-Zumino. Nous partirons de branes de dimension maximale, \cad de dimension $9+1$ car les expressions sont plus simples et capturent la physique de toute autre brane. En effet, les actions sur les branes de dimension inférieure sont obtenues par T-dualité le long des directions rendues transverses. Pour la brane non-BPS~\cite{Garousi:2000tr,Sen:1999md}, l'action effective d'une D9-brane en type IIA est~:

\begin{align}\label{eq:nonBPS}
S_{non BPS} = \sqrt 2 T_9 \, \int \di^{10}\sigma  \, e^{\Phi} V(T)\, \sqrt{\det\parent{G_{ab}+B_{ab}+2\pi \alpha' F_{ab} + \partial_a T \partial_b T}} + S_{WZ}
\end{align}

L'expression est abélienne, puisqu'il n'y a qu'une seule brane. Nous discutons de son domaine de validité dans le chapitre~\refcc{chap:mot}. Il s'étend \apriori seulement le long de condensations spatiales. Pour discuter des condensations temporelles la base d'étude est l'action tachyonique dans la jauge statique~:

\begin{align}
S_{T} = \sqrt 2 T_p \, \int \di^{p+1}\sigma  \, V(T)\, \sqrt{\eta_{ab} + \partial_a T \partial_b T}
\end{align}

Dans~\refe{eq:nonBPS}, le terme de Wess-Zumino est connu pour être de la forme~\cite{Sen:2003tm,Okuyama:2003wm,Kennedy:1999nn,Garousi:2007fk,Billo:1999tv,Kluson:2000iy}~:

\begin{align}\label{eq:chern-simons}
S_{WZ} = \mu_p \int_{p+1} W(T) dT \wedge \parent{\sum_{m\in \IIA} C_{(m)}} \wedge e^{B + 2\pi \alpha' F}
\end{align}

où $W(T) \propto V(T)$ et $B$ et $F$ respectivement les pull-back sur le volume de la brane du champ de Kalb-Ramond et le tenseur de Maxwell du champ de jauge de corde ouverte. La charge $\mu_p$ est proportionnelle à la tension de la brane. Les champs de jauge de R-R sont en type IIA les formes différentielles d'indice paire. En type IIB, ils sont d'indice impair. Notons que le tachyon n'est pas couplé minimalement aux champs de jauge des cordes ouvertes, ce qui nécessite leur confinement lors de la condensation. 

Par T-dualités successives le long des directions que l'on souhaite rendre transverses, on obtient les actions effectives des branes non-BPS de dimension inférieure donc en type IIA et IIB en fonction de leur dimension. Il faut appliquer, par exemple par T-dualité le long de la direction $X^{10}$~:

\begin{align}
&F_{a \, 10} \to \partial_a X^{10} \qquad \text{et} \qquad \partial_{10} T \to 0 \nonumber \\
&G_{a \, 10} \to A^G_a \qquad \text{et}\qquad  G_{10 \, 10} \to \phi^G \nonumber \\ 
&B_{a \, 10} \to A^B_a
\end{align}

Les champs $ A^{G,B}_a$ et $\phi^G$ sont réabsorbés respectivement dans le champ $F_{ab}$ et le dilaton $\Phi$. Si bien que l'action T-duale d'une $Dp$-brane non-BPS pour $p$ quelconque s'exprime par~:

 \begin{multline}\label{eq:nonBPS2}
S_{non BPS} = \sqrt 2 T_p \, \int \di^{p+1}\sigma  \, e^{\Phi} V(T)\, \sqrt{\det\parent{G_{ab}+B_{ab}+2\pi \alpha' F_{ab}+  \partial_a X^I \partial_b X_I + \partial_a T \partial_b T}} \\ + S_{WZ}
\end{multline}

avec $I=p+1 \ldots 9$ les dimensions transverses. \\

L'action effective du système brane-antibrane, comme expliqué dans l'introduction, est obtenue par non-abélianisation de l'action non-BPS de dimension maximale, puis projection le long des secteurs du système $D-\bar D$ commun au système non-BPS. Son domaine de validité suit celui de l'action TDBI pour la brane non-BPS, \cad $T$ de genre espace. Nous ne donnerons que la forme conjecturée pour $p=9$ en type IIB\footnote{La forme T-duale a été donnée dans le chapitre~\refcc{chap:mot} pour un fond simplifié et on pourra s'y reporter.}~\cite{Garousi:2000tr,Garousi:2007fn}~:

\begin{align}
S_{D\bar D} = \int \di^{10}\sigma \,  \Str  \, e^{\Phi} V(|T|)\, \sqrt{\det\parent{G_{ab}+B_{ab}+2\pi \alpha' F_{ab} + D_a T (D_b T)^\dagger}} + S_{WZ}
\end{align}

avec $\Str$  la trace complètement symétrique sur le groupe de jauge. Nous avons également introduit la dérivé covariante non abélienne telle que $D_a T = \partial_a T - i[A_a,T]$. Le terme de Wess-Zumino~\cite{Garousi:2008tn,Garousi:2008ge,Green:1996dd} n'est pas important pour nous, mais est très similaire à~\refe{eq:chern-simons} à une trace près. Pour obtenir l'expression de l'action dans les systèmes de dimension inférieure, nous appliquons encore une T-dualité dans les directions que l'on souhaite transverses. Par exemple, le long de  $X^{10}$ il faudrait appliquer~:

\begin{align}
&F_{a \, 10} \to D_a X^{10} \qquad \text{et} \qquad D_{10} T \to [X^{10},T] \nonumber \\ 
&G_{a \, 10} \to A^G_a \qquad \text{et}\qquad  G_{10 \, 10} \to \phi^G \nonumber \\ 
&B_{a \, 10} \to A^B_a
\end{align} 

Le champ $X^{10}$ est une matrice diagonale encodant les positions respectives de chaque brane, tandis que $T$ est donné par une matrice anti-diagonale :

\begin{align}
X^{10} = \parent{\begin{array}{cc} X^{(1) \, 10} & 0 \\ 0 & X^{(2) \, 10}\end{array}} \qquad \text{et} \qquad T = \parent{\begin{array}{cc} 0 & \tau \\ \tau^* & 0\end{array}}
\end{align}

Il apparait naturellement un couplage entre le tachyon et la position relative de la brane et de l'antibrane via $[X^{10},T]$. Ce couplage donne au tachyon interbranaire sa masse $\alpha'm^2 = \ell^2/4\pi^2\alpha' -1/2$ avec $\ell$ la distance relative entre les branes. La position du centre de masse du système est quant à elle découplée et reste un module à tachyon non nul, ce qui n'est \apriori plus le cas du champ de distance\footnote{Mais à tachyon nul il retrouve naturellement son statut de module.}. \\

Pour étudier les tachyons condensants de genre temps il faut de nouveau s'intéresser plutôt à l'action effective du tachyon seul dans la jauge statique, donc au moins dans la limite où la distance relative est nulle, \cad à la  \emph{coïncidence}~:

\begin{align}
S_{T} = \sqrt 2 T_p \, \int \di^{p+1}\sigma  \, \Str V(|T|)\, \sqrt{\eta_{ab} + \partial_a T \partial_b T^*}
\end{align}

L'étude du système $D-\bar D$ coïncident est très bien connu à l'inverse du cas à séparation non nulle que nous avons étudié dans cette thèse. Nous allons voir maintenant les diverses solutions de condensation qu'admettent les équations de mouvement de ces actions.

\subsection{Solutions de condensation spatiale~: ressaut et vortex}
\label{sec:2.2.2}

Ces solutions correspondent exactement aux ressauts -- kink -- que nous avions introduits en théorie bosonique, à la différence qu'ici il existe des contraintes supplémentaires associées à la charge des objets. Par condensation spatiale, les branes sont vues comme des solitons du champ tachyonique. Or le potentiel effectif en théorie de supercorde est symétrique $\mathbb Z_2$ ou $U(1)$ et par condensation cette symétrie est spontanément brisée. 

\subsubsection{Solutions de type ressaut}

Par conséquent, l'existence de solutions de condensation de type ressaut interpolant entre au moins deux vides stables distincts est attendue. Ces solutions ne sont pas nécessairement topologiquement non-triviales -- par exemple sur le système $D-\bar D$ le potentiel est symétrique $U(1)$ et il existe donc une transformation continue qui amène le ressaut vers le vide de corde fermée stable. Cependant, 
les solutions de ressaut sont typiquement chargées -- même si ce n'est pas toujours le cas -- du fait du changement de vide et parce que le tachyon couple aux champs de jauge par le terme de Chern-Simons~\refe{eq:chern-simons}. La construction des solitons sur le système brane-antibrane coïncident a été fait par Sen~\cite{Sen:2003tm,Sen:1998ii} exclusivement en terme de ressaut et de vortex. 

Pour y obtenir un seul ressaut, il impose la même contrainte qu'en théorie bosonique, \cad qu'il commence par compactifier une direction et il ajoute le long de celle-ci une demi-unité de ligne de Wilson. De la sorte, sa solution doit être anti-périodique autour de la coordonnée compacte et elle peut donc être une unique interpolation d'un vide $-T_0$ vers un vide $+T_0$. Or dimensionnellement, dans le terme de WZ, le tachyon ne peut pas coupler à un champ de jauge~: les dimensions du volume de la brane ne peuvent jamais \^etre remplies avec un tachyon et au moins un champ de jauge.  Cette solution n'est donc pas chargée. Or comme nous disions, elle est aussi une solution topologiquement triviale et donc très probablement instable. D'un point de vue dimensionnel, l'objet créé est de codimension 1 dans le volume du système et par conséquent, il doit s'agir d'une brane non-BPS. \\ 

Appliquons une condensation analogue le long d'une brane non-BPS. Une demi-unité de ligne de Wilson est allumée le long d'une de ses directions, supposée compacte. La solution peut alors interpoler une fois entre les deux vides $\pm T_0$ du potentiel de la brane non-BPS. Gr\^ace à l'existence d'un champ de jauge permettant par couplage au tachyon de remplir toutes les dimensions de la brane non-BPS dans le terme de WZ, l'objet obtenu est ainsi chargé. Or, à cause de la forme du potentiel que l'on sait symétrique $\mathbb Z_2$ toute solution interpolant entre les deux vides distincts est topologiquement non triviale. L'objet est donc stable. Par analyse dimensionnelle il ne peut s'agir que d'une brane BPS de codimension 1 dans le volume de la brane non-BPS. \\  

En étudiant les diverses observables du système et en particulier, le tenseur énergie-impulsion, les diverses sources et plus généralement l'état de bord du système, Sen obtient effectivement que ces solutions se présentent sous la forme de solitons de codimension 1 et bien localisés -- des murs de domaine en somme -- interpolant entre deux vides bien distincts. Tout cela peut se résumer sous la forme des relations suivantes, en tenant compte des contraintes liées à la dimension des branes BPS et non-BPS~:

\begin{align}
&\IIA~:~Dp-\overline Dp  &\longrightarrow & \qquad \text{ressaut non-BPS } D(p-1) \nonumber \\ 
&\IIB~:~\text{non BPS } Dp &\longrightarrow & \qquad \text{ressaut } D(p-1)
\end{align} 

En relaxant la condition d'anti-périodicité il est possible de construire des couples de branes par exemple dans des dimensions compactes. En l'occurrence sur la brane non-BPS au rayon auto-dual -- puis en tout rayon par marginalité exacte en $\lambda=1/2$  --  en ajoutant à la théorie de surface de corde la perturbation $\sigma^1 \otimes \lambda \oint \cos X$ nous obtenons un couple brane-antibrane séparé à distance critique -- donc stable. De plus, les branes sont réparties en des points diamétralement opposés, donc cette configuration est d'autant plus \emph{géométriquement} stable. Dans la limite de décompactification, cette déformation construit un ensemble infini de couples brane-antibrane répartis périodiquement le long de la direction concernée. Puisque tous les objets sont séparés par une distance critique, l'ensemble est tout à fait stable. 

\subsubsection{Solution de type vortex}

Dans le cadre du système brane-antibrane, il est aussi possible de construire des solutions de type vortex~\cite{Sen:2003tm} puisque le tachyon est complexe et que le potentiel est symétrique U(1). Ces solutions engagent une configuration bi-dimensionnelle, à la différence du ressaut ou du rebond qui sont purement uni-dimensionnels, car il s'agit de faire varier la phase du tachyon autour d'un défaut, une singularité, ce qui est réminiscent du cas des cordes cosmiques. De fait, nous aurons~:

\begin{align}
&\IIA \text{ ou } \IIB~:~ Dp-\overline Dp &\longrightarrow & \qquad \text{vortex } D(p-2) 
\end{align}

L'orientation, \cad la charge, de la brane dépend de l'orientation du vortex autour de la singularité par intégration le long du tachyon d'un terme de Wess-Zumino du type~\refe{eq:chern-simons} bien qu'en l'état cela ne fonctionne pas aussi trivialement -- voir plus bas. Il est clair que toute configuration de vertex est topologiquement non trivial. En effet, \emph{i)} le nombre de vortex n'est pas continuellement réductible et \emph{ii)} si un vortex est une brane BPS alors deux vortex sont deux branes BPS de même charge et ainsi de suite, ce qui constitue un système stable --  et donc irréductible -- de branes parallèles, coïncidentes et localisées au point entouré par les vortex. \\

A cause de la neutralité du système brane-antibrane initial nous faisons cependant face à un problème pour créer une brane chargée, même en tenant compte de l'éventuel couplage du tachyon aux champs R-R \apriori de la forme~\refe{eq:chern-simons}. Ce problème est clairement dimensionnel parce que le tachyon y apparaît \apriori sous la forme $dT$ et il ne peut donc pas sourcer les champs R-R couplés à une $D(p-2)$. Rappelons que cette forme était providentielle afin de construire une brane non chargée non-BPS $D(p-1)$ ci-dessus. Il a été proposé dans~\cite{Majumder:2000tt} pour palier à cette situation, de ne considérer que le cas, moins problématique, de la création de pair vortex-antivortex. Dans cet article ils proposent au passage une BCFT pour cette solution en introduisant dans l'action de surface au rayon auto-dual une déformation marginale sur le bord. 

Cependant, parce que le tachyon dépend de deux coordonnées, il doit être possible de généraliser $d\tau \to d\tau \wedge d\tau^* \propto dx \wedge dy$ avec $(x,y)$ les coordonnées du plan de condensation, de sorte que le champ de jauge R-R naturellement couplé à la $D(p-2)$-brane est effectivement sourcé. Cela se comprend d'autant mieux que par descente en construisant itérativement des solutions de ressaut depuis la pair $Dp-\bar Dp$ en passant par la brane non-BPS $D(p-1)$ on obtient proprement une et une seule $D(p-2)$ brane chargée. 

C'est exactement ce qui est conjecturé par Kennedy et Wilkins dans~\cite{Kennedy:1999nn} pour le cas particulier des branes coïncidentes et vérifié à des termes de plus grande dérivée dans~\cite{Garousi:2008tn}. Le terme de Wess-Zumino qu'ils proposent s'exprime en fonction de la superconnection~:

\begin{align}
i{\mathcal A} = \parent{\begin{array}{cc} iA^+ &  \tau^* \\ \tau & iA^- \end{array}}
\end{align}

dont ${\mathcal F}=d{\mathcal A} - i {\mathcal A} \wedge {\mathcal A}$ est la courbure et $A^\pm$ étant les champs de jauges de chacune des branes respectivement. Et nous aurions donc~:

\begin{align}
S_{WZ} = T_p \int_{p+1} C \wedge \Str e^{i 2 \pi \alpha' {\mathcal F}}
\end{align}

Nous avons noté $C=\sum C_{(m)}$ avec $m$ pair en type IIA et impair en type IIB. Kraus et Larsen~\cite{Kraus:2000nj} montrent qu'à partir de ce terme, sont retrouvées les bonnes charges R-R pour les solitons obtenus par condensation du tachyon.

\subsection{Solutions de condensation temporelle~: S-brane et solutions hybrides}
\label{sec:2.2.3}

Sen a aussi introduit dans le cadre des systèmes brane-antibrane et brane non-BPS~\cite{Sen:1998sm,Bagchi:2008et} mais aussi Larsen \emph{et al.} dans~\cite{Larsen:2002wc} des solutions de condensation de type tachyon roulant, quasiment identiques à ce que nous avons pu découvrir en théorie bosonique. En effet, il existe aussi des solutions de type S-brane \emph{complète} et \emph{demi} S-brane. De nouveau, cette dernière est plus physiquement pertinente que la première du fait qu'elle ne décrive que le mécanisme de désintégration de la brane au cours du temps et non aussi sa reconstruction\footnote{Cela impliquerait une conspiration de matière à reformer une brane, ce qui est un phénomène possible mais très improbable.}. Cependant, ici encore l'existence des solutions S-brane complètes est importante pour contraindre la forme de l'action effective associée au système. 

Sen obtient par étude des diverses observables du système -- tenseur énergie-impulsion, état de bord --  que la solution de demi S-brane s'évapore sous forme de matière tachyonique qui par dualité doit correspondre aussi à de la matière formée de cordes fermées. De nouveau, il semblerait que le confinement du champ électrique non couplé au tachyon soit à l'origine de l'identification des degrés de libertés tachyoniques à ceux d'un gaz de cordes fermées dont le profil de densité est piqué dans le volume d'univers de la brane instable initiale. Cette approche est exacte tant que l'on néglige le couplage de la brane aux champs de cordes fermées, comme nous le discutions dans le cas bosonique, mais qui devraient idéalement être aussi pris en compte ici.  \\

Nous ne nous attarderons pas sur la présentation de la solution de tachyon roulant du point de vue de l'espace-cible car d'après~\cite{Larsen:2002wc} les calculs des observables sont grossièrement identiques et les interprétations sont celles que l'on vient de donner. Enfin il peut aussi exister des solutions de condensation temporelle inhomogènes. Une telle solution est proposée dans~\cite{Larsen:2002wc} où ils calculent un certain nombre d'observables, cependant ils utilisent une factorisation qui semble un peu cavalière en présence de fermions. L'analogie au calcul en corde bosonique suggère tout de même que ce type de solutions existe de telle sorte que par condensation soit produit dans les limites de la conservation de la charge et de l'énergie, un système de brane (et antibrane suivant le système initial) correspondant à des hybrides $\nicefrac{1}{2} \, \text{S-brane} \otimes \parent{\text{vortex ou ressaut}}$. \\

\part{Tachyon roulant, syst\`emes brane-brane et brane-antibrane}
\label{part:tach_roul}

\chapter{Condensation de tachyon dans un système de branes en théorie bosonique}
\label{chap:cond_bos}

Il est toujours intéressant de tenter en premier lieu la résolution d'une théorie bosonique correspondant à la théorie supersymétrique étudiée. En effet, même si la th\'eorie bosonique est par d\'efinition pathologique\footnote{En particulier \`a cause du tachyon de corde ferm\'ee.}, les processus mis en valeur devraient aussi apparaitre dans l'extension supersymétrique mais de façon plus contrôlée. Par exemple, il existe des modes de condensation de type ressaut en th\'eorie bosonique et de fa\c con similaire nous trouverons ressaut et vortex en th\'eorie des supercordes. 

La question soulevée dans cette thèse est la suivante~: est-ce qu’un tachyon de corde ouverte tendu entre deux branes séparées spatialement, dans le secteur antidiagonal $\sigma^{1,2}\in U(2)$ peut-il condenser ? Quelles expressions du tachyon sur la surface de corde sont marginales, exactement marginales ? Correspondent-elles à des th\'eories conformes (CFT) connues, des modèles intégrables ? La séparation spatiale peut-elle être maintenue constante ou doit-elle être dynamique, ou bien encore inhomogène ?

\section{CFT du tachyon roulant dans le système de branes parallèles et séparées}
\label{sec:brane_sep}

Il est nécessaire de définir les conditions qui permettent de voir apparaître un tachyon. A cette fin, le système de branes parallèles sera étudié un peu plus en détail puis les questions de la condensation et des expressions possibles du champ dans la théorie de surface de corde seront abordées dans un deuxième temps. 

\subsection{Le système brane-brane}

Comme mentionné dans la section précédente, il existe quatre secteurs de cordes ouvertes dans un système de deux branes. Le système est représenté sur la figure~\refe{fig:branes_sep_bos}. L'ensemble est caractérisé par 4 quatre facteurs de Chan-Paton appartenant à l'algèbre $U(2)$, \cad

\begin{align}
\sigma^0 = \Big(\begin{array}{cc}1 & \\ & 1 \end{array}\Big) \qquad 
\sigma^1= \Big(\begin{array}{cc} & 1\\1 &  \end{array}\Big)  \qquad
\sigma^2= \Big(\begin{array}{cc} &-i\\i &  \end{array}\Big) \qquad
\sigma^3= \Big(\begin{array}{cc}1 &\\ & -1 \end{array}\Big) 
\end{align}

\begin{wrapfigure}{l}{.5\linewidth}
\centering
\includegraphics[scale=.5]{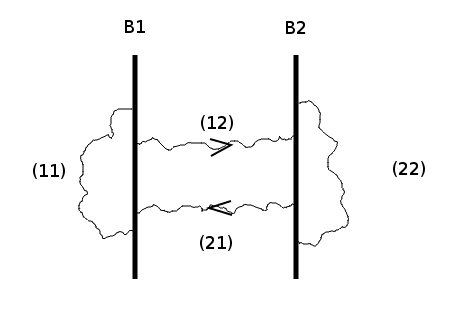}
\caption{\label{fig:branes_sep_bos} \footnotesize{Les secteurs $(11)$ et $(22)$ s'organisent en facteurs de CP $\sigma^0$ et $\sigma^3$ tandis que les secteurs interbranaires $(12)$ et $(21)$ s'organisent en $\sigma^1$ et $\sigma^2$.}}
\end{wrapfigure}

Lorsque les branes sont coïncidentes la théorie de jauge hébergée sur le volume d'univers est non-abélienne $U(2)$. Mais lorsque le système est séparé, la symétrie de jauge est spontanément brisée en $U(1)\times U(1)$~ : ce processus est identifié à un mécanisme de Higgs. Chaque $U(1)$ correspond alors au secteur hébergé sur le volume de chaque brane, ici correspondant aux secteurs diagonaux $\sigma^{0,3}$ ; les deux secteurs résultants antidiagonaux $\sigma^{1,2}$ sont associés à des champs de jauge massifs\footnote{Mais de masse nulle à la coïncidence.} donc découplés. Ces derniers constituent les secteurs interbranaires. \\

En th\'eorie bosonique, le tachyon constitue l'état fondamental d'excitation des cordes des quatre secteurs. Cependant, les secteurs interbranaires sont les plus importants. Ils sont en effet potentiellement tachyoniques aussi en supercordes, ce qui n'est jamais le cas des autres. D'après le commutateur~:

\begin{align}
[\sigma^1,\sigma^3] = 2 i \sigma^2
\end{align}  

ces derniers sont naturellement couplés au secteur $\sigma^3$ du champ de jauge $U(1)\times U(1)$ et constituent donc ensemble un champ \emph{bi-fondamental}. Ces deux tachyons sont \apriori réels mais peuvent être regroupés en un tachyon complexe et son conjugué. On parlera alors de \emph{tachyon interbranaire}. 

\subsubsection{Spectre de masse et distance critique}

Sans séparation, les spectres de masse de chaque secteur sont triviaux~:

\begin{align}
&\alpha' m^2_0 = N_0 - 1 \nonumber \\ 
&\alpha' m^2_1 = N_1 - 1 \nonumber \\ 
&\alpha' m^2_2 = N_2 - 1 \nonumber \\ 
&\alpha' m^2_3 = N_3 - 1 
\end{align}

Lorsque la séparation est ouverte, les secteurs interbranaires reçoivent un nombre d'\emph{enrou-lement} -- \emph{winding} en anglais -- proportionnel à la distance que l'on notera $\ell$. L'expression de ce terme est facile à obtenir en comparant le système à une brane de codimension 1 dont la direction transverse est compactifiée sur un cercle de rayon $R = \ell/2\pi$. Les cordes s'enroulant $1$ fois autour du cercle gagnent un terme de masse en $\alpha' m^2 \sim R^2/\alpha'$. Ainsi, on obtient~:

\begin{align}
&\alpha' m^2_0 = N_0 - 1 \nonumber \\ 
&\alpha' m^2_1 = \frac{\ell^2}{4\pi^2 \alpha'} + N_1 - 1 \nonumber \\ 
&\alpha' m^2_2 = \frac{\ell^2}{4\pi^2 \alpha'} + N_2 - 1 \nonumber \\ 
&\alpha' m^2_3 = N_3 - 1 
\end{align}

Ces formules démontrent donc qu'il existe une distance critique au-delà de laquelle le bi-fondamental $(N_1,N_2)=(0,0)$ est massif et non plus tachyonique. A la distance critique il est non-massif ; soit pour~:

\begin{align}\label{eq:crit_dist}
\ell_{cr} = 2\pi \sqrt{\alpha'}
\end{align}
 
Du point de vue de la théorie des champs, dans l'espace-cible associée au système, trois phases se distinguent~: (1) la phase massive pour laquelle on s'attend à ce que le potentiel du champ de séparation $\phi=\ell$ soit attractif en $V(\phi)\propto T^2 \phi^2$ au premier ordre ; (2) la phase non-massive critique qui est une théorie des champs d'un bi-fondamental scalaire non-massif couplé à un champ de jauge abélien $A$ et au scalaire $\phi$  ; et enfin (3) la phase tachyonique totalement dominée par la condensation \emph{classique} du champ de tachyon. 

\subsubsection{Discussion des connexions entre ces différentes phases}

L'action effective quadratique à l'ordre des arbres pour les champs $\varphi$ et $T$ peut \^etre obtenue par développement d'une action de type Garousi-TDBI~\refe{eq:TDBI} ou~\refe{eq:garousi_base} dont ne sont conservées que les contributions de la distance et du tachyon,\footnote{Nous verrons cependant dans la section~\refcc{sec:off_sep} que cette expression est probablement incomplète.}~:

\begin{align} \label{eq:quad}
S = T_p \int \di^{p+1}\sigma \, \parent{\frac{1}{2}\partial_a \phi \partial^a \phi + \frac{1}{2} \partial_a T \partial^a T^* - (\phi^2 - \ell_c^2)\frac{\module{T}^2}{8\pi^2}}
\end{align}

qui devrait être correcte tant que $T \ll 1$ et $\partial_\mu \phi \ll 1$ ainsi que $\sqrt 2 \ell_c > \phi \geq \ell_c$. En dehors de cette dernière limite, le potentiel à une boucle -- diagramme cylindrique d'échange de graviton \emph{entre autres} -- prend le relais et domine la dynamique. Tant que le tachyon est moins massif que le premier état massif de corde ouverte des secteurs $\sigma^{0,3}$, l'approximation de l'action effective par l'action à l'ordre des arbres~\refe{eq:quad} est acceptable. \\

\begin{itemize}\itemsep4pt
\item[1)] Dans le domaine massif, un potentiel effectif à une boucle de type Coleman-Weinberg est obtenu par intégration du tachyon et après renormalisation~:

\begin{align}\label{eq:coleman-Wein}
S = T_p \int \di^{p+1}\sigma \, \parent{\frac{1}{2}\partial_a \phi \partial^a \phi - \frac{1}{2} \parent{\phi^2-\ell_c^2}^2 \ln \parent{\phi^2-\ell_c^2}}
\end{align} 

Dans cette limite, le potentiel s’aplatit en $\phi = \ell_c$. Le plus important reste que le potentiel du champ $\phi$ est clairement attractif en direction de la distance critique\footnote{L'existence d'un minimum local pour $\phi>\ell_c$ est remarquable bien qu'il n'aura aucune importance pour nous.}. En outre, ce comportement attractif est confirmé par le potentiel à une boucle calculé à partir du diagramme cylindrique.

\item[2)] A la distance critique, par continuité de l'action précédente, en notant $\varphi=\ell_{c}+\phi$ et en utilisant la formule~\refe{eq:crit_dist}~:

\begin{align}\label{eq:action_bos_dist_crit}
S = T_p \int \di^{p+1}\sigma \, \parent{\frac{1}{2}\partial_a \phi \partial^a \phi + \frac{1}{2} \partial_a T \partial^a T - \frac{\phi \, T^2}{2\pi}}
\end{align}   

Le potentiel de type Yukawa domine dans la limite perturbative le terme en $\phi^2 T^2$, ce qui implique que la masse effective du tachyon dépend linéairement de la perturbation de distance. La distance critique est donc un point extrêmement instable et la solubilité de cette théorie effective est discutable. Il semblerait que cette action soit celle du modèle de Wick-Cutkorsky~\cite{Darewych:1998mb} non-massif.

\item[3)] Dans le domaine tachyonique, il pourrait \^etre attendu que le potentiel du champ $\phi$ soit attractif -- mais ce n'est pas ce qui est obtenu \emph{a posteriori} dans la section~\refcc{sec:tach_roul_CFT_bos}~ : il existe un mode de condensation temporelle pour lequel la distance reste constante. Premi\`erement, en dessous de la valeur critique, l'action~\refe{eq:coleman-Wein} n'est plus définie. D'une part, le calcul du potentiel effectif à une boucle n'est valable que dans la limite perturbative et d'autre part, \refe{eq:quad} ne reste valide que pour des valeurs de champs faibles. Or, pour $\phi<\ell_c$ une chute \emph{classique} et rapide du tachyon est in\'evitable. La limite perturbative est donc rapidement fausse~; et par conséquent, il n’est plus possible de décrire approximativement le système par l'action quadratique~\refe{eq:quad}.

Deuxi\`emement, le couplage minimal des champs de jauge (et donc aussi au champ $\phi$) au tachyon n'est \apriori valable que dans la limite où les valeurs de $T$ sont contrôlées et faibles. Un exemple d'écart à un modèle de couplage minimal causé par des corrections cordistes est connu~: dans le cadre de la production de paires par un champ électrique~\cite{Bachas:1992bh,Ferrara:1992yc} un couplage minimal cesse d'être valable lorsque la valeur du champ est augmentée -- c'est un effet purement cordiste. Lorsque  le champ critique est atteint, un tachyon apparaît. Cette situation est donc similaire \`a la n\^otre. Toutefois, l'origine de ce tachyon n'est pas exactement la même que celui qui nous intéresse, en particulier parce que le champ critique est dans un cas un champ de jauge vectoriel et dans notre cas un champ scalaire. Nous pouvons voir aussi cela en calculant le taux de production de paire de cordes interbranaires par deux branes en collision, \cad pour un champ $\phi(x^0)$ d\'ependant d'une rapidit\'e $\varepsilon$ et d'un param\`etre d'impact $b$. Nous citerons ici le calcul entre une brane et une anti-brane dont le r\'esultat est \'equivalent~ :

\begin{align}
\omega &\propto \sum_{k=1}^\infty \frac{1}{k}\parent{\frac{\epsilon}{k}}^{p/2}e^{\frac{b^2 k}{\pi \epsilon}} \left\{ - \parent{\frac{1+(-)^k}{2}}\frac{\theta_4(0|ik/\epsilon)^4}{\eta(ik/\epsilon)^{12}} + \parent{\frac{1-(-)^k}{2}} \frac{\theta_3(0|ik/\epsilon)^4}{\eta(ik/\epsilon)^{12}} \right\} \nonumber \\ 
& \stackrel{\varepsilon \to 0}{\sim} -\sum_{k \text{ even}}^\infty \frac{1}{k}\parent{\frac{\epsilon}{k}}^{p/2} e^{-\frac{b^2 k}{\pi \epsilon}} \parent{e^{\frac{\pi k}{\epsilon}}-8 + \ldots} + \sum_{k \text{ odd}}^\infty \frac{1}{k}\parent{\frac{\epsilon}{k}}^{p/2} e^{-\frac{b^2 k}{\pi \epsilon}} \parent{e^{\frac{\pi k}{\epsilon}}+8 + \ldots}
\end{align}

Le taux d\'epend du paramètre d'impact de telle sorte qu'il devient critique au-delà de la valeur~\refe{eq:crit_dist} mais ne dévie pas de la formule de couplage minimal repr\'esent\'ee par le terme dominant. Seule la valeur de la rapidit\'e est responsable de l'inclusion des termes d'ordre sup\'erieur. 
\end{itemize}

\subsubsection{Pertinence de la notion de distance à l'échelle des cordes}

Lorsque la distance entre les branes est grande, leur localisation est claire. Ainsi l'est aussi la notion de distance. Cependant, si la distance est de l'ordre de la longueur de corde $\ell_s=\sqrt {\alpha'}$ et que le tachyon roule en dehors de son potentiel, peut-on encore faire sens d'un système de branes parallèles et bien localisées ? La r\'eponse devrait \^etre affirmative puisque la finesse des conditions de Dirichlet sur les cordes ouvertes dans les directions transverses aux branes n'est pas modifiée lors de la condensation – voir par exemple~\cite{Sen:2002nu}. 

Cependant, la physique étant non-commutative à cette échelle -- les divers champs de jauge étant pris en compte -- il est possible de douter de la pertinence de la notion d'espace et de distance – voir cependant la discussion~\cite{Douglas:1996yp}. Il est tout de même surprenant que indépendamment de cette considération, les résultats montrent que le tachyon roule tout en conservant la valeur du champ $\phi$ -- du moins à l'ordre des arbres. Cela pourrait indiquer que la distance perd son statut de champ pour devenir un paramètre de la condensation et donc caractériser le vide stable atteint -- s'il existe\footnote{Cette question est r\'eserv\'ee pour des travaux ult\'erieurs, mais nous avons des raisons de penser que le vide atteint n'est pas stable.}.  De plus, le mécanisme de création des cordes fermées à partir des cordes ouvertes et des flux de champs de jauge confinés\footnote{Ce point est incertain si le tachyon n'atteint pas le vide pour lequel $V=0$ comme nous le suspectons.}, couplés aux tachyons interbranaires devrait produire des cordes fermées retenues à l'intérieur du système de branes. Donc, $\phi$ caractériserait l'épaisseur du produit final. \\ 

Il est alors impératif de comprendre la physique du système dans la phase tachyonique et quels mécanismes de condensation sont autorisés et sont les solutions des équations du mouvement dérivées de l'action effective correcte. Pour cela, il faut mettre en évidence les th\'eories conformes de bord (BCFT) dans cette phase. C'est la question \`a laquelle nous r\'epondrons en étudiant la marginalité du modèle sigma du tachyon roulant et en montrant qu'il s'agit d'une BCFT \`a distance constante.

\subsection{Tachyon roulant et marginalité}
\label{sec:tach_roul_CFT_bos}

La façon la plus directe d'étudier le champ de tachyon est probablement de s’intéresser à une théorie conforme tachyonique, \cad qui admet comme terme d'interaction sur le bord de la surface un opérateur de vertex correspondant à un tachyon on-shell. \\

Cette section traite de la théorie conforme du tachyon roulant sur un système brane-brane séparé à distance $r$ constante. Dans un premier temps, il sera démontré que ce tachyon est en général exactement marginal pour tout $r$ sauf en certaines valeurs particulières où il perd sa marginalité, contrairement à son homologue du système $D-\bar D$ ce qui sera abordé dans le chapitre~\refcc{chap:tach_cond_susy}. La cause, qui est bien spécifique au cas bosonique, sera physiquement identifiée. 

Dans un deuxième temps, le système sera analys\'e perturbativement en dehors de la CFT et en dehors de la phase tachyonique. Le groupe de renormalisation du mod\`ele sigma seront étudiés. Ce dernier sera perturbé par des déformations off-shell autour des déformations margi\-na\-les. Mais nous mettrons d'abord en évidence que le long de d\'eformations non marginales les fonctions b\^eta ne sont pas bien définies en fonction du schéma de renormalisation, tandis que le long de d\'eformations marginales tout terme contribuant aux fonctions b\^eta est universel. Leur interprétation en qualité d'équations du mouvement sera discuté en dernier lieu.

\subsubsection{Action de surface de corde du système}

Cette partie concerne la théorie de corde ouverte à l'ordre des arbres, \cad sur le disque $D^2$ ou son voisinage conforme $H_+$ le demi-plan complexe, déformée par les tachyons des secteurs interbranaires. Le fond géométrique est suppos\'e trivial, \cad avec $B_{\mu\nu}=0$ et $G_{\mu\nu}=\eta_{\mu\nu}$. La distance étant très faible, il est possible de se placer dans le référentiel inertiel sans tenir compte de la \emph{r\'eponse} des branes sur le fond géométrique.

Cette action sur le disque est la donnée pertinente pour étudier la conformalité du système puis calculer la fonction de partition à l'ordre des arbres et contraindre l'action effective. On se placera d'emblée dans la jauge unitaire sur le demi plan complexe. L'expression de cette action est\footnote{La convention $Z \propto e^{-S}$ est utilisée ici,}~:

\begin{align}
S = \frac{1}{2\pi\alpha'} \int_{H_+} \di^2 z \, \partial X^\mu \bar\partial X_\mu + \sigma^+ \otimes \frac{\lambda^+}{2\pi} \oint_{\mathbb R} \di z \, e^{i r \wt X^{p+1} + \omega X^0} + \sigma^- \otimes \frac{\lambda^-}{2\pi} \oint_{\mathbb R} \di z \, e^{-i r \wt X^{p+1} + \omega X^0}  
\end{align}

avec~:

\begin{align}
\sigma^\pm = \frac{\sigma^1 \pm i \sigma^2}{2} \qquad \text{et} \qquad \lambda^\pm \in \mathbb C
\end{align}

Donnons \`a pr\'esent les OPE des champs fondamentaux sur le bord pour $z>w$. Les champs d'indices $a$ et $b$, pour $a=0\ldots p$, vérifient des conditions au bord de type Neumann -- colinéaires à la brane -- et les champs d'indices $i$ et $j$, pour $i = p+1 \ldots D$, vérifient des conditions au bord de type Dirichlet -- transverses à la brane. Le champ dual $\wt X^{p+1}$ vérifie les conditions au bord de Neumann. Il s’agit du champ conjugué au moment d'enroulement, ici $r$.

\begin{align}
& X^i(z) X^j(w) \simeq - 2 \, \delta^{ij} \alpha' \ln |z-w|^2  \nonumber \\
& X^a(z) X^b(w) \simeq 0 \nonumber \\
& \wt X^{p+1} (z) \wt X^{p+1}(w) \simeq - 2\alpha' \ln(z-w) 
\end{align}

Par la suite, les notations seront allégées en prenant $\alpha'=1$. A condition de ne pas calculer une amplitude avec des insertions arbitraires et les champs n’ayant d'OPE non nuls qu'entre ceux de mêmes indices, il est possible d’intégrer les champs qui ne sont pas concernés par la déformation tachyonique. Il s’agit de tout $X^\mu$ tel que $\mu \neq \{0,p+1\}$. La théorie de surface de corde résultante est donc une théorie $c=2$ pour les champs $X^0$ et $X=X^{p+1}$. L'action simplifiée est~:

\begin{align}\label{eq:action_rolling}
S = \frac{1}{2\pi} \int_{H_+} \di^2 z \, \parent{ - \partial X^0 \bar \partial X^0 + \partial X \bar \partial X} + \sigma^+ \otimes \frac{\lambda^+}{2\pi} \oint_{\mathbb R} \di z \, e^{i r \wt X + \omega X^0} + \sigma^- \otimes \frac{\lambda^-}{2\pi} \oint_{\mathbb R} \di z \, e^{-i r \wt X + \omega X^0}  
\end{align}

L’objectif est d’étudier la marginalité de ce modèle $c=2$. La méthode présentée par Gaberdiel \emph{et al.} dans~\cite{Gaberdiel:2008fn} sera suivie pour étudier le groupe de renormalisation de cette théorie. Leur proposition de régularisation -- \emph{point splitting} – est souvent utilisée. Nous l'avons présentée dans l'introduction en section~\refcc{sec:mod_sig}. Les fonctions b\^eta peuvent alors être calculées dans divers schémas de renormalisation. Ceux proposés dans cet article sont particulièrement pertinents, \cad schéma de \emph{soustraction minimal} et schéma de \emph{Wilson}. 

Ce modèle est étudié on-shell \`a l'ordre dominant, \cad que toutes les déformations dans l'action~\refe{eq:action_rolling} sont imposées marginales ($h=1$). Puisque les divergences de type puissance n'empêchent pas la théorie d'être exactement marginale et que dans le schéma de soustraction minimale seules les divergences logarithmiques -- donc les résonances -- participent aux fonctions b\^eta, il est donc suffisant de se placer d'emblée dans ce dernier schéma.

\subsubsection{Fonctions b\^eta et marginalité exacte du modèle en fonction des valeurs de $r$}

Comme explicité dans l'introduction, il est possible de calculer rapidement la fonction b\^eta de chaque tachyon au premier ordre~:

\begin{align}\label{eq:beta_tach}
\beta_\pm = (1-h_{\pm}) \lambda^\pm = (1-r^2 - \omega^2) \lambda^\pm
\end{align}

Ainsi à cet ordre, il faut imposer $\omega = \pm \sqrt{1-r^2}$, avec $r$ fixé, ce qui est la condition de marginalité présupposée. L'OPE des tachyons au deuxième ordre est alors donné par la formule~:

\begin{align}\label{OPE_tach_bos}
\sigma^+\sigma^-\otimes e^{i r \wt X + \omega X^0}(z)\cdot e^{-i r \wt X + \omega X^0}(w) = \frac{e^{2\omega X^0}}{(z-w)^{4r^2-2}} + i r \frac{\partial \wt X e^{2\omega X^0}}{(z-w)^{4r^2-3}} + \ldots
\end{align}

Puisque $r^2<1$ le terme associé à l'opérateur $e^{2\omega X^0}$ sera divergent UV pour tout $r>\sqrt 3/2$ tandis que celui de l'opérateur $\partial \wt X e^{2\omega X^0}$  est divergent IR pour tout $r\neq 0$. \\

\begin{itemize}\itemsep4pt
\item L'opérateur correspondant à une perturbation de distance est $\sigma^3 \otimes \delta r\oint \partial \wt X$. Même si le terme produit dans l'OPE~\refe{OPE_tach_bos} lui ressemble, il n'y correspond pas~ : il est \emph{irrelevant}, ce qui implique qu'il ne peut pas constituer une perturbation. Ainsi, aucune perturbation de distance n'est produite à l'ordre 2 ni à aucun ordre en perturbation puisqu'à l'ordre $2n$ l'opérateur équivalent sera $\partial \wt X e^{2n\omega X^0}$ pour lequel les mêmes conclusions sont tirées. Par conséquent \`a tout ordre~:

\begin{align}
\beta_{\delta r} = 0
\end{align}

Ce résultat n’est plus vrai en $r=1$. Cette valeur se rév\`ele particulière – voir plus loin. 

\item Le premier terme de~\refe{OPE_tach_bos} est divergent pour $r>\sqrt 3/2$. Ceci implique que la perturbation correspondante, \cad $ \sigma^0\otimes \mu_1\oint e^{2\omega X^0}$ sera produite par le tachyon. Or, cette perturbation n'est autre que celle du tachyon du secteur $\sigma^0$ \cad hébergé sur le volume de chaque brane et naturellement présent dans un modèle bosonique. Sa fonction b\^eta est dans le schéma de renormalisation de Wilson~:

\begin{align}\label{eq:beta_tach_bran}
\beta_{\mu_1}=(4r^2-3)\mu_1 - \lambda^+\lambda^-
\end{align}

En $r=\sqrt 3/2$, il y a résonance car l'opérateur $e^{2\omega X^0}$ est marginal. De la sorte, la divergence est logarithmique et le terme source est universel en cette valeur. Il n'y a donc aucune ambiguïté -- reliée à la nature du schéma de renormalisation -- sur le couplage du tachyon interbranaire au tachyon du secteur $\sigma^0$ en $r=\sqrt 3/2$. 

\item Des résonances, donc des divergences logarithmiques, sont de m\^eme attendues à chaque ordre supérieur $n>1$. Puisque la dimension de l'opérateur correspondant à l'ordre $2n$ est $\Delta=4n^2\omega^2$, il y a potentiellement résonance en $\omega=1/2n$, \cad en $r=\sqrt{1-1/4n^2}$. Les calculs sont détaillés dans le cas supersymétrique pour lequel le résultat est plus intéressant et surtout plus décisif. En effet, il n'y a pas de tachyon dans le secteur $\sigma^0$ si bien que l'apparition d'une divergence logarithmique serait vraiment problématique. \\ 
\end{itemize}

Enfin, la fonction b\^eta du tachyon interbranaire est exactement~\refe{eq:beta_tach} à tous les ordres. En effet, dans l'état actuel de la théorie, le tachyon est le seul opérateur de vertex présent sur le bord. Or, toute OPE du tachyon avec lui-même est inévitablement proportionnelle au champ $e^{2 n \omega X^0}$ donc jamais égale au tachyon lui-même. En outre, compte-tenu de l'expression du tachyon du secteur $\sigma^0$ aucun terme n’est produit en $e^{\pm ir \wt X+\omega X^0}$ par OPE à un ordre quelconque. Ainsi, pour $r$ et $\omega$ fixés, à tout ordre~:

\begin{align}
\beta_\pm = 0
\end{align} 

Pour conclure, l'action complète à étudier devrait être corrigée par un certain nombre de contretermes de la façon suivante~:

\begin{multline}\label{eq:action_rolling2}
S = S_{bulk} + \sigma^+ \otimes \frac{\lambda^+}{2\pi} \oint_{\mathbb R} \di z \, e^{i r \wt X + \omega X^0} + \sigma^- \otimes \frac{\lambda^-}{2\pi} \oint_{\mathbb R} \di z \, e^{-i r \wt X + \omega X^0}  \\ + \sigma^0 \otimes \mu_{0} \oint e^{X^0} + \sigma^0 \otimes \sum_{n \geq 1} \mu_n(\lambda^+\lambda^-) \varepsilon^{4n^2\omega^2-1} \oint e^{2 n\omega X^0}
\end{multline}

Il est nécessaire d’ajouter le tachyon marginal du secteur $\sigma^0$ puisque celui-ci peut être produit en $r=\sqrt{1-1/4n^2}$ pour tout $n\geq 1$ par résonance, à travers la fusion des opérateurs de tachyon des secteurs interbranaires. En dehors de ces valeurs, le tachyon interbranaire est exactement marginal, mais il cesse de l'être en ces valeurs \emph{précisément}. Alors, le tachyon du secteur $\sigma^0$ doit résoudre une équation dont le terme source est proportionnel à $\lambda^+\lambda^-$ telle que~\refe{eq:beta_tach_bran}. Pour $r>\sqrt 3/2$, il faut aussi ajouter le contreterme en puissance du cut-off et dont l'expression dépend uniquement de $\lambda^\pm$. 

Parce que le tachyon du secteur $\sigma^0$ est physique dans la théorie bosonique, il n'y a pas de problème à l'existence de ces résonances en tout $r=\sqrt{1-1/4n^2}$. Elles impliquent simplement que dans un modèle réaliste de condensation de tachyon entre deux branes séparées, il faut aussi tenir compte du tachyon de ce secteur. On en déduit que les branes vont éventuellement se désintégrer par ce tachyon tout autant que par le tachyon interbranaire. Dans le cadre du système brane-antibrane l'absence du tachyon de secteur $\sigma^0$ et l'éventuelle existence de ces résonances pose plus de problème, parce qu'elles n’apportent pas d'interprétations physiques comme ici. \\

Il y a une limite intéressante à ce modèle qui est $r \to r_c$ avec ici $r_c=1$. Dans ce cas, le nombre de contre termes tend vers l'infini, ce qui indique que la théorie n'y est \apriori pas renormalisable en limite -- les points de résonances sont denses autour de $r=r_c$. Il convient donc de l'étudier en cette valeur, pour laquelle elle est renormalisable parce que $\omega=0$ strictement. Par contre, la théorie n'est pas \emph{exactement} marginale parce qu'à l'inverse du cas $r<1$ c'est la perturbation de distance qui est résonante.

\subsubsection{Cas particulier $r=1$}

En cette distance, par continuité de~\refe{eq:action_rolling}, le tachyon est purement statique. En effet, la déformation est simplement~:

\begin{align}
\delta S = \sigma^+ \otimes \lambda^+ \oint_{\mathbb R} \di z \, e^{i \wt X} + \sigma^- \otimes \lambda^- \oint_{\mathbb R} \di z \, e^{-i \wt X}  
\end{align}

Par commodité, nous avons red\'efini $\lambda^\pm \to 2\pi \lambda^\pm$. En supposant que $\lambda^\pm \ll 1$, cette limite peut être considérée perturbative et il est possible de développer proprement $e^{-\delta S}$. Le premier terme non trivial non nul est au second ordre~:

\begin{multline}
\frac{1}{2} \parent{\sigma^+ \otimes \lambda^+ \oint_{\mathbb R} \di z \, e^{i \wt X}(z) }~\cdot~\parent{ \sigma^- \otimes \lambda^- \oint_{\mathbb R} \di w \, e^{-i \wt X} (w) } \Theta(|z-w|-\epsilon) \Theta(L-|z-w|)  \\ + \frac{1}{2} \parent{ \sigma^- \otimes \lambda^- \oint_{\mathbb R} \di z \, e^{-i \wt X} (z) }~\cdot~\parent{\sigma^+ \otimes \lambda^+ \oint_{\mathbb R} \di w \, e^{i \wt X}(w) }\Theta(|z-w|-\epsilon) \Theta(L-|z-w|)
\end{multline}

Des cut-off UV et IR ont été ajoutés en sorte de réguler les divergences selon la m\'ethode de \emph{point-splitting}. Cette régularisation brise explicitement la symétrie conforme sur la surface de corde, mais le résultat final ne doit pas dépendre des cut-off. Donc \emph{in fine}, la sym\'etrie doit être restaurée . La formule précédente se simplifie en~:

\begin{align}
\sigma^+ \sigma^- \otimes \lambda^+\lambda^-  \int \di w \int_{w+\epsilon}^{w+L} \di z ~ e^{i \wt X}(z) e^{-i \wt X}(w)
+ \sigma^- \sigma^+ \otimes \lambda^+\lambda^- \int \di w \int_{w+\epsilon}^{w+L} \di z ~ e^{-i \wt X}(z) e^{i \wt X}(w) 
\end{align} 

Les OPE des opérateurs exponentiels sont données sur le bord par~:

\begin{align}
& e^{i \wt X}(z) e^{-i \wt X}(w) = (z-w)^{-2} \parent{1 + i(z-w) \partial \wt X + \ldots} \nonumber \\
& e^{-i \wt X}(z) e^{i \wt X}(w) = (z-w)^{-2} \parent{1 - i(z-w) \partial \wt X + \ldots}
\end{align}

La réinjection de ces résultats dans les intégrales précédentes donne~:

\begin{align}
\sigma^0 \otimes \lambda^+\lambda^- \int \di w ~ \parent{\frac{1}{\epsilon}-\frac{1}{L}} + \sigma^3 \otimes \lambda^+\lambda^- \int \di w ~  \ln \frac{L}{\epsilon} \, \partial\wt X
\end{align}

Le premier terme est divergent en puissance des cut-off mais doit simplement être ôté du calcul et ne pose pas de problème de marginalité. Par contre, le deuxième terme donne lieu à une divergence logarithmique. La perturbation de distance est par cons\'equent r\'esonante avec le tachyon interbranaire en $r=r_c$. Ce terme va donc contribuer universellement à la fonction b\^eta du couplage associé à l'opérateur de vertex $\sigma^3 \otimes \partial \wt X$. Pour le supprimer dans le cadre d'un sch\'ema de soustraction minimale, il faut ajouter à l'action un contre-terme~:

\begin{align}
S_{ct} = \sigma^3 \otimes \lambda^+\lambda^-\int \di w ~  \ln \frac{\ell}{\epsilon} \, \partial\wt X
\end{align}

avec $\ell$ l'\'echelle de renormalisation (\`a ne pas confondre de la valeur de distance $\ell$ que nous avons utilis\'ee pr\'ec\'edemment). L'ajout de l'\'echelle de renormalisation est essentiel pour compenser la dimension de $\varepsilon$ dans l'argument du logarithme. Pour étudier l'effet de ce contreterme sur la marginalité de la théorie, il faut introduire le terme de bord suivant~:

\begin{align}
\frac{\sigma^3}{2} \otimes \oint \di z \, \delta r(X) \partial \wt X
\end{align}

De manière tout à fait générale, si $\delta r$ dépend des champs $X$ il est possible de faire un développement autour des modes zéro~\cite{Tseytlin:2000mt,Tseytlin:1991bu,Tseytlin:2001ah} selon $X=x+\hat X$~:

\begin{align}\label{eq:dev_distance}
\frac{\sigma^3}{2} \otimes \oint \di z \, \parent{\delta r(x) + \partial_a \delta r(x) \, \hat X^a + \frac{1}{2}\partial_a \partial_b \delta r(x) \, \bnormal{\hat X^a \hat X^b} + \ldots} \partial \wt X
\end{align}

Pour étudier le groupe de renormalisation, il faut extraire l'échelle de renormalisation des opérateurs de vertex en appliquant $z \to \ell \, z $. En utilisant 

\begin{align} 
\bnormal{\hat X^a \hat X^b (z/\ell)} = \bnormal{\hat X^a \hat X^b (z)} - 2\eta^{ab}\ln \ell
\end{align}

le couplage $\delta r$ obtenu est le suivant~:

\begin{align}
\frac{\sigma^3}{2} \otimes \oint \di z \, \parent{\delta r(x)  + \square \delta r(x) \ln \ell + 2 \lambda^+\lambda^- \ln \ell } \partial \wt X
\end{align}

Par conséquent, la fonction b\^eta du couplage $\delta r$ au deuxième ordre est~:

\begin{align}\label{eq:beta_on_crit}
\beta_{\delta r} = - 2 \lambda^+\lambda^- - \square \delta r 
\end{align}

Une fonction b\^eta peut, \`a condition qu'elle soit indépendante du schéma de renormalisation, être interprétée comme une équation de mouvement dérivée d'une action effective $S_{eff}$ \`a des red\'efinitions des champs pr\`es. Par exemple, \`a l'ordre quadratique, en notant $\phi= \delta r$, l'action suivante~:

\begin{align}
S_{eff} \propto \int \di^{p+1} \sigma \parent{\frac{1}{2} \partial_\mu \phi \partial^\mu \phi + 2 \phi \lambda^+\lambda^-}
\end{align}

est compatible avec $\beta_{\delta r}$. Cette expression est en accord avec celle obtenue dans la formule~\refe{eq:action_bos_dist_crit} par d\'eveloppement de l'action de Garousi. En $r=r_c$, le tachyon agit donc bien comme terme source pour le champ de distance en le tirant vers $r< 1$. Ce résultat est au moins valable pour $\lambda^\pm \ll 1$.

\section{Groupe de renormalisation, fonctions b\^eta et équations du mouvement}
\label{sec:off_sep}

Dans cette section, nous étudierons le groupe de renormalisation du modèle sigma pour le tachyon et le champ de distance sur la surface de corde. Les fonctions b\^eta associées aux divers champs off-shell relevants seront calculées et rapport\'ees -- dans la mesure du possible -- \`a des équations du mouvement de cette certaine action. Cette interprétation n'est en générale pas correcte~: Tseytlin explique dans~\cite{Tseytlin:1986ws,Tseytlin:1986ti} que les équations du mouvement de chaque champ d'espace-cible sont en fait \emph{proportionnelles} aux fonctions b\^eta, à des facteurs dépendants des divers champs près, et non égales. De sorte que pour $S$ l'action effective, $\delta S/\delta \phi_i = \kappa_{ij} \beta_j$. Cet argument est justifié par la non-conservation des expressions des fonctions b\^eta par changement de schéma de renormalisation, qui sont des redéfinitions des couplages~\cite{Gaberdiel:2008fn}. Si bien que, à moins d'exprimer des fonctions b\^eta invariantes par changement de schéma, elles ne peuvent pas être interprétées directement en tant qu'équations du mouvement. 

Nous verrons que les fonctions b\^eta off-shell ne sont en g\'en\'eral pas cohérentes en tant qu'équations du mouvement. A l'inverse, lorsqu'elles sont obtenues pour des théories perturbées autour de déformations marginales, \cad sur le bord autour d'opérateurs primaires de poids $\Delta =1$ elles sont invariantes par changement de sch\'ema de renormalisation car construites \`a partir de contributions universelles.

\subsection{Phase surcritique $r>1$}

En $r>1$, le champ de tachyon primaire est de la forme $\sigma^\pm \otimes \lambda^\pm \oint e^{\pm i r \wt X \pm i \omega X^0}$. L'étude des premiers ordres des fonctions b\^eta y est justifiable pour obtenir une action effective tant que l'approximation $\lambda \ll 1$ l'est ; ce qui est bien le cas pour $r>1$. Les résultats montrent que tant que $\omega^2>r^2-1$ la déformation est relevante et constitue donc une bonne perturbation.  

\subsubsection{Schéma de soustraction minimal}

Les fonctions b\^eta de $\delta r$ et $\lambda^\pm$ sont modifiées,  en~:

\begin{align}\label{eq:beta_on}
& \beta_{\delta r} = -\square \delta r - 2 r \, \lambda^+ \lambda^- \, \delta_{r^2-\omega^2,1}\nonumber \\
& \beta_{\pm} = (1-r^2+\omega^2) \lambda^\pm - 2r \delta r \, \lambda^\pm
\end{align}

ce qui indique que si le tachyon est off-shell dans cet ansatz $T\propto e^{i\omega x^0}$, \ie si $\omega^2 < r^2-1$, alors la contribution à la fonction b\^eta de $\delta r$ est nulle. Il est un peu difficile d'interpréter cela en terme d'équations du mouvement. Sans imposer d'ansatz au tachyon, sa déformation se développe de façon similaire à celle de $\delta r(X^a)$~:

\begin{align}\label{eq:dev_tachyon}
\sigma^\pm \otimes \ell^{r^2-1} \oint \di z \, \parent{\lambda^\pm  + \hat X^a \partial_a \lambda^\pm + \frac{\partial_a\partial_b \lambda^\pm}{2}\hat X^a\hat X^b  + \ldots} e^{\pm i r \wt X}
\end{align}

Cela démontre que le tachyon est irrelevant puisque $\lambda^\pm \propto \ell^{1-r^2}$. Il est donc nécessaire d'en revenir à l'ansatz, qui peut être un peu complexifié. Les déformations suivantes sont développées

\begin{align}\label{eq:deformation_minimal}
&\sigma^\pm \otimes \ell^{r^2-\omega^2-1} \oint \di z \, \parent{\lambda^\pm  +\partial_i \lambda^\pm  \hat X^i  + \frac{\partial_i\partial_j \lambda^\pm}{2}\hat X^i\hat X^j  + \ldots} e^{\pm i r \wt X \pm i \omega X^0} \nonumber \\
& \sigma^3 \otimes \oint \di z \, \parent{\delta r + \partial_i \delta r  \hat X^i  + \frac{\partial_i\partial_j \delta r}{2}\hat X^i\hat X^j  + \ldots}\partial \wt X
\end{align}

en imposant toujours $\omega^2>r^2-1$ \cad off-shell. Dans un premier temps, la dépendance temporelle dans les couplages est négligée pour simplifier les calculs mais nous la rétablirons \infine par covariance. Les OPE utiles sont~:

\begin{align}\label{eq:OPE_irr}
& \bnormal{ e^{i r \wt X + i\omega X^0}(z)}\bnormal{ e^{-i r\wt X - i\omega X^0}(w)} = (z-w)^{2\omega^2-2r^2} \parent{1 + ir (z-w) \partial \wt X + \ldots} \nonumber \\
& \bnormal{e^{-i r \wt X - i\omega X^0}(z)} \bnormal{e^{i r\wt X+ i\omega X^0}(w)} = (z-w)^{2\omega^2-2r^2} \parent{1 - ir (z-w) \partial \wt X + \ldots} \nonumber \\ 
& \bnormal{\partial \wt X(z)} \bnormal{e^{\pm i r \wt X \pm i\omega X^0}(w)} = \frac{\pm 2 i r}{z-w} e^{\pm i r \wt X \pm i\omega X^0}(z)
\end{align}

Le premier terme des deux premières lignes est divergent. Il contribue, comme nous l'avons vu dans la section pr\'ec\'edente, \`a la fonction b\^eta du tachyon du secteur $\sigma^0$ puisque $\acomm{\sigma^+}{\sigma^-}=\sigma^0$. Toutefois, il est sans intérêt pour nous, puisque spécifique au cas bosonique. Les OPE des champs bosoniques $\hat X^a$ contribuent dans les fonctions bêta à des termes dépendants des cut-offs, leur étude est reportée dans un premier temps.

En toute rigueur, il faudrait aussi analyser la production du terme en $\partial X^0$ par les OPE des tachyons. Cette analyse ne sera pas réalisée parce que \emph{i)} seuls les relations entre le champ de distance et les tachyons sont spécifiquement étudiés et \emph{ii)} par covariantisation de l'action effective le long du groupe de jauge $U(2)$ brisé, les contributions sont exprimables sans calculs explicites – mais une vérification à l'ordre quadratique à partir des formules~\refe{eq:OPE_irr} n'est pas difficile. \\

Le second terme des deux premières lignes quant \`a lui, est divergent UV pour $\omega^2 < r^2-1$ exclusivement donc est convergent dans le domaine surcritique. En revanche, le terme obtenue en troisième ligne est clairement résonant et donne une divergence logarithmique. Les fonctions b\^eta\footnote{Celles des champs $\partial_a \delta r$ et $\partial_a\lambda^\pm$ sont simplement les dérivées des fonctions b\^eta que l'on donne.} sont obtenues en utilisant $[\sigma^+,\sigma^-]=\sigma^3$  et par soustraction des divergences UV~:

\begin{align}\label{eq:beta_off_min}
& \beta_{\delta r} = \Delta \delta r  \nonumber \\
& \beta_{\pm} = (1-r^2-\omega^2)\lambda^\pm + \Delta \lambda^\pm - 2r \delta r \, \lambda^\pm 
\end{align}

La fonction bêta de $\delta r$ est tout à fait triviale. Or, dans les formules~\refe{eq:beta_on} qui sont correctes, le long de l'ansatz $T\propto e^{\pm i\omega X^0}$ apparaissait un terme proportionnel à $r \lambda^+ \lambda^- \delta_{r^2-\omega^2,1}$. Il n'apparaît pas ici car l'exponentielle $e^{\pm i\omega X^0}$ est répartie entre les termes de dérivées multiples du tachyon que nous avons justement négligés en repoussant l'\'etude des OPE des champs $\hat X^a$. Ces derniers produisent ensemble des divergences le long de l'opérateur de vertex du champ de distance, à partir d'intégrales du type~:

\begin{align}\label{eq:integrales_derivees}
(-1)^n \frac{2^{n}}{n!} \, \frac{r}{4\pi^2} \, [\partial_{a_1}\partial_{a_2}\ldots\partial_{a_n}\lambda^+]_B \,\prod_{i}^n  \eta^{a_i b_i} \,  [\partial_{b_1}\partial_{b_2}\ldots\partial_{b_n}\lambda^-]_B ~ \int_\epsilon^\ell \di z \, \frac{\ln^{n} z}{z^{2r^2-1}} \nonumber \\ \times \sigma^3 \otimes \int \partial \wt X 
\end{align}

avec $n \in \mathbb N$. Les couplages nus ont été développés comme proposé dans le chapitre d'introduction et non sous la forme $\ell^{h-1} \mu$ ce qui est crucial dans ce schéma car, par définition, $d\mu_B/d\ell =0$. Il faut ajouter \`a l'action le contreterme~ :

\begin{align}
S_{ct} & =  (-1)^n \frac{2^{n}}{n!} \, \frac{r}{4\pi^2} \, [\partial_{a_1}\partial_{a_2}\ldots\partial_{a_n}\lambda^+]_B \,\prod_{i}^n  \eta^{a_i b_i} \,  [\partial_{b_1}\partial_{b_2}\ldots\partial_{b_n}\lambda^-]_B \nonumber \\ & \times \, 2(1-r^2) \Gamma(1+n, 2(1-r^2) \ln \frac{\ell}{\varepsilon} )  ~ \sigma^3 \otimes \int \partial \wt X 
\end{align}

avec $\Gamma(a,z)$ la fonction gamma incompl\`ete. Les contributions \`a la fonction b\^eta de $\delta r$ sont proportionnelles \`a un ensemble de facteurs d\'ependant de l'\'echelle $\ell$ selon~ :

\begin{align}
\beta_{\delta r} \sim \sum_{\alpha=1}^{n-1} C(\alpha) \varepsilon^{2(1-r^2)} \frac{ \ln^{n-\alpha} \ell}{\ell}
\end{align}

avec $C(\alpha)$ un coefficient d\'ependant de la distance et des d\'eriv\'ees des tachyons. Dans la limite IR $\ell \to \infty$ tous s'annulent, puisque $r>1$. Ils n'emp\^echent donc pas l'existence d'un point fixe infrarouge et peuvent bien \^etre n\'eglig\'es. Maintenant, d'après la formule~\refe{eq:integrales_derivees}, nous obtenons qu'on-shell, avec $\partial_0^{(n)}\lambda^\pm = (-\omega)^n \lambda^\pm$ les logarithmes se resomment sous la forme $\sum (2 \omega^2\ln (z-w))^{n} /n! = (z-w)^{2\omega^2}$ dont nous d\'eduisons la formule~\refe{eq:beta_on}. Ainsi, les fonctions b\^eta ne sont pas continues dans la transition du tachyon off-shell au tachyon on-shell~: tout se passe à un niveau pré-intégratoire, \cad dans l’intégrande, qui se resomme parfaitement à la résonance en une formule compacte. \\

Deuxièmement, l'absence du terme quadratique $r\lambda^+\lambda^-$ entre en contradiction avec la pré\-sence de son équivalent dans la fonction b\^eta du tachyon $2 r\delta r \lambda^+$ du point de vue de leur interprétation en tant qu'équations du mouvement dérivées d'une action. Cela peut suggérer que le schéma minimal n'est pas le cadre le plus adapté au calcul off-shell de l'action effective, compte-tenu du développement naturel~\refe{eq:dev_distance} et \refe{eq:dev_tachyon} des couplages sur le bord. Mais cela peut aussi suggérer que l'interprétation des fonctions b\^eta calculées off-shell en tant qu'équations du mouvement est incorrect. Cette dernière suggestion est probablement la plus raisonnable, \'etant donn\'e que l'\'etude dans le sch\'ema de Wilson, comme nous allons le voir maintenant, n'apporte aucune am\'elioration. 

\subsubsection{Schéma de Wilson}

Dans ce cadre, il faut remplacer\footnote{Voir section~\refcc{sec:mod_sig}.} dans~\refe{eq:deformation_minimal} l'échelle de renormalisation directement par le cut-off UV $\ell \to \varepsilon$ et supposer que les couplages dépendent explicitement de ce cut-off $\mu = \mu(\varepsilon)$. Les fonctions b\^eta suivantes sont obtenues~:

\begin{align}\label{eq:beta_wilson}
& \beta_{\delta r} = \Delta \delta r - 2 r \parent{\beta_+ \lambda^- + \lambda^+ \beta_-} + \ldots  \nonumber \\
& \beta_{\pm} = (1-r^2-\omega^2)\lambda^\pm + \Delta \lambda^\pm - 2r \delta r \, \lambda^\pm + \ldots 
\end{align}

Tous les termes de dérivées multiples dépen\-dants du cut-off UV en $\ln^n \epsilon$, toujours au deuxième ordre, ont été inclus dans les pointillés. Cette fois en revanche, parce que ce sont les couplages d\'ependant de l'\'echelle qui sont pris en compte dans le d\'eveloppement et \emph{non} les couplages \emph{nus}, ces termes sont directement proportionnels aux diverses fonctions b\^eta des d\'eriv\'ees de tachyon. Ils sont donc redondants, puisque la relation des fonctions b\^eta avec les équations du mouvement demande d'imposer $\beta_i=0$. Toutefois, nous avons mis en valeur le terme constant dans les cut-off. \\

Remarquons enfin qu’il n’y a pas plus de contribution en $r\lambda^+\lambda^-$ dans la fonction b\^eta de $\delta r$ que dans le schéma de soustraction minimale. Par déduction, quelque soit le schéma de renormalisation, les fonctions b\^eta off-shell ne peuvent pas être, comme présupposé, des équations du mouvement.

Pour preuve, lorsque dans un premier temps les deux équations~\refe{eq:beta_wilson} sont vérifiées, \cad $\beta_i =0$ le tachyon on-shell peut être choisi avec, par exemple, $\omega = 1/2-r^2$ et $\delta r=0$. Or, le long de cette solution, on revient dans un second temps à~\refe{eq:beta_on} pour lequel la fonction b\^eta de $\delta r$ était en fait non nulle à cause du terme $r\lambda^+\lambda^-$. Ces équations ne peuvent donc pas constituer des équations de mouvement puisque leurs solutions n'en sont pas, à l’exception de $\lambda^\pm=0$ ou $r=0$ qui sont triviales.

\subsubsection{Un ansatz plus général}

Ainsi, il semble \`a juste titre que l'utilisation des fonctions b\^eta off-shell et leur interprétation en tant qu'équations du mouvement soient sujettes à caution. Cependant, il serait intéressant de chercher des ansatz plus généraux dont les déformations seraient quasi-marginales\footnote{Des perturbations le long de d\'eformations marginales au premier ordre.} mais non exactement marginales~: elles résoudraient les équations du groupe de renormalisation \`a l'ordre dominant et permettraient éventuellement d'exprimer des fonctions b\^eta non triviales aux ordres supérieurs. En procédant de cette manière, seules des résonances apparaîtraient. Ces dernières fournissent des contributions universelles aux fonctions b\^eta et sont par conséquent indépendantes du schéma de renormalisation. Par exemple, l'ansatz~:

\begin{align}
T^+ &= \zeta_{(1)}(x^i) e^{i\omega x^0} + \zeta_{(2)}(x^i) e^{- i\omega x^0} \nonumber \\ 
T^- &= (T^+)^*
\end{align}

avec $\omega=r^2-1$ telle que la déformation avec $\zeta_{(i)}$ constante est marginale au premier ordre, donne les fonctions b\^eta suivantes~:

\begin{align}\label{eq:beta_general_massif}
& \beta_{\delta r} = \Delta \delta r - 2 r \parent{\module{\zeta_{(1)}}^2 + \module{\zeta_{(2)}}^2} + \ldots  \nonumber \\
& \beta_{{(1,2)}} = \Delta \zeta_{(1,2)} - 2r \delta r \, \zeta_{(1,2)} + \ldots 
\end{align}

Ces contributions sont obtenues uniquement à partir de divergences logarithmiques. La solution de tachyon roulant \'etudi\'ee initialement était donnée par $\zeta_{(2)} = 0$ et $\zeta_{(1)} = \lambda^+$ avec $\lambda^- = (\lambda^+)^*$. Il est clair ici que la solution à $r$ constant implose obligatoirement $\zeta_{(1,2)}=0$. Notons qu'il est permis de redéfinir les champs à des constantes près. Or, puisque $r$ est considéré constant, le tachyon peut \^etre redéfini de façon général en $\zeta \to \sqrt{f(r)} \zeta$. Si bien que les fonctions b\^eta deviennent~:

\begin{align}\label{eq:beta_onshell}
& \beta_{\delta r} = \Delta \delta r - 2 r f(r) \parent{\module{\zeta_{(1)}}^2 + \module{\zeta_{(2)}}^2} + \ldots  \nonumber \\
& \beta_{{(1,2)}} = \Delta \zeta_{(1,2)} - 2r \delta r \, \zeta_{(1,2)} + \ldots 
\end{align}

En considérant $\omega$ off-shell, même infinitésimalement, la contribution des tachyons dans $\beta_{\delta r}$ disparaîtrait. Or les deux autres équations sur les tachyons ne seraient quant-à elles pas résolues puisque $\omega$ est off-shell \emph{au premier ordre} et impose donc aussi $\zeta_{(1,2)}=0$. Ces fonctions b\^eta sont donc cohérentes off-shell et on-shell, à la différence des premières~\refe{eq:beta_off_min} ou~\refe{eq:beta_wilson}. \\

En outre, lorsque toutes les déformations constantes sont marginales, toutes les contributions aux fonctions b\^eta proviennent de résonance et sont par conséquent indépendantes du schéma de renormalisation. Cela implique que ces fonctions b\^eta peuvent éventuellement -- mais uniquement dans une limite perturbative -- constituer d'excellentes candidates au rôle d'équations du mouvement. Cet argument sera aussi valable pour $r<1$. 

\subsubsection{Equations du mouvement~: une proposition}

Pour finir, des équations du mouvement peuvent être proposées telles qu'elles seraient compatibles avec les diverses contraintes imposées par les fonctions b\^eta~\refe{eq:beta_onshell}. En supposant le regroupement en champ $\phi = r+\delta r$ nous pourrions exprimer les \'equations suivantes~:

\begin{align}\label{eq:mouvement_massif}
\frac{\delta {\mathcal L}}{\delta \phi} &= - \square \phi - \frac{\phi \, f(\phi)}{1-\phi^2} \parent{\partial_a T \partial^a T^* + (1-\phi^2) \module{T}^2} \nonumber \\ 
\frac{\delta {\mathcal L}}{\delta T^*} &= - \square T + (1-\phi^2) T  \nonumber \\ 
\frac{\delta {\mathcal L}}{\delta T} &= - \square T^* + (1-\phi^2) T^*
\end{align}

Elles redonnent bien les fonctions b\^eta obtenues dans la limite quadratique en remplaçant $\phi = r+\delta r$ ainsi que $T=T^+$ et $T^*=T^-$. Des actions compatibles avec chacune de ces équations prises séparément sont trouvables, mais il n'existe pas d'expression complète permettant de retrouver exactement~\refe{eq:mouvement_massif}. Nous verrons dans le chapite~\refcc{chap:tach_cond_susy} que cela est d\^u \`a ambiguïté\'e dans la correspondance des fonctions b\^eta aux \'equations du mouvement~ : il y a toujours la libert\'e de rajouter dans~\refe{eq:beta_onshell} des termes proportionnels \`a des fonctions b\^eta, puisqu'elles doivent \^etre nulles.

Une action de type Garousi n'est pas exactement compatible avec ces \'equations, puisque le terme de couplage dans les tachyons seraient d'apr\`es~\refe{eq:quad} en $\phi \module{T}^2$. C'est un point troublant, car elle est techniquement valable autour de tachyons dont les \'el\'ements de matrice S sont bien d\'efinis. Ce qui est le cas lorsqu'ils sont massifs. La r\'esolution de cette \'enigme tient probablement du fait que la correspondance entre la physique des champs du mod\`ele-sigma et celle des champs directement d\'efinis dans l'espace-cible n'est vraie qu'\`a des red\'efinitions de champs pr\`es – voir par exemple~\cite{Kutasov:2003er}. Or, le point important reste que les \'equations du mouvement dans chaque cas doivent simplement admettre des solutions compatibles. C'est bien le cas ici, car \`a $\phi$ constant, il faut $T=0$. Nous donnerons plus de d\'etail \`a ce propos en discutant les \'equations du groupe de renormalisation dans le chapitre~\refcc{chap:tach_cond_susy}.

\subsection{Phase sous-critique $r<1$}

Dans cette région, les fonctions b\^eta~\refe{eq:beta_off_min} et~\refe{eq:beta_wilson} restent correctes si la déformation du tachyon off-shell $e^{i\omega X^0+ir\tilde X}$ est toujours relevante, \cad si $\omega$ est réel. Toutefois, il ne s'agit pas du mode fondamental, puisque le tachyon est instable et devrait donc avoir $\omega$ imaginaire. Pour cette raison l'approximation $\lambda \ll 1$ n'est de façon générale plus valable. En effet, le tachyon marginal est de la forme  $T \propto e^{\pm ir \wt X + \omega X^0}$ avec $\omega^2 = 1-r^2$ si bien que $e^{\omega x^0} \sim 1$. Or, dans le domaine sous-critique, il est aussi possible de n'ajouter off-shell que la déformation $e^{\pm ir \tilde X}\lambda(X^a)$ puisqu'elle est relevante. Néanmoins, ça n'est pas une bonne idée de calculer les fonctions b\^eta complètement off-shell pour en extraire des équations de mouvement, car les divergences ne sont pas universels et les contributions aux fonctions b\^eta non-universelles. 

\subsubsection{Fonctions b\^eta des perturbations autour du tachyon roulant marginal}

Le développement autour de la déformation marginale au premier ordre est révisé dans ce paragraphe~:

\begin{multline}
\delta S = \sigma^+ \otimes \varepsilon^{r^2+\omega^2-1} \oint e^{ir \wt X + \omega X^0}\parent{\lambda^+   + \partial_i\lambda^+ \hat X^i + \frac{\partial_i\partial_j \lambda^+}{2} \hat X^i\hat X^j} \\ + \sigma^- \otimes \varepsilon^{r^2+\omega^2-1}\oint e^{-ir \wt X + \omega X^0}\parent{\lambda^- + \partial_i\lambda^- \hat X^i + \frac{\partial_i\partial_j \lambda^-}{2} \hat X^i\hat X^j}
\end{multline}

De nouveaux, il s’agit d’un cas particulier où toute la dépendance temporelle est supposée factorisée pour le tachyon et gelée pour le champ de distance. Les fonctions b\^eta off-shell obtenues dans le schéma wilsonien, à partir entre autres de ce qui a été calculé dans la section~\refcc{sec:tach_roul_CFT_bos} sont simplement~:

\begin{align}\label{eq:beta_wilson_tach}
& \beta_{\delta r} = -\Delta \delta r   \nonumber \\
& \beta_{\pm} = (1-r^2-\omega^2)\lambda^\pm + \Delta \lambda^\pm - 2r \delta r \, \lambda^\pm + \ldots
\end{align}

Dans le schéma minimal, les fonctions b\^eta obtenues sont identiques. L'absence du terme source dans la fonction b\^eta de la perturbation de distance est remarquable, tandis que le terme correspondant apparaît dans celle des tachyons. L'origine de cette absence est, comme nous l'avions vu dans la section pr\'ec\'edente, le facteur $e^{2 \omega X^0}$ multipliant l'opérateur $\partial \wt X$ de sorte que l'op\'erateur produit n'est pas identifi\'e \`a celui du champ de distance. L'opérateur $e^{2\omega X^0}\partial \wt X$ ne peut de plus pas être assimilé à une déformation \emph{corrigée} du champ de distance puisqu'il est irrelevant.  \\

Remarquons à nouveau que les fonctions b\^eta ne peuvent être considérées que dans la limite perturbative $\lambda \ll 1$ ou $x^0 \to -\infty$. En d'autres termes, le temps s'écoulant, les ordres $\lambda^n$ devraient tous finir par contribuer. Toutefois, l'égalité $\beta_{\delta r} = -\Delta \delta r$ est vérifiée dans tout le domaine et à tout ordre le long d'un tachyon $e^{\omega x^0}\lambda^\pm(x^i)$ à cause du facteur $e^{2n\omega X^0}$.  

Attention, ces remarques ne sont vraies que dans la limite où les dérivées successives du tachyon sont négligeables~: le long de $T^\pm \propto e^{\pm i k_i x^i \pm i \omega x^0}$ avec $\omega^2 =r^2 + k^2 -1$ et $k^2 \geq 1-r^2$, \cad tel que l'énergie est réelle, alors~\refe{eq:beta_on} est exactement retrouvée~:

\begin{align}\label{eq:beta_on_sous}
& \beta_{\delta r} = \Delta \delta r - r \lambda^+\lambda^-   \nonumber \\
& \beta_{\pm} = - 2r \delta r \, \lambda^\pm 
\end{align}

De nouveau, ce n'est donc pas une solution à moins que $\lambda^\pm=0$ ou $r=0$ et $\delta r=0$. Mais cela indique que le terme d'interaction $r \lambda^+\lambda^-$ est toujours sous-jacent dans l'équation de mouvement. \\

Etudions \`a nouveau les fonctions b\^eta des perturbations d\'efinies le long de l'ansatz plus général du tachyon, marginal au premier ordre pour $\omega^2= 1-r^2$~:

\begin{align}
T^+ &= \zeta_{(1)}(x^i) e^{\omega x^0} + \zeta_{(2)}(x^i) e^{-\omega x^0} \nonumber \\
T^- &= (T^+)^*
\end{align}

Les termes croisés contribuent à la fonction b\^eta de l'opérateur $\partial \wt X$. Après calcul des OPE, les fonctions b\^eta s'expriment selon~:

\begin{align}\label{eq:beta_onshell_sous}
& \beta_{\delta r} = \Delta \delta r - 2 r \, \parent{\zeta_{(1)}\zeta_{(2)}^*+\zeta_{(1)}^*\zeta_{(2)} } + \ldots  \nonumber \\
& \beta_{{(1,2)}} = \Delta \zeta_{(1,2)} - 2r \delta r \, \zeta_{(1,2)} + \ldots 
\end{align}

Comme précédemment, les tachyons peuvent être redéfinis par $\zeta \to \sqrt{f(r)} \zeta$. Ces formules prouvent de nouveau que le terme d'interaction doit être présent de façon implicite. Il s'annule le long du tachyon roulant, qui ici serait simplement obtenu en imposant $\zeta_{(2)}=0$. Les équations du mouvement compatibles avec ce comportement sont encore, en notant $\phi=r+\delta r$~:

\begin{align}\label{eq:mouvement_tach}
\frac{\delta {\mathcal L}}{\delta \phi} &= - \square \phi - \frac{\phi \, f(\phi)}{1-\phi^2} \parent{\partial_a T \partial^a T^* + (1-\phi^2) \module{T}^2} \nonumber \\ 
\frac{\delta {\mathcal L}}{\delta T^*} &= - \square T + (1-\phi^2) T  \nonumber \\ 
\frac{\delta {\mathcal L}}{\delta T} &= - \square T^* + (1-\phi^2) T^*
\end{align}

A la fonction $f(\phi)$ et au cas particulier de la distance critique près, il semble donc que les \'equations du mouvement sont continues entre la phase sur-critique et la phase sous-critique. C'est un point très positif. Cela sera le cas aussi en théorie des supercordes dans le système $D-\bar D$. Cependant, aucune action effective pour $r$ et pour $T$ ne peut être trouvée à partir de ces expressions. Cela sera rediscuté dans le cadre du système $D-\bar D$ pour lequel d'autres contraintes sur l'action effective existent. Toutefois, les couplages entre les champs déduits de ces équations ne sont pas compatibles avec une action quadratique aussi simple que~\refe{eq:quad}. Cela montre qu'une action de type Garousi-TDBI ne peut pas d\'ecrire ce syst\`eme, parce que les solutions seraient diff\'erentes. 

A l'instar du cas massif, les excitations tachyoniques d'\'energie r\'eelle, \cad de genre espace et pour lesquels des \'el\'ements de matrice-S sont bien d\'efinis par continuation depuis la phase massive, ne v\'erifient pas non plus d'\'equations de mouvement \'egales ou proportionnelles \`a celles d\'eriv\'ees depuis une action de type Garousi. Toutefois, elles partagent encore les m\^emes solutions, donc il y a de fortes chances qu'elles soient finalement compatibles dans ce r\'egime. 

\subsubsection{Remarque sur la solution tachyonique générale}

A partir de la formule~\refe{eq:beta_onshell_sous} nous obtenons une combinaison pour laquelle les fonctions b\^eta s'annulent telle que la d\'eformation correspondante d\'efinirait potentiellement une CFT~:

\begin{align}\label{eq:tach_gen}
T(x^0) = \zeta e^{\omega x^0} - i \frac{\lambda}{\zeta^*} e^{-\omega x^0}
\end{align}

avec $\lambda \in \mathbb R$ et $\omega^2=1-r^2$. L'interprétation physique de cette solution est incertaine car elle décrit un tachyon, de type S-brane \emph{complète}, décondensant depuis $x^0 \to -\infty$ puis recondensant en $x^0\to\infty$ mais dont la phase (complexe) subit un décalage de $\pi/2$. Cependant, cette expression n'est solution qu'à l'ordre quadratique et il faudrait vérifier que les fonctions b\^eta restent nulles aux ordres suivants -- en utilisant par exemple la méthode de~\cite{Gaberdiel:2008fn} à l'ordre 4 -- mais aussi refaire l'étude des divergences logarithmiques de la section~\refcc{sec:tach_roul_CFT_bos} en tenant compte des termes croisés et des nouveaux termes de contact. Ces calculs sont justifiables car, bien que physiquement peu pertinente, une telle solution peut contraindre l'expression de l'action effective, comme dans la méthode proposée par Kutasov et Niarchos dans~\cite{Kutasov:2003er}.

\subsection{Phase critique $r=1$}
\label{sec:phas_crit}

En comparant les fonctions b\^eta~\refe{eq:beta_general_massif},~\refe{eq:beta_on_crit} et~\refe{eq:beta_onshell_sous} et en particulier celles de $\delta r$ nous distinguons clairement une discontinuité en $r=1$. En effet~ :

\begin{align}
r>1~ :\quad & \beta_{\delta r} = \Delta \delta r - 2 r \parent{\module{\zeta_{(1)}}^2 + \module{\zeta_{(2)}}^2} + \ldots \nonumber \\
r=1~ :\quad & \beta_{\delta r} =  - \square \delta r  - 2 \module{\lambda}^2 \nonumber \\
r<1~ :\quad & \beta_{\delta r} = \Delta \delta r - 2 r \, \parent{\zeta_{(1)}\zeta_{(2)}^*+\zeta_{(1)}^*\zeta_{(2)} } + \ldots  \nonumber \\
\end{align}

avec en $r=1$ par continuité dans la d\'efinition des d\'eformations de bord $\lambda^\pm = \zeta_{(1)} + \zeta_{(2)}$. Il apparaît donc nettement qu'en la distance critique les contributions de $r>1$ s'ajoutent avec celles de $r<1$. Puisque $\module{\zeta_{(1)}}^2 + \module{\zeta_{(2)}}^2 \geq \module{\zeta_{(1)}+\zeta_{(2)}}^2$ n\'ecessairement $\zeta_{(1)}\zeta_{(2)}^*+\zeta_{(1)}^*\zeta_{(2)} \leq 0$ de sorte qu'en $r=1$ la contribution est interm\'ediaire. Il s’agit pour ces \'equations de flot d’une façon de rendre les limites $r\to r_c^+$ et $r \to r_c^-$ compatibles. Ces \'equations sont toutefois discontinues \`a la distance critique, et le terme source change de signe~: le syst\`eme y subit donc une transition de phase. La continuit\'e n'est r\'etablie que si $\zeta_{(i)}=0$ pour $i=1,2$ qui n'est \'evidemment pas une limite int\'eressante.

Cette discontinuit\'e en $r=r_c$ exprime \'egalement le fait qu'une th\'eorie des champs en $r=r_c$ n'est pas bien définie, parce que d'un côté nous avons une phase stable et de l'autre une phase instable. En outre, \`a propos de la transition sous-critique/critique, d'après notre étude sur les divergences du modèle roulant en fonction de la distance, le nombre de contretermes de type terme-de-contact à ajouter à l'action de surface tend à exploser en $r_c^-$. Ainsi, dans cette limite, la théorie est non-renormalisable, mais pas en $r=r_c$ comme nous le voyons bien. Il s'agirait d'un moyen pour le système de mener à la discontinuité, ce qui est réminiscent de la limite $c\to 1$ dans les théories de Liouville~\cite{Runkel:2001ng,Fredenhagen:2004cj,Schomerus:2003vv}. Nous savons par exemple que la continuation analytique $b\to ib$ (\ie $Q\to 0$ ou $c\to 1$) pour passer de Liouville de genre espace à Liouville de genre temps n'est pas très bien définie. Or notre modèle est très similaire dans la forme à une théorie de Liouville, mis à part les facteurs de CP. En ce sens, la transition $b\to ib$ semble similaire à la transition $\omega \to 0$ (soit $c=2 \to c=1$) dans l'opérateur du tachyon sur le bord $e^{i\sqrt{1/2-\omega^2}\wt X + \omega X^0}$. Nous n'avons pas exploré plus en détail cette relation, puisque ça n'a pas été un point crucial pour notre étude.

\section{Conclusion}
\label{sec:conclu_bos}

Dans ce chapitre, il a été démontré que le système brane-brane séparé à distance constante admettait une solution de condensation temporelle de type demi S-brane $\lambda^\pm e^{\sqrt{1-r^2} x^0}$ dans le secteur interbranaire, paramétrisée par les facteurs de Chan-Paton $\sigma^{\pm} \in U(2)$. Cette solution était représentée sur la surface de corde par une déformation de bord exactement marginale. Cette propriété était en revanche perdue en certaines valeurs de distance $r = \sqrt{1-1/4n^2}$ pour tout $n \in \mathbb N^*$. En ces valeurs, les fonctions b\^eta du tachyon du secteur $\sigma^0$ recoivent des contributions non nulles proportionnelles à $(\lambda^+\lambda^-)^n$ et cela est à des résonances entre les opérateurs $ \sigma^\pm \otimes e^{\sqrt{1-r^2} X^0 \pm i r \wt X}$ et $e^{X^0}$ \cad des divergences logarithmiques. Ces valeurs de distance brisant la marginalité sont denses autour de la valeur critique $r_c=1$ telles que la limite $r \to r_c^-$ n’est pas renormalisable. Cela est en accord avec la non-marginalité du tachyon à la distance critique, qui se couple au champ de distance et attire le système en $r=r_c^-$. 

Ainsi, dans un schéma réaliste d'évolution du système initialement posé à la distance critique, le couplage à l'autre tachyon est in\'evitable puisque les valeurs $r = \sqrt{1-1/4n^2}$ sont denses autour de $r=1$. La condensation du tachyon $\sigma^0$ \cad hébergé sur chaque brane est bien comprise -- \cf chapitre~\refcc{chap:cond_tach} -- et nous en déduisons qu'au moins chaque brane s'évapore indépendamment l'une de l'autre sous forme de cordes fermées, à la fois par confinement du champ électrique sur chaque brane quand $x^0 \to \infty$ et par couplage des cordes fermées à la brane~\cite{Lambert:2003zr}. Ainsi, il est physiquement peu probable que le système puisse persister à distance constante, mais la condensation du tachyon sur chaque brane permet d'avoir une vision assez claire de son évolution même si la vélocité relative des branes est non-nulle, menant à sa désintégration complète en cordes fermées. \\ 

En outre, le calcul de l'action effective a été initiée en étudiant le groupe de renormalisation du modèle sigma légèrement perturbée autour du tachyon roulant. Il s'agissait d'une approche pertinente, tant que le tachyon roulant était choisi marginal à l'ordre dominant dans les fonctions b\^eta. Des équations du mouvement ont été obtenues \`a partir des expressions des fonctions b\^eta invariantes\footnote{Par changement de sch\'ema de renormalisation.}, mais n'\'etaient pas compatibles avec une expression d'action effective, ni en particulier celle de type Garousi~\refe{eq:quad}. Cependant, dans le r\'egime massif et pour un tachyon de genre espace dans la phase tachyonique, au moins pour une distance constante, leurs solutions sont compatibles avec celles d\'eriv\'ees d'une action de type Garousi. Nous avons aussi mis en valeur l'existence d'une transition de phase du syst\`eme en la distance critique lorsque le tachyon est allum\'e sur le bord. \\

Cette étude sera prolongée dans le cadre du système $D-\bar D$ pour lequel d'autres contraintes sont posées. Cette autre approche montrera que les fonctions b\^eta mènent aux mêmes contraintes sur les champs que des équations du mouvement dérivées d'une action effective quadratique obtenue par une méthode indépendante également autour de la solution de tachyon roulant.

\chapter{Condensation de tachyon en supercorde et système brane-antibrane}
\label{chap:tach_cond_susy}

Dans le système brane-antibrane la situation de la coïncidence est bien assimilée et en particulier en ce qui concerne le tachyon découplé des autres champs. Cependant, nous savons que dans ce système le tachyon est couplé de façon non abélienne à ces derniers. Or comme dit dans l'introduction, le domaine de validité des propositions d'actions effectives faites par Garousi dans~\cite{Garousi:2000tr,Garousi:2007fn} concerne \apriori uniquement des tachyons de genre espace. Il faut donc les considérer avec précautions si on souhaite en étendre le domaine de validité aux tachyons condensants d\'ependants du temps. En suivant l'expression de l'action de Garousi, nous avons \emph{grosso-modo} que le tachyon est effectivement couplé minimalement à un certain nombre de champ de jauge de cordes ouvertes et que réciproquement, ces champs de jauge sont couplés au tachyon. 

En décrivant une paire $D-\bar D$ disjointe spatialement par une distance finie (disons $\ell$) pas forcément constante, nous avons que le champ de distance (disons $\phi$) est tel que sa valeur dans le "vide" est $\corr{\phi}=\ell$. Ce champ est non-massif et par T-dualité correspond à un champ de jauge minimalement couplé au tachyon. Ainsi le champ de distance (non-abélien) serait couplé au tachyon (non-abélien) par un terme du type $[\phi,T]^2$. Naturellement -- et c'est ce qu'obtient Garousi dans~\cite{Garousi:2007fn} -- la distance et le tachyon sont interdépendants de telle sorte qu'il n'existe aucune solution \emph{classique} aux équations du mouvement, condensante et à distance constante. 

Par étude des équations du mouvement, nous l'avons constaté dans le chapitre~\refcc{chap:mot}. En résolvant numériquement ces équations, nous avons observ\'e un comportement privil\'egiant une attraction vers $\ell=0$ mais qui n'y stabilise pas et oscille. Nous avons aussi vu à partir de l'expression des équations du mouvement qu'il ne pouvait pas exister de solution non triviale à distance constante. \\

Or nous allons montrer dans la section~\refcc{sec:tach_roul_CFT} qu'il existe une CFT admettant sur le bord une déformation tachyonique dépendante du temps de type demi S-brane à distance constante. Le système brane-antibrane séparé et de la distance critique est d\'efini dans l'introduction sections~\refcc{sec:BCFT} et \refcc{sec:cond_susy}. Nous utiliserons maintenant $r=\ell/2\pi$ de sorte que la distance critique dans cette variable est\footnote{Nous utilisons la convention $\alpha'=1$.} $r_c=1/\sqrt 2$. 

Bagchi et Sen~\cite{Bagchi:2008et} avaient montré que dans la partie du domaine tachyonique $r<r_c/\sqrt 2$ la déformation en question était exactement marginale. Par ailleurs, nous avons montré~\cite{Israel:2011ut} par notre étude que cette déformation était aussi une CFT dans la partie manquante $r>r_c/\sqrt 2$. Pour être plus précis nous l'avons montré pour tout $r<\sqrt{17}/6$. Cependant,  de forts indices -- supersymétrie en l'occurrence -- nous font conjecturer qu'elle doit l'être sur tout le domaine tachyonique. Ainsi les calculs -- en particulier la fonction de partition -- que l'on peut effectuer en $r<r_c/\sqrt 2$ peuvent être continués analytiquement à tout $r<r_c$. \\

Dans la section~\refcc{sec:renorm_susy} nous avons étudié le groupe de renormalisation en dehors de la CFT. Nous avons calculé les fonctions bêta associées aux divers couplages, puis nous en avons déterminé des équations du mouvement pour les champs de tachyon et de distance. Dans la section~\refcc{sec:fonc_part} nous exprimons la fonction de partition sur le disque le long de la BCFT du tachyon roulant à distance constante. Nous avons calculé la fonction de partition à l'ordre 8 dans les tachyons à la distance $r_c/\sqrt 2$ en mettant en valeur les contributions du terme de contact dans l'expression en super-espace. Nous introduisons également une méthode diagrammatique qui s'est révélée pratique d'utilisation. Enfin dans la section~\refcc{sec:act_eff_kut} nous discutons du calcul d'une action effective en utilisant la méthode de Kutasov et Niarchos et la fonction de partition à l'ordre 2 dans le tachyon. Nous obtenons une action effective quadratique dont l'expression est totalement contrainte et qui est compatible -- dans une certaine mesure -- avec les équations du mouvement que nous avions obtenu auparavant.

\section{CFT du tachyon roulant dans système brane-antibrane séparé}
\label{sec:tach_roul_CFT}

Le système brane-antibrane admet un tachyon bi-fondamental dans le spectre des cordes ouvertes du secteur interbranaire. De la même manière que dans le cas bosonique, il existe trois phases distinctes. Celle pour laquelle le bi-fondamental est massif ($r>r_c$), puis celle où il est exactement non-massif ($r=r_c$) et enfin celle où il est tout à fait tachyonique ($r<r_c$). 

\subsection{Déformation de tachyon roulant dans domaine $r<r_c$}

Comme cela est expliqu\'e dans la section~\refcc{chap:cond_bos} pour le mod\`ele bosonique \'equivalent, la théorie des champs effective pour les champs tachyoniques et non-massifs est d\'ecrite par l'action effective \`a l'ordre des arbres\footnote{Cela tient son origine dans l'instabilit\'e tachyonique de la th\'eorie \`a l'ordre des arbres qui implique la divergence de la contribution \`a l'ordre d'une boucle et plus~: les calculs quantiques perturbatives perdent tout leur sens.}. La th\'eorie des cordes doit s'exprimer dans le fond compos\'e des configurations \emph{classiques} des champs tachyoniques et non-massifs.  En ce qui concerne le fond tachyonique, cela revient \`a ajouter sur le bord, dans l'action de la surface, deux tachyons complexes conjugués l'un de l'autre dans les secteurs représentés par les facteur de CP $\sigma^+$ et $\sigma^-$ à une distance $r$. Nous nous plaçons dans le demi-plan supérieur $z \in H_+$. Les tachyons sur le bord sont représentés par les opérateurs de vertex~:

\begin{align}\label{eq:def_gen_tach}
& T= \sigma^+\otimes \lambda^+ \oint e^{i r \widetilde{X}} f(X^a,\psi^a,\widetilde \psi) \nonumber \\ 
& T^\dagger = \sigma^-\otimes \lambda^- \oint e^{-ir\widetilde{ X}}f^*(X^a,\psi^a,\widetilde \psi)
\end{align}

où nous avons regroupé tous les termes dépendant des partenaires fermioniques des bosons $X^a$ et $\wt X$ dans $f$ et $f^*$.  Nous allons préciser leur expression bientôt. Nous avons choisi une notation un peu trompeuse~: le fermion $\wt \psi$ défini sur le bord est ici le \emph{dual}\footnote{De m\^eme que $\wt X$ est le dual, conjugu\'e au moment d'enroulement, de $X$.} de $\psi$ et non le champ antiholomorphe. Cela reste suffisamment clair puisqu'il n'existe aucune distinction entre holomorphe et antiholomorphe sur le bord. Pour les fermions définis sur le bord ici, nous utilisons la convention~:

\begin{align}
\mathbb X(z) = X(z) + i \theta \psi(z) 
\end{align}

Par rapport à la définition donnée dans l'introduction~\refe{eq:convention_psi}, nous identifions donc ici $\Psi$ à $\psi$. Dans~\refe{eq:def_gen_tach}, l'exponentielle $e^{\pm ir\wt X}$ est un opérateur de twist qui permet d'inclure les conditions aux bords de façon condensée et qui donne les bons poids conformes et donc les bonnes OPE. Pour être tout à fait rigoureux, il faudrait initialement partir de l'action~\cite{Takayanagi:2000rz}~:

\begin{multline}
S = S_{bulk} + i \frac{\sigma^3}{2} \otimes \oint \phi(X^a) \, \partial \widetilde X + i \frac{\sigma^3}{2} \otimes \oint \partial_a \phi(X^a)\,\psi^a  \widetilde \psi \\ +\sigma^+ \otimes \lambda^+ \oint f(X^a,\psi^a) + \sigma^- \otimes \lambda^- \oint f^*(X^a,\psi^a)
\end{multline}

et choisir $\phi(X^a) = r + \delta r(X^a)$ avec $r$ constant. Ensuite, le terme en $\sigma^3 \otimes r \partial \wt X$ modifie les OPE des tachyons, à cause de $[\sigma^\pm,\sigma^3]\neq 0$ mais peut être réabsorbé dans la définition des champs de tachyon sous la forme de l'opérateur de twist que l'on vient de nommer. Ce point sera plus clair lorsque nous introduirons les fermions de bord. Ainsi nous avons~:

\begin{multline}\label{eq:act_duale}
S = S_{bulk} + i \frac{\sigma^3}{2} \otimes \oint \delta r(X^a) \partial \widetilde X + i \frac{\sigma^3}{2} \otimes \oint \partial_a\delta r(X^a) \psi^a \widetilde \psi \\ + \sigma^+ \otimes \lambda^+ \oint e^{i r\wt X} f(X^a,\psi^a,\widetilde \psi) + \sigma^- \otimes \lambda^- \oint e^{-i r \wt X} f^*(X^a,\psi^a,\widetilde \psi)
\end{multline}

Nous allons maintenant montrer qu'il existe une solution de condensation de tachyon roulant à distance constante, \cad~:

\begin{align}
f(X^a,\psi^a,\wt \psi) = \parent{ir \widetilde \psi + \omega \psi^0}e^{\omega X^0} \quad \text{et} \quad \delta r(X^a) = 0
\end{align}

avec $\omega = \sqrt{1/2-r^2}$. Dans la suite, $\psi^\pm = \pm ir \widetilde \psi + \omega \psi^0$. Pour montrer que cette solution existe bien, il faut s'intéresser à une action encore plus générale. En effet, une théorie des cordes est une solution des équations du mouvement si et seulement si elle est une d\'ecrite par une th\'eorie conforme sur la surface de corde. Pour cela, il faut que dans le mod\`ele sigma non lin\'eaire -- voir section~\refcc{sec:mod_sig} -- les fonds classiques constituent des configurations repr\'esent\'ees par des couplages \emph{exactement marginaux}, \cad dont les fonctions bêta du groupe de renormalisation sont nulles \`a tout ordre en perturbation. Or nous avons vu dans le chapitre~\refcc{chap:cond_bos} que les termes importants dans les fonctions bêta sont en particulier ceux qui sont universels, \cad ceux qui proviennent de résonances et qui ne peuvent pas être réabsorbés par redéfinition des champs. Il se trouve que l'action~\refe{eq:act_duale} n'est pas complète de ce point de vue, car les tachyons peuvent éventuellement entrer en résonance avec des termes du type $e^{2 n \omega X^0}$ comme nous l'avons aussi vu en théorie bosonique. Cependant, à l'inverse de ce dernier cas ces résonances sont multipliées par des coefficients dont la somme est nulle, gr\^ace aux fermions, donc gr\^ace à la supersymétrie, ce que nous montrerons maintenant. 

\subsection{Résonances et contretermes}

Nous devons dans un premier temps introduire des termes de bord du type~:

\begin{align}
\sigma^0 \otimes \sum_{n>1} \mu^n \oint e^{2 n \omega X^0 }
\end{align}

mais qui ne sont pas explicitement supersymétriques et par conséquent brisent la symétrie superconforme sur le bord. Dans le meilleur des cas, les coefficients $\mu^n$ dépendent des autres couplages et des cut-offs uniquement de telle sorte qu'ils suppriment les divergences néfastes dans les amplitudes. Même s'ils brisent explicitement la supersymétrie de surface en réalité ils l'a rétablissent \emph{in fine} dans les calculs. Rappelons que l'ajout de cut-offs UV ou IR brise explicitement la symétrie de Poincaré (et donc aussi la supersymétrie) sur la surface de corde mais qu'à la fin elle doit être recouvrée par soustraction des divergences.  

Dans le pire des cas, ils seraient indépendants des autres couplages et auraient des fonctions bêta non triviales. Alors la théorie complète serait non-marginale dans ses couplages et non-supersymétrique et par conséquent non-superconforme. Elle ne serait donc pas une solution des équations du mouvement de la SFT. \\

Ces termes ne sont associés à des résonances que si $2n\omega =1$. En effet, les tachyons étant marginaux, seule la production d'un opérateur marginal par fusion peut être associée à une résonance, puisqu'il faut $h_a + h_b -2 = h_c-1$.  En outre, ces termes entre eux n'ont clairement pas de résonance, car dimensionnellement $4(n^2+m^2-(n+m)^2)\omega^2 = 1$ devrait être résolu par $nm=-1/8\omega^2$. Or $n$ et $m$ étant positifs, cette équation est impossible à résoudre. Et donc il ne faut ultimement s'intéresser qu'au terme~:

\begin{align}
\sigma^0 \otimes \mu_0 \oint e^{X^0}
\end{align}

quand $\omega=1/2n$. Les autres entrent dans le cadre du \emph{meilleur cas} présenté plus haut et ne sont donc pas problématiques. Ce dernier peut en revanche \^etre potentiellement un \emph{pire cas} et montrer qu'il ne l'est pas n'est pas aisé ; nous y dévouerons la section suivante. \\

Puisque $n\geq 1$, alors pour tout $\omega>1/2$ la théorie est exactement marginale dans ses couplages et supersymétrique, donc superconforme. C'est la conclusion exacte de Sen dans~\cite{Sen:1998sm}. Sachant que $\omega=\sqrt{1/2-r^2}$ ce domaine correspond à tout $r<1/2$. Il nous parait étrange que la solution de tachyon roulant ne soit valide qu'à partir de cette valeur de distance, car ça n'a physiquement pas de sens contrairement au cas du modèle bosonique. C'est ce qui fait que l'on s'attend par continuité à ce que la supersymétrie rétablisse la symétrie conforme aussi pour $r_c>r \geq 1/2$. 

Comme nous l'avons suggéré plus haut, nous verrons en effet que les termes purement logarithmiques dans les cut-offs associés à cet opérateur de vertex se suppriment ensemble. Nous ne le montrerons pas pour tout $r<r_c$ car trop compliqué -- il nous est impossible de pousser le calcul jusqu'à des ordres infiniment grands -- mais pour tout $r<\sqrt{17}/6$, \cad jusqu'au sixième ordre dans les OPE des tachyons, \cad $n=3$. Nous extrairons de ces calculs les divergences UV parmi lesquels les logarithmes dont le coefficient doit s'annuler. Nous mettrons en valeur que la supersymétrie est à la source de ce mécanisme de suppression et nous argumenterons que l'on peut donc étendre ce résultat à tout $r<r_c$.

\subsection{Termes logarithmiques à l'ordre 2 et 4 dans les tachyons}

L'action de d\'epart de cette \'etude est~: 

\begin{align}\label{eq:act_base}
S= S_{bulk} + \sigma^+ \otimes \lambda^+ \oint \psi^+ e^{i r \wt X + \omega X^0} + \sigma^- \otimes \lambda^- \oint \psi^- e^{-i r \wt X + \omega X^0}  
\end{align}

Elle est définie sur le demi-plan complexe $H_+$. Ces déformations tachyoniques apparaissent à l'intérieur de l'intégrale de chemin avec la convention $Z \propto \int e^{-S}$ et sont d\'evelopp\'es suivant une s\'erie de Taylor, de la même manière que pour des interactions perturbatives en théorie des champs. Les champs fondamentaux v\'erifients les OPE suivantes~:

\begin{align}\label{eq:OPE_fond_susy}
&\wt X(z) \wt X(w) = - 2 \ln (z-w) + \ldots \nonumber \\ 
& X^0(z) X^0(w) = 2 \ln (z-w) + \ldots \nonumber \\ 
& \widetilde \psi (z) \widetilde \psi(w) = \frac{2}{z-w} + \ldots \nonumber \\
& \psi^0 (z) \psi^0 (w) = - \, \frac{2}{z-w} + \ldots
\end{align}

Nous les utiliserons pour calculer les OPE des op\'erateurs de vertex des tachyons~: perturbativement la th\'eorie fondamentale est libre et constitue une CFT. Donc toutes les amplitudes peuvent \^etre calcul\'ees \`a partir des formules des OPE fondamentales donn\'ees ci-dessus.

\subsubsection{Développement au second ordre}

Le développement de $e^{-S-\sum_i \mu^i \oint \phi_i}$ est donné conventionnellement jusqu'au deuxième ordre par la formule suivante, en utilisant des régularisations UV et IR telles qu'implici\-tement $\varepsilon\to 0$ et $L\to \infty$ et en présence de facteurs de CP~:

\begin{multline} 
{\mathcal P} e^{-S-\sum_i \sigma^i\otimes\mu^i \oint \phi_i} \\ = e^{-S}\parent{ 1 - \sum_i \sigma^i \otimes \mu^i \oint \phi_i + \sum_{i,j} \sigma^i\sigma^j \otimes \mu^i \mu^j \int \di w \int_{w+\varepsilon}^{w+L} \di z ~ \phi_i(z) \phi_j(w) + \ldots}
\end{multline}

L'opérateur d'ordre $\mathcal P$ agit sur les opérateurs en les ordonnant dans l'ordre croissant. C'est en particulier important lorsque les opérateurs sont fermioniques, tels qu'ici les tachyons~: l'ordre ne tient pas compte de leur anti-commutativité. Techniquement, il faut aussi sommer sur les facteurs de CP en ajoutant une trace dans les calculs d'amplitude mais il faut s'abstraire de cette opération ici car les divergences doivent être supprimées de telle sorte que tout calcul d'amplitude y compris avec des insertions sur le bord doit être régulier. Ainsi, au second ordre dans les tachyons, nous devons calculer~:

\begin{multline}
\sigma^+ \sigma^- \otimes \lambda^+ \lambda^- \int \di w \int_{w+\varepsilon}^{w+L} \di z ~  \bnormal{\psi^+ e^{ir\wt X+\omega X^0}(z)}\bnormal{\psi^- e^{-ir \wt X+\omega X^0}(w)} \\
+ \sigma^- \sigma^+ \otimes \lambda^+ \lambda^- \int \di w \int_{w+\varepsilon}^{w+L} \di z ~ \bnormal{\psi^- e^{-ir\wt X+\omega X^0}(z)}\bnormal{\psi^+ e^{ir \wt X+\omega X^0}(w)} 
\end{multline}

Le calcul est imm\'ediat en utilisant~\refe{eq:OPE_fond_susy} et donne~:

\begin{align}\label{eq:dev_OPE}
&\sigma^+\sigma^- \otimes \lambda^+ \lambda^- \int \di w \int_{w+\varepsilon}^{w+L} \di z ~ (z-w)^{2\omega^2 - 2 r^2 - 1} \nonumber \\ 
& \qquad\qquad \times \bnormal{e^{ir\wt X + \omega X^0}(z)e^{-ir\wt X + \omega X^0}(w)\parent{4r^2-1 + (z-w)\psi^+(z) \psi^-(w)}} \nonumber \\ 
&+ \sigma^-\sigma^+ \otimes \lambda^+ \lambda^- \int \di w \int_{w+\varepsilon}^{w+L} \di z ~ (z-w)^{2\omega^2 - 2 r^2 - 1} \nonumber\\ 
& \qquad\qquad \times \bnormal{e^{-ir\wt X + \omega X^0}(z)e^{ir\wt X + \omega X^0}(w)\parent{4r^2-1 + (z-w)\psi^-(z) \psi^+(w)}}
\end{align}

Or puisque $\omega^2 + r^2 =1/2$, alors $2\omega^2-2r^2 -1 = -4r^2$ et avec $r^2<1/2$. Par conséquent, par d\'eveloppement des opérateurs autour de $w$ seul l'ordre zéro sera éventuellement divergent UV. En effet, ce dernier est divergent tant que $r^2\geq 1/4$ tandis que l'ordre premier est divergent pour $r^2 \geq 1/2$. Nous n'étudierons pas le système en $r = r_c$ dans un premier temps car le tachyon y est non-massif, par conséquent nous ne devons garder dans~\refe{eq:dev_OPE} que l'ordre zéro. De la sorte, il vient~: 

\begin{align}
\sigma^0 \otimes \lambda^+ \lambda^- \parent{L^{1-4r^2}-\varepsilon^{1-4r^2}} ~ \int \di w~ \bnormal{e^{2\omega X^0}(w)} 
\end{align}

Le cas $r^2=1/4$ est particulier. Techniquement nous devrions avoir que l'intégrale donne un logarithme puisque l'intégrande est proportionnel à $(z-w)^{-1}$. Cependant, en cette valeur de distance, le coefficient multiplicatif qui provient directement des fermions s'annule car il est proportionnel à $4r^2-1$~:

\begin{align}
\sigma^0 \otimes \lambda^+ \lambda^- (4r^2-1) \Bigg\vert_{r^2=1/4} \ln \frac{L}{\varepsilon} ~ \int \di w~ \bnormal{e^{X^0}(w)}   = 0 
\end{align}

Pour $r> 1/2$ il faut soustraire la divergence UV tandis que pour $r<1/2$ la divergence est uniquement IR et nous pouvons la n\'egliger. Pour $r> 1/2$ la divergence est de type puissance, donc dans un schéma de soustraction minimale il faut ajouter un contreterme \`a l'action. Ce dernier n'affecte pas la marginalité de la théorie, car il n'est pas logarithmique~:

\begin{align}\label{eq:contreterme_puissance}
S_{ct} = - \sigma^0 \otimes \lambda^+ \lambda^- \varepsilon^{1-4r^2} ~ \int \di w~ \bnormal{e^{2\omega X^0}(w)} 
\end{align}

D'apr\`es la discussion de la section~\refcc{sec:mod_sig} il ne modifie clairement pas la fonction bêta d'un couplage $\mu_1$ associé à l'opérateur de vertex $e^{2\omega X^0}$~:

\begin{align}
\beta_1 = (4r^2-1) \mu_1 
\end{align}

La condition de marginalit\'e impose $\mu_1=0$ et ne subsiste donc que le contreterme dépendant de $\lambda^\pm$ et $\varepsilon$ qui ne s'oppose en rien à la marginalité de la théorie. Nous verrons dans la section~\refcc{sec:superespace} que le contreterme UV apparaît naturellement sous la forme d'un terme de contact, lorsque nous partons d'une action manifestement supersymétrique, \cad exprimée initialement dans le super-espace.  

\subsubsection{Développement au quatrième ordre}

A partir de l'action~\refe{eq:act_base}, il n'y a pas de terme résonant à l'ordre 3, ce qui est assez aisé à voir à cause du terme $e^{\pm ir\wt X + 3\omega X^0}$. Le développement à l'ordre 4 fait intervenir une OPE à 4 points du type $T^+ T^- T^+ T^-$ à cause des facteurs de Chan-Paton dont l'int\'egrale est divergente. L'extraction de toutes les divergences est une t\^ache difficile \`a cause de l'ordre d'int\'egration. Nous obtienons \apriori un ensemble de divergences de type puissance. N\'eanmoins, pour $\omega=1/4$ ou $r=\sqrt 7 /4 > 1/2$ la fonction \`a 4-points est r\'esonante et, par cons\'equent, devrait donner des divergences logarithmiques. Nous verrons dans cette section que ces divergences se suppriment ensemble, et ce, gr\^ace aux combinaisons fermioniques. Cela constitue le point principal de notre article~\cite{Israel:2011ut}.

Nous devons aussi tenir compte des corrections apport\'ees \`a l'action qui ont permis ci-dessus de r\'egulariser l'ordre quadratique. Pour $r>1/2$ il faut donc aussi consid\'erer les contributions du contreterme \refe{eq:contreterme_puissance} quadratique ajout\'e précé\-dem\-ment sous la forme $S+S_{ct}$. En effet, il n'est pas exclu qu'il produise aussi des divergences par OPE, au deuxième ordre avec lui-même, ou au troisième ordre avec les deux tachyons. Il y a cependant peu de chance qu'elles puissent être de type logarithmiques à cause du facteur en puissance des cut-offs qui le précède. Et en effet, aucune divergence logarithmique n'est produite. Au mieux, il permet de supprimer des divergences de type puissance, sous-dominantes à l'ordre 4, et dans le meilleur des cas toutes les divergences. 

Nous trouvons que le contreterme permet effectivement de supprimer une grande partie des divergences à l'ordre 4. Toutefois, nous avons obtenu qu'il laisse une divergence résiduelle associ\'ee au terme~:

\begin{align}\label{eq:div_resid}
\sigma^0 \otimes (\lambda^+ \lambda^-)^2 f(\omega) \varepsilon^{16\omega^2-1} \oint e^{4 \omega X^0}
\end{align}

avec $f(\omega)$ une fonction non-divergente dont nous n'avons pas réussi à obtenir une formulation compacte. \\

Tout d'abord, occupons-nous des OPE \`a quatre points dans les tachyons, $T^+ T^- T^+ T^-$ et $T^- T^+ T^- T^+$. Le terme correspondant est~:

\begin{multline}
\sigma^+\sigma^-\sigma^+\sigma^- \otimes (\lambda^+\lambda^-)^2 \oint \di y \int_{w+\varepsilon}^{w+L} \di z \int_{x+\varepsilon}^{x+L} \di w \int_{y+\varepsilon}^{y+L} \di x \\ ~ \bnormal{\psi^+ e^{ir\wt X+\omega X^0}(z)}\bnormal{\psi^- e^{-ir \wt X+\omega X^0}(w)}\bnormal{\psi^+ e^{ir\wt X+\omega X^0}(x)}\bnormal{\psi^- e^{-ir \wt X+\omega X^0}(y)} \\ 
+ \sigma^-\sigma^+\sigma^-\sigma^+ \otimes (\lambda^+\lambda^-)^2 \oint \di y \int_{w+\varepsilon}^{w+L} \di z \int_{x+\varepsilon}^{x+L} \di w \int_{y+\varepsilon}^{y+L} \di x \\ ~ \bnormal{\psi^- e^{-ir\wt X+\omega X^0}(z)}\bnormal{\psi^+ e^{ir \wt X+\omega X^0}(w)}\bnormal{\psi^- e^{-ir\wt X+\omega X^0}(x)}\bnormal{\psi^+ e^{ir \wt X+\omega X^0}(y)}
\end{multline}

Seules les divergences associées à l'opérateur $e^{4 \omega X^0}$ seront ici int\'eressante. La raison est à la fois que cet opérateur est le seul qui peut devenir résonant avec les tachyons pour $\omega=1/4$ et que dans la fonction de partition il est le seul à survivre dans le vide sous la forme $\int \di x^0~e^{4 \omega x^0}$. Les OPE ne sont pas difficiles à calculer, elles sont~:

\begin{multline}
\bnormal{\psi^+ e^{ir\wt X+\omega X^0}(z)}\bnormal{\psi^- e^{-ir \wt X+\omega X^0}(w)}\bnormal{\psi^+ e^{ir\wt X+\omega X^0}(x)}\bnormal{\psi^- e^{-ir \wt X+\omega X^0}(y)} \\ 
= (z-w)^{4\omega^2-1}(z-x)(z-y)^{4\omega^2-1}(w-x)^{4\omega^2-1}(w-y)(x-y)^{4\omega^2-1} \\ \times \parent{\frac{(4\omega^2-1)^2}{(z-w)(x-y)} - \frac{1}{(z-x)(w-y)}+\frac{(4\omega^2-1)^2}{(z-y)(w-x)}} \times  e^{4 \omega X^0}
\end{multline}

La combinaison qui apparaît entre parenthèses est très importante, il s'agit de la contribution des fermions que l'on a calculé en appliquant le théorème de Wick. Le fait particulier qu'il s'agisse de fermions est crucial car alors le terme central est bien soustrait et non additionné. Nous obtenons ainsi la bonne combinaison de coefficients s'annulant en $\omega =1/4$. \\

La méthode de calcul de l'intégrale est expliquée dans l'annexe de notre article~\cite{Israel:2011ut}. Nous noterons $a=4\omega^2$. Après de longs calculs, nous obtenons~:

\begin{multline}\label{ord4c}
TTTT = \\  \sigma^0 \otimes \int \di x_1 \int_{x_1-L}^{x_1-\varepsilon} \di x_2 \int_{x_2-L}^{x_2-\varepsilon} \di x_3  \int_{x_3-L}^{x_3-\varepsilon} \di x_4 \,
\bnormal{\psi^+ T^+ (x_1)}\bnormal{\psi^- T^- (x_2)}\bnormal{\psi^+ T^+ (x_3)}\bnormal{\psi^- T^- (x_4)}  \\
\sim \sigma^0 \otimes
\Bigg\{ 
	\frac{1}{2a+1} \parent{\frac{L}{\varepsilon}}^{2-2a} + \frac{a-1}{a} \parent{\frac{L}{\varepsilon}}^{1-2a} -  \parent{\frac{L}{\varepsilon}}^{1-a}  W(a)
     \displaybreak[3]\\ 
    +  \parent{\frac{L}{\varepsilon}}^{1-4a} U(a) + \frac{\parent{\frac{L}{\varepsilon}}^{1-4a}-1}{1-4a} \, V(a)
\Bigg\}  
~ L^{4a-1}  \int \di x_1 \,\bnormal{e^{4\omega ~ X_0}}(x_1)
\end{multline}

avec $U(a)$, $V(a)$ et $W(a)$ des coefficients numériques non singuliers en $a=1/4$. Il est important que $U(a)$ soit non singulier car la divergence associée est alors non logarithmique en $a=1/4$. Le coefficient $W(a)$ est explicitement~:

\begin{multline}
W(a) = \frac{2(a-1)}{3a} \Bigg(
   			\frac{\,_2\text{F}_1\left(-a,a+1,a+2,-1\right)}{a+1} + \frac{\,_2\text{F}_1\left(-a,a-1,a,-1\right)}{a-1} \\ + \frac{\,_2\text{F}_1\left(2-a,a+1,a+2,-1\right)}{a+1} 
    \Bigg)
\end{multline}

Le coefficient $V(a)$ est donné par la formule exacte~:

\begin{multline}
V(a) = (a-1)^2  \parent{\frac{\,_2\text{F}_1\left(1-2a,a-1,a,-1\right)}{a-1} + \frac{\,_2\text{F}_1\left(1-2a,2-3a,3-3a,-1\right)}{2-3a}} \\ \times 
		 	\Bigg(  
		 			\frac{\,_2\text{F}_1\left(-a,a-1,a,-1\right)}{a-1} + \frac{\,_2\text{F}_1\left(-a,1-2a,2-2a,-1\right)}{1-2a} \\ 
+ \frac{\,_2\text{F}_1\left(2-a,a+1,a+2,-1\right)}{a+1}+ \frac{\,_2\text{F}_1\left(2-a,1-2a,2-2a,-1\right)}{1-2a} 
		 	\Bigg) \\ 
			+ \parent{2(a-1)^2-1}  \parent{\frac{\,_2\text{F}_1\left(1-a,a,1+a,-1\right)}{a}
+\frac{\,_2\text{F}_1\left(1-a,1-2a,2-2a,-1\right) }  {1-2a}}    \\ \times \Bigg( \frac{\,_2\text{F}_1\left(1-2a,a,a+1,-1\right)}{a} 
+ \frac{\,_2\text{F}_1\left(1-2a,1-3a,2-3a,-1\right)}{1-3a} \Bigg)
\end{multline}

En revanche le coefficient $U(a)$ n'est connu que sous la forme d'un développement en série que nous avons vérifié comme convergeant relativement rapidement (100 itérations suffisent)~:

\begin{multline} \label{eq:Ua}
U(a) = \frac{(a-1)^2 }{4 a-1}\left(\frac{\, _2\text{F}_1(1-2 a,a-1;a;-1)}{a-1}+\frac{\, _2\text{F}_1(1-2 a,-3 a;1-3 a;-1)}{3 a}\right) \\ 
\times \Bigg(\frac{\, _2\text{F}_1(-a,1-2 a;2-2 a;-1)}{1-2 a}+\frac{\, _2\text{F}_1(-a,a-1;a;-1)}{a-1} \\ + 
\frac{\, _2\text{F}_1(2-a,1-2 a;2-2 a;-1)}{1-2 a}+\frac{\, _2\text{F}_1(2-a,a+1;a+2;-1)}{a+1} \Bigg)  \\
\displaybreak[3]+ \frac{\left(2 (a-1)^2-1\right)}{4 a-1} \left(\frac{\, _2\text{F}_1(1-2 a,1-3 a;2-3 a;-1)}{3 a-1}+\frac{\, _2\text{F}_1(1-2 a,a;a+1;-1)}{a}\right)\\ 
\times \left(\frac{\, _2\text{F}_1(1-a,1-2 a;2-2 a;-1)}{1-2 a}+\frac{\, _2\text{F}_1(1-a,a;a+1;-1)}{a}\right) \\
\displaybreak[2]+ (a-1)^2\sum _{n=0}^{\infty } \frac{ \Gamma (a+1) }{ \Gamma (n+1) \Gamma (a-n+1)(a+n-1)(3 a-n) }\\ 
\left(\frac{\, _2\text{F}_1(n-a,-2 a+n+1;-2 a+n+2;-1)}{-2 a+n+1}+\frac{\, _2\text{F}_1(n-a,-2 a+n-1;n-2 a;-1)}{-2 a+n-1}\right) \\ 
+ (a-1)^2\sum _{n=0}^{\infty } \frac{ \Gamma (a-1) }{ \Gamma (n+1) \Gamma (a-n-1)(a+n+1)(3 a-n-2) } \\ 
\Bigg(\frac{\, _2\text{F}_1(-a+n+2,-2 a+n+1;-2 a+n+2;-1)}{-2 a+n+1}+\frac{\, _2\text{F}_1(-a+n+2,-2 a+n+3;-2 a+n+4;-1)}{-2 a+n+3}\Bigg)\displaybreak[3] \\
+ 2 \left(2 (a-1)^2-1\right)  \sum _{n=0}^{\infty } \frac{\Gamma (a) }{\Gamma (n+1) \Gamma (a-n)(a+n)(3 a-n-1)}\\ \frac{\, _2\text{F}_1(-a+n+1,-2 a+n+1;-2 a+n+2;-1)}{-2 a+n+1}
\end{multline}

Puisque les séries convergent vite, $U(a)$ peut \^etre connu avec une bonne précision pour tout $a \leqslant 1/4$ (ou $\omega\leqslant 1/4$). \\

Maintenant intéressons-nous au sort des divergences logarithmiques dans l'intégrale TTTT. Puisque~:
\begin{equation}
\frac{\parent{\frac{L}{\varepsilon}}^{1-4a}-1}{1-4a} \stackrel{a\to 1/4}{\to} \log \frac{L}{\varepsilon}\, ,
\end{equation}

seul le dernier terme dans~(\ref{ord4c}) est potentiellement divergent logarithmique en $\omega=1/4$. Il se trouve que dans cette limite le coefficient $V(a)$ est exactement nul. Par comparaison de ce calcul au cas bosonique correspondant, ce sont bien les trois différentes contractions avec les bons signes relatifs qui permettent de supprimer le coefficient et donc la divergence logarithmique comme attendu. Or, le même mécanisme s'applique à l'ordre 2. Il parait naturel de conjecturer que cela s'applique aussi à tous les ordres supérieurs, car il semble bien que cela découle de la supersymétrie sur le bord.

\subsubsection{Contributions du contreterme}

Maintenant, il faut quand même s'assurer qu'aucune divergence logarithmique n'est produite dans les OPE impliquant le contreterme, que l'on notera $C(z)=e^{2\omega X^0}(z)$. En effet, celles-ci peuvent contribuer aux divergences associées à l'opérateur $e^{4\omega X^0}$ à l'ordre $(\lambda^+\lambda^-)^2$ dans les combinaisons suivantes, à l'ordre 3~:

\begin{multline}
\sigma^+\sigma^- \otimes \varepsilon^{a-1} \int \di w \int^{w-\varepsilon}_{w-L}\di x \int^{x-\varepsilon}_{x-L}\di y~\Big(C(w)T^+(x)T^-(y) + T^+(w) C(x) T^-(y)\\  + T^+(w)T^-(x)C(y)\Big) \\ 
+ \sigma^-\sigma^+ \otimes \varepsilon^{a-1}  \int \di w \int^{w-\varepsilon}_{w-L}\di x \int^{x-\varepsilon}_{x-L}\di y~\Big(C(w)T^-(x)T^+(y) + T^-(w) C(x) T^+(y) \\ + T^-(w)T^+(x)C(y)\Big)
\end{multline}

Mais aussi à l'ordre 2 sous la forme~:

\begin{align}
\sigma^0 \otimes \varepsilon^{2(a-1)}  \int\di x \int^{x-\varepsilon}_{x-L}\di y~C(x) C(y)
\end{align}

Pour les termes de type CTT, nous obtenons après quelques calculs, dont les détails sont explicités dans les annexes de notre article~\cite{Israel:2011ut}~:

\begin{multline} \label{ord4a}
 CTT+TCT+TTC \sim 
  \sigma^0 \otimes \Bigg[ 
-\frac{2 }{1+2a}~\parent{\frac{L}{\varepsilon}}^{2-2a} +  \frac{1}{a} \parent{\frac{L}{\varepsilon}}^{1-2a} \\
+ \left( \frac{L}{\varepsilon} \right)^{1-a}  \, X(a)
- \parent{\frac{L}{\varepsilon}}^{1-4a} Y(a)  
\Bigg]  
 ~ L^{4a-1}  \int \di x_1 \,\bnormal{e^{4\omega ~ X_0}}(x_1)
\end{multline}

Les coefficients $X(a)$ et $Y(a)$ sont donnés par~:

\begin{multline}\label{eq:Xa}
X(a) = \frac{2(a-1)}{3a}   \Bigg( 
\frac{\,_2\text{F}_1\left(-a,a+1,a+2,-1\right)}{a+1} + \frac{\,_2\text{F}_1\left(-a,a-1,a,-1\right)}{a-1} \\+ \frac{\,_2\text{F}_1\left(2-a,a+1,a+2,-1\right)}{a+1}
\Bigg) 
\end{multline}

et $Y(a)$ qui ne s'exprime qu'en fonction de développements en série rapidement convergeants (en $100$ itérations typiquement) ~:

\begin{multline} \label{eq:Ya}
Y(a) =  (a-1)\sum_{n=0}^\infty \sum_{s=0}^1 \frac{\Gamma(a)}{\Gamma(a-n)\Gamma(1+n)(3a-s-n)} \Bigg( 
		\frac{\,_2 \text{F}_1 (n-a,1+n-2a;2+n-2a;-1)}{1+n-2a} \\ + \frac{\,_2 \text{F}_1 (s-a,1+s-2a;2+s-2a;-1)}{1+s-2a} 
+ \frac{\,_2 \text{F}_1 (n-a,s+n-1-2a;s+n-2a;-1)}{s+n-1-2a} \\ + \frac{\,_2 \text{F}_1 (s-a,n+s-1-2a;n+s-2a;-1)}{n+s-1-2a} \Bigg) \displaybreak[2]\\ 
+ (a-1)\sum_{n,p=0}^\infty \frac{\Gamma(a) \Gamma(a-1)}{\Gamma(a-n)\Gamma(1+n)\Gamma(a-1-p)\Gamma(1+p)} 
\frac{\,_2 \text{F}_1(1-a,n+p-3a,n+p+1-3a,-1)}{3a-n-p} \\ \times \frac{\,_2 \text{F}_1(2+p-a,n+p+1-2a,n+p+2-2a,-1)}{n+p+1-2a} \\
    + (a-1) \sum_{p=0}^\infty\sum_{s,t=0}^1 \frac{\Gamma(a-1)}{\Gamma(a-1-p)\Gamma(1+p)(3a-s-t-p)} \\ \times \ \frac{\,_2 \text{F}_1(2+p-a,s+p+1-2a,s+p+2-2a,-1)}{s+p+1-2a}
\end{multline}

Nous avons vérifié que ce terme est effectivement convergeant en $\omega=1/4$. Par conséquent, la divergence associée n'est sûrement pas logarithmique. Enfin en ce qui concerne le terme du type CC, il est explicitement~:

\begin{multline}\label{ord4b}
CC\sim  \sigma^0 \otimes \Bigg(
\frac {1}{2a + 1} \left(\frac{L}{\varepsilon}\right)^{2-2a}-\left(\frac{L}{\varepsilon}\right)^{1-2a} \\- \frac{5-6a-(2a-1) 2^{2a+2} 
\,_2 \text{F}_1(1-a,-a-\frac{1}{2};-a+\frac{1}{2};\frac{1}{4})}{4(2a+1)(2a-1)}\left(\frac{L}{\varepsilon}\right)^{1-4a} \Bigg)
\\ \ \times \ L^{4a-1} \int  \di x_1 \,\bnormal{e^{4\omega ~ X_0}}(x_1)\, .
\end{multline}

Ici encore, le facteur précédant $\left(L/\varepsilon\right)^{1-4a}$ est non-singulier en $\omega=1/4$ donc la divergence est non logarithmique en cette valeur. \\

Ainsi, les contretermes ne contribuent pas logarithmiquement en $\omega=1/4$. Compte-tenu de l'expression des termes du type TTTT, nous pouvons conclure qu'il n'y a définitivement aucune divergence logarithmique à l'ordre 4 dans les tachyons. Par conséquent, la théorie est exactement marginale jusqu'à la prochaine résonance à l'ordre $6$, \cad en $r=\sqrt{17}/6$. Puisque le mécanisme de suppression de la divergence logarithmique est visiblement dû à la supersymétrie, nous pouvons conjecturer que ce mécanisme s'applique aussi aux ordres supérieurs et pour tout $r>r_c/\sqrt 2$.

En outre, il s'agit d'un comportement attendu, car à la différence du modèle bosonique, où la perte de marginalité est \emph{physiquement} justifiée par couplage au tachyon du secteur $\sigma^0$ ici cela n'a aucune justification physique.  \\

\fbox{
\begin{minipage}{.9\linewidth}
La théorie du tachyon interbranaire roulant dans le système brane-antibrane séparée est une th\'eorie conforme de bord en toute distance constante $r<r_c$.
\end{minipage}
}

\subsubsection{Remarque sur les divergences résiduelles}

Pour être bien rigoureux, il faut aussi voir ce qu'il advient des divergences de puissance et s'il persiste quelques divergences résiduelles après ressommation de toutes les contributions~: schématique\-ment au quatrième ordre  $CC + CTT + TCT + TTC + TTTT$. Par comparaison de~\refe{ord4a},~\refe{ord4b} and~\refe{ord4c}, nous trouvons que tous les coefficients placés devant chaque divergence s'annulent exactement pour tout valeur $\omega \geqslant 1/4$. De sorte que par ajout du contreterme de deuxième ordre, apparaissant en fait naturellement en exprimant la théorie de surface de manière manifestement supersymétrique -- voir section suivante -- alors la théorie est bien définie dans l'UV pour tout $\omega \geqslant 1/4$. \\

Nous n'avons pas pu obtenir de formule exacte pour le coefficient, not\'e $f(\omega)$ dans~\refe{eq:div_resid}, associé au terme d'ordre $\varepsilon^{1-4a}$ divergeant pour tout $\omega<1/4$. Son expression est en fonction de $U$ donn\'ee dans~\refe{eq:Ua} et $Y$ donn\'ee dans~\refe{eq:Ya}~:

\begin{align}
f(\omega) = U(4\omega^2) - Y(4\omega^2) - \frac{5-24\omega^2-(8\omega^2-1) 2^{8\omega^2+2} \,_2 \text{F}_1(1-4\omega^2,-4\omega^2-\frac{1}{2};-4\omega^2+\frac{1}{2};\frac{1}{4})}{4(8\omega^2+1)(\omega^2-1)} 
\end{align}

En utilisant une évaluation numérique, nous avons trouvé que ce coefficient donne une contribution finie mais non nulle pour $\omega < 1/4$. Ainsi une divergence résiduelle dans cette région persiste.
Par comptage de puissance, cette divergence non-supprimée correspond à la contribution de quatre tachyons interagissant simultanément en un même point. Ce n'est pas inattendu, puisque par nature le contreterme~\refe{eq:contreterme_puissance} correspond à la collision de deux tachyons \emph{d'abord} et par cons\'equent ne peut jamais contribuer pour une collision simultan\'ee de quatre tachyons. Puisque cette divergence n'est pas logarithmique, elle n'est pas pathologique et n'empêche pas la théorie du bord d'être conforme, mais elle doit tout de même être renormalisée. Cela indique qu'il faudrait ajouter de nouveaux termes de contact à 4 tachyons et probablement à plus aux ordres supérieurs. 

\subsection{Expression manifestement supersymétrique en super-espace}
\label{sec:superespace}

Pla\c cons-nous dans le super-espace paramétrisé par $(z,\bar z,\theta,\bar \theta)$ et tel que $z \in H_+$ et $\theta=\eta \overline \theta$ le long du bord. Ici $\eta=\pm 1$ correspond à la structure de spin que nous avons déjà introduit dans la section~\refcc{sec:BCFT}. La notation $\mathbb A$ représente un superchamp tel que~:

\begin{align}
\mathbb A(z,\bar z,\theta,\overline \theta) = A(z,\bar z) + i \theta \phi(z) + i \overline \theta \wt \phi(\bar z) + \theta \overline \theta F(z,\bar z)
\end{align}

Et sur le bord nous aurons par conséquent~:

\begin{align}
\mathbb A(z,\theta) = A(z) + i \theta \parent{\phi(z) + \eta \wt \phi(z,z)}
\end{align}

Les tachyons en super-espace sur le bord sont représentés par les opérateurs de vertex~:

\begin{align}
& \mathbb T(\mathbb X^a,\widetilde{ \mathbb X}) = \lambda^+ \oint \di z \di \theta~\Gam^+ e^{i r \widetilde{\mathbb X}} f(\mathbb X^a) \nonumber \\ 
& \mathbb T^\dagger (\mathbb X^a,\widetilde{ \mathbb X}) = \lambda^- \oint \di z \di \theta ~\Gam^- e^{-ir\widetilde{ \mathbb X}}f^*(\mathbb X^a)
\end{align}

Les facteurs de CP sont remplacés par des degrés de liberté supersymétriques définis sur le bord~\cite{Kraus:2000nj,Takayanagi:2000rz}. Ce sont des superchamps de fermi $\Gam^\pm = \eta^\pm + i \theta F^\pm$ tels que $\eta^\pm$ est un fermion de bord et $F^\pm$ un champ auxiliaire. Ils ont un terme cinétique sur le bord, tel que la quantification des champs $\eta^\pm$ donne les facteurs de Chan-Paton. Ce terme cin\'etique est $\Gam^+ D \Gam^-$ en introduisant la super-dérivée de bord $D=\partial_\theta + \theta \partial$. L'action~\refe{eq:act_duale} s'exprime directement dans le super-espace par~: 

\begin{align}\label{eq:act_super_base}
S = S_{bulk} -  \oint \di z \di \theta~ \Gam^+ {\mathcal D} \Gam^- - i\oint \di z \di \theta~\lambda^+ \Gam^+ \mathbb T^+ - i \oint  \di z \di \theta~ \lambda^-\Gam^-  \mathbb T^- 
\end{align}

avec la super-dérivée covariante ${\mathcal D} = D + i \Phi(\mathbb X^a)D\widetilde{\mathbb X}$, \cad que les fermions de bord sont chargés sous le champ de jauge $A=\Phi dX$. Le long de $\Phi(\mathbb X^a) = r$ donc pour $A$ pure jauge, l'action précédente peut \^etre r\'e\'ecrite en~:

\begin{align}\label{eq:act_super_duale}
S = S_{bulk} -  i\oint \di z \di \theta~ \Gam^+ e^{-ir \widetilde{\mathbb X}} D \parent{e^{ir\widetilde{\mathbb X}}\Gam^-} - i\lambda^+ \oint \di z \di \theta~\Gam^+ \mathbb T^+ - \lambda^- \oint  \di z \di \theta~ \mathbb T^- \Gam^- 
\end{align}

La mesure totale de l'intégrale de chemin, avec celle des fermions de bord est simplement~:

\begin{align}
\int \croch{\di \Gam^+ \di \Gam^-} \croch{\di \mathbb X}\prod_a \croch{\di \mathbb X^a} \Bigg\vert_{x_{cm}=0}
\end{align}

où nous avons indiqué que l'unique condition au bord est sur la coordonnée du centre de masse, $x_{cm}=0$ sachant aussi que les $X^a$ sont Neumann et $X$ est Dirichlet, tandis que $\Gam^\pm$ sont anti-périodiques. Les conditions aux bords précises selon les secteurs sont encodées dans le terme $\Gam^+ \Phi D \widetilde{\mathbb X} \Gam^-$. Or, tant que dans une quelconque amplitude il n'y a pas d'insertion dans le bulk dépendant explicitement de $\mathbb X$, dans~\refe{eq:act_super_duale} nous pouvons réabsorber le facteur $e^{\mp ir\widetilde{\mathbb X}}$ à l'intérieur des $\Gam^\pm$. De sorte que par redéfinition des champs et par invariance de la mesure ci-dessus,~\refe{eq:act_super_duale} devient~:

\begin{align}\label{eq:act_super}
S = S_{bulk} -  \oint \di z \di \theta~ \Gam^+  D \Gam^- - i\lambda^+ \oint \di z \di \theta~\Gam^+ e^{ir\widetilde{\mathbb X}} \mathbb T^+ - i\lambda^- \oint  \di z \di \theta~ \mathbb T^- e^{-ir\widetilde{\mathbb X}}\Gam^- 
\end{align}

Maintenant, la décomposition de cette action en composantes et après intégration de la variable de Grassmann $\theta$ est~:

\begin{multline}
S = S_{bulk} + \oint \di z ~ \Bigg(\eta^+  \partial \eta^- + \eta^+ ~\lambda^+ ~\psi^+ e^{ir\widetilde{X}+\omega X^0}  +  \eta^-~\lambda^-~\psi^- e^{-ir\widetilde{X}+\omega X^0} \Bigg) \\ + \oint \di z ~ \parent{ -F^+F^- +i F^+ \lambda^+ e^{ir\wt X +\omega X^0} +i F^- \lambda^- e^{-ir\wt X +\omega X^0} }
\end{multline}

Les fermions $\eta^\pm$ admettent un terme cin\'etique et comme attendu les champs $F^\pm$ n'en ont aucun et sont donc bien auxiliaires. Alors nous pouvons intégrer ces derniers directement.

\subsubsection{Intégration des champs auxiliaires et terme de contact}

Les champs auxiliaires ont la fonction de Green $G_F(z,w)=\delta(z-w)$ et donnent après intégration~:

\begin{multline}\label{eq:contact_irr}
S = S_{bulk} + \oint \di z ~ \Bigg(\eta^+  \partial \eta^- + \eta^+ ~\lambda^+ ~\psi^+ e^{ir\widetilde{X}+\omega X^0}  +  \eta^-~\lambda^-~\psi^- e^{-ir\widetilde{X}+\omega X^0}\Bigg) \\ + \, \lambda^+\lambda^- \oint\di z \di w ~ \delta(z-w)\bnormal{e^{ir\wt X +\omega X^0}(z)}\bnormal{e^{-ir \wt X +\omega X^0}(w)} 
\end{multline}

Ainsi par intégration des champs auxiliaires surgit un terme de contact -- à cause du $\delta(z-w)$. Or en l'état, ce terme est non-local et l'action n'est par conséquent pas bien définie. Pour obtenir une expression locale, il faudrait calculer explicitement l'OPE puis développer un des champs autour du point d'insertion de l'autre. Le problème est que~:

\begin{align}
\bnormal{e^{ir\wt X +\omega X^0}(z)}\bnormal{e^{-ir \wt X +\omega X^0}(w)}  = (z-w)^{1-4r^2} \bnormal{e^{ir\wt X +\omega X^0}(z)e^{-ir \wt X +\omega X^0}(w)}
\end{align}

n'est pas défini en $z=w$ pour $r^2>1/4$. Afin d'exprimer ce terme de contact, La solution est simplement d'appliquer, même si cela brise explicitement la supersymétrie, la régularisation UV de \emph{point-splitting} \cad que toute OPE entre deux champs quelconque $\phi_1(z)\phi_2(w)$ est restreinte au domaine tronqué\footnote{La limite IR, aussi utilisé dans les calculs précédents, est quant à elle spécifique au plan complexe et ne doit être incluse à ce niveau. Le point-splitting concerne uniquement les divergences UV.} dans l'UV par $\theta(|z-w|-\epsilon)$. Or on peut encoder cette contrainte directement à l'intérieur des fonctions de Green des fermions de bord~:

\begin{align}\label{eq:Green_ferm}
\Gam^+(z)\Gam^-(w) &= \xi(z-w) + 2 \theta_z \theta_w \delta(z-w) \nonumber \\ 
& \to \theta(z-w-\varepsilon) - \theta(w-z-\varepsilon) \nonumber \\ & \qquad + \theta_z\theta_w \,\delta(|z-w|-\varepsilon)
\end{align}

où nous avons introduit $\xi(z)=2\theta(z)-1$ la fonction signe. En somme, cette régularisation revient -- en tout cas en ce qui concerne le cut-off UV -- à répandre l'interaction autour du point de contact $z=w$. Dans la limite $\varepsilon \to 0$ on retrouve bien la formule initiale. Compte-tenu de~\refe{eq:contact_irr} il en ressort un terme de contact à petite échelle. En réinjectant dans le membre de gauche la décomposition des fermions de bords $\Gam^\pm = \eta^\pm + \theta F^\pm$, nous obtenons la fonction de Green régularisée des champs auxiliaires~:

\begin{align}
G_F(z,w) = \delta(|z-w|-\varepsilon)
\end{align}

Reprenons maintenant~\refe{eq:contact_irr} et injectons la fonction de Green ci-dessus. Nous aurons alors pour le terme de contact~:

\begin{multline}\label{eq:contact_general}
\lambda^+\lambda^- \oint\di w ~ \varepsilon^{1-4r^2} \bnormal{e^{ir\wt X +\omega X^0}(w+\varepsilon)e^{-ir \wt X +\omega X^0}(w)} \\ + \lambda^+\lambda^- \oint\di w ~ \varepsilon^{1-4r^2} \bnormal{e^{ir\wt X +\omega X^0}(w)e^{-ir \wt X +\omega X^0}(w+\varepsilon)} 
\end{multline}

L'ordre le plus bas dans le développement des opérateurs est le seul à correspondre à une divergence UV, étant donné que $r<1/2$. Par conséquent, pour $r^2>1/4$ et dans la limite où $\varepsilon \to 0$~:

\begin{align}\label{eq:contact}
\lambda^+\lambda^- \oint\di w ~ \varepsilon^{1-4r^2} \bnormal{e^{2\omega X^0}(w)}  + \ldots
\end{align}

Ce terme est dominant dans la limite $\varepsilon \to 0$. L'ensemble des autres termes tendent à s'annuler. Toutefois, l'ensemble de ces termes, tant que $\varepsilon$ est fini et non nul, devraient contribuer et en particulier pourraient fournir des termes proportionnels à des divergences sous-dominantes. C'est effectivement bien le cas, mais en comparant les calculs dans chaque cas, il revient visiblement au même d'utiliser l'expression non locale~\refe{eq:contact_general} ou l'expression~\refe{eq:contact} -- \cf notre article~\cite{Israel:2011ut}. Cela peut sembler étrange, mais il n'est pas évident que la série et l'intégrale commute~: pour preuve ici la formule suivante qui applique une simple translation de la variable d'intégration~:

\begin{align}
\int_{-\infty}^{+\infty}\di w ~ \bnormal{e^{ir\wt X +\omega X^0}(w+\varepsilon)e^{-ir \wt X +\omega X^0}(w)} = \int_{-\infty}^{+\infty} \di w ~ \bnormal{e^{ir\wt X +\omega X^0}(w)e^{-ir \wt X +\omega X^0}(w-\varepsilon)}
\end{align}

Or il est clair que les développements en séries de $e^{ir\wt X +\omega X^0}(w+\varepsilon)$ et de $e^{-ir \wt X +\omega X^0}(w-\varepsilon)$ ne sont pas égaux. Par conséquent, tant que $\varepsilon$ est fini nous ne pouvons pas développer les opérateurs et commuter la série et l'intégrale et l'action ne peut pas s'exprimer sous une forme totalement locale. En revanche dans la limite $\varepsilon \to 0$ aucune ambiguïté ne subsiste car seul le terme dominant est non nul et il est identique dans chaque membre de l'expression ci-dessus. En admettant que techniquement $\varepsilon =0$ strictement et aussi bien dans la définition de l'action que dans le calcul de toute amplitude alors il apparaît naturel de ne conserver \emph{dans l'action} que le terme local dominant.

\subsubsection{Fonctions de Green des fermions de bord et ordre de chemin}

Pour développer directement l'action~\refe{eq:act_super} exprimée dans le super-espace, il faut d'abord calculer la fonction à N-points des fermions de bord. Or, celle-ci est non nulle uniquement pour $N$ pair. Par cons\'equent, calculons donc la fonction à 2N-points des fermions de bord~:

\begin{multline}\label{eq:ordre_chemin}
\corr{\Gam^+(z_1)\Gam^-(z_2)\ldots\Gam^+(z_{2N-1})\Gam^-(z_{2N})} \\ = \sum_{\text{perm} P}^{2(n!)^2} (-1)^{P(a_1,a_2\ldots a_{2n})} \Theta(z_{a_1}-z_{a_2} + \theta_{a_1}\theta_{a_2})\Theta(z_{a_2}-z_{a_3} + \theta_{a_2}\theta_{a_3}) 
\end{multline}

avec $\Theta(z_{1}-z_2 + \theta_1\theta_2)= \Theta(z_{1}-z_2)  + \theta_1\theta_2 \delta(z_1-z_2)$ et $P$ toute permutation des indices qui conservent l'ordre $(+-+-\ldots)$ ou $(-+-+\ldots)$ donc par exemple $(a_1,a_2,a_3,\ldots)=(1,2,3\ldots)$. Le membre de gauche n'est pas \apriori ordonné, mais en utilisant le théorème de Wick avec~\refe{eq:Green_ferm} nous obtenons deux ordres distincts dans le membre de droite correspondant à $(+-+-)$ et $(-+-+)$. Donc par intégration des fermions de bord nous obtenons un \emph{ordre de chemin} dans l'intégrale de chemin, ce qui revient \`a l'effet obtenu par l'inclusion de facteurs de Chan-Paton.

Maintenant, en appliquant la régularisation précédente nous aurons par exemple pour la fonction à 2-points~:

\begin{align}
\corr{\Gam^+(z_1)\Gam^-(z_2)}_{\varepsilon} &= \Theta(z_{1}-z_2-\varepsilon + \theta_1\theta_2) - \Theta(z_{2}-z_1-\varepsilon + \theta_2\theta_1) - \theta_2\theta_1) \nonumber \\ 
 &= \Theta(z_{1}-z_2-\varepsilon) + \theta_1\theta_2 \, \delta(z_1-z_2-\varepsilon) \nonumber \\ 
    & \qquad - \Theta(z_{2}-z_1-\varepsilon) + \theta_1\theta_2 \, \delta(z_1-z_2+\varepsilon) 
\end{align}

Et dans une fonction plus grande, il n'est pas difficile d'extrapoler la formule ci-dessus. Les fonctions delta correspondent aux fonctions de Green des champs auxiliaires. Ils sont par conséquent associés aux contributions du terme de contact que nous avons calculé. Or dans toute fonction à 2N-points, entre les fonctions de Green du terme de contact s'intercalent des fonctions thêta qui ordonnent les termes de contact avec les autres opérateurs. De sorte qu'en voulant calculer les contributions du terme de contact (local ou non local) dans une amplitude, il faut imposer, de façon \emph{ad hoc}, qu'aucun opérateur ne doit approcher un autre opérateur à moins\footnote{Sur le bord de $H_+$ il faut aussi demander à ce que les opérateurs ne s'éloignent pas les uns des autres de plus de $L$.} de $\varepsilon$.

\subsubsection{Intégration des fermions et facteurs de Chan-Paton}

Pour obtenir les facteurs de Chan-Paton, il faut quantifier les champs fermioniques $\eta^\pm$ compte-tenu du terme cinétique $\eta^+\partial \eta^-$ tout en considérant les autres termes comme des perturbations. Les équations du mouvement sont triviales~:

\begin{align}
\partial \eta^\pm = 0 \qquad \text{et} \qquad \acomm{\eta^+}{\eta^-} = 1
\end{align}

Elles sont aisément résolues par $\eta^\pm = \sigma^\pm$ avec les matrices de Pauli $\sigma^\pm = (\sigma^1 \pm i \sigma^2)/2$. Il faut ensuite utiliser les identifications suivantes~:

\begin{align}
&\eta^\pm \to \sigma^\pm \nonumber \\ 
&\eta^+\eta^-(z) \to \frac{1}{2}\comm{\sigma^+}{\sigma^-} = \frac{\sigma^3}{2}
\end{align}

La deuxième identification est naturelle par anti-commutation de $\eta^\pm$. L'intégrale de chemin, compte-tenu de ce qui a été vu dans la section précédente à propos de l'ordre de chemin induit par les fermions de bord~\refe{eq:ordre_chemin} s'exprimera donc par~:

\begin{align}
\int [\di \Gam^+ \di\Gam^-][\di\mathbb X] e^{-S_{super}} \ldots = \tr \int [\di X][\di \psi] {\mathcal P} e^{-S_{decomp}-S_{contact}}
\end{align}

où dans le membre de droite nous avons utilisé l'action exprimée dans le super-espace~\refe{eq:act_super} et dans le membre de gauche l'action décomposée et exprimée en terme des facteurs de Chan-Paton. Il faut en outre ajouter la contribution du terme de contact obtenue par intégration des champs auxiliaires et qui s'exprime par~:

\begin{multline}
S_{decomp}+S_{contact} =  S_{bulk} + \oint \di z ~ \Bigg(\sigma^+ \otimes ~\lambda^+ ~\psi^+ e^{ir\widetilde{X}+\omega X^0}  +  \sigma^-\otimes ~\lambda^-~\psi^- e^{-ir\widetilde{X}+\omega X^0}\Bigg) \\ + \lambda^+\lambda^-\oint\di w ~ \varepsilon^{1-4r^2} \bnormal{e^{2\omega X^0}(w)}
\end{multline}

Nous avons aussi inclus un opérateur d'ordre ${\mathcal P}$ déjà introduit auparavant, qui applique selon~:

\begin{align}
{\mathcal P} \int \di z_1 \di z_2 \ldots \di z_N ~ \phi_1(z_1)\phi_2(z_2)\ldots \phi_N(z_N) = (2\pi)^N \int [\underset{>}{\di z_i}]_N ~ \phi_1(z_1)\phi_2(z_2)\ldots \phi_N(z_N)
\end{align}

et impose aux opérateurs $\phi_i$ d'être effectivement ordonnés, \cad en terme des facteurs de Chan-Paton qu'ils contiennent. Nous avons introduit la notation pratique que l'on réutilisera~:

\begin{align}
[\underset{>}{\di z_i}]_N = \prod_i^N \frac{\di z_i}{2\pi} \, \prod_i^{N-1} \Theta(z_i-z_{i+1})
\end{align}

\subsubsection{Action générale avec perturbation de distance et remarque sur les termes de contact d'ordre supérieurs}

En partant de l'action~\refe{eq:act_super_duale} et avec $\Phi = r + \delta r(\mathbb X^a)$, nous aurions aussi obtenu des termes de contact entre champ de distance $(\delta r \, \eta^\pm \widetilde \psi)^2 \to \delta r^2/\varepsilon$ et dont l'expression en fonction des cut-offs correspond exactement aux divergences de type Moebius~\cite{Andreev:1988cb,Andreev:1988bz,Liu:1987nz,Tseytlin:1987ww} rencontrées dans le calcul de l'action BI par exemple. Le terme proportionnel aux fermions $\eta^+\eta^-$ serait~:

\begin{align}
i \eta^+\eta^- \parent{\delta r(X^a)\partial \wt X + \partial_a\delta r(X^a) \psi^a \widetilde \psi} \to i \frac{\sigma^3}{2} \parent{\delta r(X^a)\partial \wt X + \partial_a\delta r(X^a) \psi^a }
\end{align}

Alors l'action serait très semblable à celle obtenue en théorie bosonique, mis à part le terme supersymétrique supplémentaire. 

Maintenant, en s'inspirant de l'émergence du contreterme d'ordre 2 en tant que terme de contact, il faudrait tenter de regrouper l'ensemble des contretermes dans une formulation unique manifestement supersymétrique. Autrement dit, nous voudrions une action exprimée \apriori dans le super-espace puis éventuellement décomposée de telle sorte que les opérateurs du type $\sigma^0 \otimes e^{2n\omega X^0}$ apparaissent, accompagnés comme~\refe{eq:contact} d'une expression divergente dans les cut-offs et telle qu'ils sont précisément les contretermes nécessaire à la soustraction de toutes les divergences. Par extrapolation de notre résultat à l'ordre 4, les seules divergences résultantes à soustraire seraient du type $\varepsilon^{4n^2 \omega^2-1}$. Celles-ci ne peuvent correspondre qu'à des termes de contact d'ordre $2n$. En effet, ceux-ci doivent être du type~:

\begin{align}
\varepsilon^{-1} \, \int \di^{2n-1} \vec{u} ~ \delta^{(2n)}(u^i-\varepsilon) f(\vec{u}) \int \di z \, e^{2n\omega X^0}
\end{align}

avec $\vec u$ le vecteur pris dans l'espace des distances entre opérateurs, de type $(z_1-z_2)$, dont la base est choisie de telle sorte que les distances sont indépendantes. Nous pouvons choisir par exemple $u^i=z_{i}-z_{i+1}$ avec $i \in [1,2n-1]$. La fonction $f$ est calculée par OPE des tachyons ; par exemple à l'ordre 4~:

\begin{multline}
\corr{T^+(z_1) T^-(z_2) T^+(z_3) T^-(z_4)} = f(\vec u) \\ = (z_1-z_2)^{4\omega^2-1}(z_1-z_3)(z_1-z_4)^{4\omega^2-1}(z_2-z_3)^{4\omega^2-1}(z_2-z_4)(z_3-z_4)^{4\omega^2-1}  \\ 
=(u^1)^{4\omega^2-1}(u^1+u^2)(u^1+u^2+u^3)^{4\omega^2-1}(u^2)^{4\omega^2-1}(u^2+u^3)(u^3)^{4\omega^2-1}
\end{multline}

Compte-tenu de la fonction delta, tous les facteurs sont proportionnels à $\varepsilon$. Donc à l'ordre 4 le terme de contact serait proportionnel à~:

\begin{align}
\varepsilon^{16\omega^2 -1 } \int \di z \, e^{2n\omega X^0}
\end{align}

A l'heure actuelle nous ne connaissons pas de telle formulation obtenue à l'aide des champs auxiliaires. Or ce type de terme dans son expression manifestement supersymétrique est important car il peut ajouter de nouvelles contributions finies aux calculs, en plus de soustraire les divergences. Ce devrait être un terme non-linéaire et il pourrait être par exemple de la forme  $\varepsilon^{-1} \int \di z \, F^+F^-F^+F^-(z)$ ce qui ne peut provenir que d'un terme manifestement supersymétrique de la forme $\oint \Gam^+ D\Gam^- D\Gam^+ D\Gam^-$.

Ces termes ne sont cependant pas indispensables, car une formulation supersymétrique n'exclut pas la renormalisation. En outre, le fait de rajouter des termes implique de modifier fondamentalement la théorie puisque cela revient à ajouter de nouvelles interactions sur le bord. \\

Pour conclure, en ce qui concerne le tachyon roulant à distance fixe, l'expression manifestement supersymétrique est souhaitée mais n'est pas nécessaire. Elle implique en outre d'exprimer des interactions de contacts à N-points, qu'\apriori nous ne connaissons pas. Autrement dit, le fait de devoir ajouter des contretermes de façon \emph{ad hoc} brise explicitement la symétrie superconforme dans l'action de surface de corde, mais n'empêche pas la théorie d'être exactement marginale donc de correspondre à une solution des équations du mouvement. Par conséquent trouver une expression manifestement supersymétrique n'est pas indispensable.

\section{Fonctions bêta, groupe de renormalisation et équations du mouvement}
\label{sec:renorm_susy}

Cette \'etude est similaire à celle faite en théorie bosonique dans la section~\refcc{sec:off_sep}. Elle donnera à peu de choses près les mêmes résultats. Il est donc évident qu'il faudra uniquement s'intéresser aux déformations marginales au premier ordre. Nous referons cependant le développement des trois phases massive, non-massive et tachyonique, et calculerons les fonctions bêta dans chaque cas.

Nous commencerons par étudier la phase surcritique massive ($r>1/\sqrt 2$) puis la phase sous-critique tachyonique ($r<1/\sqrt 2$) plus délicate et enfin nous \'etudierons la continuité avec la phase critique non massive ($r=1/\sqrt 2$).

\subsection{Phases surcritique $r\geq r_c$}

Dans cette phase, les opérateurs de vertex $ \oint \psi^\pm e^{\pm i r\wt X \pm i\omega X^0}$ avec $\psi^\pm = \pm (ir \wt \psi + i \omega \psi^0)$ sont marginaux en $\omega=r^2-1/2$ nous pouvons donc les inclure sur le bord, étudier les fonctions bêta résultantes et y faire correspondre des équations de mouvement de champs correspondant. Nous devons partir de l'action générale~\refe{eq:act_super_base} en décomposant $\Phi = r +\delta r$ et en absorbant $r$ constant dans les fermions de bord, de sorte que nous avons~:

\begin{multline}
S = S_{bulk} -  \oint \di z \di \theta~ \Gam^+  \parent{D + i \delta r(\mathbb X^a)D\wt{\mathbb X}} \Gam^- \\ -i \oint \di z \di \theta~\lambda^+(\mathbb X^a) \Gam^+ e^{ir\widetilde{\mathbb X}+i\omega \mathbb X^0} -i \oint  \di z \di \theta~ \lambda^-(\mathbb X^a)\Gam^- e^{-ir\widetilde{\mathbb X}-i\omega \mathbb X^0}
\end{multline}

avec à l'ordre 2 en dérivées en développant $\mathbb X^\mu = x^\mu + \hat X^\mu + \theta \psi^\mu$~:

\begin{align}
\delta r(\mathbb X^i) & = \delta r + \partial_i \delta r \parent{\hat X^i + i\theta \psi^i} + \frac{\partial_i\partial_j \delta r}{2}\parent{\hat X^i\hat X^j + 2 i \theta \hat X^i \psi^j} \nonumber \\ 
\lambda^\pm(\mathbb X^i) &= \lambda^\pm + \partial_i\lambda^\pm \parent{\hat X^i + i\theta \psi^i} + \frac{\partial_i\partial_j \lambda^\pm}{2}\parent{\hat X^i\hat X^j + 2 i\theta \hat X^i \psi^j} 
\end{align}

A l'instar du modèle bosonique nous avons choisi de complètement factoriser le comportement temporel pour ne s'intéresser qu'au comportement spatial off-shell, ce qui a le mérite de simplifier singulièrement les calculs. Nous rétablirons la covariance à la fin. Il est utile de décomposer cette action explicitement afin de bien dégager les contributions associées aux fermions de bord de celles associées aux champs auxiliaires~:

\begin{multline}\label{eq:premiere_act}
S = S_{bulk} + \oint \eta^+ \partial \eta^- - \oint F^+ F^-  \\ 
               + i \oint \eta^+\eta^- \Bigg[ \delta r' \partial \wt X + \partial_i\delta r \parent{\hat X^i \partial \wt X - \psi^i \widetilde \psi} +\frac{\partial_i\partial_j \delta r}{2}\parent{\hat X^i\hat X^j\partial \wt X - 2 \hat X^i \psi^j \widetilde \psi}  \Bigg] \\
               + \oint \eta^+ e^{ir\wt X+i\omega X^0} \Bigg[\lambda^{+\, '} \psi^+ + \partial_i \lambda^+ \parent{\hat X^i\psi^+ + \psi^i} +   \frac{\partial_i\partial_j \lambda^+}{2}\parent{\hat X^i\hat X^j\psi^+   + 2 \hat X^i \psi^j}  \Bigg]     \\ 
               + \oint \eta^- e^{-ir\wt X-i\omega X^0} \Bigg[\lambda^{-\, '} \psi^- + \partial_i \lambda^- \parent{\hat X^i \psi^- + \psi^i} +   \frac{\partial_i\partial_j \lambda^-}{2}\parent{\hat X^i\hat X^j \psi^-   + 2 \hat X^i \psi^j}  \Bigg] \\           
               + \oint  F^+ \Bigg[-\eta^- \widetilde \psi \parent{\delta r' +  \partial_i\delta r \hat X^i + \frac{\partial_i\partial_j \delta r}{2} \hat X^i\hat X^j} - i \, e^{ir\wt X +i\omega X^0}\parent{\lambda^{+\, '} + \partial_i\lambda^+ \hat X^i + \frac{\partial_i\partial_j\lambda^+}{2} \hat X^i\hat X^j} \Bigg] \\
               + \oint  F^- \Bigg[ \eta^+ \wt \psi \parent{\delta r' +  \partial_a\delta r \hat X^i + \frac{\partial_i\partial_j \delta r}{2} \hat X^i\hat X^j} -i \, e^{-ir\wt X-i\omega X^0}\parent{\lambda^{-\, '} + \partial_i\lambda^- \hat X^i + \frac{\partial_i\partial_j\lambda^-}{2} \hat X^i\hat X^j} \Bigg]
\end{multline}

Nous avons sorti explicitement le facteur d'échelle UV dans le schéma de Wilson, mais avec des déformations principalement marginales, il disparaît. Néanmoins, nous avons noté par commodité $\mu' = \mu + \ln \varepsilon \, \square \mu$ pour $\mu = \{\delta r, \lambda^\pm \}$ mais dans la suite nous ôterons le "prime" et considérerons implicite l'addition de $\ln \varepsilon \, \square \mu$. Remarquons au passage que $\beta_{\mu'}=\beta_{\mu} + \square \mu$. Par intégration des champs auxiliaires, nous obtenons un ensemble de termes de contact régularisés dans l'UV\footnote{Nous ne nous interesserons pas aux divergences IR ici.} comme il a été fait précédemment en utilisant $G_F = \delta(|z-w|-\varepsilon)$~:

\begin{multline}\label{eq:contact_gen} 
S_{contact} = - \frac{2}{\varepsilon}\croch{\delta r^2 - 2  \partial_i\delta r\partial^i \delta r  ~ \ln \varepsilon + \Bigg(\partial_i (\delta r^2)  - 4  \partial_j \delta r \partial^j\partial_i \delta r ~\ln \varepsilon \Bigg) \hat X^i + \frac{\partial_i \partial_j(\delta r^2)}{2} \hat X^i\hat X^j}    \\ 
+  \varepsilon^{-1} \Bigg[\lambda^+\lambda^- - 2 \partial_i\lambda^+\partial^i\lambda^- ~\ln\varepsilon  + \Bigg(\partial_i(\lambda^+\lambda^-) - 2 \partial_i\parent{\partial_j \lambda^- \partial^j \lambda^+}\ln \varepsilon  \Bigg) \hat X^i \\ + \frac{\partial_i\partial_j(\lambda^+\lambda^-)}{2}\hat X^i\hat X^j \Bigg] \\ 
+ i \eta^+ \widetilde \psi e^{ir \wt X +i\omega X^0}\Bigg[ \lambda^+\delta r - 2 \partial_i\lambda^+\partial^i\delta r ~\ln\varepsilon  + \Bigg(\partial_i(\lambda^+ \delta r) - 2  \partial_i\parent{\partial_j \delta r \partial^j \lambda^+}\ln \varepsilon \Bigg) \hat X^i \\  + \frac{\partial_i\partial_j(\lambda^+\delta r)}{2} \hat X^i\hat X^j \Bigg] \\ 
- i \eta^- \widetilde \psi e^{-ir \wt X-i\omega X^0}\Bigg[ \lambda^-\delta r - 2 \partial_i\lambda^-\partial^i\delta r ~\ln\varepsilon  + \Bigg(\partial_i(\lambda^- \delta r) - 2 \partial_i\parent{\partial_j \delta r \partial^j \lambda^-}\ln \varepsilon \Bigg) \hat X^i \\  + \frac{\partial_i\partial_j(\lambda^-\delta r)}{2} \hat X^i\hat X^j \Bigg]
\end{multline}

A la deuxième ligne, nous avons utilisé $2\omega^2-2r^2=-1$ dans l'OPE des tachyons, d'où le facteur $\varepsilon^{-1}$. A première vue il appara\^it un certain nombre de divergences logarithmiques qui devraient logiquement contribuer aux fonctions bêta des couplages des tachyons -- en particulier celles des deux dernières lignes -- et à la fonction bêta du "couplage" de l'opérateur unité $\oint 1$. En vérité, ça n'est pas le cas~: ces termes vont jouer le rôle de contretermes supprimant une partie des divergences apparaissant par OPE des opérateurs factoris\'es par $\eta^\pm$ dans~\refe{eq:premiere_act}. 

Dans le schéma de Wilson par exemple, les éventuels contretermes ne participent pas aux fonctions bêta des couplages car ces derniers dépendent directement du cut-off $\varepsilon$ et doivent déjà contenir toutes contributions permettant de supprimer toutes les divergences. Si ces contretermes retranchent les divergences, alors cela implique qu'il ne doit pas y avoir de contribution correspondante dans l'expression des couplages. Et donc les fonctions bêta ne sont pas modifiées. 

Dans le schéma minimal se pourrait être un peu plus ambiguë, cependant il faut tenir compte du fait que les couplages sont associés aux opérateurs de vertex \emph{avant} intégration des champs auxiliaires, par conséquent tout ce qui apparaît via l'intégration de ces derniers n'a pas à être pris en compte dans la définition des couplages. \\

Remarquons en outre que le premier terme de chacune des deux dernières lignes contribue à $\eta^\pm \widetilde \psi e^{\pm i r \wt X}$ de telle sorte que $r\to r+ \delta r$ ce qui est en accord avec l'effet attendu de la perturbation $\delta r D\wt{\mathbb X}$ censée être un opérateur de changement de distance. Ces termes de contact sont déjà d'ordre 2 dans les perturbations. Par conséquent, puisque nous ne nous intéressons qu'au deuxième ordre en perturbation dans les fonctions bêta nous pouvons les négliger dans la suite. \\

 L'action que nous allons utiliser maintenant est donnée par~\refe{eq:premiere_act} sans les termes dépendants des champs auxiliaires et sans les termes de contact. Soit donc en remplaçant les fermions de bord par les facteurs de CP~:

\begin{multline}
S = S_{bulk} + \sigma^3 \otimes \frac{i}{2} \oint \parent{\delta r \partial \wt X  + \partial_i \delta r \, (\hat X^i \partial \wt X - \psi^i \widetilde \psi) + \frac{\partial_i\partial_j \delta r}{2} (\hat X^i \hat X^j \partial \wt X - 2 \hat X^j \psi^i \widetilde \psi)} \\
+ \sigma^+ \otimes \oint \Bigg( \lambda^+ \psi^+ + \partial_i \lambda^+ (\hat X^i \psi^+ + \psi^i)  + \frac{\partial_i\partial_j \lambda^+}{2}  \parent{\hat X^i \hat X^j \psi^+ + 2 \hat X^j \psi^i } \Bigg) e^{i r \wt X+i\omega X^0} \\ 
+ \sigma^- \otimes \oint \Bigg(\lambda^-\psi^- + \partial_i \lambda^- (\hat X^i \psi^- + \psi^i)  + \frac{\partial_i\partial_j \lambda^-}{2} \parent{\hat X^i \hat X^j \psi^- + 2 \hat X^j \psi^i }\Bigg) e^{-i r \wt X-i\omega X^0}  
\end{multline}

Les OPE importantes sont ici celles des deux tachyons ensemble et celles de chaque tachyon avec les divers opérateurs associés au champ de distance. Notons d'abord l'OPE des fermions~:

\begin{align}\label{eq:OPE_ferm}
\psi^+(z)\psi^-(0) \sim \frac{1}{z}
\end{align}

Elle aura son importance lorsque nous traiterons du cas sous-critique. Les OPE des tachyons sont comme suit -- nous avons laissé implicite les facteurs de CP~:

\begin{align}\label{eq:OPE_tach_tach}
& \psi^+ e^{i r \wt X + i\omega X^0}(z)  \cdot  \psi^- e^{-i r \wt X-i\omega X^0}(w)  \simeq (z-w)^{-2} + i r (z-w)^{-1} \partial \wt X (w) + \ldots \nonumber \\ 
& (\hat X^i \psi^+ + \psi^i)e^{i r \wt X+ i\omega X^0}(z) \cdot \psi^- e^{-i r \wt X- i\omega X^0}(w) \simeq \frac{\hat X^i}{(z-w)^{2}}  + \frac{1}{z-w} \parent{ir \hat X^i \partial \wt X + \psi^i \psi^-} \nonumber \\ 
& (\hat X^i \psi^+ + \psi^i)e^{i r \wt X+ i\omega X^0}(z) \cdot (\hat X^j \psi^- + \psi^j)e^{-i r \wt X- i\omega X^0}(w) \nonumber \\ 
& \qquad\simeq \frac{1}{(z-w)^{2}} \parent{\hat X^i \hat X^j + 2\eta^{ij} } + \frac{1}{z-w} \parent{ 2ir \,\eta^{ij}\partial \wt X + \parent{ir \hat X^i \hat X^j \partial \wt X - 2\hat X^{i} \psi^{j} \psi^+ }} \nonumber \\ 
&\parent{\hat X^i \hat X^j \psi^+ + 2\hat X^{j} \psi^{i} } e^{i r \wt X+ i\omega X^0}(z) \cdot \psi^- e^{-i r \wt X- i\omega X^0}(w) \nonumber \\ 
& \qquad\simeq \frac{1}{(z-w)^{2}} \parent{\hat X^i \hat X^j } + \frac{1}{z-w} \parent{ ir \hat X^i \hat X^j \partial \wt X + 2\hat X^{j} \psi^{i} \psi^- } \nonumber \\ 
&\parent{\hat X^i \hat X^j \psi^+ + 2\hat X^{j} \psi^{i} } e^{i r \wt X+ i\omega X^0}(z) \cdot  (\hat X^k \psi^- + \psi^k)e^{-i r \wt X- i\omega X^0}(w)\nonumber \\ 
&\qquad\simeq \frac{1}{(z-w)^{2}} \parent{2\eta^{k i} \hat X^{j} + \ldots} + \frac{1}{z-w} \parent{2 ir \, \eta^{ki} \hat X^{j} \partial \wt X + \ldots} \nonumber \\ 
&\parent{\hat X^i \hat X^j \psi^+ + 2\hat X^{j} \psi^{i} } e^{i r \wt X+ i\omega X^0}(z) \cdot \parent{\hat X^k \hat X^l \psi^- + 2\hat X^{k} \psi^{l} } e^{-i r \wt X- i\omega X^0} \nonumber \\ 
&\qquad\simeq \frac{1}{(z-w)^{2}} \parent{ 4 \eta^{il}\hat X^j \hat X^k  + \ldots} + \frac{2}{z-w} \parent{ 4 i r \eta^{il}\hat X^j \hat X^k \partial \wt X + \ldots} \nonumber \\ 
\end{align}

Quelques remarques~: \\

\begin{itemize}\itemsep4pt
\item[\emph{i)}] Nous avons caché à l'intérieur des points de suspension tous les termes proportionnels à $\ln (z-w)$. Ceux du deuxième terme participeront aux fonctions bêta sous la forme $\varepsilon^{-1}\ln \varepsilon$ et sont proportionnels à des fonctions bêta. Tandis que ceux du premier terme donneront des divergences en~:

\begin{align}
\int_\varepsilon^L \di z ~ \frac{\ln z}{z^2} = \frac{1}{\varepsilon} + \frac{\ln\varepsilon}{\varepsilon} + \ldots
\end{align}

aux termes dépendants de $L$ près. Par exemple à la troisième ligne nous devrions avoir le terme complet~:

\begin{align}
\frac{1}{(z-w)^{2}} \croch{\hat X^i \hat X^j + 2\eta^{ij}(1 - \ln(z-w))} \longrightarrow \frac{\hat X^i \hat X^j - 2\eta^{ij}\ln \varepsilon}{\varepsilon}
\end{align}

En multipliant par $\partial_i \lambda^+ \partial_j \lambda^-$ et en comparant au terme de contact de la seconde ligne de~\refe{eq:contact_gen} nous retrouvons bien~:

\begin{align}
\varepsilon^{-1} 2 \ln \varepsilon ~\partial_i \lambda^+ \partial^i \lambda^- 
\end{align}

Par cons\'equent, ces deux termes s'annulent ensemble. Il en sera de même pour tous les autres termes semblables. 

\item[\emph{ii)}] Dans~\refe{eq:OPE_tach_tach} nous avons aussi caché tous les termes qui ont plus de deux indices, par exemple $\hat X^i\hat X^j \hat X^k \wt\psi$ car ils correspondent à des ordres plus élevés en dérivées des champs que nous avons ici négligés. 

\item[\emph{iii)}] Les termes int\'eressants ici sont ceux dont l'opérateur contient $\partial \wt X$ ou $\widetilde \psi$. Donc les premiers termes en $(z-w)^{-2}$ ne seront \'etudi\'es qu'en dernier lieu pour vérifier s'ils correspondent à des contretermes ou s'annulent entre eux. 

\item[\emph{iv)}]Enfin, de même que dans le développement bosonique, les termes contenant des champs de la coordonnée temporelle tels que $\psi^0$ ou $\partial X^0$ ne seront pas \'etudi\'es. \\
\end{itemize}

Les OPE entre tachyons et perturbation de distance sont donnés dans les formules suivantes. En utilisant $\sigma^3\sigma^\pm = \pm \sigma^\pm $ et en laissant les facteurs de Chan-Paton implicites tels que dans le membre de gauche il faudrait ajouter le préfacteur $\sigma^3\sigma^\pm \otimes$ et dans le membre de droite $\sigma^\pm\otimes $~:

\begin{align}\label{eq:OPE_dist_tach}
 & \partial \wt X(z)  \cdot    \psi^\pm e^{\pm i r \wt X \pm i\omega X^0}(w) \simeq \frac{\mp 2 i r}{z-w} \psi^\pm e^{\pm ir \wt X \pm i\omega X^0}(w) + \ldots \nonumber \\ 
 &\partial \wt X(z) \cdot (\hat X^i \psi^\pm + \psi^i)e^{\pm i r \wt X \pm i\omega X^0} (w) \simeq  \frac{\mp 2 ir}{z-w}  ( \hat X^i \psi^\pm + \psi^i) e^{\pm ir \wt X \pm i\omega X^0}(w) + \ldots \nonumber \\ 
 &(\hat X^i \partial \wt X - \psi^i \widetilde \psi)(z) \cdot \psi^\pm e^{\pm ir\wt X \pm i\omega X^0}(w) \simeq  \frac{\mp 2 i r }{z-w}  (\hat X^i \psi^\pm + \psi^i) e^{\pm ir \wt X \pm i\omega X^0}(w) + \ldots  \nonumber \\ 
 & (\hat X^i \partial \wt X - \psi^i \widetilde \psi)(z) \cdot (\hat X^j \psi^\pm + \psi^j)e^{\pm i r \wt X \pm i\omega X^0} (w) \nonumber \\ & \qquad \simeq \frac{\mp 2}{z-w} \Bigg[i r \parent{\hat X^i \hat X^j \psi^\pm + 2 X^{i}\psi^{j}} + \eta^{ij} \widetilde\psi + \ldots \Bigg]e^{\pm ir \wt X \pm i\omega X^0} \nonumber \\ 
 & (\hat X^i \partial \wt X - \psi^i \widetilde \psi)(z) \cdot  \parent{\hat X^j \hat X^l \psi^\pm + 2 \hat X^j \psi^l } e^{\pm i r \wt X \pm i\omega X^0}  (w) \simeq \frac{\pm 2}{z-w} \parent{2 \eta^{ik} \hat X^j\widetilde \psi + \ldots}e^{\pm i r \wt X \pm i\omega X^0}  \nonumber \\
 & (\hat X^i \hat X^j \partial \wt X - 2 \hat X^j \psi^i \widetilde \psi)(z) \cdot \psi^\pm e^{\pm i r \wt X \pm i\omega X^0}(w) \simeq \frac{ \mp 2ir}{z-w} \parent{\hat X^i \hat X^j \psi^\pm + 2 \hat X^j \psi^i}e^{\pm ir \wt X \pm i\omega X^0} \nonumber \\ 
  &  (\hat X^i \hat X^j \partial \wt X - 2 \hat X^j \psi^i \widetilde \psi)(z) \cdot (\hat X^l \psi^\pm + \psi^l)e^{\pm i r \wt X \pm i\omega X^0} (w) \nonumber \\ & \qquad \simeq \frac{\pm 2 }{z-w} \parent{2 \eta^{ac} \hat X^j \widetilde \psi + \ldots} e^{\pm i r \wt X \pm i\omega X^0} \nonumber \\ 
  & (\hat X^i \hat X^j \partial \wt X - 2 \hat X^j \psi^i \widetilde \psi)(z)\cdot \parent{\hat X^l \hat X^l  \psi^\pm + 2 \hat X^l \psi^l } e^{\pm i r \wt X \pm i\omega X^0}  (w)\nonumber \\ &\qquad \simeq \frac{\pm 2 }{z-w} \parent{4 \eta^{ad} \hat X^j \hat X^l \widetilde \psi + \ldots }e^{\pm i r \wt X \pm i\omega X^0} 
\end{align}

De nouveau nous avons caché les termes qui ne nous intéressent pas. Ainsi, au deuxième ordre en récupérant seulement les termes intéressants dans~\refe{eq:OPE_tach_tach} et \refe{eq:OPE_dist_tach} nous avons à sommer dans l'intégrale de chemin, après intégration de $z>w$~:

\begin{align}\label{eq:deuxieme_contreterme}
&1 - \sum_\pm \sigma^\pm \otimes \int \di w~\psi^\pm e^{\pm ir \wt X \pm \omega X^0}(w)\Bigg[\lambda^\pm + \parent{2 r \delta r \lambda^\pm + 2 \partial_i\delta r \partial^i \lambda^\pm} \ln \frac{\varepsilon}{\ell}\Bigg] \nonumber \\ 
& - \sigma^3 \otimes \frac{i}{2}  \int \di w~\partial \wt X \Bigg[\delta r + 2 r \parent{\lambda^+ \lambda^- + 2 \partial_i \lambda^+ \partial^i\lambda^-} \ln \frac{\varepsilon}{\ell} \Bigg] + \ldots
\end{align}
 
Nous ne nous occupons ici que des couplages $\lambda^\pm$ et $\delta r$. Les couplages $\partial_i \lambda^\pm$ et $\partial_i\delta r$ reçoivent des contributions qui sont les dérivées de celles des premiers. Il en est de même pour les dérivées d'ordres suivants. \\

Maintenant, par comparaison aux termes de contact à la troisième et quatrième ligne dans~\refe{eq:contact_gen} le terme $\pm 2 \partial_i\delta r \partial^i \lambda^+$ est exactement compensé. Si bien que ce dernier ne doit pas participer à la fonction bêta de $\lambda^\pm$ comme nous l'avons expliqué. En outre, certains termes brisent la supersymétrie de surface, comme par exemple dans l'avant-dernière ligne de~\refe{eq:OPE_dist_tach} le terme en $\hat X^i \widetilde \psi$ sans son partenaire $\psi^i$. Or, ils apparaissent aussi dans l'expression~\refe{eq:contact_gen} des termes de contacts, qui suppriment à nouveau ces contributions. Il en est de même pour toutes les contributions du premier terme de chaque ligne de~\refe{eq:OPE_tach_tach}. En calculant exactement chacun de ceux-ci, y compris ceux proportionnels à $\ln(z-w)$ qui \'etaient occultés dans les points de suspension, nous obtenons après intégration, chacun des contretermes de deuxième ordre dans les tachyons et leur dérivées. \\

Enfin, d'autres termes brisent la supersymétrie de surface mais n'apparaissent dans aucun terme de contact, tels que dans~\refe{eq:OPE_tach_tach} 
\begin{align}
2 i r\eta^{ik}\hat X^j \partial \wt X  \quad \text{ou} \quad 4 i r\eta^{il}\hat X^j \hat X^k \partial \wt X
\end{align}

Ceux-ci peuvent \^etre problématiques puisqu'impossible \`a réabsorber par des redéfinitions des champs. Bien qu'ils soient nuls après application de la trace sur le facteur $\sigma^3$ en se repla\c cant dans un contexte plus général d'une amplitude avec des insertions arbitraires alors ils peuvent contribuer de façon non triviale et effectivement briser la supersymétrie de surface dans l'amplitude~: la sym\'etrie superconforme au niveau des amplitudes n'est donc plus garantie. L'unique solution est d'imposer l'équation~:

\begin{align}\label{eq:equation_supp}
\partial_j\Bigg[\partial_i \lambda^+ \partial^i \lambda^-\Bigg] = 0 \quad \Leftrightarrow \quad \partial_i \lambda^+ \partial^i \lambda^- \propto \lambda^+\lambda^-
\end{align}

avec $\lambda^-=(\lambda^+)^*$. Il ne s'agit pas une contrainte extraordinaire. En effet, par exemple $\lambda^\pm \propto e^{\pm i k_i X^i}$ la vérifie immédiatement. On peut cependant s'interroger sur la signification physique de cette contrainte qui n'apparaît pas spécifiquement comme une équation de mouvement. Compte-tenu de la formule~\refe{eq:deuxieme_contreterme}, nous avons~:

\begin{align}\label{eq:beta_susy_1}
&\beta_{\pm} = \parent{-2r \delta r - \Delta} \lambda^\pm  \nonumber \\ 
&\beta_{\delta r} = -\Delta \delta r - 2r \parent{1+2k^2} \lambda^+\lambda^-  
\end{align}

avec $k^2$ la valeur propre de $-\Delta \lambda^+$. Par comparaison aux fonctions bêta correspondantes~\refe{eq:beta_wilson} dans le cas bosonique, nous obtenons une formule \`a peine différente. Toutefois, elles ont les mêmes solutions et~\refe{eq:beta_susy_1} peut se réécrire sous la forme~:

\begin{align}
&\beta_{\pm} = \parent{-2r \delta r - \Delta} \lambda^\pm  \nonumber \\ 
&\beta_{\delta r} = -\Delta \delta r - 2r\lambda^+\lambda^-  + r  \parent{ \lambda^+ \Delta\lambda^- + \lambda^- \Delta\lambda^+}
\end{align}

Soit à peu de chose près et au deuxième ordre~:

\begin{align}
&\beta_{\pm} = \parent{-2r \delta r - \Delta} \lambda^\pm  \nonumber \\ 
&\beta_{\delta r} = -\Delta \delta r - 2r\lambda^+\lambda^-  + r \parent{\lambda^+ \beta_- + \lambda^- \beta_+}
\end{align}

Par conséquent, à des fonctions bêta près\footnote{Soit des termes nuls par relation \`a des \'equations du mouvement.} les équations sont exactement semblables dans ce cas et dans le cas bosonique.  Maintenant, nous pouvons comme dans le mod\`ele bosonique imposer un ansatz plus général~:

\begin{align}
T^+ &= e^{ir \wt{\mathbb X}} \parent{\zeta_{(1)}(\mathbb X^i) e^{i\omega \mathbb X^0} + \zeta_{(2)}^*(\mathbb X^i) e^{- i\omega \mathbb X^0}} \nonumber \\ 
T^- &= (T^+)^*
\end{align}

avec $\omega^2=r^2-1/2$ telle que la déformation correspondante est marginale. En remarquant que seuls les termes croisés contribuent aux fonctions bêta nous obtenons~:

\begin{align}\label{eq:beta_onshell_super}
& \beta_{\delta r} = \Delta \delta r - 2 r \parent{\module{\zeta_{(1)}}^2 + \module{\zeta_{(2)}}^2} + \ldots \nonumber \\
& \beta_{{(1,2)}} = \Delta \zeta_{(1,2)} - 2r \delta r \, \zeta_{(1,2)}  
\end{align}

Elles sont exprimées également à des fonctions bêta près. A ce stade les tachyons peuvent \^etre red\'efinis par une constante dépendant de $r$ puisque la distance est constante $\zeta \to \sqrt{f(r)} \zeta$. La forme des \'equations nous suggère les équations du mouvement suivantes. En notant $\phi = r+ \delta r$ et $T^+=T=(T^-)^*$ et en rétablissant la covariance~:

\begin{align}\label{eq:mouvement_massif_super}
\frac{\delta {\mathcal L}}{\delta \phi} &= - \square \phi - \frac{\phi f(\phi)}{\frac{1}{2}-\phi^2} \parent{\partial_a T \partial^a T^* + \parent{\frac{1}{2}-\phi^2} \module{T}^2} \nonumber \\ 
\frac{\delta {\mathcal L}}{\delta T^*} &= - \square T + \parent{\frac{1}{2}-\phi^2} T  \nonumber \\ 
\frac{\delta {\mathcal L}}{\delta T} &= - \square T^* + \parent{\frac{1}{2}-\phi^2} T^*
\end{align}

Comme dans le mod\`ele bosonique, il n'existe pas d'action effective correspondant à ces équations du mouvement quadratiques. Toutefois, la solution de ces \'equations \`a distance constante reste compatible avec la solution \`a distance constante d\'eriv\'ee de l'action de Garousi d\'evelopp\'ee \`a l'ordre quadratique~\refe{eq:quad}, \cad $T=0$. En fait, elles peuvent \^etre r\'e\'ecrites en utilisant les \'equations de $T$ et $T^*$ sous la forme~:

\begin{align}\label{eq:comp_eq_Garousi_1}
\frac{\delta {\mathcal L}}{\delta \phi} &= - \square \phi - \phi f(\phi) \, \module{T}^2 -  \frac{\phi f(\phi)}{1-2\phi^2} \partial_a\partial^a \module{T}^2  \nonumber \\ 
\frac{\delta {\mathcal L}}{\delta T^*} &= - \square T + \parent{\frac{1}{2}-\phi^2} T  \nonumber \\ 
\frac{\delta {\mathcal L}}{\delta T} &= - \square T^* + \parent{\frac{1}{2}-\phi^2} T^*
\end{align}

La d\'ependance en $\zeta_{(1)}$ et $\zeta_{(2)}$ dans~\refe{eq:beta_onshell_super} sugg\`ere que les deux solutions $\zeta_{(1)}e^{i\omega x^0}$ et $\zeta_{(2)}e^{-i\omega x^0}$ sont ind\'ependantes \`a cet ordre. Par cons\'equent, pour r\'esoudre les \'equations~\refe{eq:mouvement_massif_super} il faudrait imposer l'ansatz $T= \tau_{\vec k} e^{i k^\nu x_\mu}$ qui v\'erifie $\partial_a\module{T}^2=0$. Le long de cet ansatz, ces \'equations sont compatibles avec celles d\'eriv\'ees depuis l'action quadratique~\refe{eq:quad}~:

\begin{align}\label{eq:comp_eq_Garousi_2}
\frac{\delta {\mathcal L}}{\delta \phi} &= - \square \phi - 2\phi\, \module{T}^2  \nonumber \\ 
\frac{\delta {\mathcal L}}{\delta T^*} &= - \square T + \parent{\frac{1}{2}-\phi^2} T  \nonumber \\ 
\frac{\delta {\mathcal L}}{\delta T} &= - \square T^* + \parent{\frac{1}{2}-\phi^2} T^*
\end{align}

A condition que $f(\phi)=2$. Ceci indique que l'action de Garousi est valide, au moins \`a l'ordre quadratique, pour tout champ de tachyon dans la phase sur-critique. Il faudrait \'evidemment v\'erifier que les modifications des fonctions bêta des tachyons dans~\refe{eq:beta_onshell_super} \`a des ordres sup\'erieurs sont compatibles avec les non-lin\'earit\'es des \'equations de l'action Garousi. En particulier, la propri\'et\'e d'ind\'ependance entre les solutions $\zeta_{(1)}e^{i\omega x^0}$ et $\zeta_{(2)}e^{-i\omega x^0}$ est forte et probablement fausse aux ordres sup\'erieurs. Toutefois, la fa\c con dont l'action de Garousi \`a \'et\'e d\'eriv\'ee et le fait qu'elle ai \'et\'e v\'erifi\'ee par comparaison \`a des \'el\'ements de matrice-S sugg\`ere qu'elle doit \^etre valide dans la phase surcritique. C'est ce que nous devrions obtenir en prolongeant l'\'etude aux ordres sup\'erieurs.

\subsection{Phase sous-critique $r<r_c$}

L'extension du modèle off-shell précédent au domaine sous-critique s'inspire de l'\'etude bosonique du chapitre pr\'ec\'edent, section~\refcc{sec:renorm_susy}. Puisqu'il faut des déformations de bord marginales au premier ordre à l'action, l'expression en super-espace doit \^etre~:

\begin{multline}
S = S_{bulk} -  \oint \di z \di \theta~ \Gam^+  \parent{D + i \delta r(\mathbb X^a)D\wt{\mathbb X}} \Gam^- \\ -i \oint \di z \di \theta~\lambda^+(\mathbb X^i) \Gam^+ e^{ir\widetilde{\mathbb X}+\omega \mathbb X^0} \mathbb T^+ -i \oint  \di z \di \theta~ \lambda^-(\mathbb X^i)\Gam^- \mathbb T^- e^{-ir\widetilde{\mathbb X}+\omega \mathbb X^0}
\end{multline}

avec cette fois-ci $\omega^2 = 1/2-r^2$. Les couplages sont d\'evelopp\'es \'egalement selon~:

\begin{align}
\delta r(\mathbb X^i) & = \delta r + \partial_i \delta r \parent{\hat X^i + i\theta \psi^i} + \frac{\partial_i\partial_j \delta r}{2}\parent{\hat X^i\hat X^j + 2 i \theta \hat X^i \psi^j} \nonumber \\ 
\lambda^\pm(\mathbb X^i) &= \lambda^\pm + \partial_i\lambda^\pm \parent{\hat X^i + i\theta \psi^i} + \frac{\partial_i\partial_j \lambda^\pm}{2}\parent{\hat X^i\hat X^j + 2 i\theta \hat X^i \psi^j} 
\end{align}

Par comparaison au d\'eveloppement précédent, il suffit ici de changer $\psi^\pm$ en $\psi^\pm = \pm ir\wt \psi +\omega \psi^0$. Mais~\refe{eq:OPE_ferm} devient maintenant~:

\begin{align}
\psi^+(z) \psi^-(0) \sim \frac{4r^2-1}{z}
\end{align}

Pour bien clarifier les choses il est préférable d'exprimer les parties "non-contact" et "termes-de-contact" de l'action décomposée séparément. Pour la première, nous avons~:

\begin{multline}
S = S_{bulk} + \sigma^3 \otimes \frac{i}{2} \oint \parent{\delta r \partial \wt X  + \partial_i \delta r \, (\hat X^i \partial \wt X - \psi^i \widetilde \psi) + \frac{\partial_i\partial_j \delta r}{2} (\hat X^i \hat X^j \partial \wt X - 2 \hat X^j \psi^i \widetilde \psi)} \\
+ \sigma^+ \otimes  \oint \Bigg( \lambda^+ \psi^+ + \partial_i \lambda^+ (\hat X^i \psi^+ + \psi^i)  + \frac{\partial_i\partial_j \lambda^+}{2}  \parent{\hat X^i \hat X^j \psi^+ + 2 \hat X^j \psi^i } \Bigg) e^{i r \wt X+\omega X^0} \\ 
+ \sigma^- \otimes  \oint \Bigg(\lambda^-\psi^- + \partial_i \lambda^- (\hat X^i \psi^- + \psi^i)  + \frac{\partial_i\partial_j \lambda^-}{2} \parent{\hat X^i \hat X^j \psi^- + 2 \hat X^j \psi^i }\Bigg) e^{-i r \wt X+\omega X^0}  
\end{multline}

Tandis que la partie relevant des termes de contact exclusivement est~:

\begin{multline}
S_{contact} = - \frac{2}{\varepsilon}\croch{\delta r^2 - 2  \partial_i\delta r\partial^i \delta r  ~ \ln \varepsilon + \Bigg(\partial_i (\delta r^2)  - 4  \partial_j \delta r \partial^j\partial_i \delta r ~\ln \varepsilon \Bigg) \hat X^i + \frac{\partial_i \partial_j(\delta r^2)}{2} \hat X^i\hat X^j}    \\ 
+ \varepsilon^{1-4r^2} \Bigg[\lambda^+\lambda^- - 2 \partial_i\lambda^+\partial^i\lambda^- ~\ln\varepsilon  + \Bigg(\partial_i(\lambda^+\lambda^-) - 2 \partial_i\parent{\partial_j \lambda^- \partial^j \lambda^+}\ln \varepsilon  \Bigg) \hat X^i \\ + \frac{\partial_i\partial_j(\lambda^+\lambda^-)}{2}\hat X^i\hat X^j \Bigg]e^{2\omega X^0} \\ 
+  \eta^+ \psi^+ e^{ir \wt X+\omega X^0}\Bigg[ \lambda^+\delta r - 2 \partial_i\lambda^+\partial^i\delta r ~\ln\varepsilon  + \Bigg(\partial_i(\lambda^+ \delta r) - 2  \partial_i\parent{\partial_j \delta r \partial^j \lambda^+}\ln \varepsilon \Bigg) \hat X^i \\  + \frac{\partial_i\partial_j(\lambda^+\delta r)}{2} \hat X^i\hat X^j \Bigg] \\ 
+ \eta^- \psi^- e^{-ir \wt X+\omega X^0}\Bigg[ \lambda^-\delta r - 2 \partial_i\lambda^-\partial^i\delta r ~\ln\varepsilon  + \Bigg(\partial_i (\lambda^- \delta r)  - 2 \partial_i\parent{\partial_j \delta r \partial^j \lambda^-}\ln \varepsilon \Bigg) \hat X^i \\  + \frac{\partial_i\partial_j(\lambda^-\delta r)}{2} \hat X^i\hat X^j \Bigg]
\end{multline}

Les OPE impliquant le produit de deux tachyons ne peuvent clairement pas donner de terme proportionnel à l'opérateur $\partial \wt X$ à cause du facteur $e^{2\omega X^0}$. D'autant que la dépendance UV associée sera d'ordre $\varepsilon^{4(1/2-r^2)} \to 0$. Par conséquent, similairement au cas bosonique, nous ne nous intéresserons qu'au premier terme du développement~:

\begin{align}\label{eq:OPE_tach_tach_souscrit}
&\psi^+ e^{i r \wt X+\omega X^0}(z)  \cdot  \psi^- e^{-i r \wt X+\omega X^0}(w) = \frac{4r^2-1}{(z-w)^{4r^2}} e^{2\omega X^0}(w) + \ldots \nonumber \\ 
& (\hat X^i \psi^+ + \psi^i)e^{i r \wt X+\omega X^0}(z) \cdot \psi^- e^{-i r \wt X+\omega X^0}(w) = \frac{(4r^2-1)\hat X^i}{(z-w)^{4r^2}} e^{2\omega X^0}(w) + \ldots\nonumber \\ 
& (\hat X^i \psi^+ + \psi^i)e^{i r \wt X+\omega X^0}(z) \cdot (\hat X^j \psi^- + \psi^j)e^{-i r \wt X+\omega X^0}(w) \nonumber \\ 
& \qquad = \frac{1}{(z-w)^{4r^2}} \parent{(4r^2-1)\hat X^i \hat X^j + 2 \eta^{ij} } e^{2\omega X^0}(w) + \ldots \nonumber \\ 
& \parent{\hat X^i \hat X^j \psi^+ + 2 \hat X^j \psi^i } e^{i r \wt X+\omega X^0}(z)  \cdot \psi^- e^{-i r \wt X+\omega X^0}(w) \nonumber \\ 
& \qquad = \frac{4r^2-1}{(z-w)^{4r^2}} \parent{\hat X^i \hat X^j } e^{2\omega X^0}(w) + \ldots \nonumber \\ 
&\parent{\hat X^i \hat X^j \psi^+ + 2 \hat X^j \psi^i } e^{i r \wt X+\omega X^0}(z) \cdot   (\hat X^k \psi^- + \psi^k)e^{-i r \wt X+\omega X^0}(w) \nonumber \\ 
&\qquad = \frac{2}{(z-w)^{4r^2}} \parent{2\eta^{ik} \hat X^j + \ldots}e^{2\omega X^0}(w)  + \ldots \nonumber \\ 
&\parent{\hat X^i \hat X^j \psi^+ + 2 \hat X^j \psi^i } e^{i r \wt X+\omega X^0}(z) \cdot \parent{\hat X^k \hat X^l \psi^- + 2 \hat X^l \psi^k } e^{-i r \wt X+\omega X^0}(z)  \nonumber \\ 
&\qquad = \frac{2}{(z-w)^{4r^2}} \parent{ 4 \eta^{ik}\hat X^j \hat X^l  + \ldots} e^{2\omega X^0}(w)  + \ldots \nonumber \\ 
\end{align}

Tous les termes à plus de deux indices dans les $\hat X^i$ et les termes proportionnels à $\ln\varepsilon$ sont de nouveaux cach\'es dans les pointill\'es. Après intégration ces derniers sont compens\'es exactement par les contretermes. Puis pour les OPE entre le champ de distance et le tachyon nous aurons~:

\begin{align}\label{eq:OPE_dist_tach_souscrit}
 & \partial \wt X(z)  \cdot  \psi^\pm e^{\pm i r \wt X+\omega X^0}(w)= \frac{\mp 2 i r}{z-w} \psi^\pm e^{\pm ir \wt X+\omega X^0}(w) + \ldots \nonumber \\ 
 &\partial \wt X(z) \cdot (\hat X^i \psi^\pm + \psi^i)e^{\pm i r \wt X+\omega X^0} (w)=  \frac{\mp 2 i r}{z-w} (\hat X^i \psi^\pm + \psi^i)e^{\pm i r \wt X+\omega X^0} (w) + \ldots \nonumber \\ 
 &\partial \wt X(z) \cdot \parent{\hat X^i \hat X^j \psi^\pm + 2 \hat X^j \psi^i } e^{\pm i r \wt X+\omega X^0}(w) =  \frac{\mp 2 i r}{z-w} \parent{\hat X^i \hat X^j \psi^\pm + 2 \hat X^j \psi^i } e^{\pm i r \wt X+\omega X^0}(w)+ \ldots \nonumber \\ 
 &(\hat X^i \partial \wt X - \psi^i \widetilde \psi)(z) \cdot \psi^\pm e^{\pm ir\wt X+\omega X^0}(w)=  \frac{\mp 2 i r}{z-w}  ( \hat X^i \psi^\pm + \psi^i) e^{\pm ir \wt X+\omega X^0}(w) + \ldots  \nonumber \\ 
 & (\hat X^i \partial \wt X - \psi^i \widetilde \psi)(z) \cdot (\hat X^j \psi^\pm + \psi^j)e^{\pm i r \wt X+\omega X^0} (w) \nonumber \\ &\qquad = \frac{\mp 2 i r}{z-w} \Bigg[\parent{\hat X^i \hat X^j \psi^\pm + X^{(i}\psi^{j)}} \mp \frac{\eta^{ij}}{i r} \widetilde\psi + \ldots \Bigg]e^ab{\pm ir \wt X+\omega X^0} \nonumber \\ 
 & (\hat X^i \partial \wt X - \psi^i \widetilde \psi)(z) \cdot \parent{\hat X^j \hat X^k \psi^\pm + 2 \hat X^k \psi^j } e^{\pm i r \wt X+\omega X^0}(w) =\frac{\mp 2}{z-w} \parent{\eta^{ik} \hat X^j\widetilde \psi + \ldots}e^{\pm i r \wt X+\omega X^0}  \nonumber \\
 & (\hat X^i \hat X^j \partial \wt X - 2 \hat X^j \psi^i \widetilde \psi)(z) \cdot \psi^\pm e^{\pm i r \wt X+\omega X^0}(w)= \frac{\mp 2i r}{z-w} \parent{\hat X^i \hat X^j \psi^\pm + 2 \hat X^j \psi^i}e^{\pm ir \wt X+\omega X^0} \nonumber \\ 
  &  (\hat X^i \hat X^j \partial \wt X - 2 \hat X^j \psi^i \widetilde \psi)(z) \cdot (\hat X^k \psi^\pm + \psi^k)e^{\pm i r \wt X+\omega X^0} (w) = \frac{2}{z-w} \parent{2 \eta^{ik} \hat X^j \widetilde \psi + \ldots} e^{\pm i r \wt X+\omega X^0} \nonumber \\ 
  & (\hat X^i \hat X^j \partial \wt X - 2 \hat X^j \psi^i \widetilde \psi)(z)\cdot \parent{\hat X^k \hat X^l \psi^\pm + 2 \hat X^l \psi^k } e^{\pm i r \wt X+\omega X^0}(w) \nonumber \\ & \qquad = \frac{2 }{z-w} \parent{4\eta^{ik} \hat X^j \hat X^l \widetilde \psi + \ldots }e^{\pm i r \wt X+\omega X^0} 
\end{align}

Nous obtenons donc un certain nombre de termes susceptibles de participer aux fonctions bêta de $\lambda^\pm$ mais clairement aucun pour celle de $\delta r$. Nous avons également des contraintes afin de garantir la supersymétrie de surface des amplitudes. Ces contraintes sont toutes les dérivées de l'équation suivante, à cause des termes d'ordre supérieur dans~\refe{eq:OPE_tach_tach_souscrit} qui sont à peu de choses près les mêmes que dans le cas surcritique, ce qui explique que nous ne les avons pas explicitement écrits~:

\begin{align}\label{eq:contrainte_tach}
\partial_k \Bigg[\partial_i\lambda^+ \partial^i\lambda^- + \parent{2r^2-1} \lambda^+\lambda^-\Bigg] = 0 \quad \Leftrightarrow \quad \partial_i\lambda^+ \partial^i\lambda^- \propto \lambda^+\lambda^-
\end{align}

Cette équation est bien vérifiée par la solution de tachyon roulant par exemple $\lambda^\pm \propto e^{\omega x^0}$ avec $\omega^2=1/2-r^2$ puisque la dérivée est uniquement spatiale. En fait, comme précédemment cette équation est en général vérifiée puisque les solutions les plus évidentes sont du type $e^{i k_a x^a}$ avec $k_0$ réel ou imaginaire. La contrainte de réalité du produit $\lambda^+\lambda^-$ suffit alors à extraire toute contribution spatiale, de telle sorte que $\partial_i(\lambda^+\lambda^-)=0$. Les fonctions bêta des tachyons et de la perturbation de distance obtenues sont~:

\begin{align}
&\beta_{\pm} = - \parent{ 2r \delta r + \Delta} \lambda^\pm  \nonumber \\ 
&\beta_{\delta r} = - \Delta \delta r 
\end{align}

Ce résultat est attendu, car la perturbation en $\sigma^3\otimes \oint \partial \wt X$ n'est pas produite par OPE des tachyons, ce qui appara\^it bien dans~\refe{eq:OPE_tach_tach_souscrit}. Ainsi nous obtenons le même résultat que dans le cas bosonique. 

Plaçons nous de nouveau le long de l'ansatz général~:

\begin{align}
T^+ &= e^{ir \wt{\mathbb X}}\parent{\zeta_{(1)}(\mathbb X^i) e^{\omega \mathbb X^0} + \zeta_{(2)}(\mathbb X^i) e^{-\omega \mathbb X^0}} \nonumber \\
T^-&=(T^+)^*
\end{align}

Par OPE de ces tachyons, des contributions à la fonction bêta de $\delta r$ seront produites. Pour $\omega^2=1/2-r^2$ nous obtenons~:

\begin{align}\label{eq:beta_sous-crit_susy}
&\beta_{\delta r} = -\Delta \delta r - 2 r \parent{ \zeta_{(1)}\zeta_{(2)}^* + \zeta_{(2)}\zeta_{(1)}^*  } \nonumber \\
& \beta_{{(1,2)}} = \Delta \zeta_{(1,2)} - 2r \delta r \, \zeta_{(1,2)}  
\end{align}

Elles sont exprimées à des fonctions bêta près comme dans le cas précédent\footnote{Si on tient bien compte de l'expression complète des fermions $\psi^\pm_{(1,2)}$ on obtient que le produit intervenant est $\psi^+_{(1)}(z)\psi^-_{(2)}(0) \sim 1/z$.} et à une redéfinition du tachyon près à distance constante $\zeta \to \sqrt{f(r)} \zeta$. Cela nous suggère encore les équations du mouvement suivantes, en notant $\phi = r+ \delta r$ et $T^+=T=(T^-)^*$ et en rétablissant la covariance~:

\begin{align}\label{eq:mouvement_tach_super}
\frac{\delta {\mathcal L}}{\delta \phi} &= - \square \phi - \frac{\phi f(\phi)}{\frac{1}{2}-\phi^2} \parent{\partial_a T \partial^a T^* + \parent{\frac{1}{2}-\phi^2} \module{T}^2} \nonumber \\ 
\frac{\delta {\mathcal L}}{\delta T^*} &= - \square T + \parent{\frac{1}{2}-\phi^2} T  \nonumber \\ 
\frac{\delta {\mathcal L}}{\delta T} &= - \square T^* + \parent{\frac{1}{2}-\phi^2} T^*
\end{align}

Naturellement, ces \'equations ne sont pas compatibles avec celles d\'eduites de l'action de Garousi \`a l'ordre quadratique. En effet, dans la section pr\'ec\'edente, les formules~\refe{eq:comp_eq_Garousi_1} et~\refe{eq:comp_eq_Garousi_2} \'etaient compatibles modulo le terme $\partial_a \partial^a \module{T}^2$ qui s'annule le long de l'ansatz $e^{i k_a x^a}$. En revanche ici, ce terme ne s'annule pas puisque le terme de m\'elange entre $\zeta_{(1)}$ et $\zeta_{(2)}$ dans~\refe{eq:beta_sous-crit_susy} sugg\`ere que les deux contributions de l'ansatz sont d\'ependantes~: l'ansatz n'est pas r\'eductible sous la forme $e^{i k_a x^a}$ et par cons\'equent $\partial_a \partial^a \module{T}^2 \neq 0$. 

Nous pourrons comparer ces équations à celles dérivées à partir d'une action effective obtenue par comparaison au calcul de la fonction de partition, suivant la méthode de Kutasov et Niarchos~\cite{Kutasov:2003er}. Nous présenterons dans la section~\refcc{sec:fonc_part} le calcul de la fonction de partition à distance fixe.

\subsection{Phase critique $r=r_c$}

Le passage de la phase surcritique à la phase sous-critique est discontinue en $r=r_c$, exactement comme dans le cas bosonique et nous renvoyons \`a la discussion de la section~\refcc{sec:phas_crit}. Nous referons cependant le d\'eveloppement en l'adaptant au mod\`ele pr\'esent. A la distance critique, l'opérateur de tachyon $\pm i \wt \psi e^{\pm i  \wt X/\sqrt 2}$ est marginal et on peut réduire la CFT à $c=1$ en se débarrassant de la coordonnée temporelle. En revanche, cet opérateur n'est pas \emph{exactement} marginal à cause de la fonction bêta non nulle de la perturbation de distance. Ceci implique que la brane posée à la distance critique est attirée vers le domaine tachyonique. Le long de l'ansatz $T^\pm = \pm \lambda^\pm(X^a)  (i \wt \psi/\sqrt 2) e^{\pm i \wt X /\sqrt 2}$ les fonctions bêta sont~: 

\begin{align}\label{eq:beta_onshell__crit_super}
& \beta_{\delta r} = \square \delta r - \sqrt 2 \lambda^+\lambda^- + \ldots \nonumber \\
& \beta_{\pm} = \square \lambda^\pm - \sqrt 2\delta r \, \lambda^\pm
\end{align}

La comparaison de la fonction bêta de la perturbation de distance entre les diff\'erentes phases le long de l'ansatz $T= \zeta_{(1)}e^{i\omega x^0} + \zeta_{(2)}e^{-i\omega x^0}$ avec $\omega=\sqrt{1/2-r^2}$ d\'evoile une discontinuit\'e en $r=1/\sqrt 2$ la distance critique~:

\begin{align}
r>\frac{1}{\sqrt 2}~ :\quad & \beta_{\delta r} = \Delta \delta r - 2 r \parent{\module{\zeta_{(1)}}^2 + \module{\zeta_{(2)}}^2} + \ldots \nonumber \\
r=\frac{1}{\sqrt 2}~ :\quad & \beta_{\delta r} =  - \square \delta r  - 2 \module{\lambda}^2 \nonumber \\
r<\frac{1}{\sqrt 2}~ :\quad & \beta_{\delta r} = \Delta \delta r - 2 r \, \parent{\zeta_{(1)}\zeta_{(2)}^*+\zeta_{(1)}^*\zeta_{(2)} } + \ldots  \nonumber \\
\end{align}

En $r=1/\sqrt 2$ les couplages sont $\lambda^\pm = \zeta_{(1)} + \zeta_{(2)}$ par continuité dans la d\'efinition des d\'eformations de bord. Ainsi, à l'instar du modèle bosonique, la transition surcritique/sous-critique est une transition de phase~: en omettant le terme cin\'etique, la fonction bêta change de signe, car $\module{\zeta_{(1)}}^2 + \module{\zeta_{(2)}}^2>0$ et $\zeta_{(1)}\zeta_{(2)}^*+\zeta_{(1)}^*\zeta_{(2)}<0$. La fonction bêta en la distance critique interpole entre ces deux expressions. 

Cette discontinuit\'e en $r=r_c$ exprime \'egalement le fait qu'une th\'eorie des champs en $r=r_c$ n'est pas bien définie, parce que d'un côté nous avons une phase stable et de l'autre une phase instable. En outre, \`a propos de la transition sous-critique/critique, d'après notre étude sur les divergences du modèle roulant en fonction de la distance, le nombre de contretermes de type terme-de-contact à ajouter à l'action de surface tend à exploser en $r_c^-$. Ainsi, dans cette limite, la théorie est non-renormalisable, mais pas en $r=r_c$ comme nous le voyons bien. Il s'agirait d'un moyen pour le système de mener à la discontinuité, ce qui est réminiscent de la limite $c\to 1$ dans les théories de Liouville~\cite{Runkel:2001ng,Fredenhagen:2004cj,Schomerus:2003vv}. Nous savons par exemple que la continuation analytique $b\to ib$ (\ie $Q\to 0$ ou $c\to 1$) pour passer de Liouville de genre espace à Liouville de genre temps n'est pas très bien définie. Or notre modèle est très similaire dans la forme à une théorie super-Liouville, mis à part les facteurs de CP. En ce sens, la transition $b\to ib$ semble similaire à la transition $\omega \to 0$ (soit $c=2 \to c=1$) dans l'opérateur du tachyon sur le bord $e^{i\sqrt{1/2-\omega^2}\wt X + \omega X^0}$.

\section{Fonction de partition on-shell du système séparé}
\label{sec:fonc_part}

En BSFT des cordes ouvertes introduite par Witten~\cite{Witten:1992qy,Witten:1992cr} nous avons une relation entre l'action de BSFT -- sur l'espace des théories des champs -- on-shell et la fonction de partition calculé sur le disque $D^2$ qui en théorie supersymétrique apparaît être particulièrement simple~\cite{Takayanagi:2000rz,Kutasov:2000aq,Marino:2001qc} \apriori si on suppose que la matière et les fantômes sont découplés~:

\begin{align}
S[\phi^i_{on}] = Z_{D^2}[\phi^i_{on}]
\end{align}

où l'expression de la fonction de partition est explicitement~:

\begin{align}
Z_{D^2}[\phi^i_{on}] = \tr{\mathcal P} \int [\di X][\di \psi] e^{-S_{bulk}[G_{ab},B_{ab},\Phi] - \sum_i \phi^i_{on} \oint_{S^1}V_i } 
\end{align}

Dans ce contexte, $\phi^i_{on}$ est une valeur constante correspondant à un point fixe du groupe de renormalisation pour le couplage associé à l'opérateur de vertex $\oint V_i$. L'égalité suggère par extraction du mode zéro des champs bosoniques $X^a$ que l'on peut exprimer une densité d'action, \cad un \emph{lagrangien} en fonction en la densité de fonction de partition noté $Z'$ selon~:

\begin{align}\label{eq:action_fonction_part}
\int \di^{p+1} x ~ {\mathcal L}[\varphi^i_{on}(x)] = \int \di^{p+1} x ~ Z'_{D^2}[\varphi^i_{on}(x)]
\end{align}

L'expression de $\varphi^i_{on}(x)$ est simplement donnée par l'identité $\varphi^i_{on}(x^a)= \phi^i_{on} \corr{V_i(X^a)}_0$ avec $\corr{\ldots}_0$ le corrélateur calculé en théorie libre. Par conséquent, si on veut obtenir l'action on-shell le long d'un tachyon roulant~\cite{Kutasov:2003er,Larsen:2002wc} que l'on sait être une solution des équations du mouvement dans le cas d'une seule brane non-BPS ou d'une paire de brane-antibrane séparées ou non, il faut calculer la fonction de partition sur le disque pour laquelle on ajoute une déformation supersymétrique~:

\begin{align}
\delta S = \oint_{S^1} T(X^0)
\end{align}

avec $\tau(x^0) = e^{\omega x^0}$ la forme \emph{supposée} de la solution de tachyon roulant dans l'espace-cible. Sur la paire coïncidente, Kutasov et Niarchos~\cite{Kutasov:2003er} ont pu contraindre suffisamment la forme de l'action effective, en imposant que la solution la plus générale possible est $T^+ e^{x^0/\sqrt{2}}+ T^-e^{- x^0/\sqrt 2}$. En outre, la fonction de partition est parfaitement connue~\cite{Larsen:2002wc} -- et aisée à calculer -- dans ce système précis le long du tachyon roulant de demi S-brane $e^{\omega x^0}$. Parce qu'ils n'avaient ordre par ordre qu'un seul paramètre à fixer, la seule équation~\refe{eq:action_fonction_part} suffisait à totalement déterminer l'action.  

Or pour le cas qui nous intéresse -- brane-antibrane séparées -- ce n'est pas du tout suffisant, car la solution plus générale n'existe que pour une classe de paramètres qui ne permettent pas de réduire les degrés de liberté de l'action suffisamment. Nous obtenons un système clairement sous-contraint. En outre, nous allons voir que la fonction de partition est très compliquée à calculer et que nous n'avons pu connaître pour tout $r$ que les 2 premiers ordre. Nous avons aussi obtenu les 5 premiers ordres en une distance particulière $r=1/2$, pour laquelle l'intégrande se simplifie significativement. Notons que cette distance est précisément celle à partir de laquelle le terme de contact devient important, \cad divergent. Néanmoins en cette valeur il est fini mais non nul et participe donc pleinement au calcul de la fonction de partition.   

Même si l'action n'est pas contrainte suffisamment par ce biais, il est toujours intéressant de voir si la fonction de partition est calculable perturbativement et si le développement converge ou non, bien que cela ne soit valable que pour $x^0 \to -\infty$.

\subsection{Rappels sur le système et présentation des calculs}

Je souhaite rappeler avant de rentrer plus dans les détails de la méthode de calcul. Nous regardons donc un système brane-antibrane à distance fixée que nous avons montré être pour tout $r$ une BCFT. Nous devons calculer la fonction de partition pour ce système et nous partirons de l'action définie sur le super-espace du disque suivante~:

\begin{align}
S = S_{bulk} - \oint \Gam^+ D\Gam^- - i\oint \frac{\lambda^+}{2\pi} \Gam^+ e^{ir\widetilde{\mathbb X} + \omega \mathbb X^0} - i\oint \frac{\lambda^-}{2\pi} \Gam^-  e^{-ir\widetilde{\mathbb X}+ \omega \mathbb X^0}
\end{align}

Ici le tachyon est donc choisi on-shell en une distance constante. Dans le bulk le fond est trivial, \cad que l'espace est plat et que $B_{ab}=0$ et $\Phi = \phi$ constant. Sur le disque unité, nous utiliserons que les variables sont $z=\rho e^{i t}$ avec $\rho<1$ et en particulier sur le bord, donc le long du cercle unité, nous aurons $z=e^{i t}$. La formule que nous utilisons pour calculer la fonction de partition est donc~:

\begin{align}
Z_{D^2}[\lambda^\pm,r] = \int [\di \Gam^+ \di \Gam^-][\di \mathbb X][\di \mathbb X^0] e^{-S_{bulk} - \oint \Gam^+ D\Gam^- - i\oint_{S^1}  \frac{\di z \di \theta}{2\pi}\lambda^+ \Gam^+ e^{ir\widetilde{\mathbb X} + \omega \mathbb X^0} - i\oint_{S^1}  \frac{\di z \di \theta}{2\pi} \lambda^- \Gam^-  e^{-ir\widetilde{\mathbb X}+ \omega \mathbb X^0}} 
\end{align}  

avec $\omega^2=1/2-r^2$. Nous avons intégré les autres champs $\mathbb X^{a \neq 0}$ et champs transverses $\mathbb X^{i}$, puisque comme nous le disions dans le cas bosonique, ceux-ci n'interagissent pas avec $\wt{\mathbb X}$ et $\mathbb X^0$. Nous avons aussi supposé que les fantômes sont découplés et aussi intégrés. 

\subsubsection{Remarque sur les divergences de Möbius}

Ce dernier point n'est pas complètement trivial, car \apriori nous devrions avoir des infinités de (super-)Möbius $SL(2|2,\mathbb R)$ sur le disque qu'il faudrait régulariser et supprimer en fixant un certain nombre d'opérateurs dans le super-espace. En fait, si on suit Tseytlin~\cite{Tseytlin:1987ww,Andreev:1988cb,Liu:1987nz,Andreev:1988bz} cela est finalement totalement relié au groupe de renormalisation et ce que nous avons vu précédemment. En effet, off-shell il n'y a pas de symétrie de Möbius et on-shell pas forcément suffisamment d'opérateurs à fixer surtout si on calcule une fonction de partition ; l'option qui consiste à diviser par le volume du groupe de Möbius peut donc amener à obtenir un résultat nul. Ce serait un non sens en ce qui concerne la fonction de partition. 

Tseytlin montre qu'en imposant un cut-off UV sur les fonctions de Green, la divergence de Möbius est régularisée et apparaît sous la forme d'une divergence linéaire qui peut éventuellement être réabsorbée, par renormalisation, dans les champs comme nous avons pu voir, ou simplement soustrait par une sorte de renormalisation-zeta. En supercorde, ce sont les termes de contact qui jouent ce rôle~\cite{Green:1987qu,Andreev:1988cb} et donc il n'y aurait pas d'infinités super-möbius~\cite{Andreev:1988cb}, \cad que le volume du groupe serait fini et qu'en ce qui concerne au moins les amplitudes de théorie des cordes ouvertes supersymétrique il n'est pas nécessaire de fixer la jauge.   

De ce point de vue, il peut être gênant de constater que le tachyon roulant pour une distance supérieure à $r=1/2$ produit des divergences qui ne sont, en l'état, pas supprimées par des termes de contact. C'est ce que nous voyons quand les termes du type $e^{2n\omega X^0}$ avec $n\geq 2$ sont divergent. Cependant, cela n'est pas relié au groupe de Möbius dont les divergences restent supprimées par les termes de contact proportionnels à l'opérateur unité. 

\subsubsection{Présentation des calculs}

Nous allons maintenant calculer les premiers ordres de la fonction de partition. Avant cela nous allons devoir présenter un certain nombre de règles de calculs diagrammatiques pour traiter les contributions des fermions de bord dont les termes de contact. Mais en premier lieu, nous allons simplement exprimer la fonction de partition développée. Elle est~:

\begin{multline}\label{eq:dev_fonc_part}
Z_{D^2}[\lambda^\pm,r] = \sum_{n=0}^{\infty}  \frac{(-1)^n}{n!} \int [\di \Gam^+ \di \Gam^-][\di \mathbb X][\di \mathbb X^0] e^{-S_{bulk}- \oint \Gam^+ D\Gam^-} \\ \parent{i\oint_{S^1} \frac{\di z \di\theta}{2\pi} \lambda^+ \Gam^+ e^{ir\widetilde{\mathbb X} + \omega \mathbb X^0} + i\oint_{S^1} \frac{\di z\di\theta}{2\pi} \lambda^- \Gam^-  e^{-ir\widetilde{\mathbb X}+ \omega \mathbb X^0}}^n 
\end{multline}

Nous savons que $\corr{\Gam^\pm\Gam^\pm\ldots}$ n'est non nul que s'il y a un nombre égal de $\Gam^+$ et de $\Gam^-$. Enfin, nous savons que la fonction de corrélation est telle que le résultat s'exprime comme une somme de termes du type $\hat \Theta(1,2)\hat \Theta(2,3)\ldots$ qui ordonnent les fermions selon les schémas $(+-+-\ldots)$ ou $(-+-+\ldots)$. La fonction $\hat \Theta(1,2)$ est l'extension supersymétrique de la fonction theta de heaviside sur le bord du disque, \cad $\hat \Theta(1,2) = \theta(t_1-t_2 -\theta_1\theta_2)$ pour $t_i \in [0;2\pi]$. Par anti-symétrie de redéfinition des variables d'intégration -- dans le super-espace -- nous pouvons écrire à partir de~\refe{eq:ordre_chemin}~:

\begin{align}
\corr{\Gam^+(1)\Gam^-(2)\ldots \Gam^+(2n-1)\Gam^-(2n)} = 2 (n!)^2 \hat \Theta(1,2)\hat \Theta(2,3)\ldots\Theta(2n-1,2n)
\end{align}

On regroupe les mesures d'intégration ensemble par commutation de $\di z_i$ et anti-commutation de $\di \theta_i$ avec tout $\Gam^\pm$. Nous devons donc inclure un facteur $(-1)^n$ devant la mesure suivante~:

\begin{align}
\int [\underset{>}{\di \hat t}]_n = \int_0^{2\pi} \prod_{i=1}^n \frac{\di t_i}{2\pi} \di \theta_i \prod_{i=1}^{n-1} \hat \Theta(i,i+1)
\end{align}

avec la règle d'intégration multiple de Fubini-Berezin~:

\begin{align}
\int \di \theta_1\di\theta_2 \di\theta_3 \ldots f_1(\theta_1)f_2(\theta_2)f_3(\theta_3)\ldots = \int\di \theta_1 ~ f_1(\theta_1) \int \di \theta_2~  f_2(\theta_2) \int \di \theta_3 ~ f_3(\theta_3)
\end{align}

Nous avons alors~:

\begin{align}
Z_{D^2}[\lambda^\pm,r] = \sum_{n=0}^{\infty}  \frac{2 (n!)^2}{2n!}  C_{n}^{2n} \int [\underset{>}{\di \hat t}]_{2n} ~  \corr{T^+(1) T^-(2) T^+(3)\ldots}
\end{align}

avec $T^\pm(i) = \lambda^\pm e^{\pm ir\widetilde{\mathbb X} + \omega \mathbb X^0}(z_i)$. Le coefficient combinatoire correspond au nombre de façon d'avoir dans un corrélateur autant de $\Gam^+$ que de $\Gam^-$ à partir du développement~\refe{eq:dev_fonc_part}. Le corrélateur des tachyons n'est pas difficile à calculer en utilisant la règle sur le bord pour $X$ Neumann~:

\begin{align}
\corr{\prod_i^n \bnormal{e^{ik_i \mathbb X}(\hat z_i)}} = \prod_{i<j\leq n} \module{z_i-z_j - i\sqrt{z_iz_j} \theta_i\theta_j}^{2 k_i k_j} \corr{\bnormal{\prod_i e^{ik_i \mathbb X}(\hat z_i)}}
\end{align}

Pour les impulsions réelles, la règle de conservation de l'impulsion impose techniquement $\sum k_i = 0$. En revanche, cela n'est pas nécessaire si on ne choisit pas d'intégrer sur les modes zéro, comme par exemple le long de la coordonnée temporelle $X^0$ où ici $k_0 = -i \omega$ est imaginaire pure. Remarquons que le long de la coordonnée duale $\wt X$, la conservation de l'impulsion est immédiatement vérifiée du fait des contraintes imposées par les corrélateurs des fermions de bord. Cela implique d'ailleurs que la fonction de partition ne dépend que du module carré du tachyon -- rappelons qu'il faut $\lambda^+ =(\lambda^-)^*$ par hermiticité de l'action. De sorte que nous avons~:

\begin{multline}
\corr{T^+(1) T^-(2) T^+(3e^{i n x} )\ldots T^-(2n)} \\ = \parent{\lambda^+\lambda^-}^n \prod_{1 \leq i<j}^n \module{z_{2i-1}-z_{2j-1} - i\sqrt{z_{2i-1}z_{2j-1}} \theta_{2i-1}\theta_{2j-1}} \module{z_{2i}-z_{2j} - i\sqrt{z_{2i}z_{2j}} \theta_{2i}\theta_{2j}} \\ \times \prod_{i,j=1}^n \module{z_{2i-1}-z_{2j} - i\sqrt{z_{2i-1}z_{2j}} \theta_{2i-1}\theta_{2j}}^{1 - 4r^2} ~ e^{2n \omega x^0}
\end{multline}

en appliquant $\corr{X^0}=x^0$ et en utilisant $\omega^2+r^2=1/2$. Par commodité d'écriture, nous introduirons une fonction $\hat S(i,j)$ et une autre $S(i,j)$ symétriques par permutation des arguments, telles que sur le disque~:

\begin{align}
\hat S(i,j) &= \module{z_i-z_j - i\sqrt{z_iz_j} \theta_i\theta_j} \nonumber \\ 
& = \module{2 \sin \frac{t_i-t_j}{2} - \theta_i\theta_j} \nonumber \\ 
& = \module{2 \sin \frac{t_i-t_j}{2}} - \varepsilon(i,j) \theta_i\theta_j
\end{align}

et $S(i,j)= \module{2 \sin \frac{t_i-t_j}{2}}$. La fonction $\varepsilon(i,j)\equiv \varepsilon(t_i-t_j)$ est la fonction signe que nous avons déjà introduit, anti-symétrique par permutation des arguments. Pour la fonction $S$ la symétrie par permutation des arguments est simplement une symétrie $\mathbb Z_2$ sur $t_i-t_j$. De sorte que nous pouvons écrire la fonction de partition sous la forme~:

\begin{align}\label{eq:fonct_part_super_1}
Z_{D^2}[\lambda^\pm,r] = 2 \sum_{n=0}^{\infty} \parent{\lambda^+\lambda^-e^{2\omega x^0}}^n \int [\underset{>}{\di \hat t}]_{2n} ~  \prod_{1 \leq i<j}^n \hat S(2i,2j)\hat S(2i-1,2j-1) \times \prod_{i,j=1}^n \hat S(2i-1,2j)^{1-4r^2}
\end{align}

ou encore~:

\begin{align}
Z_{D^2}[\lambda^\pm,r] = 2 \sum_{n=0}^{\infty} \parent{\lambda^+\lambda^-e^{2\omega x^0}}^n \int [\underset{>}{\di \hat t}]_{2n} ~ \prod_{i,j=1}^n \hat S(i,j) \times \prod_{i,j=1}^n \hat S(2i-1,2j)^{-4r^2}
\end{align}

Nous allons maintenant voir comment calculer les intégrales ou au moins comment les exprimer après intégrations des variables de Grassmann. Nous avons pu calculer les quelques premiers ordres uniquement pour la distance particulière $r=1/2$, où l'on voit dans~\refe{eq:fonct_part_super_1} que le deuxième facteur à la puissance $1-4r^2$ disparaît et qu'il ne reste que $\prod_{i,j=1}^n \hat S(2i-1,2j)$. Malgré cela, à cause de l'ordre d'intégration et de la présence de termes de contact dans les fonction thêta supersymétriques, l'intégration n'en est pas beaucoup plus facilité.

\subsection{Méthode diagrammatique}

Je vais définir une méthode diagrammatique pour réduire les intégrales sur les variables de Grassmann et sur les termes de contact correspondant aux $\delta(i,j)=\delta(t_i-t_j)$ dans les fonctions thêta supersymétriques. Je vais donc introduire des règles et des symboles. Dans~\refe{eq:fonct_part_super_1} nous avons deux ensembles d'indices. L'ensemble des indices pairs noté $J$ et celui des indices impairs noté $K$. Nous voyons que la première sorte de terme, d'exposant $1$ ne mélange pas chacun de ces deux ensembles tandis que, la deuxième sorte de terme, d'exposant $1-4r^2$ mélange les deux types d'indices. C'est la raison pour laquelle l'intégrale est plus complexe que dans le cas où ces termes disparaissent en $r=1/2$. Pour simplifier, mais nous verrons que l'on pourra facilement généraliser après, plaçons nous dans ce cas précis. Alors l'intégrale est simplement~:

\begin{align}\label{eq:fonct_part_super_2}
Z_{D^2}[\lambda^\pm,1/2] = 2 \sum_{n=0}^{\infty} \parent{\lambda^+\lambda^- e^{x^0}}^n \int [\underset{>}{\di \hat t}]_{2n} ~  \prod_{1 \leq i<j}^n \hat S(2i,2j)\hat S(2i-1,2j-1)
\end{align}

Deux remarques. Premièrement, car nous avons à intégrer sur les variables de Grassmann, chacune doit apparaître une et une seule fois. Ensuite, si un terme de contact $\delta(i,j)$ et une fonction $S(i,j)$ apparaissent simultanément et contiennent les mêmes variables l'intégrande s'annule car $S(i,i)=0$. \\

Maintenant, voici les règles diagrammatiques~: \\

\begin{itemize}\itemsep4pt
\item Chaque indice $i$ correspond à un vertex.
\item On sépare les indices $J$ des indices $K$ par deux colonnes distinctes.
\item Une ligne nue représente une fonction $S(i,j)$ et une ligne tiretée une fonction $\Theta(i,j)$.
\item Une ligne tiretée avec une flèche allant de $i$ vers $i+1$ représente $-\delta(i,i+1)\theta_i\theta_{i+1}$.
\item Une ligne pleine avec une flèche allant de $i$ vers $j<i$ représente $-\theta_i\theta_j \epsilon(i,j)$. \\
\end{itemize}

Il faut aussi appliquer la contrainte suivante~: \\

\begin{itemize}\itemsep4pt
\item Tout vertex doit être pointé par ou doit pointer une et une seule flèche.\\
\end{itemize}

Prenons un exemple avec $J=\{1,3,5\}$ et $K=\{2,4,6\}$. L'intégrale à calculer est explicitement~:

\begin{multline}
I = \int [\underset{>}{\di \hat t}]_{6} \;  \hat S(1,3)\hat S(1,5) \hat S(3,5)\hat S(2,4)\hat S(2,6) \hat S(4,6) \\ 
 = \int [\di t]_{6} [\di \theta]_6 \prod_{i=1}^{5} \Bigg(\theta(i,i+1)-\theta_i\theta_{i+1} \delta(i,i+1)\Bigg) \\ \times \prod_{1\leq i<j}^3 \Bigg(S(2i-1,2j-1) - \varepsilon(2i-1,2j-1) \theta_{2i-1} \theta_{2j-1}\Bigg) \Bigg(S(2i,2j) - \varepsilon(2i,2j) \theta_{2i} \theta_{2j}\Bigg) 
\end{multline}

En développant et en appliquant les règles précédemment nommées, nous trouverons entre autres les deux diagrammes suivant~:

\begin{minipage}{0.15\linewidth}
\vspace{1cm}
\shorthandoff{:}
\begin{tikzpicture}
	[vertex/.style={circle,draw=black,fill=black,inner sep=.5mm},
	midarrow/.style={postaction={decorate,decoration={markings,mark=at position .5 with {\arrow[black,line width=0.5mm]{stealth}}}}}]
	\node at (0,1) (1) [vertex,label=left:$1$]{} ;
	\node at (0,0) (3) [vertex,label=left:$3$]{} ;
	\node at (0,-1) (5) [vertex,label=left:$5$]{} ;
	\node at (1,1) (2) [vertex,label=right:$2$]{} ;
	\node at (1,0) (4) [vertex,label=right:$4$]{} ;
	\node at (1,-1) (6) [vertex,label=right:$6$]{} ;
	\draw [midarrow,dashed] (1)--(2);
	\draw [midarrow,dashed] (3)--(4);
	\draw [midarrow,dashed] (5)--(6);
	\draw [dashed] (2)--(3) ; 
	\draw [dashed] (4)--(5) ; 
	\draw (1)--(3) ; 
	\draw (1) to [bend right=60](5) ; 
	\draw (3)--(5) ; 
	\draw (2)--(4) ; 
	\draw (2) to [bend left=60](6) ;
	\draw (4)--(6) ; 
\end{tikzpicture}
\shorthandon{:}
\end{minipage}
\hfill
\begin{minipage}{0.85\linewidth}
\begin{multline}
=(-1)^3 \int [\di t]_6[\di\theta]_6  \theta_1\theta_2\theta_3\theta_4\theta_5\theta_6 \, \delta(1,2)\delta(3,4)\delta(5,6) \, \Theta(2,3)\Theta(4,5) \\ \times S(1,3)S(1,5)S(3,5)S(2,4)S(2,6)S(4,6) \nonumber
\end{multline}
\end{minipage}

\begin{minipage}{0.15\linewidth}
\vspace{1cm}
\shorthandoff{:}
\begin{tikzpicture}
	[vertex/.style={circle,draw=black,fill=black,inner sep=.5mm},
	midarrow/.style={postaction={decorate,decoration={markings,mark=at position .5 with {\arrow[black,line width=0.5mm]{stealth}}}}}]
	\node at (0,1) (1) [vertex,label=left:$1$]{} ;
	\node at (0,0) (3) [vertex,label=left:$3$]{} ;
	\node at (0,-1) (5) [vertex,label=left:$5$]{} ;
	\node at (1,1) (2) [vertex,label=right:$2$]{} ;
	\node at (1,0) (4) [vertex,label=right:$4$]{} ;
	\node at (1,-1) (6) [vertex,label=right:$6$]{} ;
	\draw [midarrow] (3)--(1);
	\draw [midarrow] (4)--(2);
	\draw [midarrow,dashed] (5)--(6);
	\draw [dashed] (1)--(2) ;
	\draw [dashed] (2)--(3) ; 
	\draw [dashed] (4)--(5) ; 
	\draw [dashed] (3)--(4) ; 
	\draw (1) to [bend right=60](5) ; 
	\draw (3)--(5) ; 
	\draw (2) to [bend left=60](6) ;
	\draw (4)--(6) ; 
\end{tikzpicture}
\shorthandon{:}
\vspace{1cm}
\end{minipage} \, 
\hfill
\begin{minipage}{0.90\linewidth}
\begin{multline}
= (-1)^3 \int [\di t]_6[\di \theta]_6 \theta_1\theta_3\theta_2\theta_4\theta_5\theta_6 \, \delta(5,6) \, \Theta(1,2)\Theta(2,3)\Theta(3,4)\Theta(4,5) \\ \times \, \varepsilon(1,3)\varepsilon(2,4) ~ S(1,5)S(3,5)S(2,6)S(4,6)\nonumber
\end{multline}
\end{minipage}

En suivant un schéma de construction identiques, nous aurons aussi les diagrammes suivant~:

\begin{minipage}{0.15\linewidth}
\vspace{1cm}
\shorthandoff{:}
\begin{tikzpicture}
	[vertex/.style={circle,draw=black,fill=black,inner sep=.5mm},
	midarrow/.style={postaction={decorate,decoration={markings,mark=at position .5 with {\arrow[black,line width=0.5mm]{stealth}}}}}]
	\node at (0,1) (1) [vertex,label=left:$1$]{} ;
	\node at (0,0) (3) [vertex,label=left:$3$]{} ;
	\node at (0,-1) (5) [vertex,label=left:$5$]{} ;
	\node at (1,1) (2) [vertex,label=right:$2$]{} ;
	\node at (1,0) (4) [vertex,label=right:$4$]{} ;
	\node at (1,-1) (6) [vertex,label=right:$6$]{} ;
	\draw [midarrow] (5)--(3);
	\draw [midarrow] (6)--(4);
	\draw [midarrow,dashed] (1)--(2);
	\draw [dashed] (2)--(3) ;
	\draw [dashed] (3)--(4) ; 
	\draw [dashed] (4)--(5) ; 
	\draw [dashed] (5)--(6) ; 
	\draw (1)--(3) ;
	\draw (1) to [bend right=60](5) ; 
	\draw (2)--(4) ; 
	\draw (2) to [bend left=60](6) ;
\end{tikzpicture}
\shorthandon{:}
\end{minipage}
\hfill
\begin{minipage}{0.15\linewidth}
\vspace{1cm}
\shorthandoff{:}
\begin{tikzpicture}
	[vertex/.style={circle,draw=black,fill=black,inner sep=.5mm},
	midarrow/.style={postaction={decorate,decoration={markings,mark=at position .5 with {\arrow[black,line width=0.5mm]{stealth}}}}}]
	\node at (0,1) (1) [vertex,label=left:$1$]{} ;
	\node at (0,0) (3) [vertex,label=left:$3$]{} ;
	\node at (0,-1) (5) [vertex,label=left:$5$]{} ;
	\node at (1,1) (2) [vertex,label=right:$2$]{} ;
	\node at (1,0) (4) [vertex,label=right:$4$]{} ;
	\node at (1,-1) (6) [vertex,label=right:$6$]{} ;
	\draw [midarrow] (5)to [bend left=60](1);
	\draw [midarrow] (6)to [bend right=60](2);
	\draw [midarrow,dashed] (3)--(4);
	\draw [dashed] (1)--(2) ; 	
	\draw [dashed] (2)--(3) ; 
	\draw [dashed] (4)--(5) ; 	
	\draw [dashed] (5)--(6) ;
	\draw (1)--(3) ; 
	\draw (3)--(5) ; 
	\draw (2)--(4) ; 
	\draw (4)--(6) ; 
\end{tikzpicture}
\shorthandon{:}
\end{minipage}
\hfill
\begin{minipage}{0.15\linewidth}
\vspace{1cm}
\shorthandoff{:}
\begin{tikzpicture}
	[vertex/.style={circle,draw=black,fill=black,inner sep=.5mm},
	midarrow/.style={postaction={decorate,decoration={markings,mark=at position .5 with {\arrow[black,line width=0.5mm]{stealth}}}}}]
	\node at (0,1) (1) [vertex,label=left:$1$]{} ;
	\node at (0,0) (3) [vertex,label=left:$3$]{} ;
	\node at (0,-1) (5) [vertex,label=left:$5$]{} ;
	\node at (1,1) (2) [vertex,label=right:$2$]{} ;
	\node at (1,0) (4) [vertex,label=right:$4$]{} ;
	\node at (1,-1) (6) [vertex,label=right:$6$]{} ;
	\draw [midarrow,dashed] (2)--(3);
	\draw [midarrow] (5)to [bend left=60](1);
	\draw [midarrow] (6)--(4);
	\draw [dashed] (1)--(2) ; 	
	\draw [dashed] (3)--(4) ; 
	\draw [dashed] (4)--(5) ; 	
	\draw [dashed] (5)--(6) ; 
	\draw (1)--(3) ; 
	\draw (3)--(5) ; 
	\draw (2)--(4) ;
	\draw (2)to [bend left =60](6) ; 
\end{tikzpicture}
\shorthandon{:}
\end{minipage}
\hfill
\begin{minipage}{0.15\linewidth}
\vspace{1cm}
\shorthandoff{:}
\begin{tikzpicture}
	[vertex/.style={circle,draw=black,fill=black,inner sep=.5mm},
	midarrow/.style={postaction={decorate,decoration={markings,mark=at position .5 with {\arrow[black,line width=0.5mm]{stealth}}}}}]
	\node at (0,1) (1) [vertex,label=left:$1$]{} ;
	\node at (0,0) (3) [vertex,label=left:$3$]{} ;
	\node at (0,-1) (5) [vertex,label=left:$5$]{} ;
	\node at (1,1) (2) [vertex,label=right:$2$]{} ;
	\node at (1,0) (4) [vertex,label=right:$4$]{} ;
	\node at (1,-1) (6) [vertex,label=right:$6$]{} ;
	\draw [midarrow,dashed] (4)--(5);
	\draw [midarrow] (6)to [bend right=60](2);
	\draw [midarrow] (3)--(1);
	\draw [dashed] (1)--(2) ;
	\draw [dashed] (2)--(3) ;
	\draw [dashed] (3)--(4) ; 
	\draw [dashed] (5)--(6) ; 
	\draw (1)to [bend right=60](5) ; 
	\draw (3)--(5) ; 
	\draw (2)--(4) ;
	\draw (4)--(6) ;
\end{tikzpicture}
\shorthandon{:}
\end{minipage}
\vspace{1cm}

Le développement en diagramme est clairement plus facile à utiliser et plus pratique en terme d'espace sur le papier que le développement analytique brutal. Nous devons maintenant intégrer les variables de Grassmann et nous pouvons trouver des équivalent graphiques à appliquer dans les diagrammes. Voici donc les règles d'intégration~: \\

\begin{itemize}\itemsep4pt
\item Si deux vertex sont raccordés par une flèche sur une ligne tiretée alors ils sont fusionnés et un facteur $1/2\pi$ apparaît devant l'intégrale résultante. 
\item Le nombre de lignes nues doit être conservé.
\item La parité (\ie $\pm 1$) du diagramme doit le multiplier et est donnée par la parité de réorganisation en ordre croissant des paires d'indices connectées par des flèches de lignes pleines -- en supposant que les indices anticommutent. Par exemple, $(35,46)$ est de parité $-1$ car il faut une seule permutation de $4$ et $5$ pour avoir $(3456)$. 
\item Il faut enfin renommer tous les vertex dans l'ordre croissant en suivant le chemin indiqué par les lignes tiretées. La parité doit être conservée.\\
\end{itemize}

Donnons tout de suite un exemple pour clarifier : 

\vspace{1cm}
\begin{minipage}{0.15\linewidth}
\shorthandoff{:}
\begin{tikzpicture}
	[vertex/.style={circle,draw=black,fill=black,inner sep=.5mm},
	midarrow/.style={postaction={decorate,decoration={markings,mark=at position .5 with {\arrow[black,line width=0.5mm]{stealth}}}}}]
	\node at (0,1) (1) [vertex,label=left:$1$]{} ;
	\node at (0,0) (3) [vertex,label=left:$3$]{} ;
	\node at (0,-1) (5) [vertex,label=left:$5$]{} ;
	\node at (1,1) (2) [vertex,label=right:$2$]{} ;
	\node at (1,0) (4) [vertex,label=right:$4$]{} ;
	\node at (1,-1) (6) [vertex,label=right:$6$]{} ;
	\draw [midarrow] (5)--(3);
	\draw [midarrow] (6)--(4);
	\draw [midarrow,dashed] (1)--(2);
	\draw [dashed] (2)--(3) ;
	\draw [dashed] (3)--(4) ; 
	\draw [dashed] (4)--(5) ; 
	\draw [dashed] (5)--(6) ; 
	\draw (1)--(3) ;
	\draw (1) to [bend right=60](5) ; 
	\draw (2)--(4) ; 
	\draw (2) to [bend left=60](6) ;
\end{tikzpicture}
\shorthandon{:}
\end{minipage}
\hfill
\begin{minipage}{0.15\linewidth}
\shorthandoff{:}
\begin{tikzpicture}
 \draw [->] (0,0) -- node[label=above:$\text{fusion}$]{} (1,0);
 \end{tikzpicture}
 \shorthandon{:}
\end{minipage}
\hfill
$-\frac{1}{2\pi}\times$
\begin{minipage}{0.15\linewidth}
\shorthandoff{:}
\begin{tikzpicture}
	[vertex/.style={circle,draw=black,fill=black,inner sep=.5mm},
	midarrow/.style={postaction={decorate,decoration={markings,mark=at position .5 with {\arrow[black,line width=0.5mm]{stealth}}}}}]
	\node at (0.5,1) (1) [vertex,label=above:$1$]{} ;
	\node at (0,0) (3) [vertex,label=left:$3$]{} ;
	\node at (0,-1) (5) [vertex,label=left:$5$]{} ;
	\node at (1,0) (4) [vertex,label=right:$4$]{} ;
	\node at (1,-1) (6) [vertex,label=right:$6$]{} ;
	\draw [midarrow] (5)--(3);
	\draw [midarrow] (6)--(4);
	\draw [dashed] (1)to [bend right=20](3) ;
	\draw [dashed] (3)--(4) ; 
	\draw [dashed] (4)--(5) ; 
	\draw [dashed] (5)--(6) ; 
	\draw (1)to [bend left=20](3) ;
	\draw (1) to [bend right=80](5) ; 
	\draw (1)--(4) ; 
	\draw (1) to [bend left=80](6) ;
\end{tikzpicture}
\shorthandon{:}
\end{minipage}
\hfill
\begin{minipage}{0.15\linewidth}
\shorthandoff{:}
\begin{tikzpicture}
 \draw [->] (0,0)--node[label=above:$\text{renommage}$]{} (1,0);
 \end{tikzpicture}
 \shorthandon{:}
\end{minipage}
\hfill
$-\frac{1}{2\pi}\times$
\begin{minipage}{0.15\linewidth}
\shorthandoff{:}
\begin{tikzpicture}
	[vertex/.style={circle,draw=black,fill=black,inner sep=.5mm},
	midarrow/.style={postaction={decorate,decoration={markings,mark=at position .5 with {\arrow[black,line width=0.5mm]{stealth}}}}}]
	\node at (0.5,1) (1) [vertex,label=above:$1$]{} ;
	\node at (0,0) (2) [vertex,label=left:$2$]{} ;
	\node at (0,-1) (4) [vertex,label=left:$4$]{} ;
	\node at (1,0) (3) [vertex,label=right:$3$]{} ;
	\node at (1,-1) (5) [vertex,label=right:$5$]{} ;
	\draw [midarrow] (4)--(2);
	\draw [midarrow] (5)--(3);
	\draw [dashed] (1)to [bend right=20](2) ;
	\draw [dashed] (2)--(3) ; 
	\draw [dashed] (3)--(4) ; 
	\draw [dashed] (4)--(5) ; 
	\draw (1)to [bend left=20](2) ;
	\draw (1) to [bend right=80](4) ; 
	\draw (1)--(3) ; 
	\draw (1) to [bend left=80](5) ;
\end{tikzpicture}
\shorthandon{:}
\end{minipage}
\vspace{1cm}

Parce que $(35,46) \to -(3456)$, ce diagramme a la parité $(-1)$. Il correspond à la réduction de l'intégrale suivante que nous avons rencontré plus tôt~: 

\begin{multline}\label{eq:int_1}
-\frac{(-1)^3}{2\pi}\int [\di t]_{5}\Theta(1,2)\Theta(2,3)\Theta(3,4)\Theta(4,5)~ \varepsilon(3,5)\varepsilon(4,6)~ S(1,2)S(1,3)S(1,4)S(1,5) \\
 = \frac{1}{2\pi}\int [\di t]_{5} \Theta(1,2)\Theta(2,3)\Theta(3,4)\Theta(4,5) ~ S(1,2)S(1,3)S(1,4)S(1,5) \\
  = \frac{1}{2\pi}\int [\underset{>}{\di t}]_{5} \symform{1}{2345} \\
\end{multline}

Nous introduisons une forme symbolique, de nouveau pour économiser de l'espace par~:

\eqna{
\symform{i_1,i_2 \ldots i_p}{j_1 j_2 \ldots j_n}=\prod_{\alpha=1}^p \prod_{a=1}^n S(i_\alpha,j_a)
}

Cette forme est complètement symétrique par échange des indices \emph{sur chaque ligne}. Maintenant, appliquons de nouveaux les règles d'intégration pour l'autre diagramme dont nous avions donné quelques détails~:

\vspace{1cm}
\begin{minipage}{0.15\linewidth}
\shorthandoff{:}
\begin{tikzpicture}
	[vertex/.style={circle,draw=black,fill=black,inner sep=.5mm},
	midarrow/.style={postaction={decorate,decoration={markings,mark=at position .5 with {\arrow[black,line width=0.5mm]{stealth}}}}}]
	\node at (0,1) (1) [vertex,label=left:$1$]{} ;
	\node at (0,0) (3) [vertex,label=left:$3$]{} ;
	\node at (0,-1) (5) [vertex,label=left:$5$]{} ;
	\node at (1,1) (2) [vertex,label=right:$2$]{} ;
	\node at (1,0) (4) [vertex,label=right:$4$]{} ;
	\node at (1,-1) (6) [vertex,label=right:$6$]{} ;
	\draw [midarrow,dashed] (1)--(2);
	\draw [midarrow,dashed] (3)--(4);
	\draw [midarrow,dashed] (5)--(6);
	\draw [dashed] (2)--(3) ; 
	\draw [dashed] (4)--(5) ; 
	\draw (1)--(3) ; 
	\draw (1) to [bend right=60](5) ; 
	\draw (3)--(5) ; 
	\draw (2)--(4) ; 
	\draw (2) to [bend left=60](6) ;
	\draw (4)--(6) ; 
\end{tikzpicture}
\shorthandon{:}
\end{minipage}
\hfill
\begin{minipage}{0.15\linewidth}
\shorthandoff{:}
\begin{tikzpicture}
 \draw [->] (0,0) -- node[label=above:$\text{fusion}$]{} (1,0);
 \end{tikzpicture}
\shorthandon{:}
\end{minipage}
\hfill
$\frac{1}{(2\pi)^3}$
\begin{minipage}{0.15\linewidth}
\shorthandoff{:}
\begin{tikzpicture}
	[vertex/.style={circle,draw=black,fill=black,inner sep=.5mm},
	midarrow/.style={postaction={decorate,decoration={markings,mark=at position .5 with {\arrow[black,line width=0.5mm]{stealth}}}}}]
	\node at (0,1) (1) [vertex,label=left:$1$]{} ;
	\node at (0,0) (3) [vertex,label=left:$3$]{} ;
	\node at (0,-1) (5) [vertex,label=left:$5$]{} ;
	\draw (1)to [bend right=20](3) ; 
	\draw (3)to [bend right=20](5) ; 
	\draw (1)to [bend left=20](3) ; 
	\draw (1) to [bend right=60](5) ; 
	\draw [dashed] (3)--(5) ; 
	\draw [dashed] (1)--(3) ; 
	\draw (1) to [bend left=60](5) ;
	\draw (3)to [bend left=20](5) ; 
\end{tikzpicture}
\shorthandon{:}
\end{minipage}
\hfill
\begin{minipage}{0.15\linewidth}
\shorthandoff{:}
\begin{tikzpicture}
 \draw [->] (0,0) -- node[label=above:$\text{renommage}$]{} (1,0);
 \end{tikzpicture}
\shorthandon{:}
\end{minipage}
\hfill
$\frac{1}{(2\pi)^3}$
\begin{minipage}{0.15\linewidth}
\shorthandoff{:}
\begin{tikzpicture}
	[vertex/.style={circle,draw=black,fill=black,inner sep=.5mm},
	midarrow/.style={postaction={decorate,decoration={markings,mark=at position .5 with {\arrow[black,line width=0.5mm]{stealth}}}}}]
	\node at (0,1) (1) [vertex,label=left:$1$]{} ;
	\node at (0,0) (3) [vertex,label=left:$2$]{} ;
	\node at (0,-1) (5) [vertex,label=left:$3$]{} ;
	\draw (1)to [bend right=20](3) ; 
	\draw (3)to [bend right=20](5) ; 
	\draw (1)to [bend left=20](3) ; 
	\draw (1) to [bend right=60](5) ; 
	\draw [dashed] (3)--(5) ; 
	\draw [dashed] (1)--(3) ; 
	\draw (1) to [bend left=60](5) ;
	\draw (3)to [bend left=20](5) ; 
\end{tikzpicture}
\shorthandon{:}
\end{minipage}
\vspace{1cm}

Ce diagramme à la parité $(+1)$ puisque qu'aucun indice n'est pointé par une flèche de ligne pleine. L'intégrale correspondante est la suivante~: 

\begin{multline}
(-1)^3 \frac{1}{(2\pi)^3}\int [\di t]_3 \Theta(1,2)\Theta(2,3) ~ S(1,2)^2 S(1,3)^2 S(2,3)^2 \\ = - \, \frac{1}{(2\pi)^3}\int [\underset{>}{\di t}]_3  \symform{1}{23}  \symform{2}{13} \symform{3}{12} 
\end{multline}

Cette intégrale est complètement symétrique par permutation des variables d'intégration. Quand un intégrande $S(1,2,3\ldots,n)$ est complètement symétrique et que l'intégrale est ordonnée, on peut enlever l'ordre d'intégration par l'identité suivante~:

\begin{align}\label{eq:sym_simpl}
\int [\underset{>}{\di t}]_n ~ S(1,2,3\ldots,n) = \int \frac{[\di t]_n}{n!} ~ S(1,2,3\ldots,n)
\end{align}
 
Ainsi, pour celle qui nous intéresse, nous aurons~:

\eqali{\label{eq:int_2}
-  \frac{1}{(2\pi)^3}\int [dt]_3 \Theta(1,2)\Theta(2,3) \symform{1}{23}  \symform{2}{13}\symform{3}{12} = -  \frac{1}{(2\pi)^3}\int \frac{[dt]_3}{3!} \symform{1}{23}  \symform{2}{13}\symform{3}{12}  
}

Les quatre autres diagrammes sont très semblables à celui qui donne l'intégrale~\refe{eq:int_1}. En fait, en suivant la méthode présentée, on trouve rapidement qu'il s'agit juste de toutes les permutations du nombre supérieur avec les nombres inférieurs dans la forme symétrique, \cad pour les cinq diagrammes que l'on somme simplement~:

\begin{align}\label{eq:int_3}
\frac{1}{2\pi}\int [\underset{>}{\di t}]_5 \parent{\symform{1}{2345}+\symform{2}{1345} +\symform{3}{1245}+\symform{4}{1235}+\symform{5}{1234}} 
\end{align}

Or nous voyons que l'intégrale ci-dessus est complètement symétrique par permutation de toutes les variables d'intégration. Par conséquent, nous pouvons la simplifier en utilisant l'identité~\refe{eq:sym_simpl}. En outre, une fois l'ordre d'intégration retiré, on voit que les cinq termes sont rigoureusement identiques à une permutation des variables d'intégration près et par conséquent égaux. Ainsi, l'intégrale finale en regroupant toutes les contributions~\refe{eq:int_2} et~\refe{eq:int_3} devient~:

\eqna{
I &=& - \frac{1}{(2\pi)^3}\int \frac{[dt]_3}{3!} \symform{1}{23}  \symform{2}{13}\symform{3}{12} +  \frac{5}{2\pi (5!)}\int [\di t]_5 \symform{1}{2345} \nonumber \\ 
}

\subsection{Application au calcul de la fonction de partition en $r=1/2$}

Nous pouvons effectivement faire explicitement le calcul en cette valeur de distance car l'intégrande est significativement simplifié par rapport au calcul en $r$ arbitraire et avec \textsc{Mathematica} nous avons pu résoudre les intégrales jusqu'à l'ordre 8 dans les tachyons. La motivation pour ce calcul est de retrouver le développement d'une fonction connue par identification des premiers ordres. Nous verrons cependant que la fonction semble trop compliquée pour être reconnue. Le calcul complet est de la forme~:

\eqna{
Z(x,y) &=& 2 \sum_{n=0}^{\infty} (\lambda^+\lambda^- e^{x^0})^n \; I_n 
}

Avec l'intégrale $I_n$ donnée par~:
\begin{align}
I_n = \int [\underset{>}{\di \hat t}]_{2n} ~  \prod_{1 \leq i<j}^n \hat S(2i,2j)\hat S(2i-1,2j-1)
\end{align}

Le calcul de $I_n$ peut être simplifié en utilisant la méthode diagrammatique, \cad en ajoutant des flèches sur ce genre de schéma,  \eg en $n=3$~: 
\begin{figure}[h!]
\centering
\begin{minipage}{0.2\linewidth}
$I_3 = \int [\underset{>}{d\hat t}]_{6}$
\end{minipage}
\hspace{0.5cm}
\begin{minipage}{0.2\linewidth}
\shorthandoff{:}
\begin{tikzpicture}
	[vertex/.style={circle,draw=black,fill=black,inner sep=.5mm},
	midarrow/.style={postaction={decorate,decoration={markings,mark=at position .5 with {\arrow[black,line width=0.5mm]{stealth}}}}}]
	\node at (0,1) (1) [vertex,label=left:$1$]{} ;
	\node at (0,0) (3) [vertex,label=left:$3$]{} ;
	\node at (0,-1) (5) [vertex,label=left:$5$]{} ;
	\node at (1,1) (2) [vertex,label=right:$2$]{} ;
	\node at (1,0) (4) [vertex,label=right:$4$]{} ;
	\node at (1,-1) (6) [vertex,label=right:$6$]{} ;
	\draw [dashed] (1)--(2);
	\draw [dashed] (3)--(4);
	\draw [dashed] (5)--(6);
	\draw [dashed] (2)--(3) ; 
	\draw [dashed] (4)--(5) ; 
	\draw (1)--(3) ; 
	\draw (1) to [bend right=60](5) ; 
	\draw (3)--(5) ; 
	\draw (2)--(4) ; 
	\draw (2) to [bend left=60](6) ;
	\draw (4)--(6) ; 
\end{tikzpicture}
\shorthandon{:}
\end{minipage}
\end{figure}

\vspace{2cm}

Comme nous venons de voir, nous avons obtenu pour ce diagramme, le résultat : 

\eqna{
I_3 &=& \frac{C_1^5}{2\pi}\int \frac{[dt]_{5}}{5!} \symform{1}{2345}-\frac{1}{(2\pi)^3}\int \frac{[dt]_3}{3!} \symform{1}{23}  \symform{2}{13}\symform{3}{12} \nonumber \\
	&=& \frac{2^{12}}{4!(2\pi)^5} -\frac{1}{(2\pi)^3} \nonumber \\
	&=& \frac{16}{3\pi^5} - \frac{1}{8\pi^3}
}

Nous avons explicitement écrit $C_1^5=5$ car le terme combinatoire apparaît réellement. En effet, il s'agit de choisir un indice dans la ligne supérieure de la forme symétrique, parmi $5$. Il est assez facile d'obtenir $I_2$ : 

\eqna{
I_2  &=& \parent{\frac{1}{(2\pi)^2}\int \frac{[dt]_2}{2!} \symform{1}{2}  \symform{2}{1} - \int \frac{[dt]_{4}}{4!} } \nonumber \\
	&=& \frac{1}{(2\pi)^2} - \frac{1}{4!} \nonumber \\
	&=& \frac{1}{4\pi^2}- \frac{1}{24}
}

Puis $I_1$ : 

\eqna{
I_1 &=& - \parent{ \frac{1}{2\pi} \int [dt]_1 } \nonumber \\
	&=& -\frac{1}{2\pi}
}

En revanche, le calcul de $I_4$ est plus complexe, mais par la méthode diagrammatique, nous obtenons finalement~: 

\begin{multline}
I_4 = \int  [\underset{>}{d t}]_8  \; \Asymform{13}{57}\Asymform{24}{68}  - \frac{1}{(2\pi)^2} \int \frac{[dt]_6}{6!} \; C_2^6 \; \symform{1}{2}\symform{2}{1} \symform{12}{3456} \\  + \frac{1}{(2\pi)^4} \int \frac{[dt]_4}{4!} \; \symform{1}{234}\symform{2}{134}\symform{3}{124}\symform{4}{123}\\
    = \frac{1}{1120}+\frac{143}{144 \pi ^6}-\frac{55}{192 \pi ^4}+\frac{13}{480 \pi ^2} -\parent{-\frac{1001}{2592 \pi ^6}-\frac{175}{432 \pi ^4}-\frac{1}{240 \pi ^2}} + \frac{1}{16 \pi^4} \\
    = \frac{1}{1120}+\frac{3575}{2592 \pi ^6}+\frac{205}{1728 \pi ^4}+\frac{1}{32 \pi ^2} + \frac{1}{16 \pi^4}
\end{multline}

Pour gagner de l'espace nous avons à nouveau introduit une forme symbolique. Ici il s'agit d'une forme complètement anti-symétrique par permutation de tous les indices~: 

\eqali{
\Asymform{ab\ldots}{cd\ldots}&= \sum_P (-1)^P \asymform{p(a)p(b)\ldots}{p(c)p(d)\ldots} \nonumber \\
							&= \asymform{ab\ldots}{cd\ldots} - \asymform{ac\ldots}{bd\ldots} + \asymform{ad\ldots}{bc\ldots} + \ldots \nonumber \\
}

pour laquelle nous avons trouvé plus pratique d'introduire une troisième forme anti-symétrique par permutation des indices de même ligne uniquement~: 

\eqali{
\asymform{abc\ldots}{def\ldots} = \varepsilon(a,b)\varepsilon(a,c)\varepsilon(b,c)\times \ldots \times \varepsilon(d,e)\varepsilon(d,f)\varepsilon(e,f)\times \ldots \times \symform{abc \ldots}{def \ldots}
}

Plus l' ordre est grand et plus la formule est compliquée, et cela à cause des multiples termes de contact et des ordres d'intégration que l'on ne peut pas toujours enlever. Résumons les formules obtenues~:

\eqna{
I_1 &=& - \parent{ \frac{1}{2\pi} \int [dt]_1 } \nonumber \\
}

\eqna{
I_2  &=& \frac{1}{(2\pi)^2}\int \frac{[dt]_2}{2!} \symform{1}{2}  \symform{2}{1} - \int \frac{[dt]_{4}}{4!} \nonumber \\
}

\eqna{
I_3 &=& \frac{1}{2\pi}\int \frac{[dt]_{5}}{5!} C_1^5 \symform{1}{2345}-\frac{1}{(2\pi)^3}\int \frac{[dt]_3}{3!} \symform{1}{23}  \symform{2}{13}\symform{3}{12} \nonumber \\
}

\eqna{
I_4 &=& \int  [\underset{>}{d t}]_8  \; \Asymform{13}{57}\Asymform{24}{68}  - \frac{1}{(2\pi)^2} \int \frac{[dt]_6}{6!} \; C_2^6 \; \symform{1}{2}\symform{2}{1} \symform{12}{3456} \nonumber \\ &&   + \frac{1}{(2\pi)^4} \int \frac{[dt]_4}{4!} \; \symform{1}{234}\symform{2}{134}\symform{3}{124}\symform{4}{123} \nonumber \\
}

En regroupant tous ces termes et leur valeur numériques nous obtenons pour la fonction de partition à l'ordre $(\lambda^+\lambda^-)^4$~: 

\begin{multline}\label{eq:partition_func_demi}
Z'(x^0) = 1 - \lambda^+\lambda^- \frac{e^{x^0}}{\pi} + \frac{\parent{\lambda^+\lambda^-}^2}{\pi^2} e^{2 x^0}\parent{1 \red{-\frac{\pi^2}{6}}} -  \frac{\parent{\lambda^+\lambda^-}^3}{\pi^3} e^{3 x^0} \parent{1 - \frac{128}{3 \pi^2}} \\+ \frac{\parent{\lambda^+\lambda^-}^4}{\pi^4} e^{4 x^0} \parent{1\red{-\frac{55}{12}+\frac{143}{9 \pi ^2}+\frac{13 \pi ^2}{30}+\frac{\pi ^4}{70}}+\frac{175}{27}+\frac{1001}{162 \pi ^2}+\frac{\pi ^2}{15}} \ldots 
\end{multline}

avec en rouge les contributions purement \emph{non-contact}. On reconnaît parmi tous ces termes un développement connu~: 

\eqna{
1 - \lambda^+\lambda^- \frac{e^{x^0}}{\pi} + \parent{\lambda^+\lambda^-}^2 \frac{ e^{2x^0}}{\pi^2} - \parent{\lambda^+\lambda^-}^3 \frac{ e^{3 x^0}}{\pi^3} + \ldots = \frac{1}{1+ \frac{\lambda^+\lambda^-}{\pi}e^{x^0}} \label{eq:triv_exp}
}

Or cette somme que nous venons de supposer infinie et exacte l'est effectivement, car en regardant l'intégrale $I_n$ de près, nous voyons que la contribution ayant le nombre maximal de termes de contact ($n$ précisément) est toujours présente et à une forme standard que l'on reconnaît être le déterminant de Vandermonde~\cite{Larsen:2002wc,Fotopoulos:2003yt}~:

\begin{align}
\Delta_{n} = \int [\di t]_n \prod_i^n \module{2\sin\frac{t_i-t_j}{2}}^2 = n!  
\end{align}

Si nous factorisons ce terme, alors nous obtenons un développement résiduel~: 

\begin{multline}
Z(x^0) = \frac{1}{1+ \frac{\lambda^+\lambda^-}{\pi}e^{x^0}} \left(1- \parent{\frac{\lambda^+\lambda^-}{\pi}}^2 \frac{\pi^2}{6}e^{2x^0} + \parent{\frac{128}{3\pi^2}-\frac{\pi^2}{6}}\parent{\frac{\lambda^+\lambda^-}{\pi}}^3 e^{3x^0} \right.\\ \left. + \parent{\frac{205}{108}+\frac{10487}{162 \pi ^2}+\frac{\pi ^2}{2}+\frac{\pi ^4}{70}}\parent{\frac{\lambda^+\lambda^-}{\pi}}^4 e^{4x^0} +\ldots \right)
\end{multline}

que l'on ne reconnaît pas provenir d'une fonction connue. Nous ne pouvons pas aller beaucoup plus loin dans cette direction. En vérité, nous ne croyons pas vraiment intéressant de calculer des termes d'ordre supérieurs car ils n'apporteraient pas grand chose du fait que le calcul devrait de toute façon être fait non-perturbativement, dans la mesure où $e^{x^0}$ croit rapidement.

\subsection{Extension de la technique à tout $r\neq 1/2$}

Il n'est pas possible de faire le calcul explicitement pour $r \neq 1/2$ mais on peut tout de même exprimer l'intégrale. En généralisant la méthode diagrammatique, nous avons pu obtenir un résultat compact pour la fonction de partition en tout $r<r_c$. En effet, nous avons alors~:

\eqali{\label{eq:sommation_func_part}
Z(r,\lambda^+,\lambda^-) &= 2 \sum_{n=0}^{\infty} (\lambda^+\lambda^-)^n e^{2n  \omega x^0} \int [\underset{>}{d\hat t}]_{2n} \prod_{i<j}^n \hat S(i,j)^{1-2r^2 \parent{1-(-)^{i+j}}}  \nonumber \\ 
& = 2 \sum_{n=0}^{\infty} (\lambda^+\lambda^-)^n e^{2n  \omega x^0} I_n
} 

L'expression générale de l'intégrale est~:

\begin{align}
I_n= \int [\underset{>}{\di \hat t}]_{2n} ~  \prod_{1 \leq i<j}^n \hat S(2i,2j)\hat S(2i-1,2j-1) \times \prod_{i,j=1}^n \hat S(2i-1,2j)^{1-4r^2}
\end{align}

avec 

\[\hat S(i,j)^{\alpha}= S(i,j)^\alpha - \alpha \,\varepsilon(i,j) S(i,j)^{\alpha-1} \theta_i\theta_j\]

 et $S(i,j)=\module{2\sin(t_i-t_j)/2}$. Pour généraliser le traitement précédent, il faut modifier un tout petit peu les règles diagrammatiques. Premièrement, il sera difficile pour des questions de clarté de conserver le visuel de l'ordre d'intégration -- les lignes tiretées. Deuxièmement, comme nous l'avons dit pour $r\neq 1/2$ il existe des termes mélangeant les indices des ensembles $J$ et $K$ et donc nous devons y faire correspondre des lignes. Puisque nous faisons la différence entre les propagateurs d'indice $J$ à indice $J$ (et d'indice $K$ à indice $K$)  et les propagateurs d'indice $J$ à indice $K$ (et inversement), nous pourrions conserver le même visuel pour les deux types, soit la ligne pleine. Cependant encore une fois nous risquons d'obtenir des diagrammes illisibles. Nous proposons donc l'ensemble de règles suivantes~: \\

\begin{itemize}\itemsep4pt
\item Chaque indice $i$ correspond à un vertex.
\item On sépare les indices $J$ des indices $K$ par deux colonnes distinctes.
\item Une ligne pleine reliant $i\in J$ avec $i \in J$ (et idem en remplaçant $J$ par $K$) représente $-\varepsilon(i,j)\theta_i\theta_{j}$.
\item Une ligne pleine reliant $i \in J$ avec $i\in K$ représente $-(1-4r^2) \epsilon(i,j) S(i,j)^{-4r^2} \theta_i\theta_j $. 
\item Une ligne tiretée reliant $i\in J$ avec $i \in K$ représente $-\delta(i,j)\theta_i\theta_{j}$.
\item Si deux vertex $(i,j)$ d'une même colonne ne sont pas connectés par une ligne pleine, alors le facteur $S(i,j)$ est implicite. 
\item Si deux vertex $(i,j)$ de deux colonnes distinctes ne sont pas connectés alors le facteur $S(i,j)^{1-4r^2}$ est implicite. 
\item Si deux vertex $(i,j)$ de deux colonnes distinctes ne sont pas connectés par une ligne tiretée et sont tels que $j=i+1$ alors le facteur $\Theta(i,i+1)$ est implicite. \\
\end{itemize}

Et une contrainte~: \\

\begin{itemize} \itemsep4pt
\item Chaque vertex ne peut être connecté qu'une et une seule fois.\\
\end{itemize}

Or nous voyons très facilement qu'il est impossible d'avoir simultanément un terme de contact $\delta(i,j)$ et $S(i,j)^{1-4r^2}$ puisque tant que $r<1/2$ le terme résultant est nul. Par contre, pour $r>1/2$ le résultat est infini et nous devrions le régulariser. Techniquement, on sait d'après les calculs que l'on a fait dans le demi-plan complexe que ces divergences ne sont jamais logarithmiques et les divergences en puissance peuvent être supprimées en ajoutant des contretermes sans conséquence pour la nature de la théorie. Nous pourrons donc appliquer une continuation analytique du cas $r<1/2$ vers le domaine $r>1/2$. Ainsi, nous nous plaçons dés à présent dans ce premier cas, et nous pouvons oublier les termes de contact et donc les traits tireté. \\

A l'ordre 3 par exemple, nous aurons~:

\begin{minipage}{0.15\linewidth}
\vspace{1cm}
\shorthandoff{:}
\begin{tikzpicture}
	[vertex/.style={circle,draw=black,fill=black,inner sep=.5mm},
	midarrow/.style={postaction={decorate,decoration={markings,mark=at position .5 with {\arrow[black,line width=0.5mm]{stealth}}}}}]
	\node at (0,1) (1) [vertex,label=left:$1$]{} ;
	\node at (0,0) (3) [vertex,label=left:$3$]{} ;
	\node at (0,-1) (5) [vertex,label=left:$5$]{} ;
	\node at (1,1) (2) [vertex,label=right:$2$]{} ;
	\node at (1,0) (4) [vertex,label=right:$4$]{} ;
	\node at (1,-1) (6) [vertex,label=right:$6$]{} ;
	\draw (1)--(2);
	\draw (3)--(4);
	\draw (5)--(6);
\end{tikzpicture}
\shorthandon{:}
\end{minipage}
\hfill
\begin{minipage}{0.85\linewidth}
\begin{multline}
= (-1)^3 \int [\di t]_6[\di \theta]_6 \theta_1\theta_3\theta_2\theta_4\theta_5\theta_6 \, \Theta(1,2)\Theta(2,3)\Theta(3,4)\Theta(4,5)\Theta(5,6) \\ \times (1-4r^2)^3 \, \varepsilon(1,2)\varepsilon(3,4)\varepsilon(5,6) \, \Big(S(1,2)S(3,4)S(5,6)\Big)^{-4r^2} ~ \\ \times  S(1,3)S(1,5)S(3,5)S(2,4)S(2,6)S(4,6) ~\\ \times  \Big(S(1,4)S(1,6)S(3,2)S(3,6)S(5,2)S(5,4)\Big)^{1-4r^2} \nonumber
\end{multline}
\end{minipage}

\begin{minipage}{0.15\linewidth}
\vspace{1cm}
\shorthandoff{:}
\begin{tikzpicture}
	[vertex/.style={circle,draw=black,fill=black,inner sep=.5mm},
	midarrow/.style={postaction={decorate,decoration={markings,mark=at position .5 with {\arrow[black,line width=0.5mm]{stealth}}}}}]
	\node at (0,1) (1) [vertex,label=left:$1$]{} ;
	\node at (0,0) (3) [vertex,label=left:$3$]{} ;
	\node at (0,-1) (5) [vertex,label=left:$5$]{} ;
	\node at (1,1) (2) [vertex,label=right:$2$]{} ;
	\node at (1,0) (4) [vertex,label=right:$4$]{} ;
	\node at (1,-1) (6) [vertex,label=right:$6$]{} ;
	\draw (1) -- (3) ; 
	\draw (2)--(4) ; 
	\draw (5)--(6) ; 
\end{tikzpicture}
\shorthandon{:}
\vspace{1cm}
\end{minipage} \, 
\hfill
\begin{minipage}{0.90\linewidth}
\begin{multline}
= (-1) \int [\di t]_6[\di \theta]_6 \theta_1\theta_3\theta_2\theta_4\theta_5\theta_6 \, \Theta(1,2)\Theta(2,3)\Theta(3,4)\Theta(4,5)\Theta(5,6) \\ \times \, \varepsilon(1,3)\varepsilon(2,4) ~(1-4r^2)\varepsilon(5,6) \, S(5,6)^{-4r^2} ~ \\ \times  S(1,5)S(3,5)S(2,6)S(4,6) ~\\ \times  \Big(S(1,2)S(1,4)S(1,6)S(3,2)S(3,4)S(3,6)S(5,2)S(5,4)\Big)^{1-4r^2} \nonumber
\end{multline}
\end{minipage}

Et nous aurons $8$ autres diagrammes similaires au dernier explicité ci-dessus et qui correspondent à choisir toutes les autres positions de la ligne transverse entre $J$ et $K$. Ils sont $(C_1^3)^2$ en tout. Mais aussi $5$ autres diagrammes correspondant aux autres configurations de 3 lignes transverses dans le premier exemple. Ils sont $3!$ en tout. Leur contributions sont très identiques à celles obtenues dans les deux exemples. 

L'intégration des variables de Grassmann est immédiate puisqu'il n'y a pas de termes de contact. Par contre, il faut toujours bien faire attention à la parité associée à l'ordre d'intégration. En étudiant tous ces diagrammes et en observant les diverses symétries de permutation, nous obtenons la formule suivante~:

\begin{multline}
I_3 = -(1-4r^2)^3 \int [\underset{>}{dt}]_{6} \symform{135}{246}^{-4r^2}  \Bigg( \symform{12}{3456} \symform{34}{56} +  \symform{12}{3645} \symform{36}{45}  + \symform{14}{2356} \symform{23}{56} \\  - \symform{14}{3625} \symform{36}{25} + \ldots \Bigg) - (1-4r^2) \int [\underset{>}{dt}]_{6}  \symform{135}{246}^{-4r^2} \Bigg(-\symform{13}{2456} \symform{24}{56} \\ - \symform{35}{4612} \symform{46}{12} + \symform{35}{2614} \symform{26}{14} + \ldots\Bigg)
\end{multline}

Les signes de parité sont données par l'ordre des paires, \cad $(12,36,45) \to (123456)$ mais $(14,36,25)\to - (123456)$. En étudiant les quelques premiers ordres, nous obtenons une formule générale~:

\begin{multline}\label{eq:integrale_compl}
I_n = (-1)^n \int [\underset{>}{dt}]_{2n} \left| \begin{array}{ccc} 1~3~5 &\ldots & 2n-1 \\ 2~4~6 &\ldots& 2n \end{array} \right|^{-4r^2} \sum_{\text{perm}~{\mathcal P}} (-1)^{P(1,2 \ldots 2n)}\parent{1-4r^2}^{u^n_P} \left|\begin{array}{c}  p(1)~p(2) \\ p(3)~p(4) \\ \ldots \\ p(2n-1) ~ p(2n)\end{array} \right| 
\end{multline}

avec 

\[u_P^n=\frac{n}{2}-\frac{1}{2} \sum_{i=1}^{n}(-1)^{p(2i-1) - p(2i)}\]

La permutation $P$ est définie de telle sorte que les pairs d'indices $(p(2i-1),p(2i))$ sont ordonnées. Il y a $(2n-1 )!!$ telles permutations. Les formes symboliques sont exactement les mêmes que celles que nous avons introduit précédemment, sauf que nous avons généralisé la forme symétrique telle que~:

\begin{align}
\left|\begin{array}{c}  a~b \\ c~d \\ \ldots \\ e ~ f\end{array} \right| = \left|\begin{array}{c}  a~b \\ c~d \end{array} \right| \ldots \left|\begin{array}{c}  a~b \\ e~f \end{array} \right| \left|\begin{array}{c}  c~d \\ e~f \end{array} \right|\ldots 
\end{align}

Après remplacement des formes symboliques par les expressions explicites dans~\refe{eq:integrale_compl}, nous obtenons exactement la formule écrite par Sen dans~\cite{Bagchi:2008et}. Donnons maintenant explicitement les 2 premiers ordres~:

\begin{align}\label{eq:int_ordre_1}
I_1 &= - \parent{1-4r^2} \int [\underset{>}{dt}]_{2} \symform{1}{2}^{-4r^2}  \nonumber \\ 
		  &= -\parent{1-4r^2} \int \frac{[dt]_{2}}{2!} \module{2 \sin \frac{t_1-t_2}{2}}^{-4r^2} \nonumber \\ 
		  &= -\frac{\Gamma(2-4r^2)}{2! \Gamma^2(1-2r^2)}  
\end{align}

en utilisant la symétrie de permutation pour enlever l'ordre d'intégration et la formule de Dixon que nous avons déjà introduite et qui est sur le cercle~\cite{Forrester_theimportance}~:

\begin{align}
\int_0^{2\pi} [dt]_{n} \prod_{i<j}\module{2 \sin \frac{t_i-t_j}{2}}^{2\gamma} = \frac{\Gamma(1+ n \gamma)}{\Gamma^n(1+\gamma)}
\end{align}

pour $\gamma>-1/2$. Nous voyons que le résultat~\refe{eq:int_ordre_1} est bien défini pour tout $r<1/\sqrt 2$, par conséquent et compte-tenu de ce que nous avons dit en début de section nous pouvons en prendre la continuation analytique pour tout $r\geq 1/2$. Or en $r=1/2$, nous aurons~:

\begin{align}
I_1(r=1/2) = - \, \frac{\Gamma(1)}{2! \Gamma^2(1/2)} = - \, \frac{1}{2\pi}
\end{align} 

qui est précisément le résultat que nous avons obtenu dans la section précédente dans~\refe{eq:partition_func_demi}. Nous voyons donc que même en l'absence de terme de contact, par continuation nous obtenons un résultat équivalent au calcul avec terme de contact. Ainsi, le fait que ce dernier donne une contribution finie est très importante pour assurer la continuité de la fonction de partition -- et sûrement aussi des amplitudes en général. On s'en assure en voyant que si nous avions fait une continuation de l'\emph{intégrande} en $r=1/2$ à cause du facteur $(1-4r^2)$ nous obtenions un résultat nul et non celui que nous avons là. \\

A l'ordre suivant, nous avons~:

\begin{align}
I_2 &= \int [\underset{>}{dt}]_{4}  \symform{1~3}{2~4}^{-4r^2} \parent{ \parent{1-4r^2}^2 \symform{1~2}{3~4} - \symform{1~3}{2~4} +  \parent{1-4r^2}^2 \symform{1~4}{2~3}} 
\end{align}

Tant que $r$ est différent de $0$ il n'y a pas de symétrie de permutation particulière et on ne peut pas simplifier cette intégrale. Nous ne connaissons pas le résultat explicite. Par conséquent nous ne pouvons pas faire de continuation analytique depuis $r<1/2$ vers $r \geq 1/2$. La formule de la fonction de partition pour tout $r<r_c$ est donc au second ordre~:

\begin{align}
Z[r,\lambda^\pm] = 2 \parent{1 - \frac{\Gamma(2-4r^2)}{2\Gamma^2(1-2r^2)} \lambda^+\lambda^-e^{2\omega x^0} + o((\lambda^+\lambda^-)^2}
\end{align}

En $r=0$ l'intégrant peut-être mis sous une forme totalement symétrique et nous obtenons la formule connue~:

\begin{align}
I_2(r=0) &= \int [\underset{>}{\di t}]_{4}  \parent{ \symform{1~2}{3~4} - \symform{1~3}{2~4} + \symform{1~4}{2~3}} \nonumber \\
&= 2^4 \int [\underset{>}{\di t}]_{4}  \sum_{\text{perm } P} \sin \frac{t_{p(1)}-t_{p(3)}}{2}\sin \frac{t_{p(1)}-t_{p(4)}}{2} \sin \frac{t_{p(2)}-t_{p(3)}}{2} \sin \frac{t_{p(2)}-t_{p(4)}}{2} \nonumber \\ 
&= 2^4 \int \frac{[\di t]_{4}}{4!}  \sum_{\text{perm } P} \sin \frac{t_{p(1)}-t_{p(3)}}{2}\sin \frac{t_{p(1)}-t_{p(4)}}{2} \sin \frac{t_{p(2)}-t_{p(3)}}{2} \sin \frac{t_{p(2)}-t_{p(4)}}{2} \nonumber \\ 
&= 3 \cdot 2^4 \int \frac{[\di t]_{4}}{4!}  \sin \frac{t_1-t_3}{2}\sin \frac{t_1-t_4}{2} \sin \frac{t_2-t_3}{2} \sin \frac{t_2-t_4}{2} \nonumber \\ 
&= \frac{3 \times 2!}{4!} = 2^{-2}
\end{align}

Avec comme contrainte que la permutation $P$ conserve l'ordre dans chaque paire regroupées en $(p(1)p(2),p(3)p(4))$ et tels que $p(1)>p(2)$ et $p(3)>p(4)$. Il y en a $(2n-1)!!$ avec ici $n=2$. En seconde ligne, nous voyons que l'intégrande est totalement symétrique et que nous pouvons donc enlever l'ordre d'intégration en divisant par $4!$. Enfin en troisième ligne, toutes les permutations sont égales après intégration, nous pouvons donc n'en conserver qu'un seul exemplaire et factoriser par $(2n-1)!!=3$. En avant-dernière ligne, le résultat est connu et donné explicitement dans~\cite{Fotopoulos:2003yt}. Et enfin nous avons appliqué la formule~\cite{Larsen:2002wc}~:

\begin{align}
\frac{(2n-1)!! n!}{2n!} = 2^{-n}
\end{align}

Nous trouvons que le même développement s'applique aux ordres supérieurs pour $r=0$ et que le résultat général est simplement~:

\begin{align}
I_n = 2^{-n}
\end{align}

qui donne après ressommation de la formule~\refe{eq:sommation_func_part} la fonction de partition~:

\begin{align}
Z'[\lambda^\pm , r=0] = \frac{2}{1+\frac{\lambda^+\lambda^-}{2} e^{\sqrt 2 x^0}}
\end{align}

C'est précisément le résultat standard de la fonction de partition du tachyon roulant sur un système brane-antibrane coïncident~\cite{Larsen:2002wc}.

\section{Discussion autour de l'action effective quadratique}
\label{sec:act_eff_kut}

Nous pouvons comparer l'approche du groupe de renormalisation à celle qui consiste à identifier l'action on-shell à la fonction de partition. L'identification de l'action off-shell à la fonction de partition off-shell renormalisée est  un peu plus délicate à cause de l’ambiguïté dans le schéma de renormalisation, tandis que l'action on-shell est admise égale à la fonction de partition calculée le long de la CFT correspondante. Nous trouvions le long de la solution de tachyon roulant $e^{\omega x^0}$ avec $\omega=\sqrt{1/2-r^2}$ la -- densité de -- fonction de partition suivante~:

\begin{align}\label{eq:fonction_part_kut}
Z'[r,\lambda^\pm] = 1 - \frac{\Gamma(2-4r^2)}{2 \Gamma^2(1-2r^2)} \lambda^+\lambda^- e^{2\omega x^0} + \ldots
\end{align} 

qui rappelons-le est valable pour tout $r<r_c$.

\subsection{L'approche de Kutasov et Niarchos}

Nous allons maintenant présenter succinctement la méthode de Kutasov et Niarchos~\cite{Kutasov:2003er} en l'appliquant à notre cas. Nous verrons que pour nous elle sera beaucoup moins puissante. L'hypothèse importante qu'ils font consiste à étudier la théorie des champs autour d'une solution de tachyon dépendant faiblement de l'espace, \cad dont les dérivées spatiales d'ordre supérieure ou égal à 2 sont négligeables. Dans cette hypothèse\footnote{Notons qu'il y a des arguments~\cite{Fotopoulos:2003yt} qui vont à l'encontre de cette hypothèse en particulier parce que les termes d'ordres $\dot T ^2$ sont de même ordre que les termes $\ddot T^2$ le long du tachyon roulant. La validité de cette hypothèse ici tient simplement au fait qu'on veut pouvoir reproduire les éléments de matrice-S quadratique dans les impulsions dans la limite $k^i \to 0$. Dans cette limite les termes d'ordres supérieurs dans les dérivées peuvent être supposés réabsorbables par redéfinition des champs dans~\refe{eq:lagrangien_base_kut}.}, on peut proposer un ansatz de lagrangien~:

\begin{align}\label{eq:lagrangien_base_kut}
{\mathcal L} = \sum_{n=0}^\infty \sum_{m=0}^\infty a^{(n)}_{m}(r) \parent{\partial_a T \partial^a T^*}^{m} \module{T}^{2(n-m)}
\end{align}

Dans cette forme les ambiguïtés de redéfinition des champs $T \to f(T,\partial T,\partial^{(2)} T ,\ldots)$ sont presque toutes fixées, à part $T \to T f(T^2)$ mais qui est équivalent à une redéfinition des coefficients $a_m^{(n)}$. Ces coefficients justement dépendent de la distance $r$ constante. Le lagrangien dépend naturellement des modules carrés par contrainte de réalité et par comparaison à la fonction de partition. Nous pouvons contraindre les coefficients inconnus $a_m^{(n)}$ en imposant aux équations du mouvement d'admettre comme solution~:

\begin{align}
T = \zeta e^{\omega x^0} + i \frac{\lambda}{\zeta^*} e^{-\omega x^0}
\end{align}

avec $\omega^2=1/2-r^2$. Nous faisons ce choix parce que nous avons montré dans la section précédente qu'à l'ordre quadratique au moins, cette expression était marginale. Nous ferons la supposition qu'elle l'est à tout ordre bien que cela appelle évidemment une vérification rigoureuse. Les équations du mouvement sont données pour tout $n$ par~:

\begin{multline}\label{eq:eom_kut}
0 = \sum_{m=1}^n m ~ a_m^{(n)} \partial^a \croch{\module{\partial_a T}^{2(m-1)} \partial_a T \module{T}^{2(n-m)} }  
						 - \sum_{m=0}^{n-1} (n-m) ~ a_{m}^{(n)} \module{T}^{2(n-m-1)} T \module{\partial_a T}^{2m} 
\end{multline}

ainsi que son complexe conjugué. En effet, les coefficients ne doivent pas dépendre de $T$ donc il faut imposer ces équations à chaque ordre $n$ dans les tachyons. Nous allons maintenant injecter la solution $T$. Remarquons tout d'abord que cette solution vérifie les équations~:

\begin{align}\label{eq:equation_spe}
&\partial_a T \partial^a T^* + \parent{\frac{1}{2}-r^2} \module{T}^2 =0 \nonumber \\ 
&\partial_a \partial^a T + \parent{\frac{1}{2}-r^2} T = 0
\end{align}

Ensuite nous écrirons par commodité de notation $T=T_+$ et $\dot T = T_-$ avec

\begin{align}
T_\pm = \zeta e^{\omega x^0} \pm i \frac{\lambda}{\zeta^*} e^{-\omega x^0}
\end{align}

Remarquons que $T_+$ et $T_-$ sont indépendants parce que $\zeta$ et $\lambda$ le sont. Maintenant, en utilisant~\refe{eq:equation_spe} dans~\refe{eq:eom_kut} le calcul est direct et nous obtenons la formule suivante~:

\begin{align}
0 = \sum_{m=0}^n \omega^{2m} a_m^{(n)}\Big(m(n-1) \, T_-^2 + (n-m+nm)\, T_+^2\Big)
\end{align}

L'indépendance des deux termes implique qu'il faut résoudre cette équation séparément pour chacun. Nous obtenons donc un système de deux équations en utilisant $T_\pm \neq 0$~:

\begin{align}
\left\{\begin{array}{l} \sum \omega^{2m} a_m^{(n)}m(n-1) = 0  \\ 
\sum \omega^{2m} a_m^{(n)} (n-m+nm) = 0
\end{array} \right.
\end{align}

Ce système est clairement sous-déterminé car nous pouvons reformuler ces deux contraintes sous la forme~:

\begin{align}\label{eq:systeme}
 \left\{\begin{array}{l} (n-1)\sum m\, \omega^{2m} \, a_m^{(n)}  = 0 \\ 
n\sum \omega^{2m} \, a_m^{(n)} = 0
\end{array} \right. 
\end{align} 

Nous voudrions réduire à chaque ordre le nombre de degré de liberté à $a_0^{(n)}$. Or il est clair que cela est impossible dans ce système pour tout $n\geq 3$. Dans l'exemple résolu par Kutasov et Niarchos en revanche, leur système était complètement déterminé par une récurrence, ce qui leur permettait d'exprimer l'ensemble des coefficients à chaque ordre $n$ en fonction de $a_0^{(n)}$. Nous n'avons pas cette chance ici à cause des équations~\refe{eq:equation_supp}. Toutefois, nous pouvons contraindre les $3$ premiers termes de l'action effective puisque pour $n<3$ le système est soluble si bien que nous pourrions obtenir une expression exacte à l'ordre quartique dans le tachyon. \\ 

Maintenant, pour déterminer les coefficients $a_0^{(n)}$ ils proposent d'utiliser la formule bien connue ${\mathcal L}_{on} = Z'$ identifiant le lagrangien on-shell à la densité de fonction de partition calculée le long d'une CFT. Rappelons que cette formule est justifiée tant qu'on peut faire sens d'éléments de matrice-S asymptotiquement, \cad au moins pour $x^0 \to \pm \infty$. Puisque nous nous plaçons tout comme eux le long d'un tachyon roulant qui est asymptotiquement nul en $x^0\to -\infty$ nous n'aurons aucun problème pour appliquer cette méthode à l'ordre quadratique.

Dans leur cas, la fonction de partition étant calculable perturbativement à tout ordre dans le tachyon, il existait une correspondance idéale entre chaque coefficient $a_0^{(n)}$ et une expression à l'ordre $\module{T}^{2n}$ dans la fonction de partition. Ainsi, l'action à l'ordre des dérivées premières dans le tachyon est totalement déterminée. Cette méthode est très intéressante et dans leur cas a été très fructueuse, puisqu'elle leur a permis d'obtenir précisément l'action effective du tachyon proposée par Sen, par exemple dans~\cite{Sen:2002in,Sen:2002nu}. Mais dans notre cas, ça ne fonctionne pas parce qu'on ne peut pas réduire le nombre de degré de liberté à 1 par ordre dans l'ansatz de l'action effective de telle sorte que l'identification à la fonction de partition ne peut pas déterminer complètement l'action.

\subsection{Action effective quadratique}

Nous allons pour notre part, simplement calculer l'action à l'ordre quadratique -- bien que nous pourrions techniquement pousser jusqu'à l'ordre quartique\footnote{C'est en projet, mais il faudra d'abord résoudre l'intégration à l'ordre 4 dans les tachyons pour tout $r$.}. Comme nous venons de voir à cette ordre l'ansatz du lagrangien à l'ordre de la dérivée première et à distance fixée est~:

\begin{align}
{\mathcal L} = a_0^{(0)}(r) + a_0^{(1)}(r) \module{T}^2 + a_1^{(1)}(r) |\partial_a T|^2 + \ldots
\end{align}

En utilisant le système~\refe{eq:systeme} précédent en $n=0,1$, nous trouvons facilement~:

\begin{align}
a_1^{(1)} = -\frac{a_0^{(1)}}{\omega^2}
\end{align}

avec $a_0^{(0)}$ et $a_0^{(1)}$ des constantes indéterminées. Nous voulons maintenant comparer l'expression de l'action obtenue, on-shell à la fonction de partition~\refe{eq:fonction_part_kut}. En utilisant la solution de tachyon roulant simple $T^\pm = \lambda^\pm e^{\omega x^0}$ et en identifiant en toute généralité $T = \kappa(r) T^+$ et $T^*= \kappa(r) T^-$ avec $\kappa(r)$ une constante éventuellement dépendante de $r$ on trouve~:

\begin{align}
\left\{ 
\begin{array}{l} 
a_0^{(0)} = 2  \\
a_0^{(1)} = \frac{\Gamma(2-4r^2)}{2 \kappa^2 \Gamma^2(1-2r^2)}
\end{array} \right.
\end{align} 

L'action quadratique à distance constante est donc finalement donnée par le lagrangien~:

\begin{align}\label{eq:lagrangien_contr2}
{\mathcal L}_Z = 2 - \frac{\Gamma(4-4r^2)}{2 \kappa^2 \Gamma^2(2-2r^2)} \parent{\parent{\frac{1}{2}-r^2} \module{T}^2 - \partial_a T\partial^a T^*} + \ldots
\end{align}

Maintenant, nous voudrions ajouter le terme cinétique du champ de distance afin d'obtenir une action pour les deux champs. On peut \apriori simplement l'addition, puisque dans la limite $T \to 0$ on sait que l'on doit retrouver le développement de l'action BI~\cite{Metsaev:1987qp,Tseytlin:1986ti}~:

\begin{align}
{\mathcal L}_{BI} = 2 ~ \sqrt{1+ \frac{1}{4} \partial_a \phi \partial^a \phi} \simeq 2 \parent{1+ \frac{1}{8} \partial_a \varphi \partial^a \varphi}
\end{align}

où on a laissé implicite le facteur $T_p e^{\Phi}$ constant. En imposant $\varphi = r/2\pi$ nous voyons donc qu'il faut ajouter le terme cinétique $\partial_a r \partial^a r /2$ à l'action~\refe{eq:lagrangien_contr2}. Soit~:

\begin{align}
\label{eq:lagrangien_contr3}
&{\mathcal L}_Z = 2\Bigg[1 + \frac{\pi^2}{2}\partial_a r \partial^a r  - f(r) \parent{\parent{\frac{1}{2}-r^2} \module{T}^2 - \partial_a T\partial^a T^*} + \ldots \Bigg] \nonumber \\
&f(r) = \frac{\Gamma(4-4r^2)}{4 \kappa^2 \Gamma^2(2-2r^2)}
\end{align}

Comme on ne connaît pas $\kappa(r)$ on peut pour l'instant conserver $f(r)$ arbitraire. Toutefois, nous pouvons déterminer exactement cette fonction en dérivant les équations du mouvement et en imposant que $r$ constant est une solution le long du tachyon~:

\begin{align}\label{eq:solution2}
T = \zeta e^{\sqrt{\frac{1}{2}-r^2} x^0} + i \frac{\lambda}{\zeta^*}  e^{-\sqrt{\frac{1}{2}-r^2} x^0}
\end{align}

A partir de la formule précédente~\refe{eq:lagrangien_contr3} on trouve que les équations du mouvement sont simplement~:

\begin{align}
0 &=-\square r - f'(r) \parent{\parent{\frac{1}{2}-r^2} \module{T}^2 - \partial_a T\partial^a T^*} + 2 r f(r) \module{T}^2 \nonumber \\
0 &= -\square T + \parent{\frac{1}{2}-r^2} T + \frac{f'(r)}{f(r)} \partial_a r \partial^a T
\end{align}

Sachant que toutes les solutions tachyoniques -- à l'ordre quadratique -- pour $r$ constant sont telles que~:

\begin{align}
\parent{\frac{1}{2}-r^2} \module{T}^2 + \partial_a T\partial^a T^* = 0
\end{align}

Nous trouvons facilement qu'il faut $f'(r)=r f(r)/(1/2-r^2)$ soit donc~:

\begin{align}
f(r) = \frac{C}{\sqrt{\frac{1}{2}-r^2}}
\end{align}

avec $C$ une constante indéterminée que l'on peut fixer à $C=1/2\sqrt 2$ sans perdre en généralité -- nous verrons pourquoi cette valeur précisément. Alors l'action effective quadratique est finalement~: 

\begin{align}\label{eq:action_quad_final}
&{\mathcal L}_Z = 2\Bigg[1 + \frac{\pi^2}{2}\partial_a r \partial^a r  + \frac{1}{2 \sqrt{1 - 2r^2}}\parent{\partial_a T\partial^a T^*-\parent{\frac{1}{2}-r^2} \module{T}^2} + \ldots \Bigg] 
\end{align}

et les équations du mouvement qui en dérivent sont~:

\begin{align}\label{eq:eom_tac_super_act}
0 &= -\square r - \frac{r }{(1-2r^2)^{3/2}}\parent{\partial_a T\partial^a T^*+\parent{\frac{1}{2}-r^2} \module{T}^2 }  \nonumber \\
0 &= -\square T + \parent{\frac{1}{2}-r^2} T + \frac{2r}{1-2r^2}\partial_a r\partial^a T
\end{align}

Cette action est valide uniquement pour $r<r_c$. Elle est discontinue en $r=r_c$ donc n'est pas continuable en $r>r_c$, d'autant que $\sqrt{1 - 2r^2}$ serait alors imaginaire. Cela signale encore que le syst\`eme subit une transition de phase \`a la distance critique. Cette discontinuit\'e permet qu'en $r>r_c$ l'action de Garousi prenne le relais~: nous avons montr\'e qu'elle \'etait compatible avec la physique interne des cordes dans le domaine surcritique.

\subsubsection{Comparaison aux équations du groupe de renormalisation}

La comparaison aux équations obtenues dans la section~\refcc{sec:renorm_susy} montre qu'il y a compatibilité entre ces équations et celles que fournissent les fonctions bêta, mais en choisissant une redéfinition de tachyon convenable et \apriori uniquement pour $r$ constant. Rappelons ici qu'elles étaient~:

\begin{align}\label{eq:mouvement_tach_super2}
\frac{\delta {\mathcal L}}{\delta r} &= - \square r - \frac{r \, h(r)}{\frac{1}{2}-r^2} \parent{\partial_a T \partial^a T^* + \parent{\frac{1}{2}-r^2} \module{T}^2} \nonumber \\ 
\frac{\delta {\mathcal L}}{\delta T^*} &= - \square T + \parent{\frac{1}{2}-r^2} T  \nonumber \\ 
\frac{\delta {\mathcal L}}{\delta T} &= - \square T^* + \parent{\frac{1}{2}-r^2} T^*
\end{align}

avec $h(r)$ une fonction de $r$ arbitraire provenant d'une redéfinition du tachyon à distance constante. En fait, si on étudie plus en détail l'ensemble des fonctions bêta, on peut voir que nous aurons aussi celle-ci~:

\begin{align}
\beta_{\Delta\lambda^\pm} &= - 2 r \Delta (\delta r\lambda^\pm) +\ldots \nonumber \\
&=  4 r \partial_i \delta r \partial^i \lambda^\pm - 2 r \lambda^\pm \Delta \delta r - 2 r \delta r \Delta \lambda^\pm
\end{align}

Soit à l'ordre quadratique simplement~:

\begin{align}
\beta_{\Delta\lambda^\pm} &= 4 r \partial_i \delta r \partial^i \lambda^\pm + o(\beta_i)
\end{align}

à des facteurs dépendant des fonctions bêta près. Puisque $r$ est constant, et que les équations du mouvement sont telles qu'il faut que les fonctions bêta soient nulles, nous aurions pu donc aussi réécrire les fonctions bêta~\refe{eq:beta_sous-crit_susy} avant d'en déduire les équations du mouvement sous la forme~:

\begin{align}
\beta_{\delta r} &= -\Delta \delta r - 2 r h(r)\parent{ \zeta_{(1)}\zeta_{(2)}^* + \zeta_{(2)}\zeta_{(1)}^*  } \nonumber \\ \beta_{{(1,2)}} &= \Delta \zeta_{(1,2)} - 2r \delta r \, \zeta_{(1,2)} + g(r) \beta_{\Delta \zeta_{(1,2)}} \nonumber \\ 
&=\Delta \zeta_{(1,2)} - 2r \delta r \, \zeta_{(1,2)} +  4 r g(r) \partial_i \delta r \partial^i \lambda^\pm + \ldots
\end{align}

Les expressions de $g(r)$ et $h(r)$ sont arbitraires, de sorte qu'il est possible de les choisir telles qu'on retrouve les équations du mouvement~\refe{eq:eom_tac_super_act}.   \\ 

Il y a donc un bon accord entre les deux calculs indépendants. Toutefois, cette \'etude montre que les fonctions bêta sont délicates à utiliser pour d\'eduire des équations du mouvement, car elles ne sont connues qu'à des termes proportionnels aux fonctions bêta près. La m\'ethode de Witten (OSFT) qui rejoint la m\'ethode de Tseytlin, semble moins ambigu\"e bien que plus complexe \`a utiliser. Nous pr\'esenterons des perspectives de recherches dans cette direction pour ce syst\`eme en conclusion, dans la partie~\refcc{part:conclu}.

\subsubsection{Contrainte sur le tachyon d'espace-cible}

Enfin, nous trouvons également qu'il faut redéfinir le tachyon de l'action effective par un facteur $\kappa(r)$ dépendant de la distance lorsqu'on l'identifie on-shell à son homologue de surface de corde. Ce n'est pas inattendu, puisque la relation entre le champ d'espace-cible et le couplage de surface n'est pas nécessairement trivial\footnote{Je pense en particulier à la correspondance entre la déformation de demi S-brane et le vide stable en $\lambda=1/2$.}. Cette constante vaut précisément~:

\begin{align}\label{eq:kappa}
\kappa(r) = \sqrt{\frac{\sqrt{1-2r^2} \, \Gamma(4-4r^2)}{2 \Gamma^2(2-2r^2)}} \quad \Rightarrow \quad T \Big\vert_{\text{espace-cible}} = \kappa(r) \, T^\pm \Big\vert_{\text{surface}}
\end{align}

Nous voyons donc quelque chose d'intéressant se produire qui est la suppression du facteur $\kappa(r)$ à la distance critique $r_c = 1/\sqrt 2$ donc du tachyon $\kappa(r) T^\pm$. En effet, les arguments des fonctions Gamma sont tous bien réguliers et non nuls en $r=1/\sqrt 2$ tandis que le facteur $\sqrt{\frac{1}{2}-r^2} $ s'annule. 
Ce n'est pas un résultat si surprenant ou inattendu car nous savons qu'à la distance critique il doit se produire un phénomène marquant une discontinuité du point de vue d'un tachyon roulant -- les fameuses limites $r \to r_c^-$ et $r_c^- \to r_c$. D'autant plus que nous ne pouvons pas calculer la fonction de partition le long d'un tachyon constant en $r=r_c$ sans tenir compte aussi de la perturbation de distance $\delta r$. Rappelons que cette perturbation a une fonction bêta non nulle en cette valeur, proportionnelle à $\lambda^+\lambda^-$. Si bien que le tachyon constant ne définit pas une CFT à moins qu'il soit nul ! Cela traduirait bien la rupture de la relation ${\mathcal L}_{on} = Z'$ pour $r$ et $T$ constants. \\

\subsection{Conclusion et comparaison à l'action effective du système coïncident}

Nous avons montré qu'il était possible de contraindre une action effective quadratique, bien que nous avons argumenté qu'il n'était pas possible de déterminer les ordres supérieurs. Nous avons obtenu l'expression de l'action en $r<r_c$ en imposant à la Kutasov et Niarchos que les équations du mouvement admettent les solutions de tachyon roulant. Puis nous avons déterminé les coefficients en comparant à la fonction de partition le long de la CFT du tachyon roulant à distance constante que nous avions calculée dans la section précédente. 

Une fois le terme cinétique du champ de distance réhabilité, nous avons vu que nous ne pouvions pas, à partir de l'expression de l'action, retrouver exactement les équations du mouvement off-shell que nous avions établi plus tôt par le biais du groupe de renormalisation. Nous trouvions cependant qu'elles étaient bien compatibles le long de $r$ constant. \\

Nous pouvons maintenant comparer l'action~\refe{eq:action_quad_final} à celle qu'obtenaient Kutasov et Niarchos dans le cas $r=0$ strictement -- \cad en gelant le champ de distance, ce qui est bien sûr discutable mais permet d'obtenir une expression exacte pour l'action effective du tachyon exclusivement. Ils trouvent précisément -- à un signe multiplicatif près~:

\begin{align}\label{eq:action_Kutasov}
{\mathcal L} &=\frac{2}{1+\frac{\module{T^2}}{2}} \sqrt{1+\frac{\module{T}^2}{2} + \partial_a T \partial^a T^*} \nonumber \\ 
  &\sim  2 \Bigg(1 - \frac{\module{T}^2}{4} + \frac{1}{2}\partial_a T \partial^a T^* +\ldots \Bigg)
\end{align}

Nous avons bien égalité entre leur formule et la nôtre, ce qui est somme toute normal par continuité de la fonction de partition en $r=0$. Nous pourrions suggérer deux formes de type TDBI correspondant en développement à l'action quadratique~\refe{eq:action_quad_final}~:

\begin{align}
{\mathcal L}^{(1)}_{TDBI} &= \frac{2}{1+\frac{\module{T}^2}{2 \sqrt{1-2r^2}}} \sqrt{1 + \partial_a r \partial^a r + \sqrt{1-2r^2}\frac{\module{T}^2}{2} + \frac{\partial_a T \partial^a T^*}{\sqrt{1-2r^2}}} \nonumber \\ 
{\mathcal L}^{(2)}_{TDBI} &= \frac{2}{1+\frac{\module{T}^2}{2 \sqrt{1-2r^2}}} \sqrt{1 + \partial_a r \partial^a r} \times \sqrt{1 + \sqrt{1-2r^2}\frac{\module{T}^2}{2} + \frac{\partial_a T \partial^a T^*}{\sqrt{1-2r^2}}} 
\end{align}

Néanmoins, une rapide étude de l'équation du mouvement de $r$ le long du tachyon roulant à $r$ constant montre dans chaque cas que ça ne fonctionne pas, dés l'ordre 4 dans les tachyons. L'équation du mouvement on-shell est~:

\begin{align}
\frac{\delta {\mathcal L}_{TDBI}}{\delta r}= \frac{r}{2} \module{T}^4  + \ldots \neq 0
\end{align}

Elle ne s'annule que pour $r=0$ car les termes d'ordres suivants sont proportionnels \`a $r$. Ces expressions de lagrangien sont donc fausses. Pour l'instant nous n'avons pas trouv\'e de formulations compatibles. Une voie de recherche consisterait \`a obtenir l'expression exacte non-perturbative de la fonction de partition en tout $r$ constant le long du tachyon roulant, mais le calcul semble dans l'imm\'ediat et \`a court terme, hors de port\'ee. Une autre option, dont nous parlerons plus longuement en conclusion, consisterait \`a \'etudier le mod\`ele sigma off-shell compos\'e d'un tachyon constant, en suivant la m\'ethode de OSFT de Witten~\cite{Witten:1992cr,Witten:1992qy,Gerasimov:2000zp}. Pour l'instant cette m\'ethode a surtout \'et\'e appliqu\'ee en th\'eorie bosonique. Puisque les physiques des syst\`emes de branes bosoniques s\'epara\'ees et de brane-antibrane s\'epar\'ees sont similaires, l'\'etude du syst\`eme bosonique reste prometteuse. Nous verrons en conclusion l'existence d'une relation entre le mod\`ele bosonique et le mod\`ele de Kondo.

\part{Conclusion et perspectives}
\label{part:conclu}

\renewcommand\theequation{\thepart.\arabic{equation}}

Au cours de cette thèse, nous avons démontré que le tachyon roulant du secteur interbranaire (anti-diagonal) $\sigma^\pm \in U(2)$ dans le système d'une brane et d'une antibrane parallèles et séparées en théorie de type IIA ou IIB, est une solution des \'equations du mouvement de la théorie des champs de cordes dans l'\emph{ensemble} du domaine tachyonique $\ell<\ell_c$ avec $\ell_c = \pi\sqrt{2\alpha'}$. Le modèle sigma incluant l'opérateur de vertex de ce tachyon roulant constitue ainsi une théorie des champs conforme. Nous nous sommes intéressés aussi au modèle bosonique correspondant qui non seulement admet un tachyon interbranaire dans son spectre de corde ouverte mais contient aussi des tachyons hébergés sur chaque brane, \cad dans les secteurs diagonaux $\sigma^{0,3} \in U(2)$. 

Dans le premier cas, nous avons montré que les résonances entre les opérateurs de tachyons roulants $\sigma^\pm \otimes \psi^\pm e^{\sqrt{1/2-r^2} X^0 \pm ir\wt X}$ avec $r=\ell/2\pi$ et l'opérateur $\sigma^0 \otimes e^{2 n \sqrt{1/2-r^2}X^0}$ n'engageaient aucune divergence logarithmique à l'ordre 2 et 4 dans les tachyons, \cad pour tout $r<\sqrt{17}/6$. En identifiant le mécanisme d'annulation de ces divergences au rôle de la supersymétrie, nous en avons déduit que cela devait s'appliquer à tout ordre et à toute distance. Nous avons vu que les divergences logarithmiques étaient la cause de la perte de marginalité en contribuant à l'expression des fonctions bêta des couplages associés aux opérateurs produit par résonance. L'absence des divergences logarithmiques nous a ainsi permis de conclure à la marginalité exacte du modèle de tachyon roulant dans le système brane-antibrane. Nous nous attendions à ce comportement par absence de raisons \emph{physiques} justifiant le contraire. 

Dans le second cas, nous avons vu en revanche que les résonances entre les tachyons roulants $\sigma^\pm \otimes e^{\sqrt{1-r^2} X^0 \pm ir\wt X}$ et l'opérateur $\sigma^0 \otimes e^{2 n \sqrt{1-r^2}X^0}$ faisaient intervenir des divergences logarithmiques -- non supprimées -- au moins à l'ordre 2 pour tout $r>r_c/\sqrt 2$. Nous en avons déduit que les contributions correspondantes dans les fonctions bêta du champ de tachyon du secteur $\sigma^0$ mettaient en valeur l'existence d'un couplage \emph{physique} non nul dans l'action effective du système entre ce tachyon et celui du secteur interbranaire. 

Ainsi, dans un cas nous obtenons une suppression des divergences logarithmiques gr\^ace à la supersymétrie, et dans l'autre cas, il n'existe aucune supersymétrie et ces divergences ne s'annulent pas. Dans le premier il n'y a aucune raison physique qui justifierait la présence de divergences logarithmiques à cause de la supersymétrie et de la projection GSO qui supprime le mode tachyonique des secteurs diagonaux. Mais dans le second, puisque le tachyon du secteur $\sigma^0$ est une excitation physique dans le modèle bosonique, nous avons une raison \emph{physique} pour expliquer la présence de divergences logarithmiques. \\

Dans le système brane-antibrane, nous avons aussi mis en valeur l'existence de termes de contact. Ces derniers apparaissent naturellement en exprimant l'action de surface de corde de façon manifestement supersymétrique dans le super-espace en introduisant des fermions de bord, puis en décomposant cette action et en intégrant sur les variables de Grassmann.  Nous avons montré que ces termes de contact jouaient le rôle de contretermes et permettaient de supprimer un grand nombre de divergences en puissance du cut-off UV $\varepsilon$ dans les amplitudes, ou au moins dans la fonction de partition sur le disque. Ce comportement relève d'une théorie manifestement supersymétrique. Or les résultats précédents ont montré que les divergences en puissance n'étaient pas \emph{toutes} supprimées et qu'il restait des divergences résiduelles. Nous avons identifiées qu'elles étaient \^otables par des termes de contact d'ordre supérieur et nous avons donc proposé de corriger l'action supersymétrique des fermions de bord en conséquence, par l'ajout de termes d'interaction à 4 points et plus. 

A distance $\ell<\ell_c$ le nombre de termes divergents à tout ordre est fini. En effet, à distance $\ell$ pour une fonction à $2N$-points des tachyons la "divergence" résiduelle est du type $\varepsilon^{4N^2(1/2 -r^2)-1} \oint e^{2 N \sqrt{1/2-r^2}X^0}$ pour tout $N <1/(2-4r^2)$. Ainsi, le nombre de contretermes à ajouter dans l'action est fini et la théorie renormalisable. Puisqu'il s'agit de divergences de puissance, la théorie reste en outre exactement marginale. Cependant dans la limite $ r\to r_c$ le nombre de divergences résiduelles tend vers l'infini. Ainsi la limite $r\to r_c^-$ ne définit pas une théorie renormalisable. La valeur $r_c$ est, aussi en ce sens, une distance critique. Nous identifions ce comportement à la perte de marginalité du tachyon roulant en $r=r_c$ à cause d'une résonance non nulle avec le champ de distance $\sigma^3 \otimes D\wt {\mathbb X}$. Cela pourrait \^etre d\^u \`a la transition d'une théorie interactive $c=2$ à une théorie interactive $c=1$ par comparaison au comportement des théories Liouville dans la limite $c\to 1$. \\

Notre étude a porté ensuite sur le groupe de renormalisation du modèle sigma perturbé autour de la déformation marginale du tachyon roulant dans le cas bosonique et dans le cas brane-antibrane. 
D'abord en théorie bosonique, nous avons inséré sur le bord la déformation correspondante à une perturbation de distance $\delta r(X^a)$ et nous avons supposé que le tachyon lui-même était une perturbation $\lambda^\pm(X^a)$. Pour étudier le groupe de renormalisation de ces perturbations et obtenir des contraintes physiques sur leur dynamique respective, nous avons développé ces fonctions des champs $X$ en décomposant ces derniers en mode zéro et modes d'oscillations $X=x+\hat X$. Nous obtenions alors des perturbations \emph{relevantes} autour des déformations marginales. De ce fait, les fonctions b\^eta obtenues pour les couplages $\delta r(x)$ et $\lambda^\pm(x)$ ne pouvaient recevoir de contributions que via des résonances, donc des termes logarithmiques et universels, \cad indépendant des schémas de renormalisation. Nous avons justifié dans cette mesure l'interprétation de ces fonctions b\^eta en tant qu'équations du mouvement. Leur expression obtenue est non triviale et prend son origine dans l'existence d'une solution de type S-brane complète à l'ordre quadratique. En l'état, ces équations ne permettent pas d'exprimer une action effective dont elles dérivent, à moins de redéfinir les champs ou les fonctions b\^eta par des termes constants ou proportionnels à d'autres fonctions b\^eta, ce que nous voyons concrètement dans la section~\refcc{sec:act_eff_kut}.

Dans le modèle brane-antibrane, nous avons exprimé une théorie équivalente, manifestement supersymétrique sur le super-espace de la surface de corde, puisque nous avions vu l'importance que revêtaient les termes de contact dans l'extraction des divergences. Les perturbations ont donc été exprimées dans le super-espace, et autour des déformations marginales, par $\delta r({\mathbb X}^a)$ et $\lambda^\pm({\mathbb X}^a)$. Après décomposition des superchamps, \cad développement par $X = x + \hat X$ et intégration des champs auxiliaires, nous avons obtenu divers termes de contact et des expressions non triviales des opérateurs \emph{relevants} couplés aux champs et leurs dérivés. Nous avons obtenu la suppression d’un certain nombre de divergences logarithmiques gr\^ace aux termes de contact. Néanmoins à partir de termes logarithmiques résiduels nous avons de nouveau calculé des fonctions bêta non triviales pour les couplages $\delta r(x)$ et $\lambda^\pm(x)$ puis des expressions pour les équations du mouvement correspondantes. Leurs expressions sont très semblables à celles du modèle bosonique, indiquant que la physique entre ces deux systèmes est très similaire, sauf dans le domaine $\ell>\ell_c/\sqrt 2$ car dans le modèle bosonique le couplage du tachyon interbranaire au tachyon du secteur $\sigma^{0,3}$ y est effectif. 

De m\^eme que dans le cas bosonique, les formules na\"ives des équations du mouvement obtenues à partir des fonctions b\^eta ne sont pas compatibles avec une action effective. Toutefois, la comparaison de ces équations à celles dérivées de l'action effective quadratique obtenue dans la section~\refcc{sec:act_eff_kut} associée à une analyse attentive des fonctions b\^eta, montre qu'en ajoutant à leurs expressions des termes proportionnels aux fonctions b\^eta et en redéfinissant certains champs, nous pouvons en réalité en dériver des équations de mouvement compatibles avec une action effective.  \\

Dans la section~\refcc{sec:act_eff_kut} nous avons utilisé la méthode proposée par Kutasov et Niarchos afin de calculer exactement l'action effective à l'ordre quadratique. Ils proposent de contraindre un ansatz d'action effective à l'ordre des dérivées premières -- ce qui est justifié le long de tachyons de type quasi-homogènes, \cad tant que leurs oscillations spatiales sont de grandes longueurs d'onde -- en imposant que ses équations du mouvement admettent la solution de S-brane complète. Cela permet de réduire considérablement, au moins dans leur modèle, le nombre de degrés de liberté de l'ansatz –- en l'occurence à un par ordre dans le tachyon.  Par comparaison à l'expression de la fonction de partition sur le disque, le long de la solution de demi S-brane, en utilisant l'identité $S_{\text{on-shell}}=Z$ l'ansatz est alors complètement contraint et l'action effective totalement déterminée. L'action obtenue par cette méthode est au moins valable autour de cette solution.

Nous avons cependant montré que dans le système brane-antibrane s\'epar\'e, il n'est pas possible de déterminer entièrement un ansatz d'action par cette méthode. En effet, la forme spécifique de la solution de S-brane complète dans ce cas précis ne permet d'exprimer qu'un système d'équations sous-déterminé pour l'ansatz. Par conséquent, nous ne pouvons pas réduire son nombre de degrés de liberté aussi significativement que dans le système étudié par Kutasov et Niarchos. Alors, l'action effective ne peut pas \^etre totalement déterminée par la formule $S_{\text{on-shell}}=Z$. Néanmoins, nous avons la possibilité de calculer exactement une action à l'ordre quartique dans le tachyon. En effet, le nombre de degré de liberté par ordre dans le tachyon étant alors réduit à un, nous pouvons les contraindre par la formule ci-dessus et obtenir une expression quartique pour l'action effective du système brane-antibrane. \\

Par ailleurs, nous avons exprimé la fonction de partition sur le disque le long du tachyon roulant. Nous avons développé une méthode diagrammatique afin de traiter analytiquement les intégrandes et réduire l'intégrale de chemin définie dans le super-espace à une intégrale de chemin définie sur le disque. L'expression finale que nous obtenons pour tout $r<r_c$ est telle que le calcul analytique de l'intégrale à tout ordre est hors de portée. Nous avons cependant pu calculer la densit\'e de fonction de partition, not\'ee $Z'$, à l'ordre 8 dans les tachyons pour la distance $r=r_c/\sqrt 2$. Nous espérions obtenir une expression connue du développement, mais les résultats ordre par ordre semblent se complexifier à mesure que l'ordre augmente. Nous ne reconnaissons aucune série connue dans sa formule~:

\begin{multline}
Z'(x^0) = 1 - \lambda^+\lambda^- \frac{e^{x^0}}{\pi} + \frac{\parent{\lambda^+\lambda^-}^2}{\pi^2} e^{2 x^0}\parent{1 -\frac{\pi^2}{6}} -  \frac{\parent{\lambda^+\lambda^-}^3}{\pi^3} e^{3 x^0} \parent{1 - \frac{128}{3 \pi^2}} \\+ \frac{\parent{\lambda^+\lambda^-}^4}{\pi^4} e^{4 x^0} \parent{1-\frac{55}{12}+\frac{143}{9 \pi ^2}+\frac{13 \pi ^2}{30}+\frac{\pi ^4}{70}+\frac{175}{27}+\frac{1001}{162 \pi ^2}+\frac{\pi ^2}{15}} \ldots 
\end{multline}

Nous avons aussi pu calculer exactement l'ordre quadratique ce qui nous a permis de déterminer au moins une action effective à l'ordre quadratique par la méthode présentée ci-dessus. L'expression de l'action effective à l'ordre quadratique pour le tachyon et la distance est~:

\begin{align}\label{eq:act_eff_conclu}
&{\mathcal L} = 2\Bigg[1 + \frac{\pi^2}{2}\partial_a r \partial^a r  + \frac{1}{2 \sqrt{1 - 2r^2}}\parent{\partial_a T\partial^a T^*-\parent{\frac{1}{2}-r^2} \module{T}^2} + \ldots \Bigg] 
\end{align}

Cette expression comme celle de l'action obtenue par Kutasov et Niarchos n'est valable que le long de la solution de référence. Donc son domaine de validité concerne les perturbations des champs autour du tachyon roulant à distance constante. Cette expression est nettement différente de celle correspondant au développement quadratique de l'action de Garousi. Ce qui n’est pas surprenant, puisque nous argumentons en introduction que l'action de Garousi n'est \apriori valable que pour des tachyons de genre espace, \cad non dynamique. En outre, l'action de Garousi n'admet pas de solution de tachyon roulant à distance constante comme nous l’avons démontré. 

Comme nous le disons plus haut, la dérivation des équations du mouvement à partir de~\refe{eq:act_eff_conclu} a montré après une analyse attentive que les équations obtenues par le biais du groupe de renormalisation étaient compatibles avec cette action. L'absence à premier abord d'un terme d'interaction entre la dérivée du tachyon et la dérivée du champ de distance rendait la correspondance impossible. Toutefois nous avons pu identifier ce terme à une contribution proportionnelle à la fonction bêta de $\Delta \lambda^\pm$ à l'ordre quadratique. Or nous savons que les fonctions bêta sont soumises à cette ambiguïté~\cite{Tseytlin:1986zz} lorsque nous voulons les interpréter en tant qu'équations du mouvement. La compatibilité entre ces deux développements semble donc plutôt correcte. Par cons\'equent, la formule de l'action quadratique que nous avons d\'eriv\'ee est en accord avec la physique interne des cordes dans le fond de champs off-shell. \\

Le sch\'ema final concernant les actions effectives le long des diff\'erentes phases du champ bi-fondamental interbranaire semble privil\'egier la distribution suivante~: (1) dans la phase surcritique $r>\sqrt 2 r_c$ la physique du syst\`eme est domin\'ee par l'attraction coulombienne r\'esultant de l'\'echange de cordes ferm\'ees entre les deux branes~; (2) dans la phase surcritique $\sqrt 2 r_c>r>r_c$ le syst\`eme est d\'ecrit par l'action de Garousi et le potentiel est \`a l'ordre d'une boucle attractif et de type Coleman-Weinberg~; (3) dans la phase critique elle-m\^eme la d\'efinition d'une th\'eorie des champs semble non pertinente \'etant donn\'e l'instabilit\'e du syst\`eme en cette position~; et (4) dans la phase sous-critique, la physique est d\'ecrite par une action dont le d\'eveloppement quadratique en de faibles valeurs de tachyon est~\refe{eq:act_eff_conclu} du moins autour d'un mode de condensation \`a distance constante et d\'ependant du temps de type \emph{tachyon roulant}. Dans cette phase, l'action de Garousi est probablement pertinente en ce qui concerne la description des modes de condensation de genre espace uniquement -- tels que ressaut ou vortex. \\

En d\'emontrant la marginalit\'e exacte du mod\`ele de tachyon roulant \`a distance constante, la voie \`a l'\'etude de la condensation en elle-m\^eme et de la d\'etermination de son issue est ouverte. La question importante concerne en particulier la nature du vide de condensation~: s'il existe bel et bien, est-il stable ou instable~? Un vide \'etant simplement une configuration de l'espace-temps, par exemple, la solution de ressaut dans le syst\`eme brane-antibrane co\"incident est un vide, certes d\'ependant des coordonn\'ees d'espace, mais instable car consiste en une brane non-BPS de co-dimension $1$. A l'inverse, la solution de vortex dans ce m\^eme syst\`eme est un vide stable car il consiste en une brane BPS de co-dimension $2$. De m\^eme le vide global de condensation, \cad celui minimisant le potentiel du tachyon qui peut \^etre consid\'er\'e globalement constant, est soit stable soit instable mais constitue n\'eanmoins dans chaque cas une issue de condensation pour le tachyon. Le calcul exact de la fonction de partition le long du tachyon roulant devrait donner cette information, comme dans~\cite{Sen:2002nu} et permettre de d\'eterminer l'\'etat de bord du syst\`eme brane-antibrane condensant. Si le vide final est un vide de corde ferm\'ee, alors dans la limite $x^0 \to \infty$ l'\'etat de bord de la brane doit s'annuler. D'apr\`es les \'etudes de Sen~\cite{Sen:2002nu} et de Lambert~\cite{Lambert:2003zr} dans ce cas, la condensation est accompagn\'ee d'une \'evaporation de la brane en corde ferm\'ee et \`a un confinement des flux \'electriques le long de chaque brane transformant les cordes ouvertes interbranaires en cordes ferm\'ees contraintes \`a circuler dans l'espace d\'elimit\'e par les deux branes. Cela reste coh\'erent car la s\'eparation est de l'ordre de la longueur de corde. Toutefois nous avons vu que le calcul perturbatif de la fonction de partition le long du tachyon roulant est tr\`es complexe et est pour l'instant inconnu. De plus il est fort probable que l'expression perturbative compl\`ete soit difficile \`a resommer en une forme compacte. Dans cette direction, il faudrait donc id\'ealement concentrer les recherches vers un calcul non-perturbatif de la fonction de partition.

\begin{wrapfigure}{l}{.5\linewidth}
\centering
\includegraphics[scale=.4]{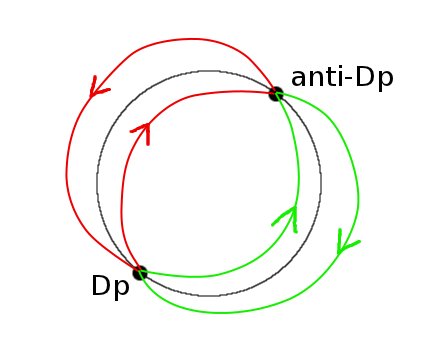}
\caption{\label{fig:systeme_comp} \small{Syst\`eme $Dp-\overline Dp$ diam\'etralement s\'epar\'e le long d'une direction compacte en forme de cercle $S^1$ de rayon $R$. Les cordes ouvertes interbranaires se s\'eparent en deux secteurs et en deux ensembles ind\'ependants \emph{droit} (en vert) et \emph{gauche} (en rouge).}}
\end{wrapfigure}

Une autre direction de recherche, que nous explorons actuellement, consiste en la construction d'un mod\`ele de condensation \'equivalent \`a celui du syst\`eme s\'epar\'e. L'utilisation d'une des nombreuses dualit\'es de la th\'eorie des cordes pourrait \'egalement se r\'ev\'eler utile. L'avancement actuel de nos recherches dans cette voie est le suivant. L'\'etablissement de l'universalit\'e du potentiel de tachyon par Sen dans~\cite{Sen:1999xm} sugg\`ere que l'issue de condensation du tachyon interbranaire dans le syst\`eme brane-antibrane ne d\'epend pas non plus de la g\'eom\'etrie dans laquelle le syst\`eme est ins\'er\'e. En particulier, il peut \^etre \'etudi\'e dans un espace compact. Ainsi, nous pouvons \'etudier la configuration repr\'esent\'ee dans la figure~\refe{fig:systeme_comp}~: une brane et une antibrane diam\'etralement s\'epar\'ees dans une direction compacte de rayon $R$. La distance s\'eparant les branes est $\ell = \pi R$. Le syst\`eme admet de nouveaux deux secteurs interbranaires, repr\'esent\'es par les facteurs de Chan-Paton $\sigma^+$ et $\sigma^-$. Pour $R<\sqrt 2$ il existe deux tachyons complexes \apriori ind\'ependants correspondant aux cordes du c\^ot\'e droit et \`a celles du c\^ot\'e gauche. Nous parlerions de tachyons droit et gauche, mais sans rapport avec une quelconque notion de chiralit\'e.

En pr\'esence de tachyons gauches et droits constants et \'egaux, le syst\`eme est g\'eom\'etriquement stable, \cad que les branes restent en leur localisations respectives. Cette derni\`ere configuration est int\'eressante car elle ouvre la possibilit\'e d'\'etudier le potentiel effectif du tachyon \`a distance constante, celui auquel nous nous int\'eressons pour conna\^itre l'issue de condensation du syst\`eme s\'epar\'e en espace plat. En effet, cela reste \`a prouver, et ce serait un point majeur dans cette \'etude, mais nous pourrions conjecturer que le potentiel du tachyon interbranaire en syst\`eme compact soit \'egal -- ou au moins \'equivalent -- \`a celui du tachyon interbranaire en espace plat. Pour le moins, le produit de condensation devrait \^etre semblable~: la topologie de l'espace n'est pas pertinente en ce qui concerne un ph\'enom\`ene local. En effet, un voisinage de $S^1$ contenant la brane et l'antibrane, par exemple le c\^ot\'e droit, ne contient qu'un seul tachyon complexe. L'existence de l'autre tachyon -- suppos\'e \'egal au premier -- contenu dans le voisinage compl\'ementaire de gauche permet de conserver la stabilit\'e g\'eom\'etrique du syst\`eme mais il ne devrait pas influencer la nature du produit local de condensation. 

Or, nous pouvons identifier ce produit qui constitue pour un tachyon constant le vide de condensation. En $R=\sqrt 2$ la d\'eformation de bord associ\'ee \`a un tachyon interbranaire \emph{constant} et pour lequel le syst\`eme est g\'eom\'etriquement stable est la suivante~:

\begin{align}
\delta S = \parent{\begin{array}{cc} 0 & \lambda \\ \lambda^* & 0 \end{array}} \otimes \oint \frac{\wt \psi}{\sqrt 2} \cos \frac{\wt X}{\sqrt 2}
\end{align}

Nous avons not\'e $X$ la dimension compacte. Du point de vue de la fonction de partition sur le disque et \`a cause des facteurs de CP, cette d\'eformation est \'equivalente \`a~:

\begin{align}\label{eq:deform_comp}
\delta S = \module{\lambda} \sigma^1 \otimes \oint \frac{\wt \psi}{\sqrt 2}  \cos \frac{\wt X}{\sqrt 2}
\end{align}

Elle correspond exactement -- \`a une T-dualit\'e et des fermions pr\`es -- \`a celle \'etudi\'ee par Sen~\cite{Sen:1999mh,Sen:2002nu} dans le syst\`eme bosonique brane-brane co\"incidentes. Pour le cas des supercordes, Sen a d\'etermin\'e la th\'eorie conforme associ\'ee \`a la  solution de ressaut~\cite{Sen:1998tt} puis Majumder et Sen ont \'etudi\'e la d\'eformation associ\'ee \`a la formation d'un vortex~\cite{Majumder:2000tt}. La d\'eformation \refe{eq:deform_comp} est exactement marginale pour tout $\lambda$. Une T-dualit\'e transforme $R\to 1/R$ et $(\wt X,\wt \psi) \to (X,\psi)$ de sorte que la d\'eformation T-dual repr\'esente la formation d'un ressaut par condensation de tachyon sur un syst\`eme $D(p+1)-\overline D (p+1)$ co\"incident enroul\'e autour de la dimension compacte de rayon $\wt R = 1/\sqrt 2$. Pour $\module\lambda=1/2$ le ressaut prend la forme concr\^ete d'une brane non-BPS localis\'ee en $x=\pi \wt R$. Par T-dualit\'e inverse, la th\'eorie de surface d\'eform\'ee par~\refe{eq:deform_comp} en $\module\lambda=1/2$ devrait donc d\'ecrire une $D(p+1)$-brane non BPS enroul\'ee autour de la dimension compacte. 

En outre, nous pouvons g\'en\'eraliser cette construction pour tout $R<\sqrt 2$~: Sen montre, dans le cas bosonique~\cite{Sen:1999mh} et dans le cas $D-\overline D$~\cite{Sen:1998tt} que le ressaut est identifi\'e \`a une brane de codimension 1 pour tout $\wt R$ tant que $\module\lambda=1/2$. Ceci implique que la d\'eformation

\begin{align}
\delta S = \frac{\sigma^1}{2}  \otimes \oint \frac{R \, \wt \psi}{2}  \cos R \frac{\wt X}{2}
\end{align}

est exactement marginale pour tout $R<\sqrt 2$. Par cons\'equent, le fond constant $\module\lambda=1/2$ est une solution des \'equations de la SFT~: il s'agit donc d'un minimum du potentiel tachyonique, ce qui s'appelle un \emph{vide}. Il consiste en une $D(p+1)$-brane non BPS enroul\'ee autour de la direction compacte -- voir figure~\refe{fig:dual_ressaut}. Cela implique qu'en condensant, le tachyon se rassemble en une \emph{mati\`ere branaire} pour recomposer cette brane non-BPS. 

\begin{figure}[h]
\centering
\includegraphics[scale=.35]{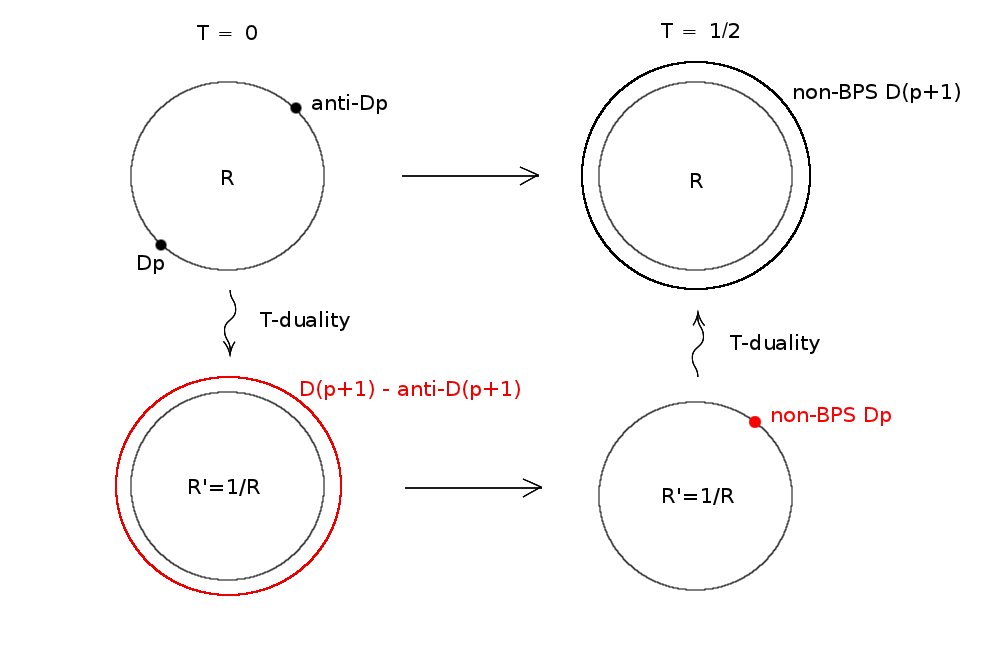}
\caption{\label{fig:dual_ressaut} \small{Condensation du tachyon interbranaire dans le syst\`eme $Dp-\overline Dp$ diam\'etralement s\'epar\'e, par T-dualit\'e avec la solution de ressaut dans le syst\`eme $D(p+1)-\overline D (p+1)$ co\"incident.}}
\end{figure}

D'apr\`es Sen, le potentiel du tachyon doit s'\'ecrire sous la forme $V(T)=2 T_p f(T)$ avec $f$ une fonction d\'ecroissante interpolant entre les valeurs extr\^emes d\'etermin\'ees par la composition de l'espace-cible en ces points. Ici, parce que la brane cr\'e\'ee est de dimension sup\'erieure \`a la brane initiale, nous avons la contrainte~:

\begin{align}
V(1/2) = \int_0^{2\pi R} \di x \, \sqrt 2 \, T_{p+1} = \sqrt 2 R \, T_p
\end{align}

Par cons\'equent, les valeurs extr\^emes de $f$ sont $f(0)=1$ et $f(1/2)=R/\sqrt 2$. Pour $R=\sqrt 2$ le potentiel est bien plat, ce qui est compatible avec un "tachyon" non-massif. Pour $R<\sqrt 2$ le potentiel est d\'ecroissant mais la valeur extr\^eme en $T=1/2$ est non nulle \`a la diff\'erence du cas co\"incident, et est d\'etermin\'ee par la distance s\'eparant initialement les branes $\ell=\pi R$.

\begin{figure}[h]
\centering
\includegraphics[scale=.3]{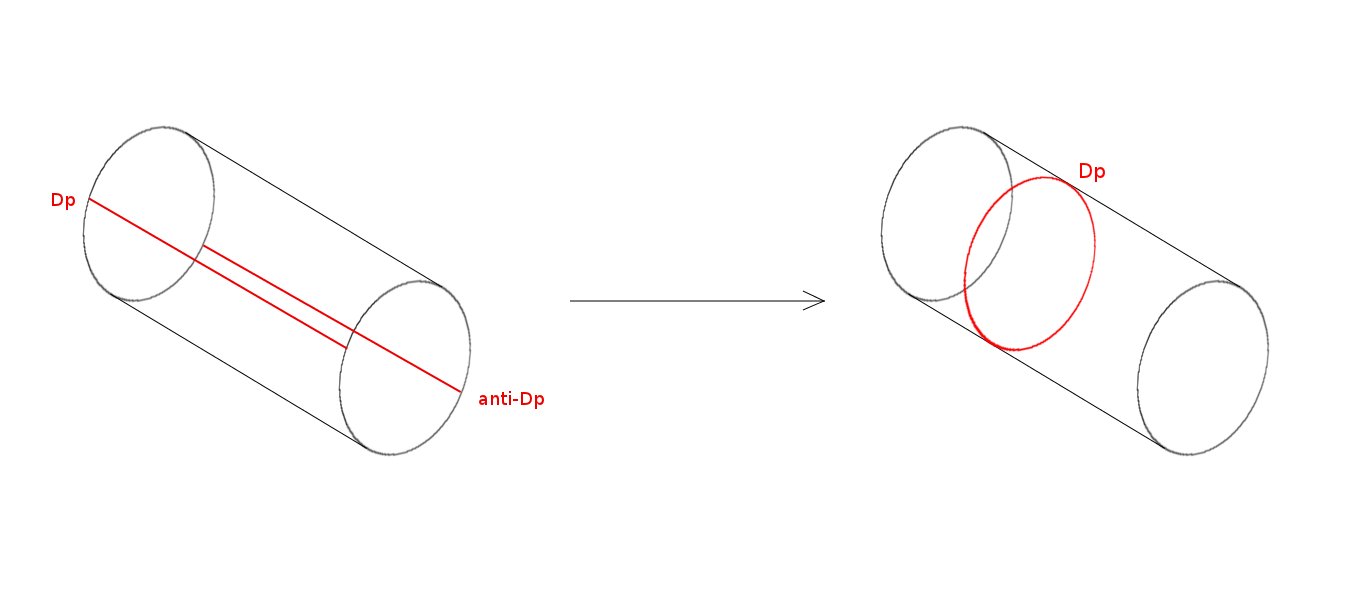}
\caption{\label{fig:ressaut_nonBPS} \small{Condensation du tachyon dans le syst\`eme brane-antibrane s\'epar\'ee, par formation d'un vortex dans le syst\`eme T-dual.}}
\end{figure}

N\'eanmoins, une brane non-BPS dans cette g\'eom\'etrie n'est pas stable puisqu'elle admet un tachyon dans le spectre de ses excitations de cordes ouvertes. Ainsi, le vide $\module\lambda=1/2$ est en v\'erit\'e instable et le tachyon de la brane non-BPS est appel\'e \`a condenser, temporellement ou sous la forme d'un ressaut. Ce dernier mode de condensation est \'equivalent \`a la formation d'un vortex -- une $Dp$-brane BPS -- directement depuis le syst\`eme T-dual de la paire brane-antibrane s\'epar\'ee. Il est repr\'esent\'e sur la figure~\refe{fig:ressaut_nonBPS}. \\

Ce d\'eveloppement sugg\`ere que le syst\`eme $Dp-\overline Dp$ s\'epar\'e par la distance $\ell$ dans la limite de d\'ecompactification tend, par condensation, vers un vide compos\'e d'une mati\`ere branaire non-BPS et instable remplissant l'espace d\'elimit\'e par les deux branes -- voir figure~\refe{fig:brane_decomp}. 

\begin{wrapfigure}{l}{.5\linewidth}
\centering
\includegraphics[scale=.4]{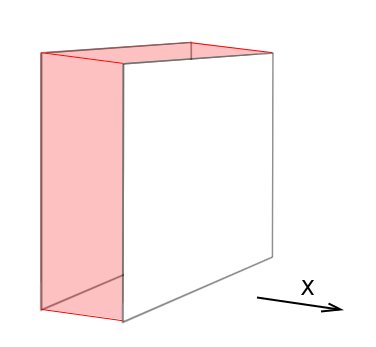}
\caption{\label{fig:brane_decomp} \small{Syst\`eme $Dp-\overline Dp$ s\'epar\'e le long d'une direction $X$ tranverse. Le produit de condensation serait une mati\`ere branaire non BPS et instable remplissant l'espace d\'elimit\'e par la surface des deux branes. Il est repr\'esent\'e ici en rouge.}}
\end{wrapfigure}

Cette mati\`ere est instable pour deux raisons, d'abord intrins\`equement \`a cause du tachyon de corde ouverte contenu dans le volume d'univers d'une brane non-BPS et ensuite g\'eom\'etriquement \`a cause de la tension non nulle de l'objet qui tend \`a r\'eduire l'\'epaisseur du syst\`eme. Il faudrait \'etudier les constantes de temps associ\'ees \`a ces deux effets pour d\'eterminer lequel est dominant. \\

Le passage du syst\`eme compact au syst\`eme d\'ecompactifi\'e n'est pas trivial et il s'agit d'un des points importants \`a d\'evelopper dans cette direction de recherche. Une confirmation directe serait apport\'ee par le calcul du potentiel associ\'e \`a un seul des tachyons droit ou gauche dont la d\'eformation de bord serait de la forme~:

\begin{align}
\delta S = \sigma^+ \otimes \lambda \oint \frac{R \wt \psi}{2} e^{i \frac{R}{2} \wt X} -  \sigma^- \otimes \lambda^*\oint \frac{R \wt \psi}{2} e^{-i \frac{R}{2} \wt X}
\end{align}

Cela pourrait \^etre effectu\'e en utilisant la relation de ce mod\`ele sigma avec le mod\`ele de Kondo -- que nous pr\'esentons plus bas. N\'eanmoins, la distinction g\'eom\'etrique nette entre les tachyons droit et gauche sugg\`ere tr\`es fortement que la valeur $\module\lambda=1/2$ pour chacun de ces tachyons correspond \`a un vide dans lequel chaque tachyon s'agr\`ege sous la forme d'une mati\`ere branaire.

A premi\`ere vue, nous pourrions opposer \`a ce raisonnement que la fonction bêta du champ de tachyon interbranaire constant est possiblement non nulle. En effet, \emph{i)} l'op\'erateur de vertex $e^{ir \wt X}$ n'est pas marginal et \emph{ii)} des termes suppl\'ementaires en $\module{\lambda}^{2n}$ devraient y \^etre ajout\'e \`a cause de la divergence des OPE du type $\parent{\wt \psi e^{ir \wt X} \cdot \wt \psi e^{-ir \wt X}}^n$. Par cons\'equent, le mod\`ele sigma correspondant n'est \apriori pas une CFT, de sorte que le fond constant $\module\lambda=1/2$ n'est \apriori pas un vide de la th\'eorie. Toutefois, la formule finale de cette fonction bêta pourrait aussi bien se ressommer non-perturbativement en une fonction $g(\module\lambda)$ nulle en $\module\lambda=1/2$. L'analyse de la relation de ce mod\`ele sigma au mod\`ele Kondo pourrait fournir des r\'eponses. \\

Nous conclurons en pr\'esentant les similitudes que revêtent les modèles de tachyon du secteur interbranaire et les modèles de Kondo bosoniques -- et par conséquent aussi supersymétriques. Le modèle de Kondo est défini par l'hamiltonien suivant~\cite{LeClair:1997sd} sur le demi-plan complexe avec $z=x+iy$~:

\begin{equation}\label{eq:kondo_hamil}
H_K = \frac{1}{2} \int_{-\infty}^{+\infty} \di x ~ \parent{ \parent{\partial_x \phi}^2 + \parent{\partial_y \phi}^2 } + \omega_0 \parent{S^+ e^{i \beta \phi(0,y)} + S^- e^{-i \beta \phi(0,y)}}
\end{equation}

Les matrices $S^\pm$ agissent dans une représentation de spin $j/2$ de $SU(2)_q$ avec $q=e^{i\pi \beta^2}$. Voir~\refe{eq:algebre_q} plus loin. Par comparaison, le lagrangien du modèle de tachyon interbranaire constant est donné par~:

\begin{equation}\label{eq:kondo_lag}
S = \frac{1}{2 \pi} \int \di{x} \di{y} ~ \parent{ \parent{\partial_x \phi}^2 + \parent{\partial_y \phi}^2 } + \oint \di{y} ~ \parent{ \lambda^+ \sigma^+ e^{i \beta \phi(0,y)} + \lambda^- \sigma^- e^{-i \beta \phi(0,y)}}
\end{equation}  

où $\sigma^\pm$ appartiennent à la représentation $1/2$ de $SU(2)$ en fait équivalente à une représentation $1/2$ de $SU(2)_q$. Nous voyons donc par transformation de Legendre de~\refe{eq:kondo_lag} que la correspondance est ici donnée par $\omega_0 = - \lambda^\pm$ et $S^\pm = \sigma^\pm$. La relation du modèle de Kondo au modèle de sine-Gordon de bord (BSG) est bien connue -- voir par exemple~\cite{Fendley:1995kj}. En l’occurrence, les fonctions de partition respectives vérifient certaines relations comme nous voyons un peu plus bas dans la formule~\refe{eq:partition_kondo_BSG}. En outre, le modèle Kondo est intégrable à condition que les matrices $S^\pm$ appartiennent bien à la représentation $j/2$ de $SU(2)_q$ et vérifient l'algèbre~:

\begin{align}\label{eq:algebre_q}
\comm{S^+}{S^-} = \frac{q^{S_z}-q^{-S_z}}{q-q^{-1}} \qquad \comm{S_z}{S^\pm}=\pm 2 S^\pm
\end{align}

qui est effectivement la même que $SU(2)$ le long d'une représentation $1/2$. Le modèle a deux régimes autour de la limite de Toulouse $\beta\to 1/\sqrt 2$ qui, par comparaison à l'étude du tachyon roulant dans le modèle bosonique, est celle à partir de laquelle le couplage aux tachyons du secteur $\sigma^0$ est effectif. Pour $\beta< 1/\sqrt 2$ on parle de régime attractif et pour $1>\beta> 1/\sqrt 2$ de régime répulsif. Dans ce dernier régime la fonction de partition sur le disque a des pôles pour tout $\beta =\sqrt{ (1+2n )/(2+2n)} $. Dans le premier la fonction de partition est analytique. Pour donner un exemple de relation entre les modèles BSG et de Kondo citons la relation montr\'ee par Fendley, LeSage et Saleur~\cite{Fendley:1995kj} le long de la représentation de spin $1/2$ et pour $\beta<1/\sqrt 2$~:

\begin{align}\label{eq:partition_kondo_BSG}
Z_{1/2}[(q-q^{-1})x]= \frac{Z_{BSG}(q x)+Z_{BSG}(q^{-1} x)}{Z_{BSG}(x)}
\end{align}

avec ici $x=2\pi \omega_0$ et au moins pour $q$ racine de l'unité, \cad $\beta=1/\sqrt{n}$. La fonction de partition de sine-Gordon est connue jusqu'à de larges ordres en perturbation et exactement sous la forme~\cite{Fendley:1995kj}~:

\begin{align}
Z_{BSG}(x) = 1+ \sum_{n=1}^\infty \frac{x^{2n} }{\Gamma(\beta^2)^{2n}} \sum_{{\bf m}} \prod_{i=1}^n \parent{\frac{\Gamma(m_i+\beta^2(n-i+1))}{\Gamma(m_i+\beta^2(n-i)+1)}}^2
\end{align}

avec la somme faite sur tous les ensembles (tableaux de Young) ${\bf m}=(m_1,m_2,\ldots,m_n)$ pour des entiers $m_i$ tels que $m_1 \geq m_2 \geq \ldots \geq m_n$. Il existe une méthode itérative pour calculer ces sommes ordre par ordre. La fonction de partition du modèle de Kondo dans la représentation de spin $1/2$ paraît donc calculable, au moins perturbativement, dans le régime attractif. En utilisant la BSFT de Witten, nous pensons possible d'exprimer une action effective off-shell, sachant que $\beta_\pm = (1-r^2)\lambda^\pm$ et toutes les autres fonctions bêta s'annulent, car la limite UV n'est plus divergente. La formule habituelle de relation de l'action effective et de la fonction de partition off-shell est en théorie bosonique~\cite{Witten:1992cr,Gerasimov:2000zp,Tseytlin:1986zz,Tseytlin:2000mt}~:

\begin{align}
S[T] = (1+\beta_T \partial_T )Z[T]
\end{align}

Ici, nous aurions cependant $Z[T,r]$ avec $\beta =r$ identifiée à la distance. Et le long de champ constant, l'action $S[T,r]$ serait simplement identifi\'ee au potentiel effectif, dont nous souhaitions obtenir la formule un peu plus haut. Nous devons encore exploiter cette direction d'étude. Mais les calculs semblent à portée de main au moins dans le cas bosonique. En outre, par similitude du modèle bosonique au modèle brane-antibrane dans le régime attractif, et par comparaison au développement quadratique obtenu dans notre \'etude, nous pourrions probablement conjecturer une forme d'action effective du système brane-antibrane séparé.

\part{Annexe~: article publi\'e}
\label{part:annexe}

\includepdf[pages={1-40}]{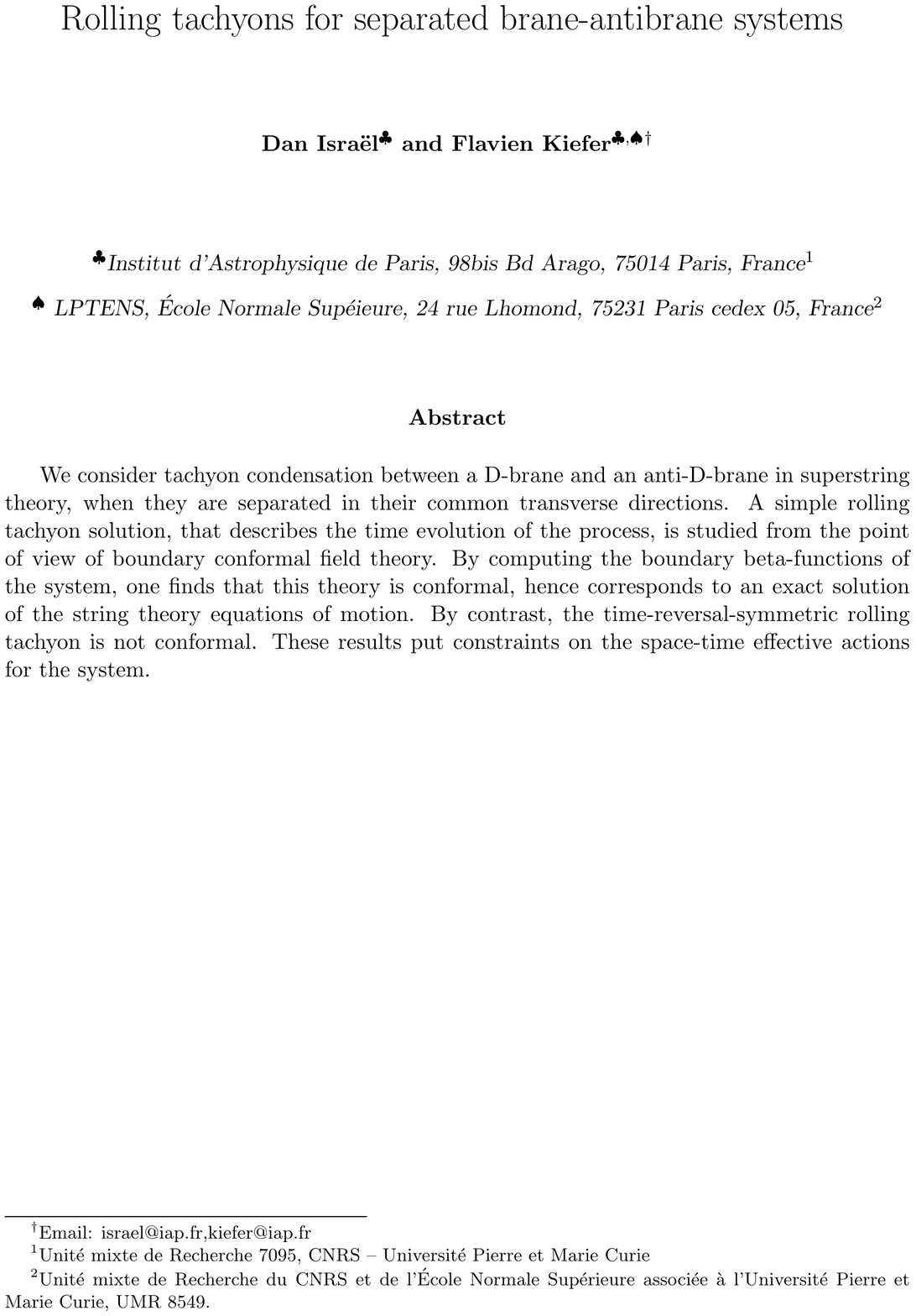}

\bibliographystyle{abbrv}
\bibliography{these_bib}

\end{document}